\documentclass[10pt,twocolumn]{revtex4-2}

\usepackage{hyperref}
\usepackage{url} 

\usepackage{array}
\usepackage{amsmath,amssymb}
\usepackage[utf8]{inputenc}
\usepackage{dcolumn}
\usepackage{bm}
\usepackage{color}
\usepackage{xcolor}
\usepackage{mathtools}
\usepackage{natbib}
\usepackage{subfigure}
\usepackage{url}
\usepackage{graphicx}
\usepackage{enumitem}
\usepackage{epstopdf}
\usepackage{ulem}
\usepackage{footnote}
\usepackage{mathrsfs}
\usepackage{physics}

\usepackage{bm}	
\renewcommand{\vec}[1]{\bm{#1}}
\renewcommand{\l}{\left(}
\renewcommand{\r}{\right)}
\renewcommand{\bra}[1]{\langle#1|}
\renewcommand{\ket}[1]{|#1\rangle}
\newcommand{\hc}{\text{h.c.}}

\renewcommand{\H}{\hat{\mathcal{H}}}

\renewcommand{\a}{\hat{a}}
\renewcommand{\b}{\hat{b}}
\newcommand{\cd}{\hat{c}^\dagger}
\newcommand{\ad}{\hat{a}^\dagger}
\newcommand{\bd}{\hat{b}^\dagger}

\renewcommand{\P}{\hat{\mathcal{P}}}

\newcommand{\vc}[1]{\bm{\mathrm{#1}}}

\hypersetup{
    colorlinks=true,
    linkcolor=blue,
    filecolor=magenta,      
    urlcolor=purple,
    citecolor=purple
}

\DeclareUnicodeCharacter{2212}{\textendash}

\begin{document}

\title{Impurities and polarons in bosonic quantum gases: a review on recent progress}

\author{ F. Grusdt$^{1,2}$, N. Mostaan$^{1,2}$, E. Demler$^3$,  L. A. P. Ardila$^{4,5}$}
\affiliation{ $^1$Department of Physics and Arnold Sommerfeld Center for Theoretical Physics (ASC), Ludwig-Maximilians-Universit\"at M\"unchen, Theresienstr. 37, D-80333 M\"unchen, Germany}
\affiliation{ $^2$Munich Center for Quantum Science and Technology (MCQST), Schellingstr. 4, D-80799 M\"unchen, Germany}
\affiliation{ $^3$Institut f\"ur Theoretische Physik, ETH Zurich, 8093 Zurich, Switzerland}
\affiliation{ $^4$School of Science and Technology, Physics Division, University of Camerino, Via Madonna delle Carceri, 9B-62032 (MC), Italy}
\affiliation{ $^5$Dipartimento di Fisica, Universit\`a di Trieste, Strada Costiera 11, I-34151 Trieste, Italy}
\email{luisaldemar.penaardila@units.it}

\begin{abstract}
This review describes the field of Bose polarons, arising when mobile impurities are immersed into a bosonic quantum gas. The latter can be realized by a Bose-Einstein condensate (BEC) of ultracold atoms, or of exciton polaritons in a semiconductor, which has led to a series of experimental observations of Bose polarons near inter-species Feshbach resonances that we survey. Following an introduction to the topic, with references to its historic roots and a presentation of the Bose polaron Hamiltonian, we summarize state-of-the-art experiments. Next we provide a detailed discussion of polaron models, starting from the ubiquitous Fr\"ohlich Hamiltonian that applies at weak couplings. Already this highly simplified model allows insights into ultra-violet (UV) divergencies, logarithmic and power-law, that need to be properly regularized. To capture the physics near a Feshbach resonance, two-phonon scattering terms on the impurity as well as phonon-phonon interactions need to be included. We proceed by a survey of concurrent theoretical methods used for solving strongly interacting Bose polaron problems, ranging from Lee-Low-Pines mean-field theory, Chevy-ansatz, Gross-Pitaevskii-equation to diagrammatic Monte Carlo approaches. The subsequent sections are devoted to the large bodies of work investigating strong coupling Bose polarons, including detailed comparisons with radio-frequency (RF) spectra obtained in ultracold atom experiments; to investigations of universal few-body and Efimov states associated with a Feshbach resonance in atomic mixtures; to studies of quantum dynamics and polarons out of equilibrium; Bose polarons in low-dimensional 1D and 2D quantum systems; induced interactions among polarons and bipolaron formation; and to Bose polarons at non-zero temperatures. We end our review by detailed discussions of closely related experimental setups and systems, including ionic impurities, systems with strong light-matter interactions, and variations and extensions of the Bose polaron concepts e.g. to baths with topological order or strong interactions relevant for correlated electrons. Finally, an outlook is presented, highlighting possible future research directions and open questions in the field as a whole.
\end{abstract}

\maketitle


%
%

\tableofcontents

\newpage
\clearpage

\section*{Notation and Nomenclature}

\begin{tabular}{p{1.75cm}p{10cm}p{1cm}}
$N_B$ & Number of bosons in the host bath.\\
$n$ & Density of the host bath.\\
$a_0$ & Bohr radius.\\
$a_\mathrm{BB}$ & Boson-Boson scattering length.\\
$a$ or $a_{\rm IB}$ & Boson-impurity scattering length.\\
$m_{\rm I}$ & Impurity mass.\\
$m_{\rm B}$ & Boson mass.\\
$m_{\rm p}$ & Polaron effective mass.\\
$m_\mathrm{red}$ & Impurity-boson reduced mass.\\
$g_\mathrm{IB}$ & Impurity-boson coupling strength.\\
$g_\mathrm{BB}$ & Boson-boson coupling strength.\\
$\textit{d=1,2,3}$ & Dimensionality.\\
$\omega_\textrm{RF}$ & Radiofrecuency pulse.\\
$\Gamma$ & Polaron decay rate.\\
\end{tabular}

\newpage
\clearpage
\section{Introduction}
\label{secIntro}

Highly imbalanced mixtures of quantum degenerate gases, cooled by lasers and trapped in optical potentials \cite{Bloch2008}, have opened up the possibility of studying mobile impurities and their resulting polaron physics with ultracold atoms. In such settings the minority atom, which takes the role of the impurity, turns into a new composite object, that can have properties very different from the original bare impurity. Namely, it becomes dressed by a cloud of excitations of the majority component, turning it into a single particle with renormalized energy, effective mass and single-particle wavefunction at low energies. Understanding how -- and under which conditions -- this \emph{quasiparticle formation} takes place, and making quantitative predictions of the resulting quasiparticle properties, remains an overarching motivation in the study of quantum degenerate mixtures.

Among the reasons why quantum impurity problems play a central role in modern physics, is their special connection to foundational problems in physics. On the one hand, being constituted by just one impurity particle allows to establish direct connections with single-particle quantum mechanics; this allows for simplified theoretical descriptions in terms of a single-particle formalism and intuitive interpretation of many experiments. On the other hand, the interactions of impurities with a surrounding bath renders such problems genuinely many-body in nature; this means many-body quantum mechanics has to be evoked for a complete microscopic description and experiments can directly observe emergent phenomena in these systems that cannot be explained by single-particle quantum mechanics alone. Along the same vein, the coupling of the impurity to a many-body bath leads to decoherence, which plays a central role for understanding how classical phenomena emerge in the context of many-body quantum mechanics. This constituted a central motivation for the seminal work by Caldeira and Leggett~\cite{caldeira1981influence}, who explored the coupling of a two-level system to a bosonic bath that appears in many polaron problems today. 

Polaron, or more generally quantum impurity, problems constitute a natural stepping stone in the quest for realizing even more complex -- and potentially novel -- phenomena in quantum many-body systems whose properties are dominated by strong correlations. This is true both for experiments and theory: The former can be benchmarked and brought to technical perfection by studying a theoretically accessible setting; The latter relies equally on experimental data and benefits from the many possible angles of attack that different theoretical techniques can utilize to make significant advances. Indeed, among the existing many-body problems studied in the context of correlated quantum systems, the impurity problems discussed here are considered to be relatively tractable -- though far from trivial. One of the most prominent examples how polaronic effects can help understand more complex emergent phenomena is the BCS theory of superconductivity, where the same impurity-phonon interactions causing polaron formation constitute the pairing mechanism between electrons in a conventional superconductor.

In this review, we provide an overview of the field and report on recent progress made in the study of mobile impurities immersed in a surrounding bosonic bath. This setting is broadly referred to as the \emph{Bose polaron problem}. Experimentally, significant progress has been made in realizing strongly interacting Bose polarons in the vicinity of Feshbach resonances, where several emergent phenomena can also be expected to occur, such as multi-particle or Efimov-type bound states or quasiparticle breakdown through so-called orthogonality catastrophe. How these effects are influenced by the finite-density bath of bosons remains subject of active research, in theory and experiment. Another interesting aspect of the problem that we will discuss, is the use of mobile impurities as probes that can provide important information about the underlying bath (e.g., its equation of state, its temperature or properties of its collective excitations).

\begin{figure}
	\centering
	\includegraphics[width=\linewidth]{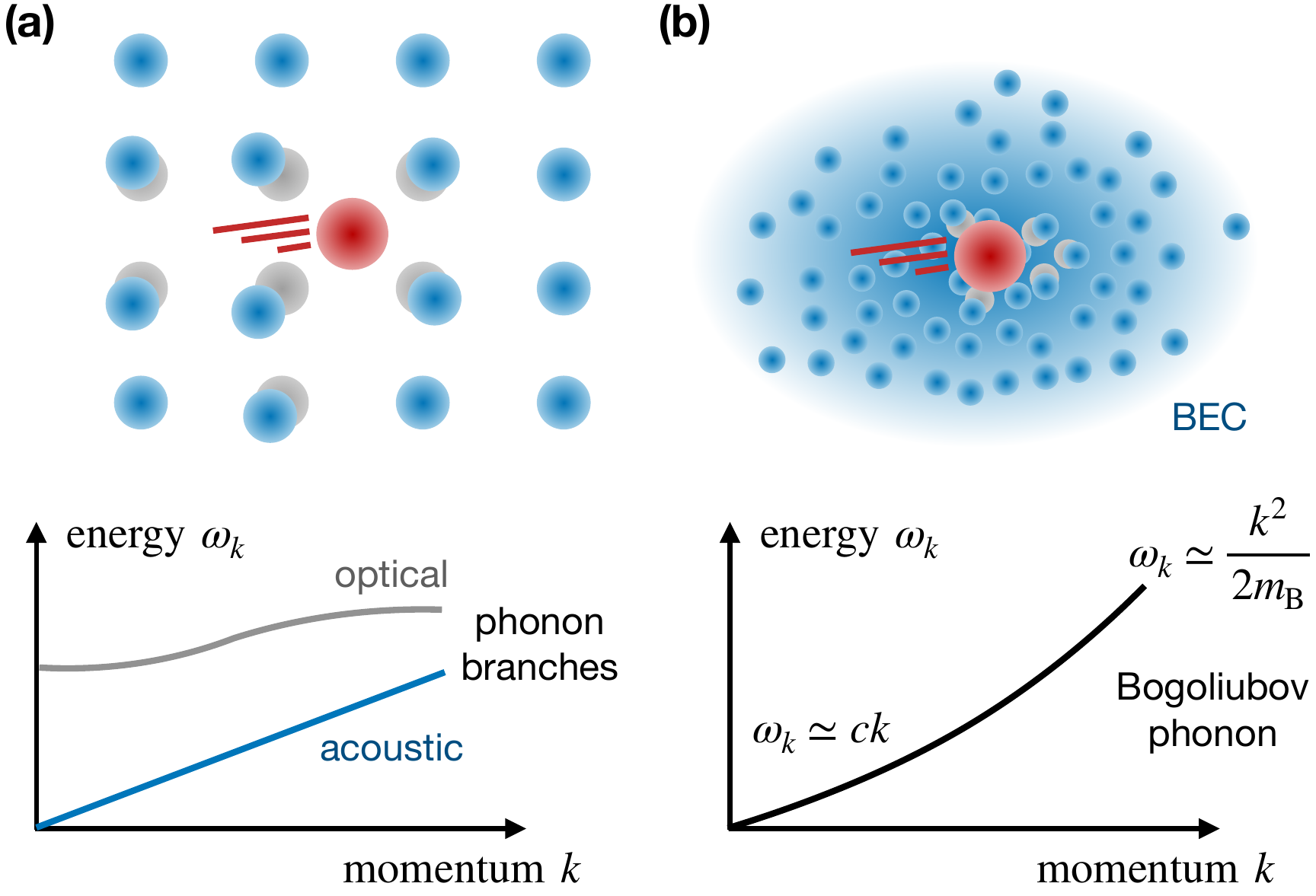}
	\caption{\textbf{Bose polarons} in (a) solids and (b) a Bose-Einstein condensate (BEC). At low energies Bogoliubov phonons correspond to phase fluctuations and their dispersion relation $\omega_k \simeq c k$ resembles that of acoustic phonons in solids. At high energies, in contrast, Bogoliubov phonons are particle-like excitations with a massive dispersion $\omega_k~\simeq k^2/(2m_{\rm B})$. At strong coupling, as realized in the vicinity of an impurity-boson Feshbach resonance, the particle-like nature of the bosonic bath gives rise to emergent few-body effects, significantly enriching the phenomenology of Bose polarons in a BEC beyond what is observable in solids.}
	\label{figIntro}
\end{figure}

\subsection{Short history of Bose polarons}
The Bose polaron problem was originally studied in the context of individual conduction electrons  -- treated as mobile impurities -- interacting with collective vibrations -- phonons -- in their host solid, see Fig.~\ref{figIntro} (a). To describe this system, Landau and Pekar \cite{Landau48} introduced the concept of a quasiparticle: They suggested that electron-phonon interactions merely \emph{renormalize} the effective mass $m_{\rm I} \to m_{\rm p}$ of the electrons, while the so-formed polarons retain their free-electron nature, signified by a non-vanishing overlap $Z(\vec{k}) = |\bra{\psi(\vec{k})} \cd_{\vec{k},\sigma} \ket{0}|^2$ (the quasiparticle weight or residue) of the polaron ground state $\ket{\psi(\vec{k})}$ with the non-interacting electron wavefunction. This laid the foundation of modern many-body theory, including the rigorous mathematical description of Landau's Fermi liquid describing strongly interacting electrons as effectively free particles at low energies.

It was also realized that dressing mobile impurities with phonon excitations can potentially lead to more dramatic effects, such as self-trapping where the enhanced polaronic mass is orders of magnitude heavier than the bare impurity mass, eventually leading to complete localization of the impurity in its phonon cloud \cite{Landau48}. Such effects depend strongly on the nature of -- and the interactions with -- the involved phonons, and typically require strong impurity-phonon coupling. This has made direct experimental realizations of polaronic self-trapping difficult. Strong mass enhancement due to polaron formation and phononic dressing of electrons is a ubiquitous phenomenon however, and plays an important role e.g. in semiconductors or for understanding transport properties of organic materials \cite{alexandrov2009advances}.

The mathematical foundation of modern polaron physics in the context of solids was provided by Fr\"ohlich \cite{Frohlich1954}, who derived the famous Fr\"ohlich Hamiltonian. We will discuss it in more detail later in this review. To include lattice effects, the model was extended to the Holstein Hamiltonian \cite{Holstein1959}. To solve these classes of models theoretically, Landau's and Pekar's original strong-coupling solution \cite{Landau48} was quickly complemented by a weak-coupling theory by Lee, Low, and Pines \cite{Lee1953}, and afterwards extended by an all-coupling solution due to Feynman \cite{Feynman1955} as one of the first applications of the newly developed path integral formalism. We provide brief reviews of these methods later on.

Polarons also appear in completely different physical settings, for example when surface electrons are levitated over liquid helium \cite{Shikin1974,Jackson1981,Tress1996}. Another, newer, class of so-called Fermi polaron problems involves impurities dressed by particle-hole excitations around a Fermi surface. These quasiparticles were originally discovered -- and explored in great detail \cite{Schirotzek2009,Kohstall2012,Koschorreck2012,Chevy2010,Massignan2014} -- in mixtures of ultracold atomic gases. More recently they were also observed in semiconductor heterostructures \cite{Sidler2017}, where excitonic impurities are interacting with a valance or conduction band Fermi sea; notably, this setting was analyzed even earlier in the semi-conductor community, see e.g. work by Suris~\cite{Suris2003}. Returning to bosonic baths, polaronic effects also play an important role in strongly correlated electron systems, where the charge carriers -- taking the role of impurities -- interact with collective bosonic modes. These include most prominently (para-) magnons~\cite{Miyake1986,Scalapino1986,Kane1989,Sachdev1989,Monthoux1991,Bruegger2006} and phonons~\cite{Alexandrov1994,Devereaux1994,Johnston2010,Mishchenko2011,Alexandrov2011}, indicating the existence of (the vicinity to) a symmetry-broken, ordered state. Examples include electrons forming nearly anti-ferromagnetic Fermi liquids~\cite{Schrieffer1989,Su1988,Millis1990,Schmalian1998,Abanov2003} and extend to non-Fermi liquid states e.g. near quantum critical points~\cite{Varma1989,Lee1998,Senthil2004,Lee2008a,Sachdev2016}.

In this review, we will focus on one of the most actively researched topics involving polarons: Mobile impurity atoms interacting with a surrounding Bose-Einstein condensate (BEC), see Fig.~\ref{figIntro} (b). Quickly following the rise of ultracold atomic mixtures as a new experimental platform, it was theoretically proposed that Bose polarons can arise in such settings \cite{Mathey2004,Kalas2006,Cucchietti2006,Tempere2009}. Initially they were thought of being akin to the Fr\"ohlich polarons in a solid, but as we will review in details below the particle-like nature of Bogoliubov excitations at intermediate- to high energies renders them considerably richer when the underlying impurity-boson interactions are strong. Compared to its fermionic counterpart, the Bose polaron remains more challenging: while Pauli blocking suppresses many potential scattering processes in Fermi polarons and allows for a relatively simple variational description due to Chevy~\cite{Chevy2010}, bosons can more easily accumulate around the impurity in the Bose polaron case leading to strongly non-linear effects. 

Inspired by the successful exploration of Fermi polarons, the excellent control of interactions via Feshbach resonances, and the rich toolbox of experimental probes such as radio-frequency spectroscopy and Ramsey interferometry, experimentalists have also set out to explore Bose polarons in BECs. By now several breakthroughs have been achieved and the study of the strongly interacting regime has started. However, significant challenges remain and many key quantities, such as the renormalized polaron mass, have not been studied in sufficient detail experimentally. The latest experimental tools, including box-shaped optical potentials and improved imaging techniques, are promising new insights that will allow for more direct and exhaustive comparison with theoretical predictions.

\subsection{The Bose polaron Hamiltonian}
The problem we consider consists of an impurity particle of mass $m_{\rm I}$, described by a field $\hat{\phi}(\vec{r})$, interacting with a surrounding Bose gas. Bosons are assumed to have a mass $m_{\rm B}$ and will be described by a field operator $\hat{\Psi}(\vec{r})$. In $d$ spatial dimensions this system can be modeled by the following Hamiltonian,
\begin{equation}
    \H = \H_{\rm B} + \H_{\rm I} + \H_{\rm IB}
    \label{eqHmic}
\end{equation}
with the bare Hamiltonian of the Bose gas
\begin{equation}
    \H_{\rm B} =  \int d^d \vec{r} ~ \hat{\Psi}^\dagger(\vec{r}) \frac{- \nabla^2}{2 m_{\rm B}} \hat{\Psi}(\vec{r}) + \frac{g_{\rm BB}}{2}  \left( \hat{\Psi}^\dagger(\vec{r})\right)^2 \left(\hat{\Psi}(\vec{r}) \right)^2,
    \label{eqHB}
\end{equation}
the free impurity Hamiltonian
\begin{equation}
    \H_{\rm I} =  \int d^d \vec{r} ~ \hat{\phi}^\dagger(\vec{r}) \frac{- \nabla^2}{2 m_{\rm I}} \hat{\phi}(\vec{r}),
    \label{eqHI}
\end{equation}
and the impurity-boson interactions
\begin{equation}
    \H_{\rm IB} =  g_{\rm IB} \int d^d \vec{r} ~ \hat{\Psi}^\dagger(\vec{r}) \hat{\Psi}(\vec{r}) \hat{\phi}^\dagger(\vec{r}) \hat{\phi}(\vec{r}).
    \label{eqHIB}
\end{equation}

Here $g_{\rm BB}$ and $g_{\rm IB}$ are the effective boson-boson and impurity-boson interaction strengths, respectively. These must be related to the corresponding measurable ($s$-wave) scattering lengths $a_{\rm BB}$ and $a_{\rm IB}$ in an appropriate way (see Sec.~\ref{subsecUVdiv}) and can be tuned using atomic Feshbach resonances \cite{Chin2010}. In all realistic experimental configurations, additional trapping potentials, harmonic or box-shaped, will be present and can be included in the microscopic Hamiltonian. Throughout this review the density of the Bose gas will be denoted by $\langle \hat{\Psi}^\dagger(\vec{r}) \hat{\Psi}(\vec{r}) \rangle = n_0$. In the presence of a trap the boson density becomes inhomogeneous and $n_0$ refers to its largest value (peak density).  

\subsection{Key concepts}
\label{secKeyConcepts}
Before we start reviewing the extensive literature on Bose polarons, we briefly summarize key theoretical concepts that are used to analyze them. A central quantity, that is directly accessible experimentally and to many theoretical methods, is the impurity spectrum. It can be defined from the impurity Green's function, $\mathcal{G}(\vec{r},t) = -i \langle \mathcal{T} \hat{\Psi}(\vec{r},t) \hat{\Psi}^\dagger(\vec{0},0) \rangle$, where $\mathcal{T}$ denotes the time-ordering operator. Taking Fourier transforms in space and time, the momentum ($\vec{k}$) and frequency ($\omega$) resolved spectral function is
\begin{equation}
    A(\vec{k},\omega) = \frac{1}{\pi} {\rm Im} \mathcal{G}(\vec{k},\omega-i0^+).
\end{equation}

The spectral function is directly proportional to the transfer rate of a weak radio-frequency (RF) pulse at frequency $\omega$ from a non-interacting to an interacting impurity state, rendering it of prime experimental interest. Using Raman lasers the momentum transfer $\vec{k}$ can also be controlled~\cite{Dao2007}. The significance of the spectral function becomes apparent from its spectral decomposition,
\begin{equation}
    A(\vec{k},\omega) \propto \sum_n |\bra{\psi_n} \hat{\Psi}_{\vec{k}}^\dagger \ket{\phi_0}|^2 ~ \delta(\omega - E_n + E_0)
\end{equation}
where $E_n$ ($\ket{\psi_n}$) are many-body eigenenergies (states) of the Bose polaron Hamiltonian and $E_0$ ($\ket{\phi_0}$) is the energy (state) of the Bose gas without an impurity. Hence, $A(\vec{k},\omega)$ probes the entire many-body spectrum.

The polaron spectrum $A(\vec{k},\omega)$ has a generic form, consisting of a low-energy quasiparticle peak of weight $Z(\vec{k})$: the \emph{quasiparticle residue}. Relative to the peak corresponding to a non-interacting impurity, an energy shift $E(\vec{k})$ is observed: the \emph{polaron energy}. At small momenta, the dispersion of the latter is quadratic $\sim \vec{k}^2/(2m_{\rm p})$ where $m_{\rm p}$ is: the \emph{effective polaron mass}. These are key characteristics of Bose polarons, which strongly depend on the underlying impurity-boson interactions as will be discussed in great detail throughout this review.

\subsection{Outline}
This review is organized as follows. In Sec.~\ref{secOverviewExp} we discuss several recent experiments involving Bose polarons that provide an overview of the current state of the field. We provide references to additional studies. While our discussion cannot capture all contributions that define this field today, we hope that it will serve as a useful summary or as a starting point for experienced researchers or those who are new to the topic. In Sec.~\ref{secEffHamiltonians} we review effective Hamiltonians and standard theoretical approaches currently used for their solution. In Sec.~\ref{secStrongCplgPolaron} we discuss Bose polarons in three dimensions in the strong coupling regime, which are at the focus of most concurrent experiments. In Sec.~\ref{secFewBodyEfimov} we discuss universal few-body and Efimov-type effects arising in the vicinity of the impurity-boson Feshbach resonance. In Sec.~\ref{secQuantDynamics} we focus on non-equilibrium polaron physics, which also play a central role for many experiments. Sec.~\ref{secLowDimPolarons} is concerned with Bose polarons in reduced dimensions $d<3$. In Sec.~\ref{secInducedInteractions} we review progress in describing multi-polaron effects and bath-induced interactions among impurities. Sec.~\ref{secFiniteT} is devoted to effects of thermal fluctuations on the polaron properties. In Sec.~\ref{secOtherSettings} we provide a summary of Bose polarons in experimentally relevant settings beyond ultracold atoms, and notable other problems with close relations to Bose polarons, including Bose polarons involving ionic impurities. We close by an outlook and a discussion of open problems in the field in Sec.~\ref{secOutlook}.

\section{Overview of experiments}
\label{secOverviewExp}
The exploration of mobile impurity atoms immersed in a surrounding Bose-Einstein condensate started in the late 1990s, when the first two-component Bose-Bose mixtures where systematically explored~\cite{Hall1998}. In the early 2000s, researchers at MIT studied the motion of microscopic impurities through a BEC, directly revealing its superfluid properties~\cite{Chikkatur2000}. A more systematic exploration of polarons in ultracold atomic gases started in 2009 when the Zwierlein group observed Fermi polarons~\cite{Schirotzek2009}. In 2011, the same group created a Bose-Fermi mixture of $^{41}\mathrm{K}-^{40}\mathrm{K}$ in the polaronic regime~\cite{ChengHsunWu2011}; in 2012 they used a Bose-Fermi mixture of $^{23}\mathrm{Na}-^{40}\mathrm{K}$ to spectroscopically detect Feshbach molecules in the vicinity of an inter-species Feshbach resonance
~\cite{Wu2012}, finding first signatures of an interaction-broadened impurity spectrum -- a hallmark of polaron formation. 

Following this brief historic perspective, we start the review by providing a brief overview of a selection of some recent experimental achievements, all of which had a strong impact on the parallel theoretical developments that we will discuss later. In subsections \ref{subsecFlorenceExp} - \ref{subsecCambridgExp} we proceed chronologically, starting from the $d=1$-dimensional experiment by Catani et al.~\cite{Catani2012} and covering important spectroscopic studies of $d=3-$  dimensional Bose polarons: back-to-back by Jorgensen, \textit{et al}.~\cite{Jorgensen2016} (Aarhus experiment) and Hu \textit{et al}.~\cite{Hu2016} (JILA experiment), by Yan \textit{et al}.~\cite{Yan2020} (MIT experiment), and most recently by Etrych \textit{et al.}~\cite{Cambridgde2023} (Cambridge experiment). In subsection \ref{subsecOthExp} we summarize notable other experiments in which impurity physics and signatures of Bose polarons have been observed.

\subsection{$^{87}\mathrm{Rb}$ and $^{41}\mathrm{K}-$ in Florence}
\label{subsecFlorenceExp}
One of the first systematic experimental studies of mobile impurities in a bosonic quantum gas was performed in quasi-1D systems by Catani et al.~\cite{Catani2012} in 2012. Using a species-selective dipole potential, impurity atoms of $^{41}\mathrm{K}$ were loaded into arrays of one-dimensional Bose gases of $^{87}\mathrm{Rb}$ atoms. The host atoms were cooled down using microwave evaporation and the impurities were sympathetically cooled in a magnetic trap at low temperatures $T = 1.5\mu \mathrm{K}$. A Feshbach resonance between the impurity and the host atom was ramped, changing the inter-species coupling $g_{\mathrm{K,Rb}}$ while keeping the intra-species coupling $g_{\mathrm{Rb,Rb}}$ in the bath essentially fixed. The experiment was performed with approximately 6000 atoms of $^{41}\mathrm{K}$, while the bath contained $1.8\times 10 ^{5}$ atoms. The average number of atoms per tube was 1.4 and 180 for the impurity and the bath respectively.

The main probe tracking polaron formation in the experiment was constituted by releasing the impurity atoms from a tight dipole trap that initially confined them to the center of the Bose gas. In the subsequent dynamics, impurity atoms expanded in a much shallower dipole trap and their interactions with bath atoms were observed to slow their expansion. The experiments directly measured the axial size of the impurity density distribution, $\sigma(t)=\sqrt{\left\langle x^{2}\right\rangle}$ as a function of expansion time $t$ for different values of the dimensionless coupling strength $\eta=\frac{g_{\mathrm{K,Rb}}}{g_{\mathrm{Rb,Rb}}}$. The first important key observation was that the initial amplitude — during roughly the first ms — of $\sigma(t)$ decreased as the coupling $\eta$ was increased, while the frequency of the oscillations did not change with $\eta$. This observation was interpreted as a direct manifestation of an increase in the effective mass due to the presence of interactions, indicating a renormalization of the single-particle properties. The series of measurements were taken by ramping the relative coupling strength from $\eta=0$ (no interactions) to the strongly interacting regime $\eta=30$.

The authors partially explained the experimental observations by using a simple but robust quasiparticle model consisting of a damped quantum harmonic oscillator. The resulting Langevin equation gives the equation of motion for the position and momentum of the impurities coupled to a thermal bath. In this picture, the attenuation in the oscillation amplitude of the averaged impurity squared-displacement $\sigma(t)$ was linked to the increase of the effective mass $m_{\mathrm{K}} \to m_{\mathrm{K}}^{\star}$ of the underlying quasiparticle, namely $\max_{t} \sigma(t) \propto \sqrt{m_{\mathrm{K}} / m_{\mathrm{K}}^{\star}}$. We will review the dynamical sequence of the experiment in Sec.~\ref{subSecPolaronOscillations}.

From the theoretical point of view, the authors also derived a 1D version of the Fr\"ohlich Hamiltonian, assuming that host bosons form a Tomonaga-Luttinger liquid with density fluctuations around the unperturbed background density. The  effective polaron mass $m_{\mathrm{K}}^{\star}$ was computed using a Feynman variational method~\cite{Tempere2009}. This description is adequate in the weakly interacting regime, where the experiment was reasonably well described by theory. However, deviations were found in the more strongly interacting regime, and a more sophisticated analytical model needed to be employed. We will return to this experiment in our discussion of low-dimensional polaronic systems, see Sec.~\ref{Sec:1DBosePolaron}.

\subsection{$^{39}\mathrm{K}$  in Aarhus}
\label{subsecExpAarhus}
\begin{figure}
	\centering
	\includegraphics[width=\linewidth]{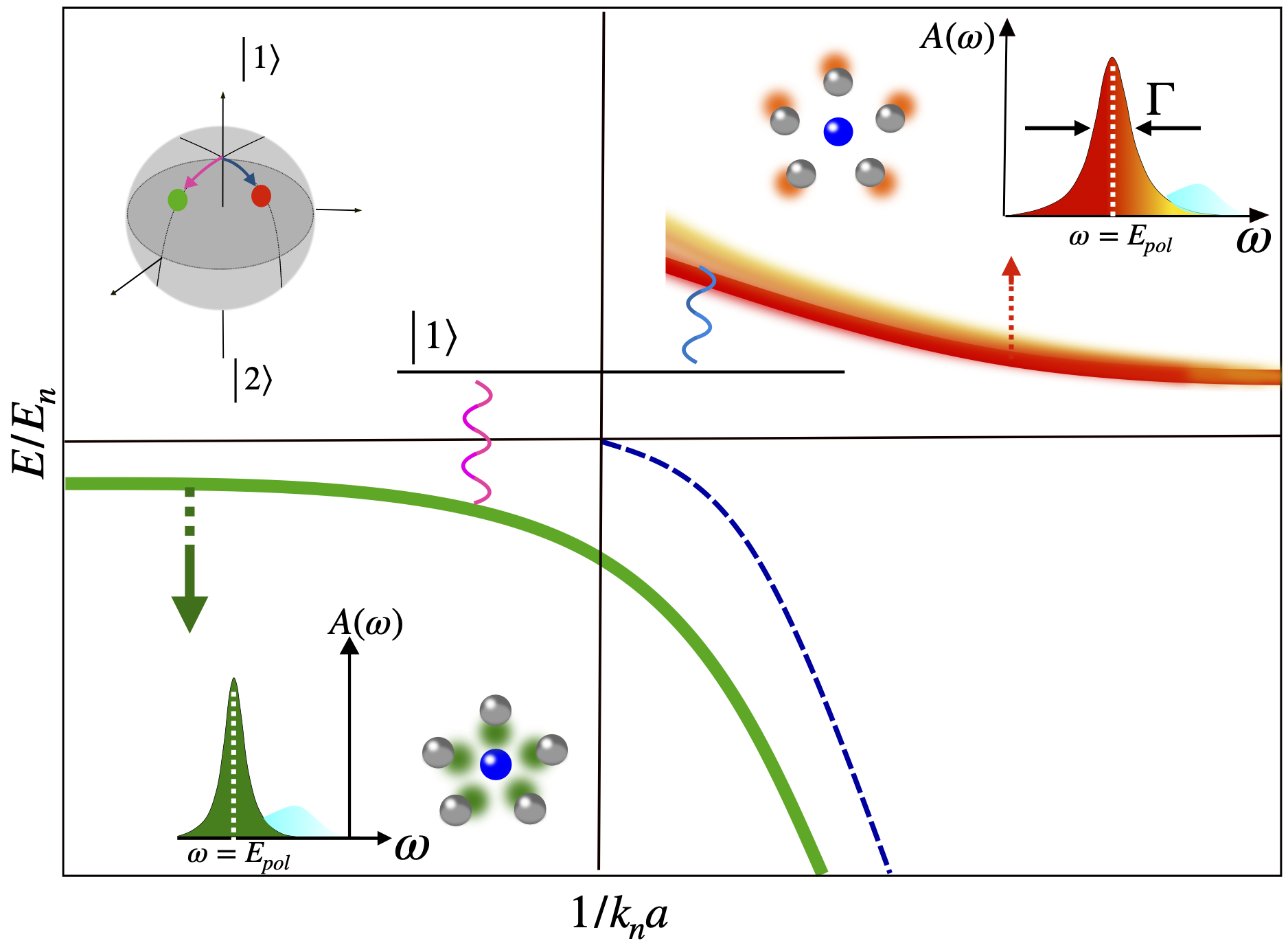}
	\caption{\textbf{Bose polaron spectroscopy around a Feshbach resonance.} (a) A BEC was created in the majority hyperfine state $\left|1\right\rangle$. A small fraction of atoms was transferred from the majority state \cite{Jorgensen2016} -- or a weakly interacting internal impurity state \cite{Hu2016} -- to another hyperfine state $\left|2\right\rangle$ interacting with the majority atoms by using a short RF pulse. The polaron exhibits a many-body ground state termed the attractive branch (green) and a meta-stable repulsive branch (red) when the coupling strength $1/(k_{n} a_{\rm IB})$ is varied. The blue dashed line corresponds to the energy of the two-body molecular impurity-boson-bound state in vacuum. The effective interactions between the impurity and the bath are attractive and repulsive, respectively, for $1/(k_{n}a)<0$ and $1/(k_{n}a)>0$. Figure adapted from reference ~\cite{Jorgensen2016}.}
	\label{figRFspec}
\end{figure}

In 2016, two breakthrough experiments were able to measure the spectral function of Bose polarons in a $d=3$-dimensional BEC \cite{Hu2016,Jorgensen2016}. The first we will discuss was performed in Aarhus by Jorgensen et al.~\cite{Jorgensen2016}, in which a single component BEC of $^{39}\mathrm{K}$ atoms was loaded in a harmonic trap. The condensed atoms were initially in a hyperfine state $\left|1\right\rangle =\left|F=1,m_{F}=-1\right\rangle$ and impurities were created by transferring a small fraction of atoms to a second hyperfine state $\left|2\right\rangle =\left|F=1,m_{F}=0\right\rangle$, using a radio frequency (RF) pulse with a duration on the order of 100 microseconds. When tuning the frequency $\omega_{\rm RF}$ of the RF pulse, the observed resonance was shifted with respect to the natural transition line between the two hyperfine states as a result of impurity-boson interactions, see Fig.~\ref{figRFspec}. 

Importantly, the transfer of atoms to the impurity hyperfine state was performed slowly, i.e. employing sufficiently weak Rabi frequency $\Omega$ in the linear-response regime. Thus the rate of transferred atoms is proportional to the polaron spectral function probed at very low impurity momentum, namely
\begin{equation}
    \frac{dN}{dt}=-\Omega^{2}n_{0}A(\mathbf{k}=0,\omega=\omega_\mathrm{{RF}}),
\end{equation}
where $A(\vec{k},\omega)$ denotes the spectral function, see Sec.~\ref{secKeyConcepts}. In this particular experiment, the RF-protocol realized was injection spectroscopy, where impurities are transferred from a weakly to a strongly interacting state with the BEC~\cite{Vale&Zwierlein2021}.

Boson-boson and impurity-boson interactions are $s$-wave and tunable.  In particular, the ground state properties of the polaron were probed by changing the impurity-bath scattering length $a$ across the Feshbach resonance located around $B_0 = 113.8 \mathrm{G}$ \cite{Jorgensen2016,Lysebo2010,Tanzi18}. Experiments were performed at very low temperatures $T\ll T_{c}$, such that thermal fluctuations should be negligible. Moreover, impurities were created with low momentum, commonly assumed to satisfy $\left|\mathbf{P}\right|\ll m c$; Here $c$ and $T_c$ are the speed of sound and critical temperature of the condensate respectively. In the experiment, the BEC was formed by $2\times10^{4}$ atoms and an average density on the order of $2.3 \times10^{14} \mathrm{cm}^{-3}$, whereas the boson-boson scattering length was fixed to $a_{\mathrm{BB}}=9a_{0}$; therefore, the average gas parameter of the BEC is on the order of  $na_{\mathrm{BB}}^{3}\approx 2.5\times10^{-8}$.

At least at weak coupling and low temperatures, important length and energy scales are set by the inter-particle distance 
\begin{equation}
    k_{n}^{-1}=(6\pi^{2}n)^{-1/3}
    \label{eq:kn}
\end{equation}
and the energy
\begin{equation}
   E_{n}=\frac{\hbar^{2}k_{n}^{2}}{4m_{\mathrm{red}}}, 
   \label{eq:En}
\end{equation}
where $m_{\rm red}$ is the reduced mass of the impurity-boson system. This allows to conveniently parameterize impurity-boson interactions by the dimensionless coupling strength $1/\left(k_{n}a\right)$. Another notable length scale in the problem that will be discussed further below is the healing length of the condensate,  $\xi=\left(8\pi na_{\mathrm{BB}}\right)^{-1/2}$, which also depends on the boson-boson scattering length $a_{\rm BB}$.

Atoms transferred from the BEC to the impurity state by the RF pulse undergo three-body recombination processes, in which two atoms in state $\left|1\right\rangle$ can collide with one in state $\left|2\right\rangle$. Because of energy and momentum conservation, such collisions allow for recombination into a molecular state and a free atom with high energy, all of which are no longer confined by the optical potential. Close to the impurity-boson Feshbach resonance, this corresponds to a dominant loss mechanism, in which eventually all impurities were lost in the experiment and the number of atoms in the condensate reduced accordingly. For a given coupling strength $1/(k_{n}a)$ an RF pulse was used to scan the frequency $\omega_{\rm RF}$ of the pulse and map out transitions $\left|1\right\rangle \rightarrow\left|2\right\rangle$. To this end, the impurity atoms were held in the BEC for a variable time following the RF pulse until \emph{all} impurity atoms were lost in three-body recombination processes. The desired spectroscopic signal measured in the experiment was taken to be the loss of the majority of atoms: This method increases the overall signal since each formed polaron corresponds to three lost bosons overall. 

\begin{figure*}
	\centering
	\includegraphics[width=0.95\linewidth]{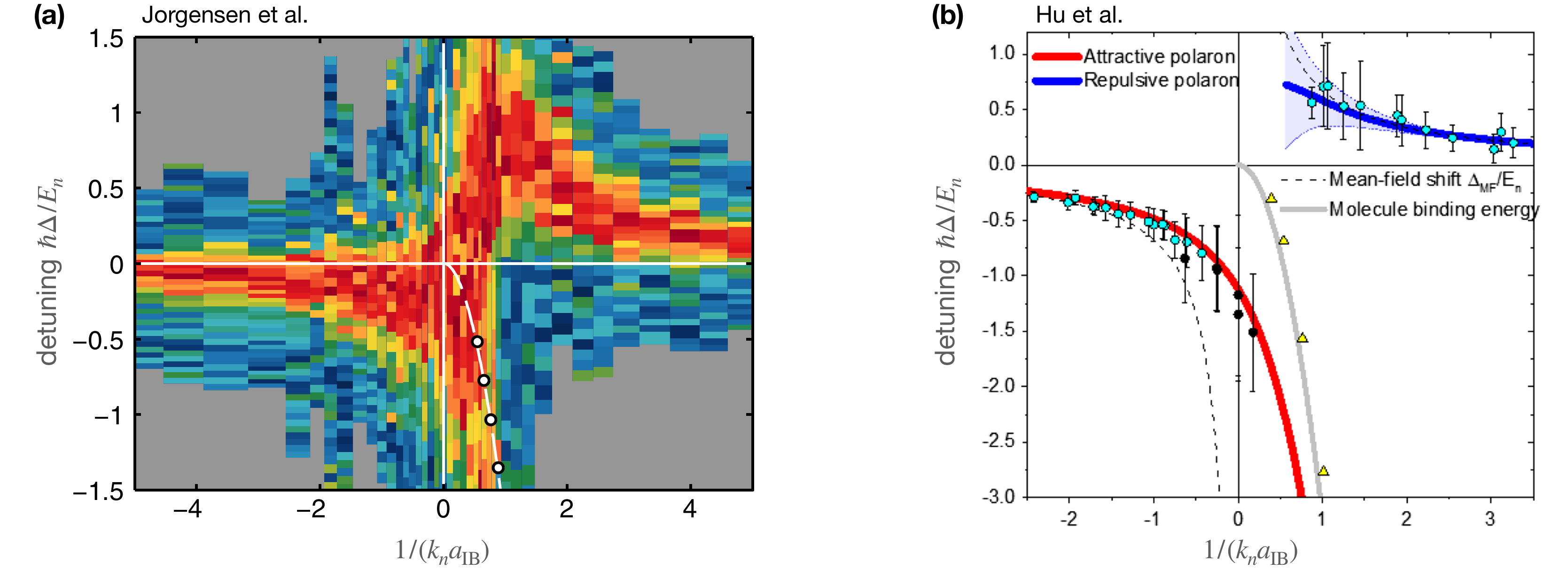}
	\caption{\textbf{Bose polaron spectra} were measured across a Feshbach resonance by Jorgensen et al.~\cite{Jorgensen2016} (a) and Hu et al.~\cite{Hu2016} (b). In (a) the white dots correspond to independently measured energies of the molecular impurity-boson bound state in the absence of a BEC. In (b) the solid red and blue lines correspond to theory predictions at zero temperature by a variational Chevy-type state \cite{Rath2013,Li2014}; the solid grey line (triangles) corresponds to the energy of the molecular impurity-boson bound state (measured independently at low densities) in vacuum without a BEC. Figure adapted from Refs.~\cite{Jorgensen2016} (a) and \cite{Hu2016} (b).}
	\label{figRFspecExp}
\end{figure*}

By extracting the peak positions in the spectrum, see Fig.~\ref{figRFspecExp} (a), the experiment determined the zero-momentum many-body polaron energies as a function of the coupling strength. An attractive and a repulsive branch characterize the energy landscape, as depicted in Fig.~\ref{figRFspec}. In the weakly interacting regime, $1/\left|k_{n}a\right|\gg1$, the polaron energy is determined exactly by the mean-field interactions experienced by the impurity when it scatters off the condensate. There the polaron energy follows a simple expression depending only on the inter-particle distance and the impurity-boson coupling strength, $E/E_{n}=\frac{4}{3\pi}k_{n}a$. 

For attractive interactions extending to the unitarity regime, $1/(k_{n}a)\rightarrow0^{-}$, the polaron energy remains well-defined and finite. The attractive branch was found experimentally to extend smoothly further across the Feshbach resonance $1/(k_{n}a) > 0$ before the spectrum broadens significantly, see Fig.~\ref{figRFspecExp} (a). In absence of a BEC, a universal two-particle molecular bound state is  formed by one impurity and one boson and it exists on the repulsive side, $1/(k_{n}a) > 0$. The energy of this in-vacuum molecular state was measured independently in the experiment and shown as a reference to compare to the energy of the attractive polaron branch in this regime, see Fig.~\ref{figRFspecExp}. It was argued in the Aarhus experiment that the universal bound state may be adiabatically connected to the attractive polaron branch, but the precise characterization of this connection remains a subject of debate. We will return to this question later in the review.

Instead, the repulsive polaron branch was observed only for $1/(k_{n}a) > 0$ and corresponds to a meta-stable excited state of the system. Its energy $E/E_{n}>0$ is positive and the state has a finite lifetime: the quasiparticle state can decay into lower energy states, such as the attractive polaron branch, by releasing energy into the surrounding bath. The spectral function is well defined in the weakly interacting regime, $1/(k_n a_{\rm IB}) \gg 1$, where it may be approximated by a Lorentzian function. Yet, in the regime of strong interactions, the repulsive polaron couples to the many-body excitation spectrum strongly and a significantly broadened spectral response was observed. Understanding the many-body character of repulsive polarons in both Bose-Bose and Bose-Fermi mixtures constitute an intriguing and rich current research direction, see \cite{scazza2022repulsive} for a comprehensive review of the recent progress.

In addition to the polaron energy, the measured spectral response also unveiled information regarding the quasiparticle lifetime, extracted experimentally by analyzing the width $\Gamma$ of the quasiparticle peaks. Besides the three-body losses discussed above, collisions between impurity and condensate atoms result in momentum broadening; the inhomogeneity of the trap and the finite duration of the RF pulse also broaden the spectrum. These effects were analyzed by Jorgensen et al.~\cite{Jorgensen2016} by comparing against theoretical models of broadening mechanisms. The experiment found a wide regime of parameters for which the three-body recombination processes, used directly in the measurement protocol, take place on longer time scales compared to the average polaron lifetime. A detailed review of the theoretical analysis, its range of validity and theoretical interpretations of the measured polaron spectra will be devoted to Sec.~\ref{secStrongCplgPolaron}.

\subsection{$^{40}\mathrm{K}$ and $^{87}\mathrm{Rb}$ in JILA}
\label{subsecExpJILA}
The second of the two 2016 breakthrough experiments was performed in JILA by Hu \textit{et al.}~\cite{Hu2016}. In contrast to the Aarhus experiment, this experimental realization used two different atomic species to create a mass- and particle-number imbalanced mixture. They realized minority fermionic $^{40}\mathrm{K}$ (impurity) atoms immersed in a majority BEC of $^{87}\mathrm{Rb}$, corresponding to a mass ratio $m_{\mathrm{I}}/m_\mathrm{B}\approx 0.46$. In this setting, RF spectroscopy was performed by driving a transition from a weakly to a strongly interacting hyperfine state of the $^{40}\mathrm{K}$ impurities. 

The ultra-dilute gas of fermionic impurities is ruled by Fermi-Dirac statistics. For small impurity concentrations, the impurity's statistics played no appreciable role in the polaron properties. Nevertheless, due to the Pauli pressure exerted on the fermionic impurities, the spatial extent of the fermionic cloud in the experiment was considerably larger than that of the condensate. This caused experimental challenges in the detection of polaron formation since only impurities interacting with the BEC can contribute to the polaronic signal. 

To probe polaron formation, the JILA experiment performed injection spectroscopy. First, a weakly interacting Fermi gas was prepared in the hyperfine state $\left|\uparrow\right\rangle =\left|F=9/2,m_F=-7/2\right\rangle$ of $^{40}\mathrm{K}$. Subsequently, an RF pulse was applied to drive the atomic transition to the $\left|\downarrow\right\rangle =\left|F=9/2,m_F=-9/2\right\rangle $ impurity state. The interaction $1/(k_{n}a)$ of the latter with the BEC in the $\left|F=1,m_F=1\right\rangle$ hyperfine state of $^{87}\mathrm{Rb}$ can be tuned all the way from weak to strong coupling by a broad inter-species Feshbach resonance found around a magnetic field $B_0=546.62 {\rm G}$ \cite{Klempt2008}. 

Due to the large extent of the fermionic cloud, the spectral response was dominated by fermionic atoms with almost negligible overlap with the condensate. To extract only the polaronic signal originating from fermions interacting with the BEC, the data was post-selected to frequencies $\omega_{\rm RF}$ where the root-mean-square size of the cloud of excited $\left| \downarrow \right\rangle$ impurity atoms was within the size of the BEC. Further post-selection of data was applied by using an inverse Abel transformation to extract the central density of $\left| \downarrow \right\rangle$ impurities, which avoids excessive inhomogeneous broadening of the signal due to the spatially varying BEC density. Finally, the data plotted over RF frequency $\omega_{\rm RF}$ was fitted to a Gaussian to extract the central peak position as a function of the impurity-boson interaction strength. The results are shown here in Fig.~\ref{figRFspecExp} (b).  The qualitative features observed by Hu \textit{et al.}~\cite{Hu2016} are in agreement with the findings by Jorgensen et al.~\cite{Jorgensen2016}. In particular, a repulsive and an attractive polaron branch were identified, also in agreement with early theoretical expectations \cite{Rath2013,Li2014}, see Fig.~\ref{figRFspecExp} (b). The attractive branch extends across unitarity into the regime $1/(k_{n}a)>0$, where a two-body $\mathrm{K}-\mathrm{Rb}$ bound state exists in the absence of a BEC. 

In the JILA experiment, a $^{87}\mathrm{Rb}$ BEC of $N=2\times10^5$ atoms were trapped in a harmonic potential. The peak density of the BEC  was $n_0=1.8\times10^{14}~\mathrm{cm}^{-3}$, while the peak density of the impurities was $n_{\rm K}=2\times10^{12}~\mathrm{cm}^{-3}$. This corresponds to a concentration of impurities in the center of the trap around $1.1\%$, which justifies comparisons with idealized theoretical scenarios assuming a single isolated impurity interacting with the surrounding BEC. The background boson-boson scattering length was $a_{\mathrm{BB}}=100a_{0}$, yielding a gas parameter on the order of $\bar{n}a_{\mathrm{B}}^{3}=2.3\times10^{-5}$ -- here $\bar{n}$ is the average probed density. Thus, the gas parameter is almost three orders of magnitude larger than the one in the Aarhus experiment \cite{Jorgensen2016}. In contrast, the temperature and average density of the bath were of the same order in both experiments.
  
By introducing a variable hold time following an RF pulse close to the polaron peak frequency and measuring the decay of the interacting impurity state, the JILA experiment quantified the lifetime of the polaron state. To this end, the averaged central $\left| \downarrow \right\rangle$ density was measured and fitted to an exponential decay $\propto \exp(-\Gamma t)$, where $\Gamma^{-1}$ is the obtained lifetime. This protocol was performed for different interaction strengths.  For the attractive branch, the decay rate $\Gamma$ was found to increase when the coupling $1/(k_{n}a_{\textrm{IB}})$ approaches unitarity, in agreement with theoretical considerations~\cite{D'Incao2005}. Instead,  at strong coupling, where the lifetime is shortest, three-body recombination processes are expected to be relevant. For the repulsive branch, the decay rate was observed to be of the same order of magnitude as for the strongly coupled attractive polaron, $\hbar\Gamma/E_{n}\sim10^{-2}$, and it remained almost independent of the coupling strength for increasing $1/(k_{n}a_{\textrm{IB}}) > 0$. This observation was in contrast with the observation of an interaction-dependent decay rate on the repulsive side in the Aarhus experiment \cite{Jorgensen2016}.  Hu \textit{et al.}~\cite{Hu2016} speculate that when repulsive polarons are created, they may escape to the outside regions of the trap where there are no bosons. Such escape does not happen for attractive polarons, which stay in the central region after they are created. Hence the observed difference in the lifetimes may not be an intrinsic property of the two types of polarons, but arises from the inhomogeneous density of BEC atoms used in experiments. A theoretical analysis of this scenario was later performed in Ref.~\cite{Ardila2019}.
 
\subsection{$^{40}\mathrm{K}$ and $^{23}\mathrm{Na}$ in the MIT}
\label{sec:MITExp}
The first experiments on Bose polarons, reviewed in Secs.~\ref{subsecFlorenceExp} - \ref{subsecExpJILA} above were performed at very low temperatures. These early studies were complemented in 2019 by another cornerstone experiment performed at MIT by Yan et al.~\cite{Yan2019}, which investigated finite-temperature effects on Bose polaron spectra. Moreover, the experiment characterized the breakdown of the quasiparticle description in the vicinity of quantum criticality associated with the condensation of the host bosons. This study revealed an interesting interplay between quantum and thermal fluctuations and will be reviewed next.

The MIT experiment by Yan \textit{et al.}~\cite{Yan2020} characterized the Bose polaron properties by employing fermionic $^{40}\mathrm{K}$ impurities immersed in a Bose gas of $^{23}\mathrm{Na}$ atoms at different coupling strengths and temperatures. In contrast to earlier experiments, attractive Bose polarons were initially prepared in their interacting equilibrium state by ramping up the magnetic field close to the used interspecies Feshbach resonances and letting them equilibrate within $2 {\rm ms}$ -- well above the expected equilibration time scale, $\hbar / E_n \sim 4 \mu {\rm s}$ at unitarity, and before three-body losses limit the lifetime of the mixture around $4 {\rm ms}$. 

In this experiment, an alternative protocol called ejection spectroscopy was used, where the initially strongly interacting impurity state is driven to a weakly interacting hyperfine state \cite{Vale&Zwierlein2021} -- opposite to the injection spectroscopy sequence chosen by the earlier experiments~\cite{Jorgensen2016,Hu2016}. This way, the experimental team was able to measure the polaron energy and spectral width as a function of temperature and impurity-boson coupling strength. A direct advantage of ejection spectroscopy is that the polaron energy can be directly extracted from the peak position rather than the energy onset in the case of injection spectroscopy \cite{Ardila2019}. Moreover, high-energy tails in the ejection spectra provide access to short-range correlations by measuring the contact parameter $C$. A detailed discussion about the connection between
injection and ejection spectroscopy, as well as the relation to impurity thermodynamics such as the contact and free energy is presented in Liu et al.~\cite{liu2020theory}.

At small temperatures $T\simeq 0.1 T_c$, well below the critical temperature $T_c$ of the BEC, a well-defined quasiparticle was observed, in agreement with theoretical expectations and earlier experiments~\cite{Jorgensen2016,Hu2016}. At high temperatures around or above $T_c$, the quasiparticle picture was found to break down. Particular emphasis was put on analyzing unitarity-limited phenomena and signatures of quantum criticality, e.g. in the observed lifetime, as will be reviewed in more detail shortly. 

The experiment realized an ultracold imbalanced Fermi-Bose mixture, where $^{40}\textrm{K}$ fermions take the role of impurities embedded in a BEC of $^{23}\textrm{Na}$. This corresponds to an impurity-to-boson mass ratio of $m_{\rm I}/m_{\rm B}=1.75$. The peak densities were $n_{\mathrm{Na}}=6\times10^{13}\mathrm{cm}^{-3}$ and $n_{\mathrm{K}}=2\times10^{11}\mathrm{cm}^{-3}$ for the BEC and the impurities, respectively, corresponding to an impurity concentration of $\approx0.3\%$. This is well in the regime where impurity atoms can be considered independently. The impurity-boson scattering length $a_{\rm IB}$ was tuned using inter-species Feshbach resonances of the two hyperfine states $\left|F=1,m_F=1\right\rangle$ for $^{23}\textrm{Na}$ and $\left|F=9/2,m_F=-9/2\right\rangle$ for $^{40}\textrm{K}$, located between $75 {\rm G}$ and $93.2 {\rm G}$ \cite{Park12}. The equilibrium gas parameter realized in the experiment was $n_{\mathrm{Na}}a_{\mathrm{BB}}^{3}=1.25\times10^{-6}$, similar to the JILA experiment \cite{Hu2016}.

\begin{figure}
\includegraphics[width=\linewidth]{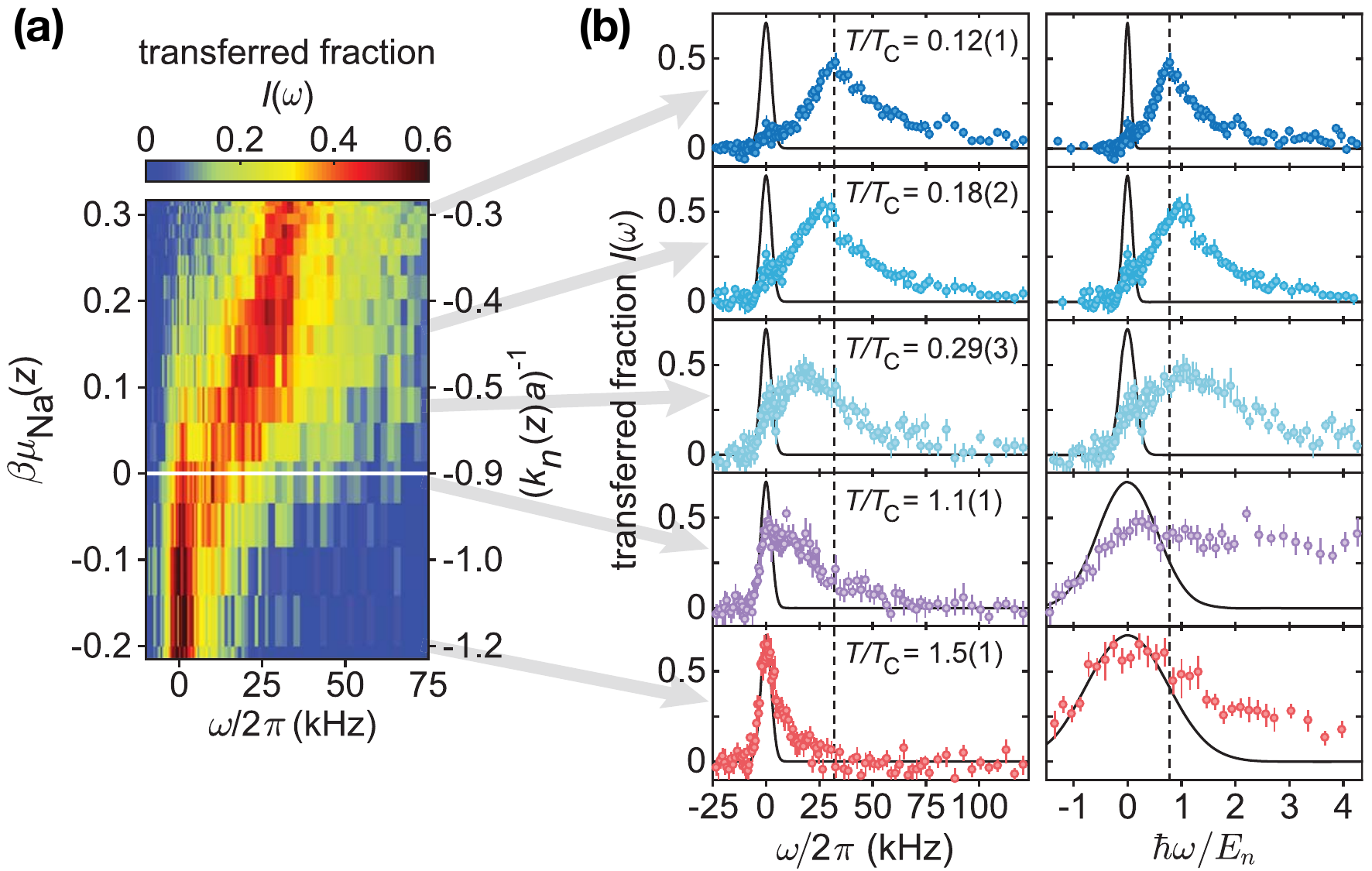}
\caption{\textbf{Temperature-dependent Bose polaron spectra} extracted from a single setup using the local density approximation in the MIT experiment by Yan et al.~\cite{Yan2019}. (a) The spatially (along $z$) resolved spectroscopic signal (ejection) is plotted over the majority ($^{23}\textrm{Na}$) chemical potential $\mu_{\rm Na}(z)$. (b) Different $\mu_{\rm Na}$ correspond to different critical temperatures $T_c(z)$ of the BEC, allowing to extract  temperature dependence of Bose polaron spectra. In the right column, frequency is shown in units of $E_n$. The experiment was performed close to unitarity, at $1/(k_n a)=-0.3$ with $a=a_{\rm IB}$ in the center. This figure was adapted from~\cite{Yan2020}.}
\label{figMITSpectrocopy}
\end{figure}

As described above, $^{40}\textrm{K}$ impurities were prepared initially in the interacting state $\left|F=9/2,m_F=-9/2\right\rangle$, for three different interaction strengths: $1/\left(k_{n}a_{\textrm{IB}}\right)=-1.7$ (weak coupling), $-0.7$ (strong coupling) and $-0.3$ (near-resonant). The impurities were then driven to the hyperfine state $\left|F=9/2,m_F=-7/2\right\rangle$ of $^{40}\textrm{K}$, which is not (or only very weakly) interacting with the condensate. The resulting transfer rate into the weakly interacting impurity state directly constitutes the spectroscopic signal $I(\omega,z)$, and could be probed with good spatial resolution along the axial direction $z$ in the experiment, see Fig.~\ref{figMITSpectrocopy} (a).  

The subsequent experimental analysis relied on the local density approximation. Since the local chemical potential $\mu_{\rm Na}(z)$ -- and correspondingly the density $n_{\rm Na}(z)$ -- of the bosons in the trap depend on the position $z$, the dimensionless interaction strength $1/(k_n(z) a_{\rm IB})$ varies across the trap. For the strongest interactions, it ranges from $1/(k_n(z) a_{\rm IB}) = -0.3$ in the center to $1/(k_n(z) a_{\rm IB})\sim -1.0$ which is still a significant interaction strength around the edge, see Fig.~\ref{figMITSpectrocopy} (a). More importantly, the critical temperature $T_{c}(z)=3.31\frac{\hbar^{2}}{k_{B}m_{\rm Na}}\left|n_{\rm Na}(z)\right|^{2/3}$~\cite{DalfovoRMP99} also depends on the position $z$ in the trap through the BEC density and vanishes at the edge of the cloud. This allowed to study the Bose polaron ejection spectrum around the BEC phase transition at $T=T_c$, see Fig.~\ref{figMITSpectrocopy} (b).

The measured spectra for near-resonant interactions $1/\left(k_{n}a_{\textrm{IB}}\right)=-0.3$ revealed a pronounced shift of the polaron peak to lower energies with increasing temperature $T<T_c$, in accordance with earlier theoretical predictions \cite{Guenther2018}. Formulated differently, this corresponds to the disappearance of the negative frequency shift of attractive polarons, because the atomic hyperfine states are such that the non-interacting energy state is higher in energy. At the critical temperature $T=T_c$, the peak position shifts back sharply and becomes centered around the bare atomic transition when $T>T_c$. For weak interactions, in contrast, the polaron energy is found to depend only weakly on temperature, while it always reduces to the bare atomic frequency above $T_c$.

The experiment also measured the polaron lifetime $1/\Gamma$ by extracting the width $\Gamma$ of the spectral peak. The most interesting behavior was found for near-resonant interactions, $1/\left(k_{n}a_{\textrm{IB}}\right)=-0.3$, where $\Gamma \simeq 8.1(5) k_{\rm B} T /\hbar$ scales linearly with temperature below $T_c$, quite substantially at the Planckian scale $k_{\rm B} T$. On the one hand, this suggests a well-defined quasiparticle with a vanishing spectral width in the zero-temperature limit. On the other hand, near $T_c$ the RF spectral width was found to increase beyond the measured energy of the polaron $E$. This feature characterizes the breakdown of the quasiparticle picture, which is consistent with quantum critical behavior expected around $T_{\mathrm{c}}$ and for strong impurity-boson interactions. 

From the high-frequency tail of the spectral function one can gain further insights into the short-range correlations, as quantified by the thermodynamic parameter called contact $C$. Experimentally, the contact was extracted from the universal scaling of high-energy tails in the spectral function as $A(\omega)=A_{0}C/\omega^{3/2}$, where $A_0$ is a known function of the peak Rabi coupling $\Omega$, the Gaussian $e^{-1/2}$ width of the RF pulse and fundamental constants \cite{Yan2020}. By relating these tails to universal two-body correlations \cite{Tan2008,Braaten2010}, it can be shown that the contact describes the probability that the impurity is in close vicinity to the host bosons.  For a fixed temperature but below $T_{\mathrm{c}}$, the contact was found to increase monotonically with the coupling strength, in agreement with theoretical expectations \cite{Ardila2015} at zero temperature. Instead, as a function of temperature but below $T_{\mathrm{c}}$, the contact $C$ decreased with increasing $T$ at weak couplings. The observable $C(T)$ remained approximately constant and large for near-resonant interactions, still below $T_{\mathrm{c}}$, corresponding to an average number of host bosons in a sphere of radius $d$ given by the inter-particle spacing on the order of one. Above the critical temperature $T_c$ the contact was found to drop sharply. This is consistent with the fact that the contact is inversely proportional to temperature for a thermal gas of bosons~\cite{Fletcher2017}.


\subsection{$^{39}\mathrm{K}$ in Cambridge}
\label{subsecCambridgExp}
\begin{figure}
\centering
\includegraphics[width=0.85\linewidth]{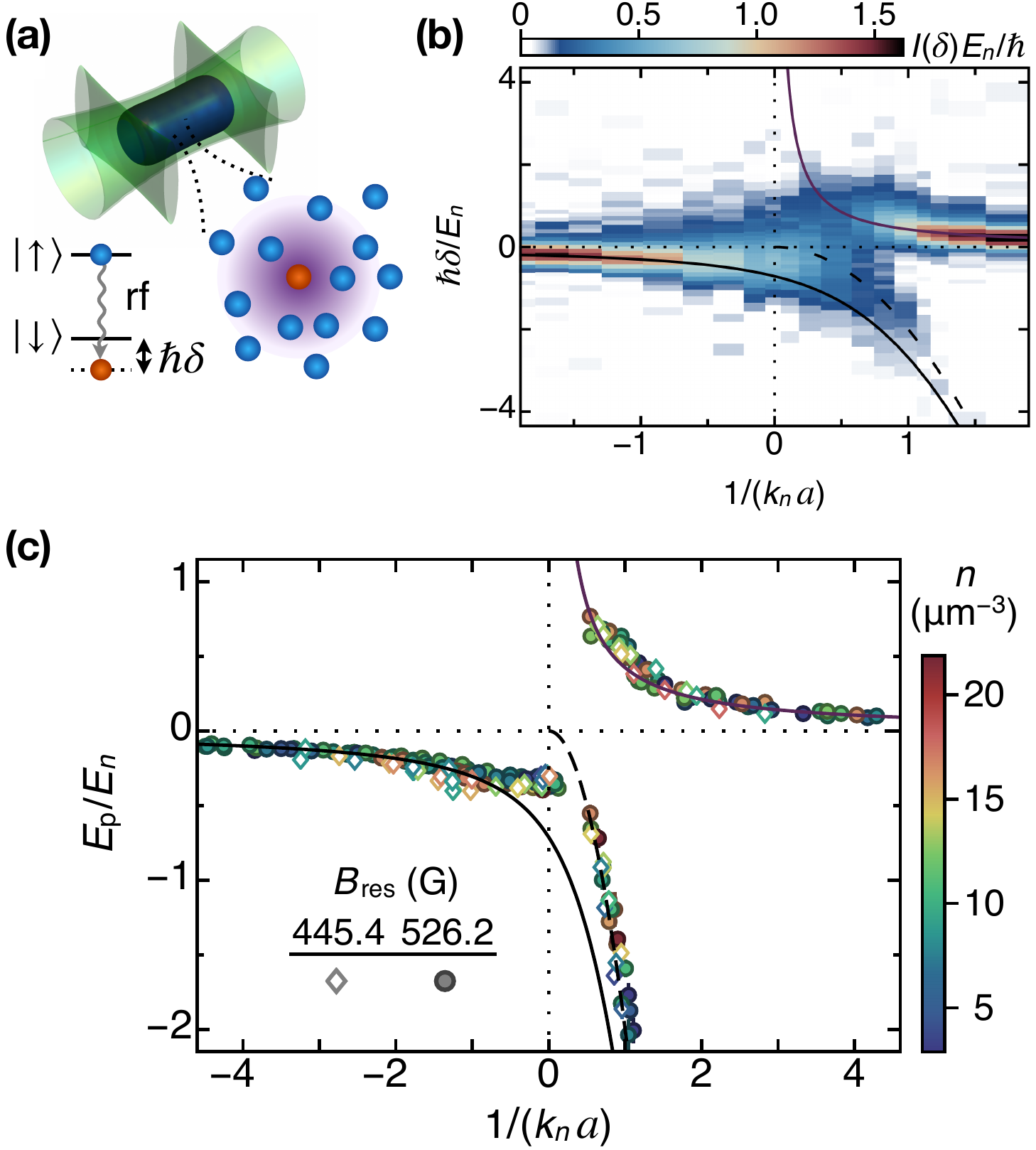}
\caption{\textbf{Universality of Bose polarons, measured in a box trap} realizing a quasi-uniform BEC of host bosons (a). Using injection RF spectroscopy in the linear-response regime, Bose polaron spectra with strongly reduced inhomogeneous broadening were measured by Etrych et al.~\cite{Etrych24} (b), revealing the molecular branch co-existing with the repulsive polaron. (c) The peaks in the Bose polaron spectra, measured for different boson-boson interactions $a_{\rm BB}$ realized around the impurity-boson Feshbach resonances at different magnetic fields $B_{\rm res}$, reveal a near-perfect universal dependence on $k_n a_{\rm IB}$ when energy $E_n$. This figure was adapted from~\cite{Etrych24}.}
\label{figCambridge}
\end{figure}

The experiments reviewed so far all used harmonically trapped BECs. In a recent experiment in Cambridge, UK, Etrych \textit{et al.}~\cite{Etrych24} were able to overcome this limitation: they realized a Bose-Bose mixture of two hyperfine states of $^{39}\mathrm{K}$ trapped in an optical box potential~\cite{Navon2021,Gaunt13}, see Fig.~\ref{figCambridge} (a). One hyperfine state was initialized in a quasi-homogenous BEC, whereas the other served as an impurity species with tunable impurity-boson interactions. 

The experiment used two combinations of hyperfine states to realize impurity-boson Feshbach resonances at two different values of the corresponding boson-boson interaction $a_{\rm BB}$: Around magnetic fields $B_{1}=526.16(3)\mathrm{G}$ and $B_{2}=445.42(3)\mathrm{G}$, respectively. This leads to bath states $\ket{F=1,m_F=-1}$ and $\ket{F=1,0}$ whose intra-bath interactions $a_{\rm BB}$ differ by a factor of $\approx 3$ around the resonances; correspondingly, the compressibility of the bath changes. Moreover, the density of the quasi-uniform BEC was varied in the range $(3-22)\mu \textrm{m}^{-3}$ and therefore typical gas parameters on the order of $na_{\rm BB}^{3}=(1.5-11)\times10^{-9}$ and $na_{\rm BB}^{3}=(4-30)\times10^{-8}$ were realized for magnetic fields around $B_{1}$ and $B_{2}$ respectively.

A first main achievement of the experiment by Etrych \textit{et al.}~\cite{Etrych24} was the high quality of their measured injection RF spectrum, see Fig.~\ref{figCambridge} (b). In contrast to earlier experiments, the box trap largely avoids inhomogeneous broadening, probing the polaron at a quasi-uniform BEC density, and thus giving access to the intrinsic spectral widths resulting from interaction effects. As in earlier experiments, the authors recorded the loss rate of the majority atoms due to impurity-induced three-body losses, although they did so after switching the magnetic field onto the resonance directly following the RF pulse, in order to avoid further atom losses. This also ensures that the loss mechanism is the same for all $1/(k_n a_{\rm IB})$, so that the recorded area under the spectra has a clear physical meaning.

As a result, the experiment revealed coexisting spectral features on the repulsive side, associated with the repulsive polaron and the molecular branch at positive and negative energies, respectively. This is also visible in Fig.~\ref{figCambridge} (b). The authors then analyzed the molecular branch further in their injection spectra and observed signatures of additional many-body states below the molecular peak. Their observation was further corroborated by a differential interferometry technique proposed by Etrych \textit{et al.}~\cite{Etrych24} that allows to isolate the molecular-branch spectrum in the regime where it overlaps with the repulsive-polaron one.

Finally, Etrych \textit{et al.}~\cite{Etrych24} varied the boson-boson interaction strength $a_{\rm BB}$ as described above, as well as the bath density $n$. For various such parameter sets they recorded RF spectra and analyzed their peak positions $E_{\rm p}$. When measured in units of the characteristic length and energy scales $k_n^{-1}$ and $E_n$, a near-perfect data collapse was observed, see Fig.~\ref{figCambridge} (c). This is true for all observed branches, attractive and repulsive polarons as well as the molecular branch. Thereby the experiment demonstrates a remarkable universality of Bose polaron properties, which were found to be almost completely independent of the gas parameter characterizing the BEC. The only notable exception corresponds to intermediate interactions on the attractive side, possibly observed experimentally around $1/(k_n a_{\rm IB})=-1.5$ as a weak residual dependence of the polaron peak position $E_p/E_n$ on $a_{\rm BB}$ -- diamonds vs. circles in Fig.~\ref{figCambridge} (c).

\subsection{Notable others}
\label{subsecOthExp}
The experimental realization of mobile quantum impurities in ultracold bosonic gases faced many challenges before Bose polarons were first observed. For instance, cooling down impurity atoms to reach thermal equilibrium with the host gas involves some special trapping techniques. In addition, the detection of low densities of impurity atoms, essentially at the single-atom level, posed a significant experimental challenge~\cite{Weber10}. 

Another relevant issue inherent to the bosonic system are 3-body recombination processes. They are a blessing and a curse in the current experiments with Bose polarons. For example, in the Aarhus~\cite{Jorgensen2016} and Cambridge~\cite{Etrych24} experiments, injection spectroscopy employed losses for inferring the impurity fraction. Conversely, in the case of ejection spectroscopy such as in the MIT experiment~\cite{Yan2020}, equilibrium states with strongly interacting impurities are challenging to achieve due to three-body losses which prevent adiabatic preparation schemes from working. In the case of injection spectroscopy, three-body losses are also an issue when measuring the photoexcited states, causing unwanted spectral broadening. 

Thus, engineering clever techniques for cooling a highly imbalanced quantum bosonic mixture, high precision in measuring impurity population and avoiding the 3-body loss rate were among the most important questions addressed in earlier experiments that we summarize now.

\subsubsection{$^{161}\mathrm{Cs}$  and  $^{87}\mathrm{Rb}$ in Bonn/ Kaiserslautern --}
Among earlier studies on quantum impurities that ultimately paved the way for the realization of Bose polarons was an experiment in 2012 by Spethmann et al.~\cite{spethmann2012dynamics}. They reported a controllable doping of $^{133}\mathrm{Cs}$ impurities in an ultracold $^{87}\mathrm{Rb}$ quantum gas~\cite{spethmann2012dynamics}. In this experiment, $\mathrm{^{87}Rb}$ atoms were precooled and prepared in a crossed dipolar trap in the hyperfine state $\left|F=1,m_{F}=0\right\rangle$, while $\mathrm{Cs}$ atoms were trapped in close vicinity by a high-gradient magneto-optical trap (MOT). Both species were then loaded into separate sites of a one-dimensional optical lattice, which was ramped down adiabatically to immerse individual $\mathrm{Cs}$ atoms into the $^{87}\mathrm{Rb}$ gas in a controlled manner.

The temperature of $\mathrm{Cs}$ was on the order of $\mu \mathrm{K}$, which is too high with respect to the temperature of the Rubidium buffer bath. Therefore, further cooling of the $\mathrm{Cs}$ atoms was achieved by adiabatically ramping down the lattice, keeping both species "living" in the original crossed dipole trap of the buffer bath. Thus, both species began to undergo elastic two-body collisions, achieving thermalization after 50 ${\rm ms}$. The impurity atoms reached the same temperature as the buffer gas ($\approx 250 \mathrm{nK}$) well into the regime where $s-$ wave scattering dominates. The experiment also took advantage of the inelastic 3-body collisions (e.g. $\mathrm{Rb-Cs-Rb}$) that are important in bosonic systems as those are related to the lifetime of each component. The authors measured the survival probabilities event by event. While the survival probability of the buffer gas remained unaffected by the presence of a few impurity atoms, the survival probability for the impurities decayed exponentially with time, which is related to the dominant mechanism of losses: molecule formation through three-body recombination.

Later experiments using the same mixture $\mathrm{Cs-Rb}$ addressed two fundamental questions regarding using impurities as a quantum probe and how to measure the coherence of the mixture. In the former case, spin-changing collisions in the internal impurity states were used as a "probe" to extract information about the hot bath. In particular, in the experiment~\cite{Bouton2020}, the authors used $\mathrm{Cs}$ impurities to measure the temperature of the buffer $\mathrm{Rb}$ gas. In the experiment~\cite{Adam21}, the goal was to perform Ramsey spectroscopy on the internal states of individual $\mathrm{Cs}$ atoms to characterize the coherent superposition of the internal states of $\mathrm{Cs}$ with the host ultracold bath of $\mathrm{Rb}$ atoms.

\subsubsection{$^{87}\mathrm{Rb}$ Mixture in Munich --}
Bose polarons can also arise in settings relevant to quantum magnetism, as explored in an experiment by Fukuhara et al.~~\cite{Fukuhara2013}. To this end, individual spin excitations -- which can also be viewed as magnons -- were created and their dynamics was analyzed in an ultracold quantum gas in an optical lattice. 

The experiment started with a $d=2$ dimensional degenerate Bose gas of  $^{87}\mathrm{Rb}$ atoms prepared in the hyperfine state $\left|\uparrow\right\rangle =\left|F=1,m_{F}=-1\right\rangle$ from which $d=1$ arrays of Mott insulator were created; each chain contained roughly ten atoms~\cite{Endres2011}. In order to create the impurity, two procedures were applied: off-resonant light first created a negative potential for a specific lattice site. Subsequently a microwave pulse was used to flip the spin of an $\left|\uparrow\right\rangle$ atom on that site, hereby creating the impurity atom: $\left|\uparrow\right\rangle \rightarrow\left|\downarrow\right\rangle$. This technique is known as single-site addressing~\cite{Weitenberg2011}. 

The experiment aimed to study the quantum dynamics of the spin $\left|\downarrow\right\rangle$ impurity propagating along the 1D spin chain $\left|\uparrow\right\rangle$. The impurity dynamics were studied as a function of the spin-independent single-particle tunneling rate  $J$, and the onsite interaction energy $U$. 
In the Mott insulator regime $J\ll U$, the impurity dynamics was governed by a single-particle superexchange process between the impurity and the neighbouring spins. The probability (that depends only on the superexchange coupling) of finding the impurity at a specific position and time was measured and compared with the exact result obtained from the homogeneous spin-$1/2$ Heisenberg model at zero temperature.  Moreover, temperature effects were also investigated, although finite temperature fluctuations were found not to affect impurity diffusion. Deep into the Mott regime, $J\ll U$, the impurity exhibited a diffusive behavior where the measured velocity is proportional to the superexchange coupling. However close to the Mott-insulator transition $\left(J/U\right)_{\mathrm{c}}\approx0.3$ the velocity was reduced and it was found to be almost constant as $J/U$ increases.

In the opposite regime, namely the superfluid regime, $J\gg U$, the impurity was coupled to the low-energy excitations of the superfluid, forming a polaron. The measured velocity of the dressed impurity was nearly half of the free particle. The velocity reduction in this regime was due to the strong dressing between the spin impurity and the bath, which manifests as an increase in the impurity’s effective mass.

Related impurity dynamics in a one-dimensional system was also investigated by other groups, for instance, Palzer et al.~\cite{Palzer2009} in Cambridge (UK). They studied impurity dynamics in the Tonks-Girardeau gas in one-dimensional tubes and without a lattice along the direction of motion, but applied an external force.

\subsubsection{$^{6}\mathrm{Li}$ and  $^{23}\mathrm{Na}$ in Heidelberg --} 
One of the first experiments testing decoherence and relaxation dynamics of impurities in a Bose-Einstein condensate was the experiment by Scelle et al.~\cite{Scelle2013}, using a technique known as motional spin-echo spectroscopy. The main scope of the experiments was to study the decoherence experienced by a two-level quantum system, realized by motional states of the impurity, interacting with a Bose-Einstein condensate. The interferometric protocol facilitates precise extraction of dephasing mechanisms resulting exclusively from interactions between impurities and the host gas.

The experiment consisted of $^{6}\mathrm{Li}$ impurities loaded in a species selective $d=1$-dimensional optical lattice, confining the impurities into "pancakes" created by two intercepting laser beams. Additionally, the fermionic impurities were embedded in a BEC of $^{23}\mathrm{Na}$ ultracold gas. The experiment worked in a regime where tunneling along the lattice direction was negligible; i.e. in the longitudinal direction the impurities can be described as harmonic oscillators (one oscillator per potential well). Tunneling in transverse direction was weak, which justified averaging over transverse degrees of freedom of the impurities.

Decoherence was extracted experimentally from the interference fringes between excited and ground states obtained after applying the interferometric protocol. In order to achieve this, two protocols were employed: standard Ramsey interferometry and spin-echo spectroscopy. In the first case, the lattice was "shaken" with a $\pi/2$ pulse in the longitudinal direction such that Li atoms were prepared in an equal superposition between the motional ground and the excited state in longitudinal direction; after a free evolution, another $\pi/2$ pulse was applied and decoherence was extracted from the final population of the excited state. However, due to inhomogeneities of the trap frequencies, a rapid dephasing for time scales on the order of $\sim 1 {\rm ms}$ was observed. 

In order to resolve decoherence arising from the effects of interaction between the impurities and atoms, a spin-echo protocol was used. The difference with respect to the first Ramsey protocol is that a $\pi$ rephasing pulse is applied between the two $\pi/2$ pulses, allowing to discern different sources of decoherence. Indeed, this allowed to measure coherent Ramsey fringes up to much longer times $\sim 10 {\rm ms}$ and detect the bath-induced decoherence of the impurity motion. 
The relaxation time, related to the population decay from higher motional states to the ground state, was also measured and compared with the decoherence time: the latter was found to be slightly larger.

\subsubsection{$^{23}\mathrm{Na}$ and $^{40}\mathrm{K}$  in Munich  --} 
Controlling the interaction between atomic species opens rich possibilities to study complex systems, such as the formation of molecules from individual atoms. In their experiment, Duda et al.~\cite{Duda2023} prepared a binary quantum mixture formed by $^{\ensuremath{23}}\mathrm{Na}$ bosonic and $^{\ensuremath{40}}\mathrm{K}$ fermionic species. 

\begin{figure}
\centering
	\centering
\includegraphics[width=6.5cm]{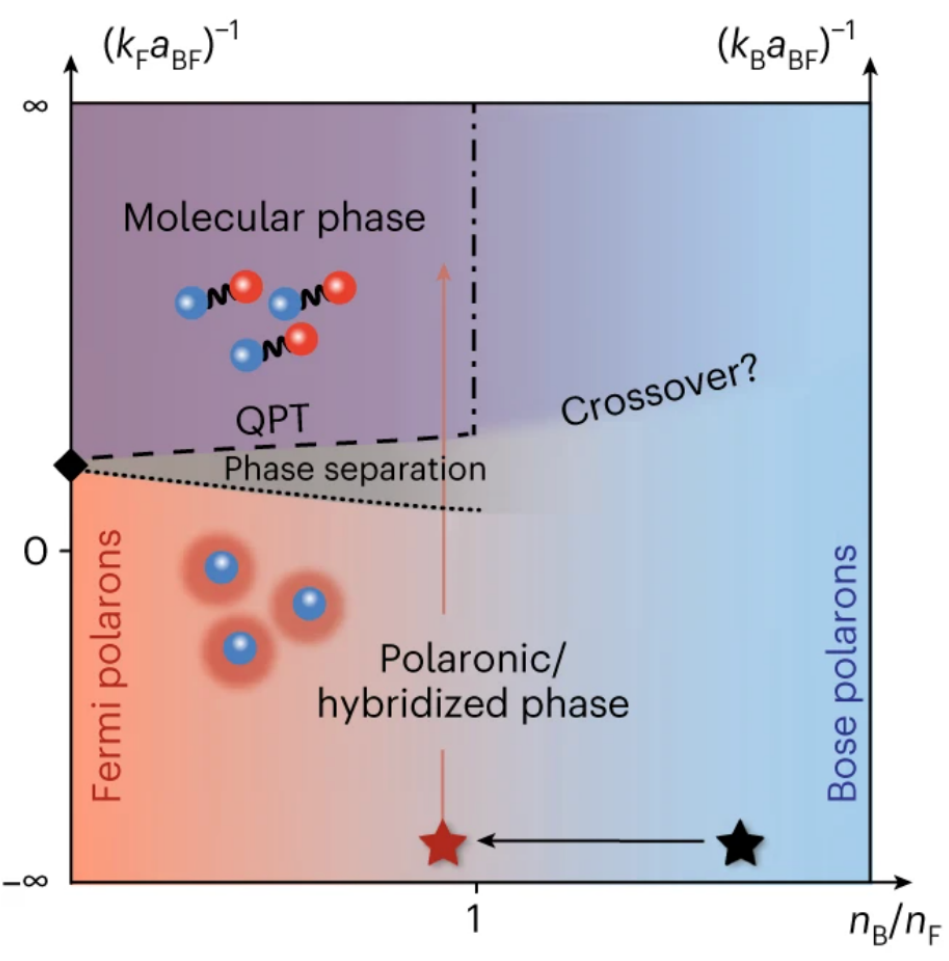}
	\caption{\textbf{Phase diagram of the $^{23}\mathrm{Na}$ and $^{40}\mathrm{K}$} as a function of the density ratio $n_{\mathrm{B}}/n_{\mathrm{F}}$ and the dimensionless interaction strength $\left(k_{i}n_{\mathrm{BF}}\right)^{-1}$. Figure taken from~\cite{Duda2023}.} \label{ExperimentInducedInteraction1}
\end{figure}

The main goal of the experiment was to explore the rich phase diagram of a Bose-Fermi mixture at very low temperatures, see Fig.~\ref{ExperimentInducedInteraction1}. Starting from the impurity limit to the highly mass-balanced case where fermions and bosons are highly correlated, forming thus Feshbach molecules. In the experiment,  the phase diagram was investigated by changing two parameters, first the relative concentration of densities $n_{\mathrm{B}}/n_{\mathrm{F}}$ being $n_{\mathrm{B}}$ and $n_{\mathrm{F}}$ the bosons and fermion density respectively; and second, the coupling strength between fermions and bosons 
$(k_{i}a_{\mathrm{FB}})^{-1}$ with $k_{i}=(6\pi^{2}n_{i})^{1/3}$ and  the index $i=\mathrm{F,B}$ denoting fermions and bosons  respectively. In addition, $a_{\mathrm{BF}}$ is the boson-fermion scattering length. The latter did drive the transition from a weakly interacting mixture to a degenerate molecular gas when interactions were strong.

The phase diagram as a function of both the density ratio 
$n_{\mathrm{B}}/n_{\mathrm{F}}$ 
and the Bose-Fermi coupling strength $(k_{i}a_{\mathrm{BF}})^{-1}$ explored two regimes, (i) polaronic phase: In the weakly interacting regime $(k_{i}a_{\mathrm{FB}})^{-1}\rightarrow-\infty$, the mixture was formed by an uncoupled sea of quasiparticles, extrapolating between the Fermi polaron for 
$n_{\mathrm{B}}/n_{\mathrm{F}}\ll1$ to the Bose polaron limit, $n_{\mathrm{B}}/n_{\mathrm{F}}>1 $. Increasing the interspecies interaction  may lead to induced interaction effects between quasiparticles; (ii) molecular phase: In the strongly  interacting regime 
$(k_{i}a_{\mathrm{BF}})^{-1}\rightarrow +\infty$ and $n_{\mathrm{B}}<n_{\mathrm{F}}$, a binding between bosons and fermions instead leads to the formation of a degenerate gas of molecules. In the extreme limit $n_{\mathrm{B}}\ll n_{\mathrm{F}}$ and for zero temperature, the transition is very well-known as the polaron-molecule quantum phase transition (QFT) for the Fermi polaron problem explored theoretically ~\cite{Prokofev2008, Mora2009, Punk2009, Combescot2009, Schmidt2011, Ness2020} and observed experimentally~\cite{Schirotzek2009,Koschorreck2012,Ness2020,Fritsche2021}. However, in the current experiment where $n_{\mathrm{B}}\leq n_{\mathrm{F}}$, the transition was predicted to be a first-order phase transition accompanied by phase separation or a second-order one~\cite{LudwigD2011,Bertaina13}. 

The biggest challenge in the experiment was to achieve an almost mass-balanced mixture  $n_{\mathrm{B}}\sim n_{\mathrm{F}}$  to study the regime where the mixture is highly correlated. The difficulty arose when both components were loaded in the MOT and subsequently cooled down by using radiofrequency evaporation, yet the total density of bosons exceeds the fermionic density below the critical temperature; thus, the system is prone to undergo losses as the excess unbound bosons will collide with the Feshbach molecules while they form. However, this problem was circumvented by using first a species selective dipole potential that moves the mixture out of the MOT, and when the trap was ramped down, a combination of two laser beams was ramped on, creating a local trap for the mixture. The wavelength of the trap was chosen to coincide with the D lines of the $\mathrm{K}$ atoms, making the polarizabilities of the Feshbach and ground-state molecules similar. Once the optical trap was ramped down, a density-matched mixture was obtained where the densities of fermions and bosons were roughly equal. When the boson-fermion coupling strength was tuned to resonance, the formation of heteronuclear $\mathrm{NaK}$ molecules was stable, and it was accompanied by a BEC depletion. The phase-separated transition occurred when the number of depleted bosons  (or order parameter) vanishes. Finally, a series of ramps in the magnetic field permitted to tuning the Bose-Fermi interaction, maintaining molecular losses at the minimum. The crossover from the polaronic phase to the molecular phase was fully characterized by measuring the in-situ density profiles and the number of bosons depleted, which was linked to the order parameters of the transition. In addition, absorption and dissociation optical density in time of flight measurements exemplify the formation and dissociation of Feshbach molecules, respectively. The latter, in particular, is associated with heating effects, yet the partial coherence of the condensate was retrieved, where the condensed and thermal fraction of bosons were tracked.

\section{From weak-coupling Fr\"ohlich theory to strong-coupling Bose polarons}
\label{secEffHamiltonians}
Polarons were first introduced as quasiparticles forming when electrons (treated as mobile impurities) in a solid get dressed by deformations of the underlying ionic lattice \cite{Landau48}. This situation can be theoretically described by the electron-phonon interaction, as first derived by Fr\"ohlich \cite{Frohlich1954}, which takes the form
\begin{equation}
    \H_{\rm F} = \frac{\hat{\vec{p}}^2}{2m_{\rm I}} + \int d^d\vec{k} \left[ \omega_{\vec{k}} \ad_{\vec{k}} \a_{\vec{k}} +  V_{\vec{k}} \l \a_{\vec{k}} + \ad_{- \vec{k}}  \r e^{i \vec{k} \cdot \hat{\vec{r}}} \right].
    \label{eqHFroh}
\end{equation}
Here $\hat{\vec{p}}$ and $\hat{\vec{r}}$ denote first-quantized electron coordinates and $\a_{\vec{k}}^{(\dagger)}$ annihilates (creates) a phonon with momentum $\vec{k}$. The phonon dispersion relation is given by $\omega_{\vec{k}}$, which can be an optical $\omega_{\vec{k}} \sim \omega_0$ or acoustical $\omega_{\vec{k}} \sim c |\vec{k}|$ phonon. The electron-phonon scattering is described by $V_{\vec{k}} \propto \sqrt{\alpha}$, where $\alpha$ is a dimensionless coupling strength.  

The Hamiltonian in Eq.~\eqref{eqHFroh} constitutes the celebrated \emph{Fr\"ohlich Hamiltonian}, which describes a remarkable range of physical phenomena. While some features, like the existence of a coherent quasiparticle ground state (i.e. a stable polaron), are universal, others depend on the details of the phonon dispersion $\omega_{\vec{k}}$ and the scattering amplitude $V_{\vec{k}}$ in the Hamiltonian. Especially in the strong coupling limit $\alpha \gg 1$, where perturbation theory in the coupling strength $\alpha$ is no longer justified, describing the polaron ground state becomes challenging and different theoretical approaches often predict vastly different results. For example, some theories predict that a phase transition separates the localized from the self-trapped regime where the effective polaron mass becomes excessively large \cite{Peeters1985,Tempere2009}, while others predict a smooth cross-over \cite{Grusdt2015RG}. The relative simplicity of the Fr\"ohlich polaron model has made it popular in the mathematical physics community too, and indeed rigorous results on the non-existence of phase transitions exist for certain classes of models \cite{Gerlach1991}.

Before discussing extensions of the Fr\"ohlich model to ultracold atoms, we mention a powerful technique allowing to simplify general, translationally invariant Hamiltonians involving a single impurity, such as Eq.~\eqref{eqHFroh}. By applying a unitary so-called Lee-Low-Pines~\cite{Lee1953} transformation $\hat{U}_{\rm LLP}$ that we describe in more detail in Sec.~\ref{subsubsecLLPmeanField}, a reference frame co-moving with the impurity can be chosen. In the case of the Fr\"ohlich Hamiltonian~\eqref{eqHFroh} one obtains
\begin{multline}
    \hat{U}_{\rm LLP}^\dagger \H_{\rm F} \hat{U}_{\rm LLP} = \frac{1}{2 m_{\rm I}} \left( \vec{K}_{\rm tot} - \hat{\vec{P}} \right)^2 +\\
    +\int d^d\vec{k} \left[ \omega_{\vec{k}} \ad_{\vec{k}} \a_{\vec{k}} +  V_{\vec{k}} \l \a_{\vec{k}} + \ad_{- \vec{k}}  \r  \right],
\end{multline}
where $\vec{K}_{\rm tot}$ is the total conserved system momentum and $\hat{\vec{P}} = \int d^d\vec{k}~\vec{k} \ad_{\vec{k}} \a_{\vec{k}}$ is the total phonon momentum operator. Notably, the impurity degrees of freedom have been completely eliminated from the problem. This comes at a price, however: The Hamiltonian is no longer quadratic in phonon operators; instead, phonons are now experiencing an impurity-mediated interaction $\sim 1/m_{\rm I}$. This is a conceptually useful trick that will be utilized in various incarnations by the theoretical methods reviewed below.

The Fr\"ohlich Hamiltonian, Eq.~\eqref{eqHFroh}, can also be applied to describe the interaction of a mobile impurity atom immersed in an ultracold Bose gas, colloquially referred to as the Bose polaron problem in the cold atom community. While Bose polarons were first predicted in cold one-dimensional quantum gases \cite{Mathey2004}, the connection to the Fr\"ohlich Hamiltonian was pointed out soon after \cite{Cucchietti2006,sacha2006self}, kicking off a flurry of theoretical activity on the problem. An early motivation of the field was the study of strong coupling Fr\"ohlich polarons - i.e. Eq.~\eqref{eqHFroh} with large $\alpha \gg 1$ that was hoped could be realized by tuning an atomic Feshbach resonance \cite{Tempere2009}. However, subsequently it became clear that the Fr\"ohlich model can only quantitatively capture the regime of weak impurity-boson interactions in a quantum gas since additional two-phonon terms \cite{Rath2013} and phonon non-linearities need to be added to capture the physics in the vicinity of an impurity-boson Feshbach resonance. In the context of Bose polarons in a BEC, this region is nowadays referred to as the strong coupling regime. 

In the following, we will review the Fr\"ohlich model for an impurity in a weakly interacting BEC, valid at weak impurity-boson couplings. Then we will describe the various additional terms that need to be included to capture strong-coupling Bose polarons and the physics around the impurity-boson Feshbach resonance.

\subsection{The Bogoliubov -- Fr\"ohlich Hamiltonian}
\label{subsecBogoFroh}
The collective excitations of a weakly interacting BEC can be described within standard Bogoliubov theory \cite{Pitaevskii2003}. This allows to replace the Hamiltonian of the Bose gas, Eq.~\eqref{eqHB}, by a linearized phonon Hamiltonian,
\begin{equation}
    \H_{\rm B} \rightarrow \int d^d\vec{k}~ \omega_{\vec{k}} \ad_{\vec{k}} \a_{\vec{k}}.
    \label{eqHBogo0}
\end{equation}
Here $\omega_{\vec{k}} = c k \sqrt{1 + \xi^2 k^2 / 2}$ is the dispersion of the Bogoliubov phonons $\a_{\vec{k}}^{(\dagger)}$, with the speed of sound $c=\sqrt{g_{\rm BB} n_0 / m_{\rm B}}$ and the healing length of the BEC $\xi = 1/\sqrt{2 m_{\rm B} g_{\rm BB} n_0}$.

The impurity-boson interactions can be simplified by replacing $\hat{\Psi}(\vec{k}) \to \sqrt{n_0} \delta(\vec{k}) + u_k \a_{\vec{k}} - v_k \ad_{-\vec{k}}$, taking into account the macroscopic occupation of the $\vec{k}=0$ mode in the BEC. Here $u_k$ and $v_k$ denote the mixing angles of the Bogoliubov transformation that diagonalizes the free phonon Hamiltonian in Eq.~\eqref{eqHBogo0}. Keeping only terms linear in $\a_{\vec{k}}^{(\dagger)}$ yields the Fr\"ohlich interaction in second quantization
\begin{equation}
    \H_{\rm IB} \to  \int d^d\vec{k} \int d^d\vec{r}~ V_{k} \l \a_{\vec{k}} + \ad_{- \vec{k}}  \r e^{i \vec{k} \cdot \vec{r}} \hat{\phi}^\dagger(\vec{r}) \hat{\phi}(\vec{r}).
    \label{eqHIBFroh}
\end{equation}
Some cumbersome but straightforward algebra finally yields the scattering amplitude / interaction strength
\begin{equation}
    V_{k} = \sqrt{n_0} (2 \pi)^{-d/2} g_{\rm IB} \l \frac{(k \xi)^2}{2 + (k \xi)^2} \r^{1/4}.
    \label{eqVkDef}
\end{equation}
Note that we assumed canonical commutation relations with a Dirac-delta function, $[\a_{\vec{k}},\ad_{\vec{k}'}] = \delta^{(d)}(\vec{k} - \vec{k}')$. If a discrete Kronecker-delta function $\delta_{\vec{k},\vec{k}'}$ is used instead, normalization in Eq.~\eqref{eqVkDef} changes and $V_k$ acquires a factor one over volume \cite{Grusdt2015Varenna}.

Combining the simplified results in Eqs.~\eqref{eqHBogo0} and \eqref{eqVkDef} with the impurity kinetic term from Eq.~\eqref{eqHI} and switching back to first quantized notation, a Fr\"ohlich Hamiltonian as in Eq.~\eqref{eqHFroh} is obtained. Because we had to linearize in the phonon operators, this so-called \emph{Bogoliubov-Fr\"ohlich} model is only valid for weak boson-boson and impurity-boson interactions and represents a weak-coupling theory of the Bose polaron problem. Below we will discuss extensions of the Fr\"ohlich Hamiltonian, valid beyond the weak coupling regime.

\subsubsection{Mean-field saddle-point solution.}
\label{subsubLLPmf}
To obtain some intuition for the polaron ground state, it is intuitive to study the case of an infinitely heavy impurity $m_{\rm I} \to \infty$, which is thus localized at, say, $\vec{r}=0$. The resulting Hamiltonian consists of displaced harmonic oscillators -- one for each $\vec{k}$-mode $\a_{\vec{k}}$. Its ground state corresponds to coherent states
\begin{equation}
    \ket{\Psi; m_{\rm I}=\infty} = \prod_{\vec{k}} e^{\alpha_{\vec{k}} \ad_{\vec{k}} - \hc} \ket{0},
    \label{eqCohStateAnsatz}
\end{equation}
with amplitudes $\alpha_{\vec{k}} = - V_k / \omega_k$, see e.g.~\cite{Devreese2013}. A similar mean-field ansatz of coherent phonons can be made for a mobile impurity, if combined with a Lee-Low-Pines (LLP) transformation \cite{Lee1953}: we introduce this method in Sec.~\ref{subsubsecLLPmeanField} below. Readers not already familiar with the LLP approach can simply focus on the infinite-mass case, $m_{\rm I} \to \infty$, in the following discussion. 

From Eq.~\eqref{eqCohStateAnsatz} one obtains a non-linear variational energy landscape $\mathscr{H}[\alpha_{\vec{k}}]$, which can be minimized with respect to the coherent amplitudes $\alpha_{\vec{k}}$, see e.g.~\cite{Devreese2013,Grusdt2015Varenna} for pedagogical discussions. It is also useful to note that the coherent states can be used to define a coherent-state path integral description, which highlights why analyzing the variational energy landscape $\mathscr{H}[\alpha_{\vec{k}}]$ - dictating the resulting effective action - is useful for developing an intuitive understanding of the problem. Indeed, the resulting mean-field ground state corresponds to the solution of a set of saddle point equations:
\begin{equation}
    \frac{\delta}{\delta \alpha_{\vec{k}}} \mathscr{H}[\alpha_{\vec{k}}] = 0, \qquad \forall \vec{k}.
\end{equation}

Within the Fr\"ohlich theory, the saddle point is always \emph{stable}, corresponding to a global minimum of the variational energy $\mathscr{H}[\alpha_{\vec{k}}]$ \cite{Shashi2014RF}. The solutions at $\pm g_{\rm IB}$ correspond to the attractive ($g_{\rm IB} < 0$) and repulsive $g_{\rm IB} > 0$ polaron, and -- within Fr\"ohlich theory -- are related by a gauge transformation $\alpha_{\vec{k}} \to - \alpha_{\vec{k}}$. These solutions describe the wings $1 / | k_n a_{\rm IB}| \gg 1$ in Fig.~\ref{figRFspec}.

\subsubsection{Ultra-violet divergences and Lippmann-Schwinger equation.}
\label{subsecUVdiv}
In order to work with the above results, the impurity-boson interaction strength $g_{\rm IB}$ of the contact pseudopotential needs to be expressed in terms of the impurity-boson scattering length $a_{\rm IB}$ that can be experimentally measured in low-energy collisions. Their relation depends on dimensionality $d$ and can be obtained by solving the Lippmann-Schwinger equation to all orders, see e.g. \cite{Shchadilova2016}. The latter describes exactly the two-body scattering problem, involving one boson and one impurity. 

For $d > 2$ ($d=2$) this yields power-law (logarithmically) divergent terms with the ultra-violet (UV) cut-off $\Lambda_{\rm UV}$ in the relation between $a_{\rm IB}$ and $g_{\rm IB}$. For example, in $d=3$ the exact solution of the Lippmann-Schwinger equation yields
\begin{equation}
    \frac{1}{g_{\rm IB}} = \frac{m_{\rm red}}{2 \pi a_{\rm IB}} - \int^{\Lambda_{\rm UV}} \frac{d^3\vec{k}}{(2 \pi)^3} \frac{2 m_{\rm red}}{k^2},
    \label{eqLSEfull}
\end{equation}
where $m_{\rm red} = [m_{\rm I}^{-1} + m_{\rm B}^{-1}]^{-1}$ is the reduced mass. To leading order in the interaction strength this gives
\begin{equation}
    a_{\rm IB} = \frac{g_{\rm IB}}{2 \pi} m_{\rm red} - \frac{4 g_{\rm IB}^2}{(2 \pi)^3} m_{\rm red}^2 \Lambda_{\rm UV} + \mathcal{O}(g_{\rm IB}^3),
    \label{eqLSEresult}
\end{equation}
including UV-divergent terms. However, these divergences with $\Lambda_{\rm UV}$ disappear in any consistent treatment of the Bose-polaron problem: the solution of the polaron problem itself parallels the resummation in the Lippmann-Schwinger equation, although in a many-body context, and leads to the cancelation of the leading UV divergent terms. As a result, the Bose polaron energy can be meaningfully calculated from the measured value of $a_{\rm IB}$.

The resummation of interaction processes in a many-body environment can, however, lead to remaining sub-leading UV divergences, even on the level of the weak-coupling Fr\"ohlich model. For example, in $d=3$ the power-law divergence with $\Lambda_{\rm UV}$ in Eq.~\eqref{eqLSEresult} disappears from the energy, but a sub-leading logarithmic UV divergence remains. This was first revealed in a renormalization group analysis of the problem \cite{Grusdt2015RG} and at the same time observed in numerically exact diagrammatic Monte Carlo simulations (DiagMC)~\cite{Vlietinck2015}. Subsequently, rigorous mathematical proofs were provided by perturbative analysis \cite{Christensen2015} and later even non-perturbatively \cite{Lampart2020}. Similar logarithmic corrections -- so-called Lee-Huang-Yang terms \cite{Lee1957,Lee1957a} -- are well known to occur in interacting Bose gases. In order to obtain physically meaningful results, the effective range $r_{\rm eff}$ of the underlying interactions must be used for regularization; this is the case even when the more accurate microscopic Hamiltonians are used \cite{Grusdt2017} which we review next.

\subsection{Beyond Fr\"ohlich I: Two-phonon terms}
\label{secExtdBogoFroh}
As a first step beyond the weak-coupling Fr\"ohlich Hamiltonian, the Bogoliubov approximation, Eq.~\eqref{eqHB}, is kept but the impurity-boson interactions are treated to all orders in the phonon operators $\a_{\vec{k}}$. In addition to the Fr\"ohlich terms, Eq.~\eqref{eqHIBFroh}, this introduces two-phonon terms, as first considered by Rath and Schmidt in \cite{Rath2013}. Using the same notation as above, the additional terms read
\begin{multline}
    \H_{\rm 2ph} = \int d^d\vec{k}d^d\vec{k}' \int d^d\vec{r} ~ e^{i (\vec{k} - \vec{k}' ) \cdot \vec{r}} \hat{\phi}^\dagger(\vec{r}) \hat{\phi}(\vec{r}) \\
    \times \frac{g_{\rm IB}}{(2\pi)^d} \l u_k \ad_{\vec{k}} - v_k \a_{- \vec{k}} \r \l u_{k'} \a_{\vec{k}'} - v_{k'} \ad_{-\vec{k}'} \r.
    \label{eqH2ph}
\end{multline}
To relate $g_{\rm IB}$ to the underlying scattering length $a_{\rm IB}$, the full Lippmann-Schwinger equation must be used, see Eq.~\eqref{eqLSEfull}. The resulting model Hamiltonian $\H_{\rm F} + \H_{\rm 2ph}$ is sometimes referred to as the \emph{extended Bogoliubov-Fr\"ohlich} model.

Unlike the Bogoliubov-Fr\"ohlich theory, including two-phonon terms allows to capture strong impurity-boson interactions. In particular, the two-particle case with one impurity scattering on just one boson is correctly captured. This situation corresponds to a vanishing BEC density $n_0=0$, where the Fr\"ohlich interaction Eq.~\eqref{eqHIBFroh} vanishes identically. As a consequence, the physics of the Feshbach resonance, $a_{\rm IB} \to \infty$, can be captured, including for $n_0 \neq 0$. This was first achieved by Rath and Schmidt in \cite{Rath2013} within the self-consistent T-matrix approximation. 

\subsubsection{Mean-field saddle-point structure.}
\begin{figure}
	\centering
	\includegraphics[scale=0.3]{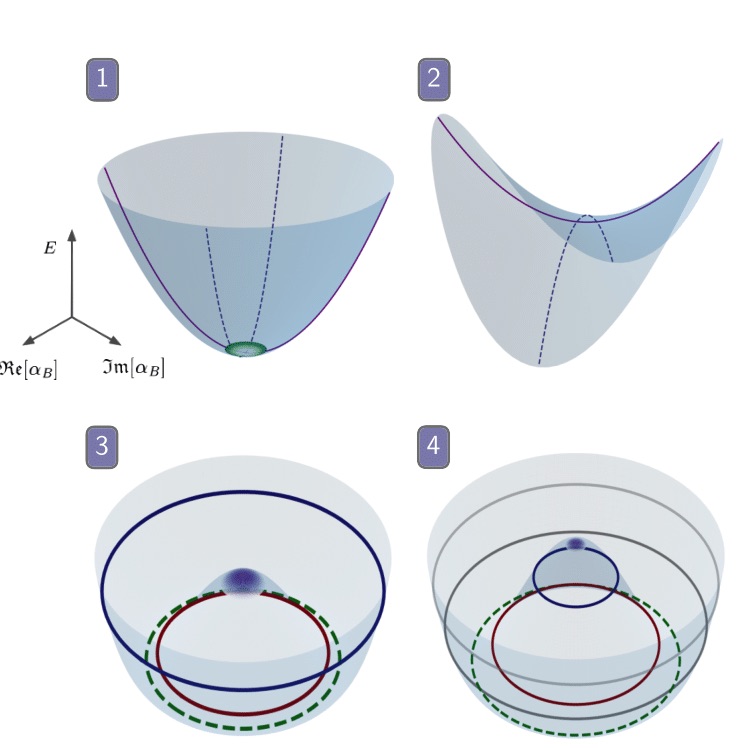}
	\caption{\textbf{Saddle-point structure of Bose polarons.} Many characteristics of strong coupling Bose polarons can be understood from the change of their saddle-point structure. To this end, coherent mean-field states of phonons are assumed and the resulting variational energy landscape is explored. This point of view can be formalized when making a coherent state path-integral analysis of the polaron Hamiltonian. The variational energy landscape contains the following features: (1) The attractive polaron corresponds to a stable saddle-point. (2) Around the Feshbach resonance, a dynamical instability can occur where the variational energy decreases along one direction in phase space. (3) The repulsive polaron corresponds to an unstable saddle-point around which the energy decreases along two dimensions $({\rm Re}[\alpha_{\rm B}],{\rm Im}[\alpha_{\rm B}])$ in phase space; including phonon-phonon interactions eventually leads to a rise of the variational energy for larger $|\alpha_{\rm B}|^2$, which can realize one (3) or multiple (4) meta-stable molecular bound states below the unstable repulsive polaron saddle-point. Figure adapted from Ref.~\cite{Mostaan2023}.}
	\label{FIGSaddlePntStrct}
\end{figure}

After adding the two-phonon term in Eq.~\eqref{eqH2ph} to the Hamiltonian, the mean-field theory (see \ref{subsubLLPmf}) can still be applied, as first done by Shchadilova et al. in \cite{Shchadilova2016}. The structure of the saddle-point solutions was worked out shortly afterwards in \cite{Grusdt2017} for the case of $d=3$ dimensions: for most values of the scattering length $a_{\rm IB}$ one still obtains a saddle-point solution, which reduces to the Fr\"ohlich solution when $|a_{\rm IB}| \to 0$ is in the weak-coupling regime. However, unlike in the Fr\"ohlich case, only the attractive polaron ($a_{\rm IB} < 0$) corresponds to a stable solution, i.e. a global minimum of the variational energy. The repulsive polaron ($a_{\rm IB}>0$), in contrast, corresponds to an \emph{unstable} saddle-point: Here the variational energy is unbounded from below. This, however, is an artifact of the Bogoliubov approximation, which neglects phonon-phonon non-linearities (discussed next) that ultimately bound the variational energy from below. See Fig.~\ref{FIGSaddlePntStrct} for an illustration.

Around the Feshbach resonance, at $a_{\rm IB} \to \infty$, the unstable saddle-point solution (repulsive polaron) disappears, and the stable saddle-point solution (attractive polaron) re-appears on the other side. Unless the BEC is non-interacting this does not happen directly, however: instead, an intermediate dynamically unstable regime was predicted \cite{Grusdt2017} where the variational energy remains unbounded from below, see Fig.~\ref{FIGSaddlePntStrct}. Furthermore, many-body effects lead to a renormalization of the two-body Feshbach resonance \cite{Grusdt2017}, which is shifted from $a_{\rm IB} \to \infty$ to a different value $a_{{\rm IB},+} > 0$ which depends on the BEC density $n_0$.

\subsubsection{Molecular bound states.}
The description of the repulsive polaron after adding two-phonon terms \eqref{eqH2ph}, as an \emph{unstable} saddle-point, can be associated with the simultaneous emergence of molecular bound states at lower energies, see Fig.~\ref{FIGSaddlePntStrct}. The first among such bound states corresponds to the universal two-body impurity-boson dimer, which disappears at the Feshbach resonance. In the two-body limit, i.e. without renormalization effects from the medium, the dimer appears at $1/a_{\rm IB} > 0$ and has a binding energy $E_{\rm bdg} = - 1 / (m_{\rm red} a_{\rm IB}^2)$ around $1/a_{\rm IB} \approx 0$, as indicated by the dashed line in Fig.~\ref{figRFspec}.

In the presence of the surrounding Bose gas, $n_0 \neq 0$, the two-body dimer itself can be viewed as an impurity which can then form a different kind of Bose polaron. The underlying interactions with the BEC shift the energy of the dressed two-body dimer, changing at which two-body scattering length $a_{\rm IB}$ this bound state emerges from the scattering continuum. This leads to the above-mentioned renormalization of the in-medium Feshbach resonance. Direct numerical evidence for the dressed two-body dimer was first obtained by Shchadilova et al. \cite{Shchadilova2016}, who predicted a corresponding sharp peak in the impurity-spectral function below the repulsive polaron. 

In addition to the two-body dimer, Shchadilova et al. \cite{Shchadilova2016} predicted multi-body bound states at even lower energies on the repulsive side. These states correspond to bound states of the impurity to $N_{\rm B}$ bosons, all of which occupy the same universal two-body dimer orbital. Since the Bogoliubov theory neglects interactions between the phonons, i.e. it treats the Bose-Bose interactions on a mean-field level, the two-body bound state described by the two-phonon term Eq.~\eqref{eqH2ph} can be occupied by any number of bosons $N_{\rm B}$, with energies $E_{\rm bdg}(N_{\rm B}) \simeq - N_{\rm B} / (m_{\rm red} a_{\rm IB}^2)$ unbounded from below. Indeed, Shchadilova et al. \cite{Shchadilova2016} numerically found equally spaced spectral lines corresponding to $N_{\rm B}+1$-body bound states. While the study in \cite{Shchadilova2016} focused on $d=3$ dimensions, similar multi-body bound states are expected to exist in any $d$. To study how Bose-Bose interactions affect these highly regular lines, phonon non-linearities must be included, as discussed next in \ref{subsecFrohPhonPhon}.

Furthermore, the introduction of two-phonon terms in Eq.~\eqref{eqH2ph} leads to the emergence of multi-body \emph{Efimov-states}, an effect of intrinsic three-body correlations which is limited to $d=3$ dimensional systems. Such Efimov states were analyzed systematically in the presence of a BEC within the extended Bogoliubov-Fr\"ohlich model Eqs.~\eqref{eqHIBFroh}, \eqref{eqH2ph} in Refs.~\cite{ChristianenPRL_2022,ChristianenPRA_2022} and will be discussed further below in Sec.~\ref{secFewBodyEfimov}.

\subsection{Beyond Fr\"ohlich II: Phonon non-linearities}
\label{subsecFrohPhonPhon}
So far we used the Bogoliubov rotation, assuming a macroscopically occupied BEC and replacing the underlying bosonic field operators by phonon operators $\hat{\Psi}(\vec{k}) \to \sqrt{n_0} \delta(\vec{k}) + u_k \a_{\vec{k}} - v_k \ad_{-\vec{k}}$, and combined this with the assumption that terms involving more than three phonons can be ignored. Notably, the first step is mathematically rigorous and can be viewed as a unitary basis transformation. Hence by performing the described Bogoliubov rotation while including terms with arbitrary numbers of phonon operators $\a_{\vec{k}}^{(\dagger)}$ yields an \emph{exact} representation of the Bose-polaron Hamiltonian \eqref{eqHmic}:
\begin{equation}
    \H = \H_{\rm F} + \H_{\rm 2ph} + \H_{\rm ph-ph}.
    \label{eqHBogoFull}
\end{equation}

Here $\H_{\rm F}$ is the Bogoliubov-Fr\"ohlich Hamiltonian described in Sec.~\ref{subsecBogoFroh}; $\H_{\rm 2ph}$ describes two-phonon scattering terms on the impurity, see Eq.~\eqref{eqH2ph}; and $\H_{\rm ph-ph}$ includes all higher-order phonon-phonon terms. The latter do not involve any impurity operators, and due to the underlying two-body nature of Bose-Bose interactions $\propto g_{\rm BB}$ in \eqref{eqHB} they consist of exactly three or four phonon operators $\a_{\vec{k}}^{(\dagger)}$.

Expressing the full Hamiltonian \eqref{eqHmic} as in Eq.~\eqref{eqHBogoFull} in terms of Bogoliubov phonons has the advantage that connections to the canonical polaron theories reviewed above can be easily made. Numerically, this form can be advantageous or disadvantageous depending on the method used -- ultimately at strong-coupling Eq.~\eqref{eqHBogoFull} describes a strongly interacting system which is hard to solve even with advanced numerical tools, and the more compact form Eq.~\eqref{eqHmic} often leads to more tractable analytical expressions.

Next we discuss which new physical aspects the phonon-phonon interactions add, and when they are important. Early analysis of this question was performed in one dimension in Ref.~\cite{Grusdt2017RG1D}, but similar considerations apply for any $d$. While it is tempting to assume that weak Bose-Bose interactions $a_{\rm BB} \to 0$ justify neglecting the phonon-phonon terms, this is not a valid approach sufficiently close to the impurity-boson Feshbach resonance, $a_{\rm IB} \to \infty$: Namely, on the attractive side, bosons tend to accumulate around the impurity. While their number $N_{\rm B}$ can become large, their mutual interaction energy scaling as $a_{\rm BB} N_{\rm B}^2$ will eventually overcome any impurity-boson attraction scaling as $a_{\rm IB} N_{\rm B}$. This simple argument implies that even for infinitesimal but non-zero $a_{\rm BB} \neq 0$, boson-boson interactions play a role in the strong impurity-boson coupling regime. It moreover highlights how the case of strictly vanishing $a_{\rm BB} = 0$ constitutes an exceptional point, where for a localized impurity with $a_{\rm IB}>0$ \emph{all} bosons condense in the universal impurity-boson dimer orbital, with a total energy $- \infty$ and a vanishing quasiparticle residue (as shown in \cite{Guenther2021}, this orthogonality catastrophe remains present even when $a_{\rm IB}<0$ if $a_{\rm BB} = 0$). Non-vanishing boson-boson repulsion will avoid such collapse.  

\subsubsection{Mean-field saddle-point structure and bound states.}
\label{subsubsecMFsaddlePointAndBoundStates}
Including the non-linearities $\H_{\rm ph-ph}$ in the mean-field analysis regularizes the variational energy landscape $\mathscr{H}[\alpha_{\vec{k}}]$ \cite{Mostaan2023}. On the attractive side where the energy was previously bounded from below, only quantitative changes take place. On the repulsive side, however, the previously unbounded energy becomes bounded from below. As a result, a stable saddle-point solution emerges energetically below the unstable repulsive polaron branch, which corresponds to an extension of the attractive polaron to the repulsive side of the Feshbach resonance, see Fig.~\ref{FIGSaddlePntStrct}. Its quasiparticle weight is heavily suppressed, however, because the molecular impurity-boson bound state is occupied by a fluctuating number of bosons \cite{Mostaan2023} -- making it invisible in the spectral function.

Likewise, the energies of the few-body impurity-boson bound states first predicted by Shchadilova et al. \cite{Shchadilova2016} are strongly affected by the phonon non-linearities. They are no longer equally spaced, and as a result of the competition between impurity-boson attraction and boson-boson repulsion, the lowest-energy state has a finite number of bosons $N_{\rm B}$ bound to the impurity. The latter depends on the values of $a_{\rm IB}$ and $a_{\rm BB}$. Remarkably, the lowest-energy multi-body bound state resembles the extended attractive polaron revealed on the repulsive side $a_{\rm IB}>0$ of the Feshbach resonance -- establishing an interesting direct connection between the repulsive and attractive Bose polarons \cite{Mostaan2023}.  

The fate of universal Efimov states after including phonon-phonon non-linearities remains unsettled. Encouraging theoretical work exists \cite{Levinsen2015,Yoshida2018PRX}, based on even more detailed two-channel models of the impurity-boson Feshbach resonance going beyond Eq.~\eqref{eqHBogoFull}, which suggests that Efimov states should still be stable in the Bose-polaron settings. See Sec.~\ref{secFewBodyEfimov} for more details.

\subsection{Theoretical methods}
Many powerful theoretical methods have been developed for solving polaron problems, starting in the context of electrons interacting with lattice vibrations in a solid. Because the polaron problem is sufficiently tractable -- constituted by just one impurity particle -- while being a genuine many-body problem involving arbitrary numbers of bosons in an extensive number of spatial modes, it has played a significant role in developing the modern formalism that essentially defines the field of quantum many-body theory. For example, the concept of a quasiparticle has been introduced by Landau and Pekar \cite{Landau48} in their path-breaking studies of polarons. 

Since then, a variety of theoretical approaches have been developed, ranging from simplified but fully analytical tools, semi-analytical variational methods, perturbative (diagrammatic) theories, renormalization group methods, fully numerical variational approaches, to different variations of all-out Monte-Carlo procedures. Being challenging, yet simple enough to have a variety of methods at hand, the Bose polaron remains a key battle ground for theorists to develop, improve and benchmark theoretical approaches of various kinds.

While some methods are rather general and can be applied for a large range of polaron problems, others are specifically designed, e.g. through the simplifying assumptions they make, for the Bose polaron problem at hand. In the following we will provide an overview of the different methods currently in use, without explaining any of them in detail -- we will refer to original work or previous reviews instead. We focus exclusively on the Bose polaron problem constituted by a free-space impurity atom in a weakly interacting free-space BEC, without discussing in detail how a given method may or may not be generalized to other classes of problems. 

The approaches summarized below are sorted roughly according to their level of complexity, which correlates strongly with their predictive power while also leading to increasing computational costs down the list. More in-depth discussions of a selection of the following approaches can also be found in the reviews \cite{Devreese2013,Grusdt2015Varenna}.

\subsubsection{Perturbative analysis, self-consistent Born and T-matrix approach.}
The conceptually simplest method to solve the polaron problem at weak couplings is to apply Rayleigh-Schr\"odinger perturbation theory, expanding in powers of the impurity-boson interaction strength $g_{\rm IB}$. To leading order this yields compact analytical expressions for the polaron energy, its effective mass, the quasiparticle residue, etc. As a starting point it requires an analytically tractable non-interacting system for $g_{\rm IB}=0$, making the method applicable to the Bogoliubov-Fr\"ohlich and the extended Bogoliubov-Fr\"ohlich Hamiltonians discussed in Secs.~\ref{subsecBogoFroh}, \ref{secExtdBogoFroh} based on non-interacting phonons. Pedagogical derivations for the Fr\"ohlich model can be found in \cite{Devreese2013,Grusdt2015Varenna}.

The formally most compact way to formulate perturbation theory, is to use Feynman diagrams in order to expand the self-energy. This procedure has been thoroughly worked out up to third order for the extended Bogoliubov-Fr\"ohlich Hamiltonian from Sec.~\eqref{secExtdBogoFroh} in $d=3$ by Christensen et al.~\cite{Christensen2015}, confirming the logarithmic UV divergence of the energy with the UV cut-off and providing valuable analytical expressions for the self-energy. 

Extensions of the perturbative summation include self-consistent treatments of the self-energy, by solving the corresponding Dyson equation. This is a standard improvement of the lowest-order Born approximation, leading to the self-consistent Born result - see e.g.~\cite{Grusdt2015Varenna} for a derivation. While an infinite number of diagrams is included in the re-summation, the error of the result is not controlled anymore. This method has been applied in the context of Bose polarons, e.g. to describe Fr\"ohlich polarons in a dipolar condensate \cite{Kain2014}. 

\begin{figure}
	\centering
	\includegraphics[scale=0.3]{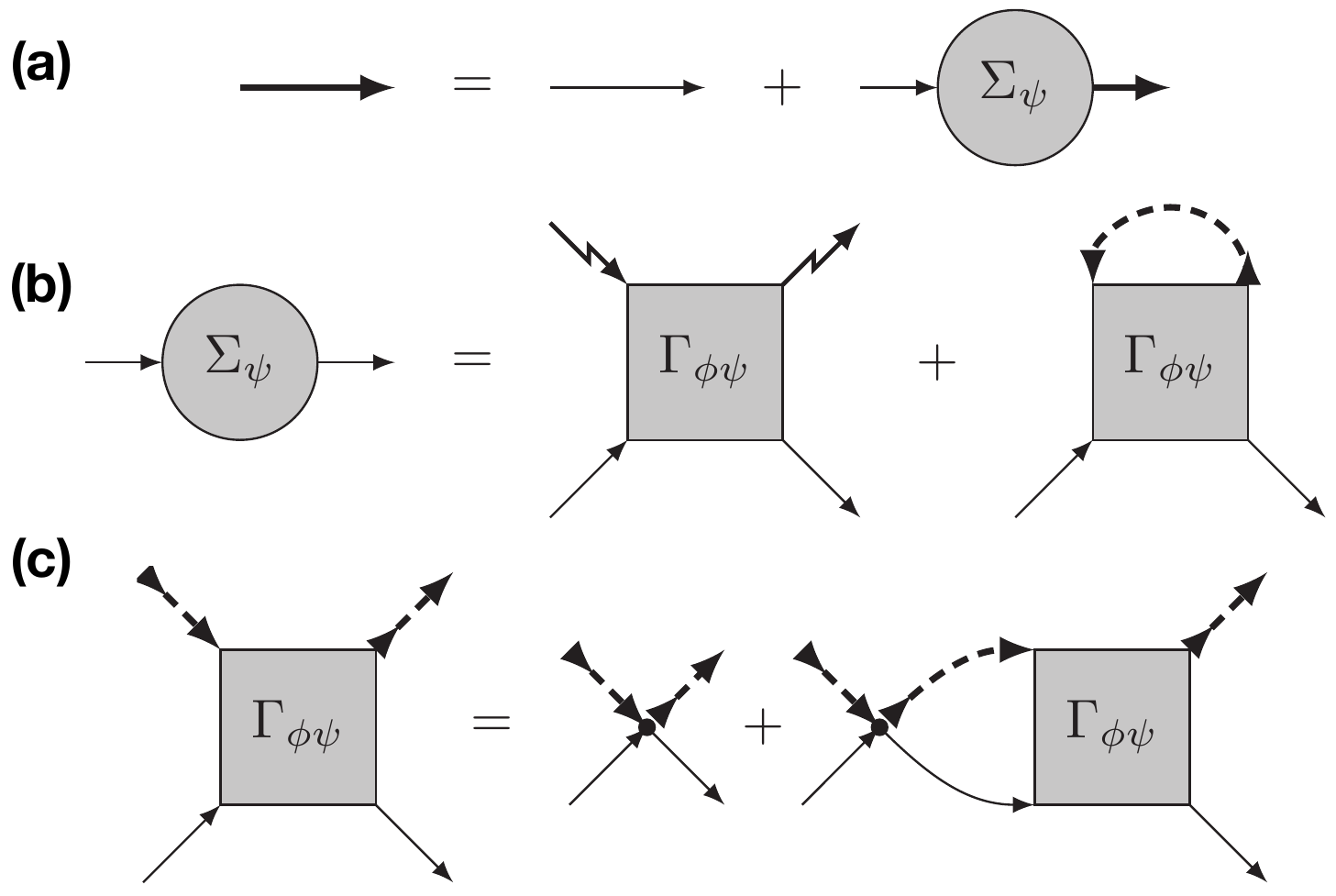}
	\caption{\textbf{The $\mathrm{T}$-matrix approach} to the Bose polaron problem is based on the following diagrams: (a) Dyson equation for the impurity Green's function (solid lines); (b) Impurity self-energy $\Sigma$, involving the $\mathrm{T}$-matrix $\Gamma$, condensate particles (zig-zag lines) and non-condensed bosons (dashed lines); (c) In-medium $\mathrm{T}$-matrix equation for the non-self consistent case. In the self-consistent $\mathrm{T}$-matrix approach the impurity propagator in the particle-particle loop in (c) is replaced by the dressed propagator from (a). Figure adapted from Ref.~\cite{Rath2013}.}
	\label{FIGTmatrixDiagrams}
\end{figure}

Further extensions include perturbative modifications of the scattering $\mathrm{T}$-matrix of the impurity on the BEC, corresponding to vertex corrections in the diagrams, see Fig.~\ref{FIGTmatrixDiagrams}. This approach was pioneered for Bose-polarons in a BEC by Rath and Schmidt \cite{Rath2013}. In the non-self consistent $\mathrm{T}$-matrix approach, one works with the bare lowest-order self-energy; In the more accurate self-consistent $T$-matrix approach, the self-energy used for the perturbative calculation of the $\mathrm{T}$-matrix is replaced by the self-consistent solution of the Dyson equation. 

Methods based on the self-energy have the added advantage to directly yield the impurity spectral function. They also have known disadvantages, however; For example, both the self-consistent and non-self-consistent $\mathrm{T}$-matrix approaches predict an unphysical gap separating the polaron ground state from the phonon continuum \cite{Rath2013} and cannot be used to describe the quasiparticle breakdown when impurities move through the BEC at super-sonic speeds \cite{Shashi2014RF}.
 
\subsubsection{Chevy-type variational analysis: Truncated numbers of excitations.}
A popular variational approach, which is formally closely related to the non-self-consistent $\mathrm{T}$-matrix approximation, is constituted by the so-called \emph{Chevy wavefunction} \cite{Chevy2006}. Inspired by first-order perturbation theory, an ansatz is made consisting of a free impurity in a superposition with an impurity dressed by one excitation,
\begin{equation}
    \ket{\psi_{\rm Chevy}(\vec{k})} = \sqrt{Z_{\vec{k}}} \hat{\psi}^\dagger_{\vec{k}} \ket{0} + \int d^d\vec{q}~ \phi_{\vec{q}} ~ \hat{\psi}^\dagger_{\vec{k}-\vec{q}} \ad_{\vec{q}} \ket{0}.
    \label{eqPsiChvy}
\end{equation}
Here $\vec{k}$ denotes the total momentum of the state, and the quasiparticle weight $Z_{\vec{k}}$ along with the complex one-phonon wavefunction $\phi_{\vec{q}} \in \mathbb{C}$ serve as variational parameters. States of this type were first studied in the context of Bose polarons in \cite{Li2014,Levinsen2015,Jorgensen2016}.

While the analytic structure of Eq.~\eqref{eqPsiChvy} is that of a first-order perturbative result, making it asymptotically exact at weak interactions, the self-consistent determination of the variational parameters can render the ansatz useful even at strong couplings. Originally this wavefunction was introduced by Chevy \cite{Chevy2006} to describe the Fermi-polaron problem \cite{Chevy2010}, where it enjoys remarkable success. In the case of Bose polarons its accuracy is less clear, and the ansatz is known to fail in capturing some situations, e.g. the orthogonality catastrophe of Bose polarons in $d=1$ dimension \cite{Grusdt2017RG1D} or for a non-interacting BEC in $d=3$ \cite{Guenther2021}. But without doubt it remains a competitive variational state in the regime of strong coupling in $d=3$ dimensions.

Generalizations of the ansatz in Eq.~\eqref{eqPsiChvy} are possible. One option is to truncate the series only after a larger number of excitations is included. This approach has been successfully applied, for example to describe three-body Efimov states involving an impurity in a BEC \cite{Levinsen2015}. Another option includes the use of time-dependent variational parameters, which allows to calculate dynamical properties and the Bose polaron spectrum \cite{Jorgensen2016}.

\subsubsection{Strong-coupling variational approach and Gross-Pitaevskii equation.}
\label{subsecSCthy}
Another popular class of variational approaches is based on a mean-field factorization of impurity and bath coordinates, assuming further that the bath can adiabatically adjust to slow changes in the impurity wavefunction. This is justified in strong coupling (SC) regimes and becomes exact when the impurity mass is large, $m_{\rm I} \to \infty$. Formally one makes an ansatz,
\begin{equation}
    \ket{\psi_{\rm SC}} =  \int d^d\vec{r} ~ \phi_{\rm I}(\vec{r}) \hat{\phi}^\dagger(\vec{r}) \ket{0}_{\rm I} \otimes \ket{\psi_{\rm B}(\vec{r})},
    \label{eqPsiSC}
\end{equation}
where $\phi_{\rm I}(\vec{r})$ denotes the impurity, or polaron, wavefunction and $\ket{\psi_{\rm B}(\vec{r})}$ describes the state of the bath depending on the instantaneous position of the impurity. In this way, strong impurity-boson correlations can be easily incorporated. This is sometimes formalized by a Lang-Firsov transformation, see e.g.~\cite{Bruderer2007}, which realizes a particular case of Eq.~\eqref{eqPsiSC}.

The SC ansatz in Eq.~\eqref{eqPsiSC} was originally introduced by Landau and Pekar \cite{Landau48} who studied self-localization of impurities. They worked with coherent states of phonons $\ket{\psi_{\rm B}(\vec{r})}$ and determined the impurity wavefunction $\phi_{\rm I}(\vec{r})$ variationally by minimizing the energy of the effective impurity Hamiltonian, $\H_{\rm I,eff} = \bra{\psi_{\rm B}(\hat{\vec{r}})} \H \ket{\psi_{\rm B}(\hat{\vec{r}})}$. If this energy is minimized by a localized Gaussian wavepacket, this indicates self-localization of the impurity inside the phonon cloud created in the bath. 

The SC approach was applied to the Bose-polaron problem in $d=3$ dimensions by Cuccietti and Timmermans \cite{Cucchietti2006}, who predicted that self-localization can occur for an impurity immersed in a BEC. They essentially worked on the level of the Bogoliubov-Fr\"ohlich Hamiltonian since they linearized in the bosonic fields. Similar self-localization was subsequently predicted within the same SC formalism but without linearization in $d=3$ \cite{Kalas2006,Bruderer08} and $d=1$ dimensional \cite{sacha2006self} Bose-polarons. Working with the Bogoliubov-Fr\"ohlich Hamiltonian again, Casteels et al.~\cite{Casteels2011} applied the SC theory to Bose polarons in $d=3$ and calculated metastable excited states of the impurity within its self-trapping potential. Blinova et al.~\cite{blinova2013single} extended the analysis to predict bubble-polaron states, in which the impurity becomes impenetrable to the condensate that is completely expelled from the vicinity of the impurity.

More recently, the SC approach has gained renewed attention in $d=1$ \cite{Volosniev2017} and $d=3$ \cite{Drescher2020,Guenther2021,Schmidt2022} because it constitutes one of the most tractable approaches that allows to fully treat the boson-boson interactions, i.e. phonon non-linearities, in the bath (discussed in Sec.~\ref{subsecFrohPhonPhon}). Instead of making a self-localization ansatz, a plane-wave impurity wavefunction $\phi_{\rm I}(\vec{r}) \simeq e^{i \vec{k} \cdot \vec{r}}$ is assumed, where $\vec{k}$ describes the total impurity momentum. Moreover, the recent works \cite{Guenther2021,Schmidt2022} assumed a \emph{Gross-Pitaevskii} type mean-field ansatz for the Bose gas,
\begin{equation}
    \ket{\psi_{\rm B}(\vec{r})} = \prod_{n=1}^N \psi(\vec{x}_n-\vec{r}),
    \label{eqPsiGPE}
\end{equation}
with $n=1...N$ labeling the bosons which macroscopically occupy the single-particle orbital $\psi(\vec{x})$. Alternatively, in second quantization, products of coherent states of bosons $\prod_{\vec{x}} \ket{\psi(\vec{x}-\vec{r})}$ with complex amplitudes $\psi(\vec{x}-\vec{r}) \in \mathbb{C}$ defined relative to the impurity can be used, see e.g. \cite{Guenther2021}. 

Minimizing the variational energy of states in Eq.~\eqref{eqPsiGPE} yields a non-linear Gross-Pitaevskii equation (GPE) for $\psi(\vec{x})$ with the original boson mass $m_{\rm B}$ replaced by the reduced mass $m_{\rm red}$ \cite{Guenther2021},
\begin{equation}
    \l - \frac{\hbar^2 \nabla^2_{\vec{r}}}{2m_{\rm red}} + U_{\rm IB}(\vec{r}) + g_{\rm BB} |\psi(\vec{r})|^2 - \mu \r \psi(\vec{r}) = 0,
    \label{eqGPEimpurity}
\end{equation}
where $U_{\rm IB}(\vec{r})$ denotes the impurity-boson interaction potential and $\mu$ the chemical potential. This equation can be interpreted as the GPE defined in the co-moving frame with the impurity \cite{Drescher2020}; indeed, the GPE~\eqref{eqGPEimpurity} can be derived directly by making a LLP transformation~\cite{Lee1953} and treating the interacting Bose gas on a mean-field level. The ansatz in Eq.~\eqref{eqPsiGPE} was further extended to include additional boson-boson correlations by the inclusion of Jastrow factors in the variational wavefunction $\ket{\psi_{\rm B}(\vec{r})}$ \cite{Drescher2020}. 

Recent theoretical analysis showed that the GPE ansatz allows to derive several universal aspects of strong-coupling Bose polarons~\cite{Massignan2021}. It was further pointed out in Ref.~\cite{Drescher2020} that the GPE variational approach requires some care when treating the impurity-boson interaction beyond Born approximation by including effective-range effects; this can lead to the accumulation of bosonic density around the impurity beyond regimes where the mean-field GPE ansatz is valid. A detailed study of the GPE in the heavy-impurity limit was performed by Yegovtsev et al.~\cite{yegovtsev2023_2}. By comparison to diffusion Monte Carlo (DMC) calculations their study demonstrated that GPE predictions are accurate provided that the gas parameter remains small everywhere, including in the vicinity of the impurity.

\subsubsection{Lee-Low-Pines (LLP) mean-field analysis.}
\label{subsubsecLLPmeanField}
One of the most commonly employed theoretical tools is constituted by the Lee-Low-Pines (LLP) transformation \cite{Lee1953}. It corresponds to an exact basis transformation to a co-moving reference frame with the impurity in its center, and can be cast in the form of a unitary rotation
\begin{equation}
    \hat{U}_{\rm LLP} = \exp \left[ - i \hat{\vec{r}} \cdot \hat{\vec{P}}_{\rm B} \right],
\end{equation}
where $\hat{\vec{r}}$ is the first-quantized impurity position operator and $\hat{\vec{P}}_{\rm B}$ denotes the total momentum operator of the bosonic bath. As a direct consequence of the overall translational invariance of the impurity-plus-bath system, the transformed Hamiltonian $\H_{\rm LLP} = \hat{U}^\dagger \H \hat{U}_{\rm LLP}$ becomes explicitly block diagonal in the total conserved momentum $\vec{K}_{\rm tot}$. Thereby, the impurity-degree of freedom is effectively removed from the problem. This trick is routinely used and works on any microscopic Hamiltonian that satisfies the condition of translational invariance.

In the canonical LLP mean-field treatment of the Bose-polaron problem \cite{Lee1953}, one first performs a LLP transformation and second assumes coherent states $\ket{\alpha_{\vec{k}}}$ of phonons, with $\alpha_{\vec{k}} \in \mathbb{C}$. The overall variational wavefunction in the original frame,
\begin{equation}
    \ket{\Psi_{\rm LLP-MF}(\vec{K}_{\rm tot})} = \hat{U}_{\rm LLP} ~ \ket{\vec{K}_{\rm tot}}_{\rm I} \otimes \prod_{\vec{k}} \ket{\alpha_{\vec{k}}},
    \label{eqPsiLLPmf}
\end{equation}
includes non-trivial correlations due to the use of the non-Gaussian LLP transformation. The ability to explicitly choose $\vec{K}_{\rm tot}$ allows one to easily access the momentum-dependence of the polaron ground state.

As we discussed in the context of the effective Hamiltonians in Sec.~\ref{secEffHamiltonians}, the resulting variational energy landscape $\mathscr{H}[\alpha_{\vec{k}}] = \prod_{\vec{k}} \bra{\alpha_{\vec{k}}} \bra{\vec{K}_{\rm tot}} \H_{\rm LLP} \ket{\vec{K}_{\rm tot}} \ket{\alpha_{\vec{k}}}$ provides valuable insights into the properties of mean-field solutions and their low-energy excitations, by searching for saddle-point solutions. By considering time-dependent variational parameters $\alpha_{\vec{k}(t)}$, dynamical properties can also be calculated within this class of variational states \cite{Shashi2014RF,Shchadilova2016,Dzsotjan2020}.

In the context of Bose-polarons in a BEC, the LLP mean-field approach was successfully applied to the Bogoliubov-Fr\"ohlich Hamiltonian to study the subsonic-to-supersonic transition \cite{Shashi2014RF}; These studies were later extended and performed also for the extended Bogoliubov-Fr\"ohlich Hamiltonian \cite{Seetharam2021arXiv}. Since the variational states include infinite numbers of phonons, this approach correctly predicts the superflow around the impurity in the BEC to break down when the velocity of the polaron reaches the speed of sound $c$ in the BEC. This is not consistently captured by other variational approaches including truncated numbers of phonon excitations only. On the other hand, a weakness of the mean-field approach is its difficulty in capturing non-Gaussian, e.g. number states of phonons believed to play a role around the Feshbach resonance.

In some cases, the LLP mean-field approach is equivalent to the strong-coupling description reviewed in Sec.~\ref{subsecSCthy}. For example, the GPE approach with coherent boson number states can be cast in the form of a LLP mean-field state \cite{Guenther2021}; in the limit of a localized impurity, $m_{\rm I} \to \infty$, the SC ansatz for a Fr\"ohlich model also becomes equivalent to the LLP mean-field description.

\subsubsection{Feynman path-integral method.}
To describe the transition from weak- to strong coupling polarons, Feynman introduced a variational method based on the path-integral formalism \cite{Feynman1955}. In it, one replaces the true action of the impurity obtained after integrating out phonons by a simplified trial action with an effective potential seen by the impurity. Parameters in the latter can be optimized to minimize the overall free energy and obtain a variational result at non-zero or zero temperature. For pedagogical derivations see e.g.~\cite{Devreese2013,Grusdt2015Varenna}.

Feynman's approach was applied to study the Bogoliubov-Fr\"ohlich model at strong coupling, where Tempere et al.~\cite{Tempere2009} predicted a sharp self-localization transition to occur. Although subsequent work clarified that the sharp transition was an artifact of the variational method and becomes a smooth cross-over \cite{Vlietinck2015,Grusdt2015RG} (consistent with general expectations \cite{Gerlach1991}), the path-integral approach remains a powerful tool and was used early on to compare theory and experiment~\cite{Catani2012}. It allows to make accurate quantitative predictions for key properties, such as the quasiparticle energy and effective mass, while capturing exactly the weak- and strong-coupling limits. This makes it more powerful than the LLP mean-field or strong-coupling approaches, which only correctly capture the weakly and strongly interacting regimes, respectively. None of these theories can correctly capture the logarithmic UV divergences (see Sec.~\ref{subsecUVdiv}) in $d=3$ dimensions, however.

The path-integral approach can also be used to calculate spectral responses, see e.g. Ref.~\cite{Casteels2012} which considered Bogoliubov-Fr\"ohlich models in reduced dimensions. More recently, Feynman's path integral approach has also been applied to the extended Bogoliubov-Fr\"ohlich Hamiltonian including two-phonon terms (see Sec.~\ref{secExtdBogoFroh}) \cite{Ichmoukhamedov2019}, where it yielded accurate results. 

Overall, the strength of the path-integral method is its predictive power and immediate inclusion of finite-temperature effects. Drawbacks include computational difficulties when more complicated bosonic baths are considered, e.g. the inclusion of boson-boson interactions remains a challenging open task.

\subsubsection{(Non-) Gaussian variational states.}
\label{sec:NonGaussianState}
In the LLP mean-field theory, the non-Gaussian LLP transformation is combined with a Gaussian coherent state mean-field ansatz. A natural extension of this approach is to replace the coherent states by the most general Gaussian states, including squeezing of the phonon fields: 
\begin{multline}
    \ket{\Psi_{\rm LLP-Gauss}(\vec{K}_{\rm tot})} =   \hat{U}_{\rm LLP} ~ \ket{\vec{K}_{\rm tot}}_{\rm I} \otimes \\
    \exp\left[ \int d^d\vec{k}~\hat{\Phi}^\dagger_{\vec{k}} \sigma^z \Phi_{\vec{k}} \right] \times \\
    \exp \left[ \frac{i}{2} \int d^d\vec{k}d^d\vec{k}'~ \hat{\Phi}^\dagger_{\vec{k}} \Xi_{\vec{k},\vec{k}'} \hat{\Phi}_{\vec{k}'}  \right] \ket{0}.
\end{multline}
Here $\hat{\Phi}_{\vec{k}} = (\a_{\vec{k}},\ad_{\vec{k}})^T$ is the Nambu vector constituted by the phonon annihilation and creation operators at momentum $\vec{k}$; $\sigma^z$ is the Pauli matrix acting on Nambu space and $\ket{0}$ is the phonon vacuum. The numbers $\Xi_{\vec{k},\vec{k}'}$ determine the covariance matrix encoding two-point correlations and $\Phi_{\vec{k}} \in \mathbb{C}$ is the coherent (mean-field) displacement amplitude; Together they serve as variational parameters.

This approach was applied to the Bogoliubov-Fr\"ohlich model and shown to significantly improve the mean-field variational energy of the Fr\"ohlich polaron \cite{Shchadilova2016PRA,Kain2016}, yielding results close to Feynman's path integral method \cite{Kain2016}. Later it was applied to the extended Bogoliubov-Fr\"ohlich model in $d=1$ dimension \cite{Kain2018} and in $d=3$ dimensions to study Efimov states \cite{ChristianenPRL_2022,ChristianenPRA_2022}. Notably, the required three-body correlations are captured by the combination of the non-Gaussian LLP transformation with general squeezing operations on the bosons. 

The above strategy can be applied more broadly, a technique pioneered by Shi et al.~\cite{TaoShi2018}: By combining the LLP with other non-Gaussian unitary transformations, additional correlations can be built into the wavefunction. In the last layer it is always useful to allow the most general Gaussian states.

\subsubsection{Renormalization group and flow-equation approaches.}
Another strategy to explore beyond LLP mean-field effects is to include fluctuations around the mean-field saddle-point systematically in a renormalization group (RG) procedure. To this end, fast and slow phonon degrees of freedom, corresponding to large and small momenta respectively, are decoupled shell by shell using infinitesimal (non-)Gaussian transformations, allowing to include (non-)Gaussian fluctuations. In each step this leads to a renormalization of the effective low-energy Hamiltonian, described by RG flow-equations capturing how the effective coupling constants change. At the same time one keeps track of the flow of observables like the ground state energy, and the converged results after the RG flow describe the polaronic state \cite{Grusdt2015RG,Grusdt2016B}.

The directly related flow-equation method performs similar sequences of unitary transformations which step-by-step approximately diagonalize the entire Hamiltonian \cite{Wegner1994,Kehrein2006}. The main technical difference between the general flow-equation approach and the RG described above is that an abstract energy parameter is used to define the flow instead of momentum shells in the RG, which leads to some formal differences of the two methods. The general flow-equation method was first applied to Bose polarons in $d=1$ by Brauneis et al.~\cite{Brauneis2021}, and allowed to fully include boson-boson interactions.

The RG procedure for polarons was first introduced as an all-coupling theory to describe the Bogoliubov-Fr\"ohlich polaron in Ref.~\cite{Grusdt2015RG}. Subsequently it was technically improved to obtain quantitatively more accurate results \cite{Grusdt2016B}. The method proved powerful by analytically predicting logarithmic UV divergencies (see Sec.~\ref{subsecUVdiv}) \cite{Grusdt2015RG} and providing accurate predictions for the ground state energy of the Bogoliubov-Fr\"ohlich polaron all the way from weak to strong coupling, clarifying the nature of the associated self-trapping as occuring in a smooth cross-over \cite{Grusdt2016B}. 

The RG method was subsequently applied to study the extended Bogoliubov-Fr\"ohlich polaron in $d=3$ \cite{Grusdt2017} and $d=1$ dimensions \cite{Grusdt2017RG1D}. This allowed to map out the saddle-point structure and revealed different unstable regimes around strong coupling, some of which are also present in LLP + Gaussian variational descriptions \cite{Kain2018}. In $d=1$, the RG procedure is necessary to properly regularize an unphysical logarithmic infrared (IR) divergence of the LLP mean-field polaron energy that simpler variational approaches with truncated numbers of excitations cannot capture in the first place \cite{Grusdt2017RG1D}.

Recently, the RG analysis has been extended by working with path-integrals and an effective action. This allows to apply the powerful tools of the \emph{functional renormalization group} (FRG) analysis. Like the RG procedures above, its main advantage is the ability to include fluctuations around simpler saddle-point solutions corresponding to mean-field states, and its accuracy depends on the quality of the ansatz made for the effective low-energy action. The formalism has been applied to the full microscopic Bose-polaron Hamiltonian by Isaule et al.~\cite{Isaule21} and yielded accurate results in $d=3$ and $d=2$ dimensions. In fact, the formalism is powerful enough to capture also more complicated atomic mixtures exhibiting the physics of Bose polarons and beyond \cite{Milczewski2022}.

\subsubsection{Truncated basis and multi-configuration Hartree methods.}
All approaches discussed so far are largely based on physically motivated approximations. A complementary strategy is to start from the full many-body problem, i.e. Eq.~\eqref{eqHmic}, and make computationally practicable simplifications. 

A powerful paradigm to reduce computational complexity is to work in truncated bases and treat the many-body system projected to the retained basis states using exact numerical tools. Depending on the problem at hand, different truncation procedures can be applied, and this procedure is generally easier to work with in low-dimensional settings or if the number of orbitals is limited. To treat free-space problems one either works with discrete orbitals or discretizes the underlying single-particle wavefunctions. 

In the context of Bose polarons, or more generally atomic mixtures, the multi-layer multi-configuration Hartree method \cite{Cao2017} has become a method of choice, see e.g. Refs.~\cite{Mistakidis2019,Mistakidis2019a}. In it, the single-particle orbitals underlying the truncated many-body basis can be optimized in order to minimize the overall variational energy. This approach has been extended to include time-dependence, optimizing the single-particle orbitals in every time step \cite{Cao2017}.

As a main advantage of the multi-configuration Hartree methods, their accuracy can be systematically controlled, and in the limit of a few particles essentially exact results are obtained. The entanglement between subsystems, e.g. between impurity and bath, can be readily extracted, and used to analyze systematically the level of entanglement building up within the system. Reaching large system sizes in higher dimensions remains a challenge for this type of method.

\subsubsection{Diagrammatic Monte Carlo.}
The most powerful and numerically best-controlled methods to date are constituted by different Monte Carlo approaches \cite{Ardila2022NR}. The first is the \emph{diagrammatic Monte Carlo} (DiagMC), which starts out from a diagrammatic expansion of the impurity Green's function. Instead of making approximations which diagrams are kept, one samples from all possible diagrams in a suitable way until the imaginary time Green's function can be determined within a controlled statistical error \cite{Prokofev1998,Mishchenko2000}.

The main advantage of the approach is its ability to produce numerically exact results (within controlled statistical error bars). However, it has to be adapted separately for every microscopic Hamiltonian, which ultimately determines the structure of the allowed Feynman diagrams. As a result, the method becomes more challenging to apply for more complicated polaron models. Indeed, in the context of Bose polarons in a BEC, the diagrammatic method has only been applied to the Bogoliuvov-Fr\"ohlich Hamiltonian \cite{Vlietinck2015}, but with great success in that case.

\subsubsection{Wavefunction and diffusion Monte Carlo methods.}
Another class of powerful Monte Carlo methods directly consider the many-body wavefunction in first quantization and sample configurations in order to estimate the corresponding energy. The simplest, so-called \emph{variational Monte Carlo} (VMC), variant makes a variational ansatz $\ket{\Psi}$ for the many-body ground state and samples its energy. The most commonly used type of ansatz uses Jastrow factors, explicitly incorporating one- and two-body correlations:
\begin{equation}
    \ket{\Psi_{\rm VMC}} = \prod_{i<j}^N \psi_{\rm BB}(\vec{x}_i-\vec{x}_j) \prod_{j=1}^N \psi_{\rm IB}(\vec{x}_j - \vec{r}).
    \label{eqPsiJastrow}
\end{equation}
Here $\vec{x}_j$ denote the coordinates of the $i,j=1,...,N$ bosons and $\vec{r}$ is the impurity coordinate. The two wavefunctions $\psi_{\rm BB}(\vec{x})$ and $\psi_{\rm IB}(\vec{x})$ can be used as variational parameters and are typically initialized in a way ensuring that the two-body limits are correctly recovered. For application of the Jastrow ansatz to Bose polarons in a BEC, see e.g.~\cite{Ardila2015,Ardila2016} (in $d=3$) and \cite{Parisi2017,Grusdt2017RG1D} (in $d$=1).

To obtain the ground state wavefunction, a variational state, e.g. Eq.~\eqref{eqPsiJastrow}, can be used as a starting point which is propagated in imaginary time $\tau$ by applying $e^{- \H \tau}$ for $\tau \to \infty$. Formally this requires solving the imaginary-time Schr\"odinger equation $\partial_\tau \ket{\psi(\tau)} = - \H \ket{\psi(\tau)}$ for the many-body ground state wavefunction $\ket{\psi_0} \propto \ket{\psi(\tau\to \infty)}$ where $\ket{\psi(\tau)} = e^{- \H \tau} \ket{\Psi_{\rm VMC}}$. Being able to sample from the time-evolved wavefunction $\ket{\psi_0}$ one can estimate, up to statistical errors, the true ground state energy, see e.g.~Refs.~\cite{Boronat1994,Toulouse2016,ThesisPOL} for a pedagogical explanation of the details of this method. 

In practice the quality and efficiency of the approach depends on how accurately one can sample from the wavefunction evolved in imaginary time. In practice this is done by applying the Green's function to an initial sample drawn according to the trial wavefunction. In the \emph{diffusion Monte Carlo} (DMC) method the Green's function itself is calculated for infinitesimal time steps by Trotterizing the imaginary time evolution. In order to reduce numerical noise, importance sampling according to the trial wavefunction $\ket{\Psi_{\rm VMC}}$ is used, which can already be a good approximation to the true ground state in the Bose polaron problem \cite{Grusdt2017RG1D}. It should be emphasized, however, that unless the initial trial state is strictly orthogonal to the true ground state, the DMC method yields numerically exact results; the VMC state merely controls its efficiency. In this review, the use of both variational and diffusion Monte Carlo will be denoted as quantum Monte Carlo methods (QMC).  

Different algorithms to compute the effective mass, quasiparticle residue and pair-correlation function in the ground-state are provided in references~\cite{Ardila2015,Ardila2020,Grusdt2017}.

Beyond the Bogoliubov-Fr\"ohlich Hamiltonian, the DMC method is the most accurate method to date. It has been successfully applied to analyze the Bose polaron all the way from weak to strong coupling~\cite{Ardila2022NR}, including around the Feshbach resonance, in $d=1$ \cite{Parisi2017,Grusdt2017RG1D}, $d=2$ \cite{Ardila2020} and $d=3$ dimensions \cite{Ardila2015,Ardila2016}, and - as will be reviewed in the next sections - compares very well with experimental results \cite{Ardila2019}.

\section{Strong coupling Bose polaron}
\label{secStrongCplgPolaron}
Chapter 2 provided a comprehensive overview of the experimental realizations of Bose polarons in ultracold gases; chapter 3 independently summarized the theoretical and numerical machinery that serves to characterize the ground-state and out-of-equilibrium dynamics of Bose polarons. In this chapter, we will bring these topics together and discuss how the experimental results compare with different theoretical predictions, ranging from perturbation theory to ab initio numerical techniques such as quantum Monte-Carlo methods. In particular, comparisons are performed for the polaron energy, namely the attractive and the repulsive polaron branches, whereas quantities such as the effective mass and quasiparticle residue remain limited to only a few experimental realizations. As we will discuss, the plethora of available analytical, semi-analytical and numerical methods in the literature predicts different behaviors, especially in the strongly interacting regime that we will mainly focus on.

The coupling of impurities to BECs has been a topic of extensive research, mainly when strong correlations arise between the impurity and the bosons. One of the seminal works in this field is the study by Cucchietti et al.~\cite{Cucchietti2006}, which first focused on the problem of a neutral impurity atom immersed in a dilute BEC. The coupling between the impurity and the BEC was predicted to create a localized state where the impurity distorts the density of the surrounding BEC so strongly that a self-trapping potential for the impurity emerges. As a result, the renormalized polaron mass was predicted to also increase sharply in the self-localized state (see Sec.~\ref{subsecSCthy}). 

Today, in the context of beyond-Fr\"ohlich descriptions of Bose polarons at strong coupling, self-localization is viewed as an artifact stemming from the specific choice of the employed variational wavefunction and the underlying model Hamiltonian. Instead, recent progress in the development and control of new experimental techniques, such as radio-frequency spectroscopy and Ramsey interferometry, opened up the possibility of probing the strong coupling Bose polaron spectroscopically. As we will review, this has led to a less exotic picture of strong coupling Bose polarons with a focus shifting away from questions concerning mass renormalization while triggering new research directions.

\subsection{Bose polaron energy}

\begin{figure}[t]
	\centering
 \includegraphics[width=8.3 cm]{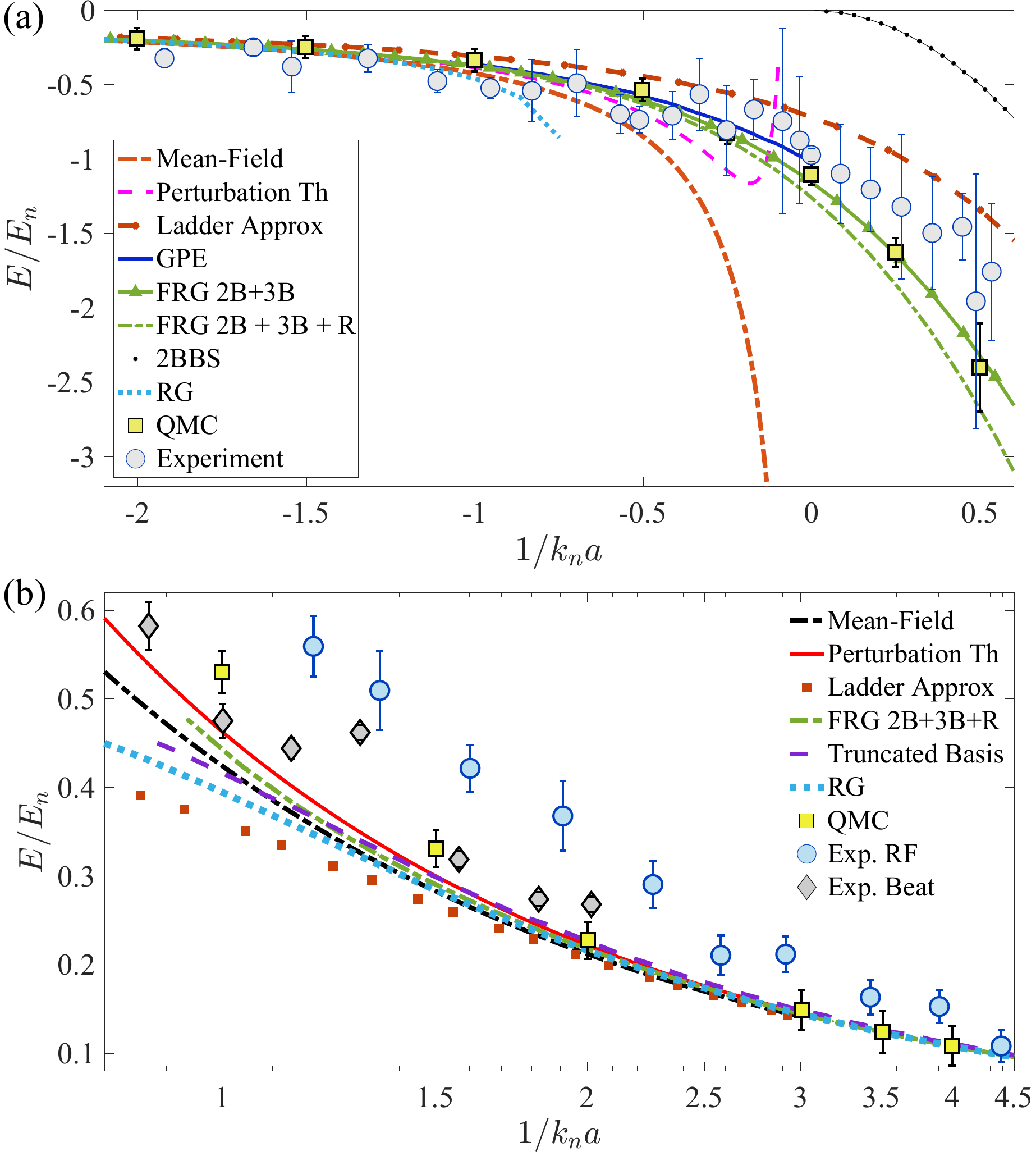}
	\caption{\textbf{Comparison to theory: Aarhus experiment.} Experimental measurements by Jorgensen et al.~\cite{Jorgensen2016} of the ground-state polaron energy using RF spectroscopy are compared to analytical and numerical techniques, for the zero momentum (a) attractive and (b) repulsive polaron branches. The methods depicted are mean-field theory $E/E_{n}=\frac{4}{3\pi}k_{n}a$, second-order perturbation theory~\cite{Ardila2015,Christensen2015}, $\mathrm{T-}$matrix (ladder approximation) approach~\cite{Ardila2019,Rath2013},  Gross-Pitaevskii (GPE) approach~\cite{Guenther2021,Massignan2021}, functional renormalization group (FRG): using only two and three-Bogoliubov excitations (2B+3B)with zero range and finite range (2B+3B+R) theory~\cite{Isaule21}, truncated basis variational method including three-body correlations~\cite{Jorgensen2016} (repulsive branch only) and renormalization group theory (RG)~\cite{Grusdt2017}. In addition, Quantum Monte Carlo predictions~\cite{Ardila2019} and the results of the Ramsey spectroscopy experiment by Morgen et al.~\cite{Morgen2023} (Exp. Beat) on the repulsive side are shown. For comparison, the energy of the two-body impurity-boson bound state (2BBS) are shown in (a).}\label{fig:ArhusExp}
\end{figure}

\begin{figure}[t]
	\centering
 \includegraphics[width=8.3 cm]{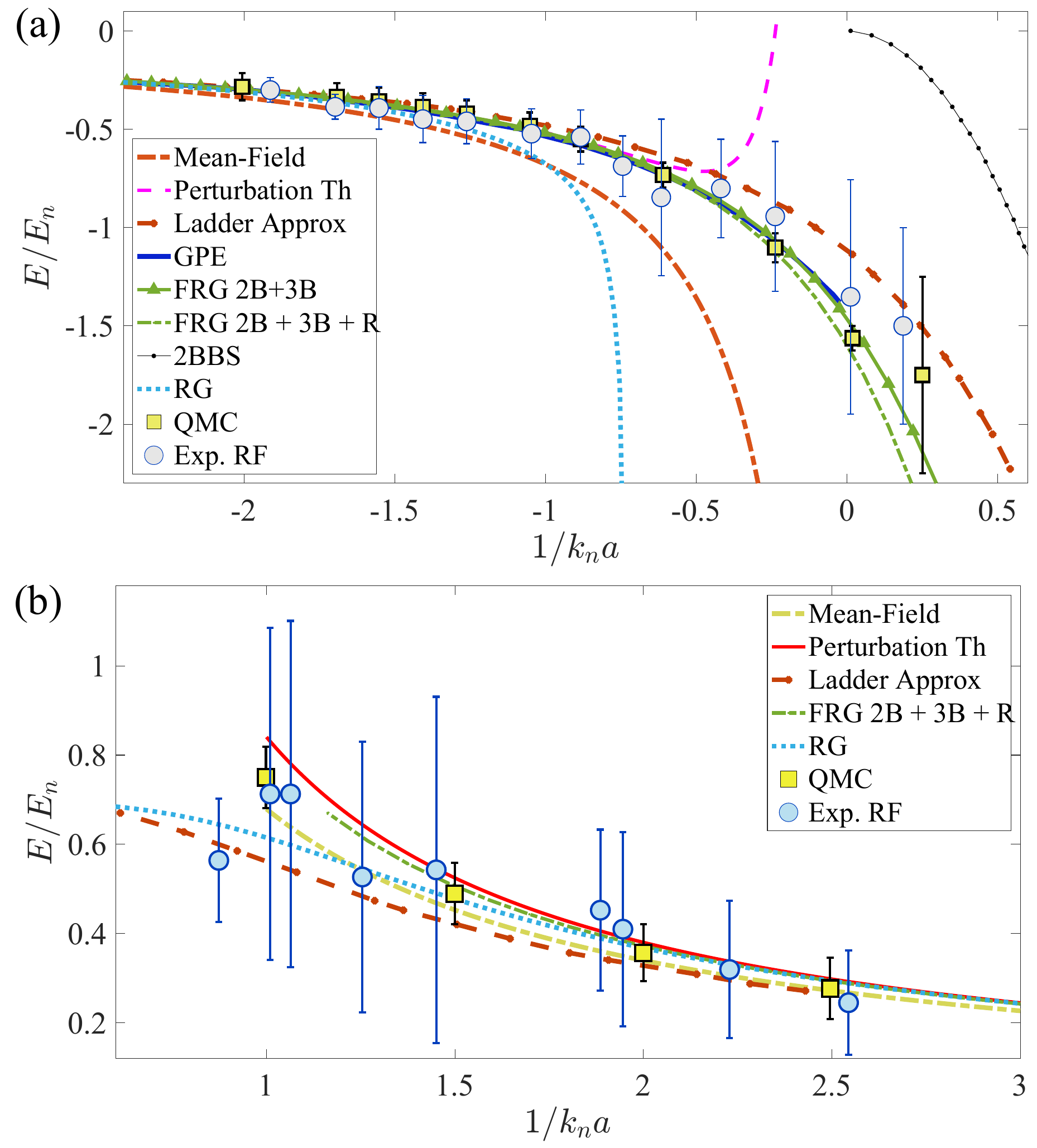}
	\caption{\textbf{Comparison to theory: JILA Experiment.} Experimental measurements by Hu et al.~\cite{Hu2016} of the ground-state polaron energy using RF spectroscopy are compared to analytical and numerical techniques, for the zero momentum (a) attractive and (b) repulsive polaron branches. The methods are mean-field theory $E/E_{n}=\frac{4}{3\pi}k_{n}a$, second-order perturbation theory~\cite{Ardila2015,Christensen2015}, $\mathrm{T-}$ matrix (ladder approximation) approach~\cite{Rath2013}, Gross-Pitaevskii (GPE) approach~\cite{Guenther2021,Massignan2021}, functional renormalization group theory (FRG): using two and three-Bogoliubov excitations (2B+3B) with zero range and finite range (2B+3B+R) theory~\cite{Isaule21} and renormalization group theory (RG)~\cite{Grusdt2017}. In addition, Quantum  Monte Carlo~\cite{Ardila2019} results are shown. For comparison, the energy of the two-body impurity-boson bound state (2BBS) are shown in (a).}
	\label{fig:JILAExp}
\end{figure}

The polaron energy is plotted as a function of the coupling strength $(k_{n}a)^{-1}$ comparing different theoretical, semi-analytical and numerical approaches against experimental measurements for both the Potassium mixture in Aarhus, Fig.~(\ref{fig:ArhusExp})(a--b) [see also subsec.~\ref{subsecExpAarhus}] and  the Potassium-Rubidium mixture at JILA in Fig.~\ref{fig:JILAExp} (a-b) [see also subsec.~\ref{subsecExpJILA}]. The experimental data is first compared to mean-field and perturbation theory, which are asymptotically exact in the weakly interacting regime, $\left|\left(k_{n}a\right)^{-1}\right|\gg1$. For intermediate and strong coupling, different techniques have been employed to compute the ground state polaron energy, for example, $\mathrm{T-}$matrix approaches within the ladder approximation~\cite{Rath2013}, renormalization group (RG)~\cite{Grusdt2017}, functional renormalization group (FRG)~\cite{Isaule21}, truncated basis method~\cite{Jorgensen2016}, Gross - Pitaevskii (GPE)~\cite{Guenther2021,Massignan2021} and quantum-Monte Carlo (QMC) techniques~\cite{Ardila2019}. 

In the weakly interacting regime, the mean-field energy can be computed analytically, namely $E/E_{n}=\left(4/3\pi\right)k_{n}a$ ($E_n$ and $k_n$ defined in Eq.~(\ref{eq:kn}) and Eq.~(\ref{eq:En}); $a=a_{\rm IB}$ is the impurity-boson scattering length). It depends only on the density and the impurity-boson coupling strength. Both predictions for the attractive and repulsive branches measured in the experiments are plotted in Fig.~\ref{fig:ArhusExp} and Fig.~\ref{fig:JILAExp} for both experiments. Beyond-mean-field energy contributions can be calculated by using the Rayleigh-Schr\"odinger perturbation theory. This perturbative expansion is performed in two ways: (i) at the Fr\"ohlich level, starting from Eq.~(\ref{eqHIBFroh}) where the impurity couples to a single mode, or (ii)  on the full Hamiltonian Eq.~(\ref{eqHmic}), by truncating up to lower-order; in both cases, corrections to the mean-field energy are added to the parameter $k_{n} a$, turning to be the small parameters in the perturbative series.

At second-order, these corrections are known as quantum fluctuations and depend on the bath compressibility accounted in the gas parameter $na_{\mathrm{BB}}^3$ and the coupling strength $a$. Quantum fluctuations are repulsive as they scale as $\propto (a/a_{\mathrm{BB}})^2$ and increase the polaron energy with respect to the mean-field energy shift as observed in Fig.~\ref{fig:ArhusExp} and Fig.~\ref{fig:JILAExp}. The third order can be retrieved from the perturbative expansion on the full Hamiltonian~\cite{Christensen2015}. At the Fr\"ohlich level, high-order scattering processes, such as phonon-phonon interactions, are neglected, whereas the perturbative expansion on the full Hamiltonian is only accurate for low-order by definition.

Mean-field and perturbative expansion provide a valuable benchmark for all the theoretical methods and experiments at weak coupling. This is corroborated in both Fig.~\ref{fig:ArhusExp} and Fig.~\ref{fig:JILAExp} where a nearly perfect collapse is observed for all methods onto the mean-field and perturbative prediction when $|k_n a|^{-1} \gg 1$. In contrast, for larger coupling strength, the mean-field and perturbative results deviate from the experiment as expected. Note that, for the repulsive branch, the regime of validity for the perturbative results roughly coincides with the regime where the polaron remains metastable before decaying into lower-energy states.

Near the resonance and on the attractive branch, all non-perturbative methods including phonon-phonon interactions agree with both experiments within statistical uncertainty. The RG and LLP mean-field approaches based on the extended Bogoliubov-Fr\"ohlich Hamiltonian without phonon-phonon interactions, in contrast predict too low energies around unitarity, $1/|k_n a| \approx 0$. The $\mathrm{T-}$matrix calculation is closely related to a Chevy-type truncated basis calculation, which includes no more than one bosonic excitation. This resembles the effect of a strong phonon-phonon non-linearity, and explains its finite variational energy around unitarity. Compared to the FRG and GPE approaches, the $\mathrm{T-}$matrix (ladder approximation) predicts slightly higher energy at strong coupling on the attractive polaron branch.

QMC methods provide the numerically exact ground-state energy of the attractive polaron, by including an unlimited number of excitations, finite range, and back-action effects due to the local deformation of the condensate density by the impurity. Additionally, other methods, such as GPE and FRG, incorporate finite range corrections and density inhomogeneities, and they compare quite well with QMC. FRG also discerns between the inclusion of two-body and three-body correlations. Indeed, including extra correlations lowers the polaron energy, as seen for the attractive branch in both experiments. Nonetheless, the error bars in both cases do not resolve the quantitative difference. In fact, for the current experiments in Fig.~(\ref{fig:ArhusExp}) and Fig.~\ref{fig:JILAExp}, deviations between GPE, FRG, and QMC are very small and do agree within the experimental uncertainties. 

As previously mentioned, the variational ansatz, which includes only one Bogoliubov excitation, is equivalent to the non-self-consistent $T$-matrix approach~\cite{Chevy2010,Rath2013,Li2014}. In reference ~\cite{Field2020}, the authors extend the variational ansatz to include three-body correlations (the impurity and two Bogoliubov excitations) and four-body correlations (the impurity and three Bogoliubov excitations). The former case agrees with Quantum Monte Carlo (QMC) results and experimental data in the regime $(n^{1/3}a)^{-1} \lesssim -1$. In fact, three-body correlations are the minimal number required to capture beyond-mean-field corrections. However, Field et al. ~\cite{Field2020} demonstrated that four-body correlations are necessary near the resonance. Perfect agreement with QMC was achieved when the effective range was adjusted such that the ground-state Efimov trimer coincided with the one predicted by QMC. Furthermore, the variational ansatzes also show remarkable agreement with the Aarhus experiment~\cite{Jorgensen2016} when the effective range from the experiment is used, corresponding to $n^{1/3}R^{\star} \approx 0.02$.

It is worth mentioning that a good agreement at resonance between these theories and the experiment is naively expected because the local deformations caused by the impurity are small -- but not perturbatively small. This is a result of several factors, such as high dimensionality, mass balance, and the absence of Efimov physics. A new generation of experiments is needed to assess the validity of the current state-of-the-art methods.

The situation is less clear for the repulsive side in both experiments. Regarding the first generation of experiments in Aarhus using RF spectroscopy, most theoretical approaches predict a lower energy than measured experimentally, as shown in Fig.~\ref{fig:ArhusExp}(b). Although QMC achieves the closest agreement with the experiment, a systematic discrepancy with theory was measured. In reference~\cite{Ardila2019}, this discrepancy was partially resolved by taking into account the effect of the incoherent part of the spectrum when extracting the repulsive polaron energy from the latter; moreover, a simplified model of the impurity dynamics was employed in order to quantify the probability of three-body loss processes at different densities, which become relevant for strong repulsive interactions. This additional analysis improved the agreement between theory and experiment, as shown by Ardila et al.~\cite{Ardila2019}. Further improvement was later achieved in a second generation of experiments in Aarhus using Ramsey interferometry to extract the polaron energy more directly~\cite{Morgen2023}: These experimental measurements (grey diamonds in Fig.~\ref{fig:ArhusExp} (b) are in very good agreement with QMC predictions. In the JILA measurements of the repulsive polaron energy, larger error bars are present; in this case, all theoretical methods agree within error bars, see Fig.~\ref{fig:JILAExp}(b).

\begin{figure}
	\centering
 \includegraphics[width=8.3 cm]{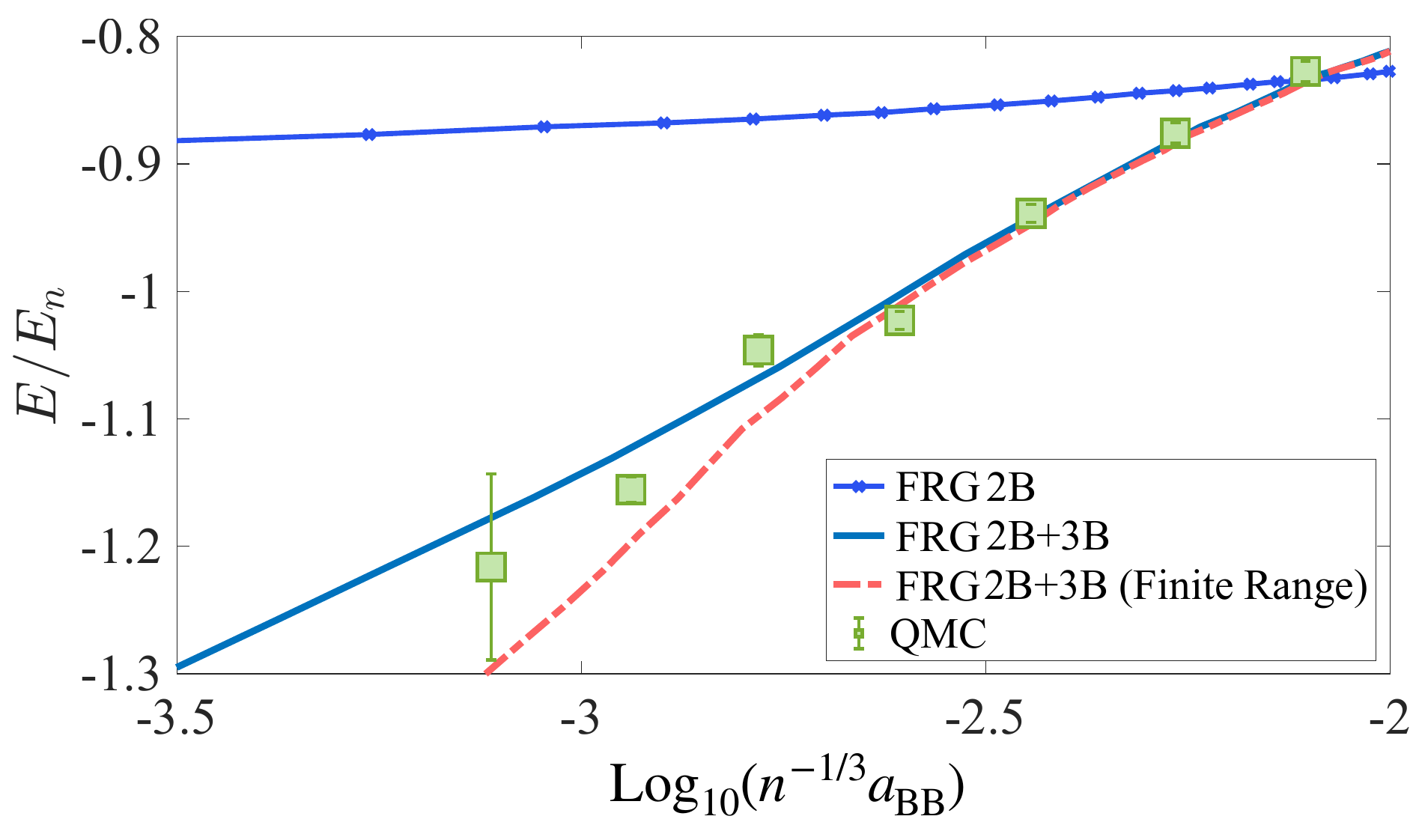}
\caption{\textbf{Theoretical prediction of the polaron energy as a function of the parameter $(na_{\mathrm{BB}}^{3})^{1/3}$} for parameters as in the Aarhus experiment. The methods are functional renormalization group theory (FRG)~\cite{Isaule21}, using only 2 Bogoliubov excitations (2B), three Bogoliubov excitations (2B+3B), finite range effects $R$ and Quantum Monte Carlo (QMC) data extracted from~\cite{Ardila2019}. Figure adapted from~\cite{Isaule21}.}
	\label{fig:DensityDependence}
\end{figure}

Another interesting question concerns the effects of a modified compressibility of the bath on Bose polaron properties at strong coupling. In particular, at the resonance it was expected that the polaron properties would depend only on the interparticle distance $n^{1/3}$, since $a_{\rm IB} \to \infty$ disappears as a lenght scale in the problem. In Fig. \ref{fig:DensityDependence} we compare the polaron energy at unitarity using QMC method and the functional renormalization group (FRG). The agreement between both theories is remarkable. However, as mentioned earlier, the quasiparticle properties depend non-universally on the specific model parameters. We will return to this question in the following chapters of this review, where we discuss polarons in different dimensions and with different types of impurity-atom interactions.

\begin{figure*}
	\centering
	\includegraphics[width=\linewidth]{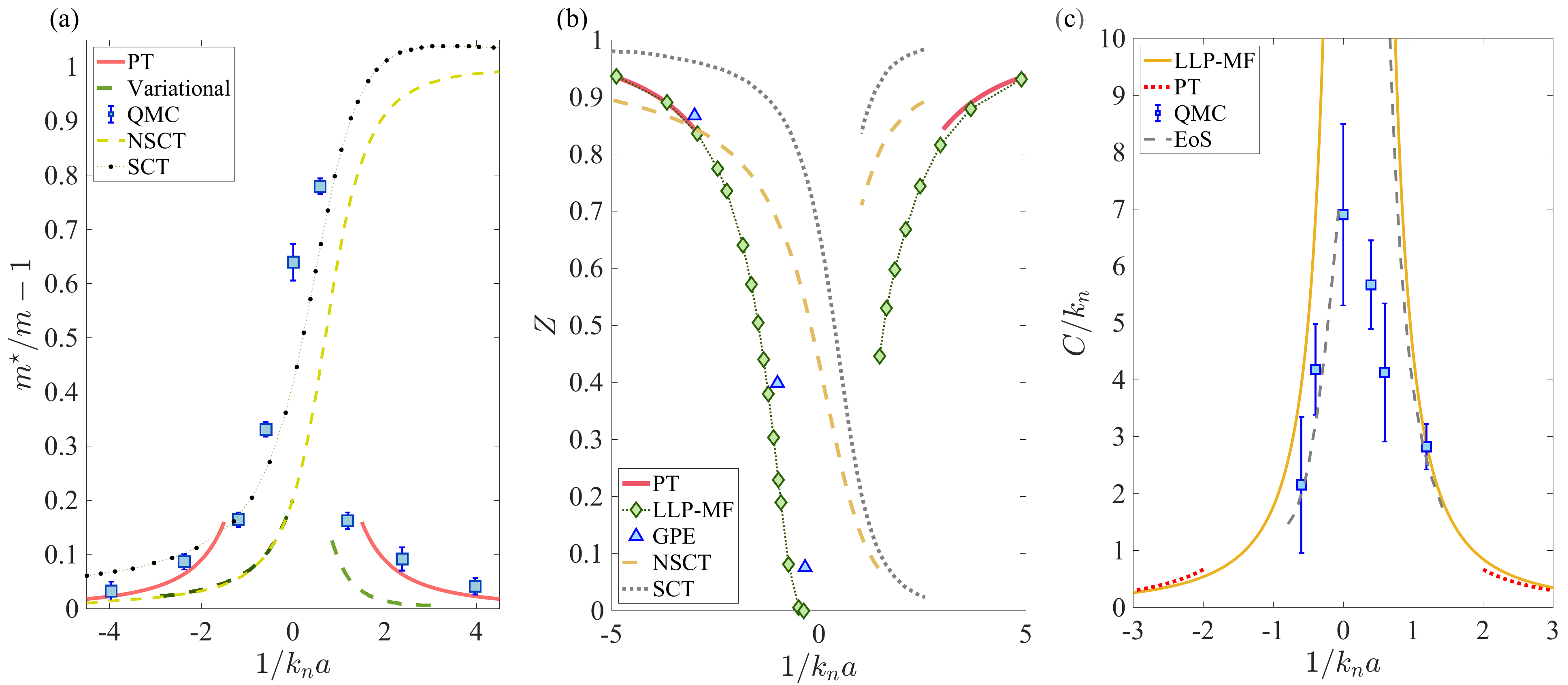}
	\caption{\textbf{Bose polaron characteristics.} (a) Effective mass for $na_{\mathrm{BB}}^{3}=1\times 10^{-5}$ and for the case $m_{\mathrm{B}}=m_{\mathrm{I}}$ using second-order perturbation theory (PT), variational ansatzs from~\cite{Li2014} and both  non-self-consistent (NSCT) and self-consistent (SCT) T-matrix  approaches~\cite{Rath2013} and in both cases $na_{\mathrm{B}}^{3}\rightarrow 0$. (b) Quasiparticle residue  prediction for the Aarhus experiment with parameters  $na_{\mathrm{BB}}^{3}=4.28\times10^{-8}$ and $m_{\mathrm{B}}=m_{\mathrm{I}}$. Here we plotted the second-order perturbation theory (PT) prediction for the residue (PT), the Lee-low Pines mean-field theory (LLP-MF)~\cite{Grusdt2016}, the Gross-Pitaevskii (GPE) approach~\cite{Guenther2021} as well as the results for the NSCT and SCT with the parameters in (a), namely $na_{\mathrm{B}}^{3}\rightarrow 0$ ~\cite{Rath2013}. (c) Contact parameter for $na_{\mathrm{BB}}^{3}=1\times 10^{-5}$ and $m_{\mathrm{B}}=m_{\mathrm{I}}$. The contact is extracted from the short-distance behavior of the impurity-boson pair correlation function using QMC~\cite{Ardila2015}. In addition, the contact can be extracted from the equation of state from QMC data (EoS)~\cite{Ardila2015} and from the equation of state computed with perturbation theory (PT) or LLP-MF ~\cite{Grusdt2016}.}
	\label{fig:3}
\end{figure*}

By comparing different theoretical predictions to experimental data, the most disagreement occurs close to the resonance. Moreover, for $1/(k_n{a}) > 0$, despite the abundance of experimental data and theoretical results for the repulsive polaron, little is known about the fate of the ground state and further few-body resonances below the impurity-boson scattering threshold. This energy range on the repulsive side of the Feshbach resonance is fascinating since the few-body spectrum of the impurity-boson system is known to possess a wide variety of resonances, such as Efimov trimers, the associated tetramers, and larger body clusters, whose properties depend on many parameters of the impurity-boson system such as inter-particle interactions and mass ratios. Even in the case of an infinitely heavy impurity coupled to non-interacting bosons, where no Efimov trimer state is expected, theories predict the emergence of exotic resonances known as "many-body bound states". The detailed discussion of this rich landscape of few and many-body states is the subject of the next chapter of the review.

\subsection{Bose polaron effective mass and quasiparticle residue}
In Fig. \ref{fig:3} we show theoretical predictions for the effective mass $m^{\star}$, the quasiparticle residue $Z$ and the contact parameter $C$ of the Bose polaron as a function of $1/(k_na)$. These all constitute central characteristics of Bose polarons that allow to compare the predictive powers of different theoretical approaches. 

The effective mass has been measured in 1D by Catani's experiments~\cite{Catani2012}, but measurements have still not been reported in the 3D experiments. Nonetheless, different theoretical and numerical approaches have been employed to predict this quantity. Within the Fr\"ohlich Hamiltonian valid for $1/\left|k_{n}a\right|\gg1$, the effective mass is computed from perturbation theory. For finite and small momentum, the total energy of the system can be written as 
\begin{equation}
E(\mathbf{P})\approx E_{\mathrm{B}}+E+\frac{\mathbf{P}^{2}}{2m^{\star}} + \mathcal{O}(\mathbf{P}^4),
\end{equation}
where $E_{\mathrm{B}}$ is the ground-state energy of the host bath, $E$ is the polaron energy discussed in the previous section and $m^{\star}$ is the renormalized polaron mass. 

Within perturbation theory one obtains a small shift of the bare mass of the impurity $\frac{m^{*}}{m_{\rm I}}=1+F(m_{\mathrm{BB}}/m_{\mathrm{I}})\frac{a^{2}}{a_{\mathrm{BB}}\xi}+G(m_{\mathrm{B}}/m_{\mathrm{I}})\frac{a^{3}}{a_{\mathrm{BB}}\xi^{2}},$ where the functions $F$ and $G$ depend on the mass ratio and are explicitly computed in ~\cite{Christensen2015}. On the other hand, variational theories such as the Chevy ansatz~\cite{Chevy2006,Li2014}, self-consistent and non self-consistent T-matrix approaches~\cite{Rath2013} and quantum Monte Carlo methods allow to estimate the effective mass in the strongly interacting regime~\cite{Ardila2015,Ardila2016}. These non-perturbative approaches are in good agreement with perturbation theory in the weak coupling limits, yet considerable deviations are observed in the strongly correlated regime. 

In Fig.~\ref{fig:3}(a), we observe that the second-order perturbative theory agrees quantitatively better with the QMC, while the results computed within the variational approaches underestimate the effective mass renormalization. Around the resonance, the perturbative result is meaningless and diverges, although many theories predict polaron self-localization~\cite{blinova2013single,sacha2006self,Bruderer08}. Instead, the $\mathrm{T}$-matrix approaches agree qualitatively with the QMC results, and a better agreement is obtained when employing the self-consistent $\mathrm{T-}$ matrix approach. The polaron mass increases from one up to approximately two, in units of the bare impurity mass, around the resonance, which is roughly the mass of the dimer formed for $1/a\rightarrow0^{+}$. It was also shown that, despite the high compressibility of the bath, the results do not change considerably when the density of the host bosons is varied~\cite{Ardila2015}. Crossing the resonance, a large increase in the effective mass is expected due to the formation of many-body bound-states in the scattering sector.

The quasiparticle residue $Z$ describes how much free-impurity character the quasiparticle retains. Within a Chevy-type truncated basis ansatz $Z$ can be related to the normalization of the polaron wavefunction~\cite{Chevy2006}, which includes the perturbative limit of very weak interactions. In this case, the unperturbed wavefunction of the system is written as $\left|\mathbf{P},0_{\mathbf{k}}\right\rangle$, representing an impurity of momentum $\mathbf{P}$ and the vacuum of phonons. Thus, the lowest-order polaron wavefunction within perturbation theory can be written as, $\left|\psi\right\rangle=\left|\mathbf{P},0_{\mathbf{k}}\right\rangle +\sum_{\mathbf{k\neq0}}\frac{\left\langle \mathbf{k}\left| \hat{\mathcal{H}}_{\mathrm{Frohlich}}\right| 0\right\rangle }{E_{0}^{(0)}-E_{\mathbf{k}}^{(0)}} \mathbf{\left|k\right\rangle } + \mathcal{O}(g_{\rm IB}^2)$. The quasiparticle residue, defined as the overlap squared with the non-interacting impurity state $\left|\mathbf{P},0_{\mathbf{k}}\right\rangle $, is then computed as
\begin{equation}
    Z = \left| \langle \psi | \mathbf{P},0_{\mathbf{k}} \rangle \right|^2.
\end{equation}
By normalizing the perturbative polaron wavefunction, $\ket{\psi} \to \ensuremath{\sqrt{Z}\left|\psi\right\rangle}$, the leading-order contribution to the residue can be calculated. Perturbatively one obtains $Z^{-1}=1+C(m_{\mathrm{B}}/m_{\mathrm{I}})\frac{a^{2}}{a_{\mathrm{B}}\xi}+D(m_{\mathrm{BB}}/m_{\mathrm{I}})\frac{a^{3}}{a_{\mathbf{\mathrm{BB}}}\xi^{2}}$, where the functions $C$ and $D$  were computed in Ref.~\cite{Christensen2015}. Note that the third-order shift for the energy, effective mass and quasiparticle residue depends on the factor $na_{BB}^{3}(a/a_{BB})^{3}$, which is much smaller than 1 in the experimentally relevant cases.

In Fig.~\ref{fig:3}~(b), we summarize different predictions for the quasiparticle residue. The results are plotted using perturbation theory, the Gross-Pitaevskii (GPE) approach and the Lee-Low Pines mean-field (LLP-MF) for the parameters relevant in the Aarhus experiment, $na_{\mathrm{BB}}^{3}=4.28\times10^{-8}$, while for both $\mathrm{T}$-matrix approaches the limit $na_{\mathrm{BB}}^{3}\rightarrow0$ is taken. In all cases, 
$m_{\mathrm{B}}=m_{\mathrm{I}}$. QMC results for the quasiparticle residue have not been reported in the literature so far. In the strongly interacting regime, GPE and LLP-MF agree qualitatively with each other, rendering a quasiparticle residue of the attractive polaron almost zero in the vicinity of the resonance. This signals a breakdown of the quasiparticle picture around unitarity. $\mathrm{T}$-matrix approaches find a similar drop of the quasiparticle residue, but further on the attractive side of the resonance, and predict quasiparticle residues between 0.5 and 0.6 at unitarity. Experimentally the quasiparticle residue can be measured as the contrast in a Ramsey interferrometric sequence at long times, on the order of the polaron formation time scale; see Sec.~\ref{secRFandDynamics} for a detailed discussion. However, experimental decoherence sources such as finite RF pulse, polaron lifetime, trap and temperature effects play an important role in inhibiting an accurate measurement of this quantity thus far.

The contact $C$ is a thermodynamic quantity that estimates the average number of bosons in the neighborhood of the impurity. In ejection spectroscopy experiments, the high-frequency tail of the spectral function provides direct information about such short correlations: The corresponding spectral function features a universal scaling as $C/\omega^{3/2}$ for $\omega \to \infty$, see e.g. Refs.~\cite{Punk2007,Shashi2014RF}. Knowing the two-body universal dynamics allows one to estimate the asymptotic value of the high-frequency tail (see also Sec.~\ref{Sec:1DBosePolaron}). Another way to estimate the contact is through the Hellmann-Feynman theorem~\cite{SanchezBaena22}, which connects the contact and the polaron energy via 
\begin{equation}
C=\frac{8\pi m_{\mathrm{red}}}{\hbar^{2}}\frac{\partial E}{\partial\left(a^{-1}\right)}.
\label{eqCHellmannFeynman}
\end{equation}

In Fig.~\ref{fig:3} (c), we show predictions for the contact (in units of $k_n$) as a function of the impurity-boson coupling strength. We compare QMC results from Ref.~\cite{Ardila2015} to calculations based on LLP-MF following Ref.~\cite{Shchadilova2016} and predictions within perturbation theory for $na_{\mathrm{BB}}^{3}=1\times10^{-5}$ and $m_{\mathrm{B}}=m_{\mathrm{I}}$. In QMC, the contact $C$ is extracted directly from the pair-correlation function; however, in the other aforementioned approaches, this quantity was extracted via Eq.~\eqref{eqCHellmannFeynman}. As expected, the results agree well for weak coupling. To extract the contact from QMC, the relation of the high-frequency behavior of the spectral function to the many-body wavefunction at short distances can be used: the contact is determined by the behavior of the impurity-boson correlation function $\gamma_{\mathrm{IB}}(\mathbf{r})=\left\langle \rho_{\mathrm{I}}\left(\mathbf{r}^{\prime}\right)\rho_{\mathrm{B}}\left(\mathbf{r}^{\prime}+\mathbf{r}\right)\right\rangle$, at short distances $R_{0}<r\ll n^{-1/3}$ and is estimated as $C=\lim_{r\rightarrow0}\gamma_{\mathrm{IB}}(r)\frac{r^{2}}{a_{\mathrm{BB}}^{2}}(na_{\mathrm{BB}}^{3})^{2/3}$~\cite{Ardila2015}.

In the strong coupling regime and for attractive interactions, the contact found in QMC from short-range correlations has a finite maximum. In contrast, LLP-MF predicts significantly larger results. On the repulsive side, $1/(k_na)>0$, LLP-MF closely follows the Hellmann-Feynman result obtained from the QMC equation of state of the polaron, however both deviate from the contact computed directly from $\gamma_{\mathrm{IB}}(r)$. Note that the QMC estimator for the contact via the short range of the pair-correlation function is more accurate than the derivative of the equation of state. In the MIT experiment~\cite{Yan2020}, the contact in the neighborhood of the resonance is on the order of 3, while in the QMC case, it tends to $4\pm 1$. This is consistent with the agreement observed in polaron energies from the experiment and QMC. Hence, theory and experiment conclude that only a few bosons gather around the impurity at strong coupling, which is consistent with the prediction by QMC that the polaron effective mass remains small in $d=3$ around unitarity.

\section{Few-body effects and polaron physics}
\label{secFewBodyEfimov}

One of the primary motivations for studying Bose polarons comes in connection with the long-standing goal of realizing unitary Bose gases. The unitary Bose gas is a strongly correlated phase of matter in which bosons at low energies interact at the unitarity limit, where the two-particle total s-wave scattering length $a_{\rm IB}$ diverges and the scattering cross section reaches its maximal value. While realizing a stable Fermi gas at unitarity is possible due to the effective Pauli repulsion, a unitary Bose gas is unstable toward decay due to the absence of a mechanism to prevent the gas from collapsing. A more controllable situation arises when the interaction among each pair of indistinguishable bosons is non-resonant while they resonantly interact with another distinguishable species. When the density of the second species is much smaller than the boson density, the minority species behaves as non-interacting, single impurities immersed in the Bose gas. In such settings, Bose polaron models give plausible descriptions of the physics. Thus, investigating Bose polarons can provide insights into the physics of unitary Bose gases.

\subsection{Few-body Efimov states}
When bosons interact with an impurity particle close to unitarity but via a short-range potential, an exciting situation can arise where two bosons can bind to the impurity, even without any impurity-boson bound states. Here, the emergence of an effective impurity-mediated attractive interaction among bosons leads to the formation of three-body bound states similar to ``Efimov trimers" of three identical bosons. More specifically, Efimov trimers are bound states among three identical bosons that interact via short-range, nearly resonant interactions. Although the interactions are short-ranged, an effective, attractive interaction is mediated among the bosons via exchanging the third boson. The range of this effective interaction is of the order of the scattering length, which is much larger than the inter-particle potential range due to the resonant nature of the interactions. Such an effective interaction has remarkable implications on the structure and energy of the Efimov trimers, such as their Borromean character, discrete scale invariance, and universality \cite{Naidon2017}. 

The Borromean character refers to the presence of Efimov trimers in the few-body spectrum even when inter-particle interaction strengths are not strong enough to support two-body bound states. In this sense, Efimov trimers fall apart by removing a particle from the bound state, analogous to a similar property of Borromean rings. 

The other notable aspect of Efimov states is their universality, meaning that, independent of microscopic details of inter-particle interactions, the properties of Efimov states, such as binding energy or size, are controlled by a few universal parameters. This universality results from the large separation of length scales involved in low-energy scattering processes. Importantly, in addition to the scattering length $a$ and effective range $r_{\mathrm{eff}}$ which characterize the two-particle scattering amplitude, the Efimov effect depends on another universal parameter $a_{-}$, also known as ``three-body parameter". There are different ways to define the three-body parameter. Still, a standard definition is the negative scattering length at which the ground state Efimov trimer emerges from the three-body scattering continuum. The three-body parameter is also intimately related to the hyper radial boundary condition of the three-body wavefunction and, in this way, is connected to $r_{\mathrm{eff}}$, although $|a_{-}|$ is larger than $|r_{\mathrm{eff}}|$ by few orders of magnitude \cite{Naidon2017}.

Interestingly, Efimov trimers obey a discrete scaling symmetry, whereby the energy $E^{(n)}$ and three-body parameter $a^{(n)}_{-}$ of a trimer is related to its immediate excited counterpart by $E^{(n+1)}=\lambda^{-2}_{0}\,E^{(n)}$ and $a^{(n+1)}_{-}=\lambda_0 \, a^{(n)}_{-}$, where $\lambda_0$ is the Efimov scaling factor. Although the ground state and first few excited trimers typically do not follow this scaling symmetry due to the influence of microscopic details, it is obeyed by the larger excited trimers.

Similar to the few-body case, understanding the universal aspects of Bose polaron physics is of prime importance. In principle, Bose polaron energy can depend on parameters such as length scales $n^{-1/3}$ and the BEC healing length $\xi$ associated with the Bose gas, two-body scattering parameters $a$ and $r_{\mathrm{eff}}$, few-body length scales like $a_{-}$, as well as other microscopic details that are non-universal. It is therefore crucial to understand the influence of various few- and many-body phenomena and the associated length scales on Bose polaron properties. 

In the weak coupling regime, $k_n a \!\ll\! 1$, it is well understood that polaron energy $E/E_n$ universally depends on $k_n a$, where $k_n\!=\!(6 \pi^2 n_0)^{1/3}$ and $E_n\!=\!\hbar^2 k^2_n/2m$ are analogues of Fermi momentum and energy associated to the Bose gas with condensate density $n_0$, respectively. However, close to unitarity and in the highly dilute regime, few-body physics is expected to have dominant effects. The critical question, in this respect, concerns the dependence of polaron energy and other quasiparticle properties on quantities related to Efimov physics, such as $a_{-}$, or even further microscopic details, and to identify parameter regimes where Efimov physics is relevant. It is tempting to expect that Efimov physics plays a crucial role in the extremely dilute regime where $n^{1/3} |a_{-}| \lesssim 1$, and close to the unitarity of impurity-boson interaction strengths, where the conditions for Efimov physics are satisfied \cite{Naidon2017}. Another interesting question is how the properties of few-body states carry over to the Bose polaron.

\subsection{Efimov states in the Bose polaron problem}
The effect of Efimov correlations on Bose polarons was first studied by Levinsen et al. in Ref.~\cite{Levinsen2015} for equal mass boson-impurity systems, employing a trunctated basis variational (TBV) approach to account for three-body correlations exactly. Considering a weakly-interacting BEC and a two-channel model to describe impurity-boson interactions results in the Hamiltonian
\begin{equation}\label{eq:BosePolaronTowChannelModel}
\begin{split}
\hat{H} & = \sum_{\vc{k}} \, \Big[ E_{\vc{k}}\hat{\beta}^{\dagger}_{\vc{k}} \hat{\beta}_{\vc{k}} + \epsilon_{\vc{k}} \hat{c}^{\dagger}_{\vc{k}} \hat{c}_{\vc{k}} + (\epsilon^{\mathrm{d}}_{\vc{k}}+\nu_0) \hat{d}^{\dagger}_{\vc{k}}\hat{d}_{\vc{k}}\Big] \\
& + g_{\rm IB}\sqrt{n_0} \, \sum_{\vc{k}} \, \big( \hat{d}^{\dagger}_{\vc{k}} \hat{c}_{\vc{k}} + h.c. \big) \\ & + g_{\rm IB} \sum_{\vc{k},\vc{q}} \, \big( \hat{d}^{\dagger}_{\vc{q}} \hat{b}_{\vc{q}-\vc{k}} \hat{c}_{\vc{k}} + h.c. \big) \, .
\end{split}
\end{equation}
The first three terms in Eq.~\eqref{eq:BosePolaronTowChannelModel} are the kinetic energies of Bogoliubov excitations of the BEC, the impurity and the closed-channel dimer respectively, with corresponding annihilation operators $\hat{\beta}_{\vc{k}}$, $\hat{c}_{\vc{k}}$ and $\hat{d}_{\vc{k}}$, and dispersion relations $E_{\vc{k}}$, $\epsilon_{\vc{k}}$ and $\epsilon^{\mathrm{d}}_{\vc{k}}$, respectively. The remaining terms describe recombination processes of the impurity and a boson into a molecular state, where $\hat{b}_{\vc{k}}$ denotes a bare boson excitation that can be expressed through Bobogliubov operators $\hat{\beta}_{\vc{k}}$ and $\hat{\beta}^\dagger_{-\vc{k}}$.

\begin{figure}
    \centering
    \includegraphics[scale=0.5]{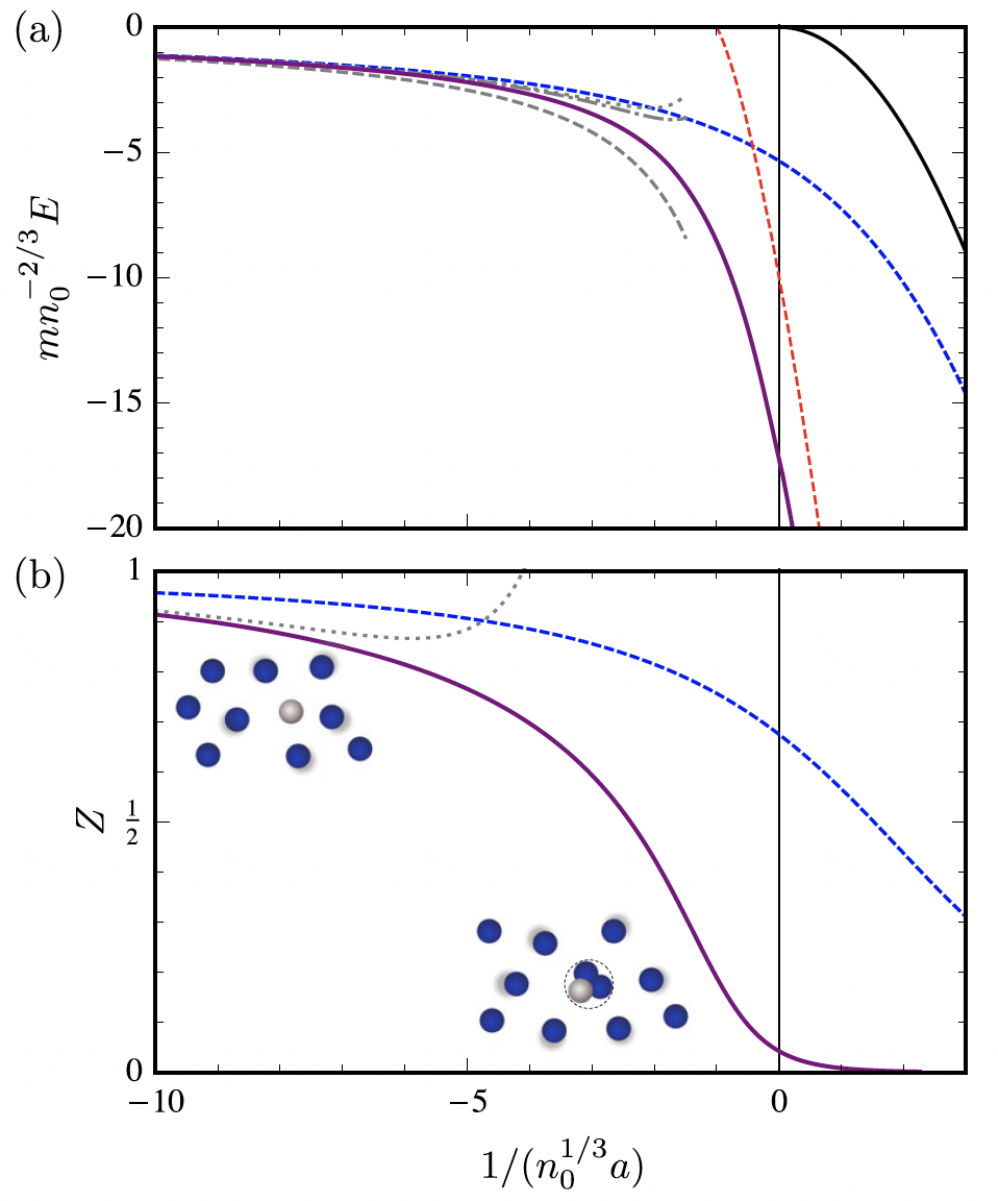}
    \caption{\textbf{Effect of Efimov correlations on Bose polarons.} (a) Energy spectrum and (b) residue $Z$ of the Bose polaron for $n_0^{1/3} a_{-} = - 1$ and $a_{\mathrm{BB}}/a_{\rm IB} = − 1/50$; note that $a_{\rm IB} \equiv a$ is denoted by $a$ in the figure. In (a) the results of variational calculations including three-body correlations (purple solid line) is compared to a Chevy-type variational ansatz which does not include three-body and effective range effects (dashed blue line). The perturbative energy results including mean-field (dashed line), second order (dotted-dashed line), and third order (dotted line) are displayed in grey. The energy of the ground state Efimov trimer in vacuum (red dashed line), and for $a > 0$ the atom-dimer continuum threshold given by the impurity-boson bound state energy $-\Big( \sqrt{1-2 r_{\mathrm{eff}}/a} - 1 \Big)^2/(2m_{\mathrm{red}}r^2_{\mathrm{eff}})$ (black solid line) are also shown. In (b), the corresponding residues $Z$ of the variational states are shown. The insets illustrate the crossover from a polaronic state to a dressed Efimov trimer. Figure is reprinted from Ref.~\cite{Levinsen2015}.}
    \label{fig:polaronEfimov}
\end{figure}

The bare parameters of the two-channel model -- the impurity-boson coupling constant $g_{\rm IB}$, the boson (impurity) mass $m_{\rm I} = m_{\rm B} \equiv m$, the closed-channel dimer detuning $\nu_0$ and a high energy momentum cut-off $\Lambda_0$ -- are tuned to reproduce the physical scattering length $a_{\rm IB}\equiv a$ and effective range $r_{\rm eff}$ via the relations
\begin{equation}\label{eq:twoChannelParams}
\begin{split}
    a = \frac{mg_{\rm IB}^2}{4 \pi} \frac{1}{\frac{m g_{\rm IB}^2 \Lambda_0}{2 \pi^2} - \nu_0} \, , \quad r_{\mathrm{eff}} = -\frac{8 \pi}{m^2 g_{\rm IB}^2} \, .
\end{split}
\end{equation}

The TBV ansatz used by Levinsen et al.~\cite{Levinsen2015} includes up to two Bogoliubov excitations of the condensate interacting with the impurity, resulting in the variational state
 \begin{equation}\label{eq:varState}
\ket{\psi}=\ket{\psi_0} + \ket{\psi_1} + \ket{\psi_2} \, ,
 \end{equation}
 where the components $\ket{\psi_{N}}$ have $N$ number of Bogoliubov excitations and take the forms
\begin{equation}\label{eq:psiComp}
\begin{split}
    & \ket{\psi_0} = \alpha_0 \hat{c}^{\dagger}_0 \ket{\emptyset} \, , \\
    & \ket{\psi_1} = \Big ( \sum_{\vc{k}} \, \alpha_{\vc{k}} \hat{c}^{\dagger}_{-\vc{k}} \hat{\beta}^{\dagger}_{\vc{k}} + \gamma_0 \hat{d}^{\dagger}_{0} \Big ) \ket{\emptyset} \, , \\
    & \ket{\psi_2} = \bigg ( \frac{1}{2}\sum_{\vc{k},\vc{q}} \, \alpha_{\vc{k}\vc{q}} \, \hat{c}^{\dagger}_{-\vc{k}-\vc{q}} \,\hat{\beta}^{\dagger}_{\vc{k}} \hat{\beta}^{\dagger}_{\vc{q}} + \sum_{\vc{k}} \, \gamma_{\vc{k}} \hat{d}^{\dagger}_{-\vc{k}}\hat{\beta}^{\dagger}_{\vc{k}} \bigg ) \ket{\emptyset} \, .
\end{split}
\end{equation}
The variational approximations to the ground state energy are obtained by minimizing $\bra{\psi}\big(\hat{H}-E\big)\ket{\psi}$. The resulting ground state spectrum and quasiparticle residue $Z$ are depicted in Fig.~\ref{fig:polaronEfimov} (a) and (b), respectively. First, it is evident from Eqs.~\eqref{eq:psiComp} that $\ket{\psi_2}$ completely includes three-body correlations of the impurity with two Bogoliubov excitations, and in the few particle limit $n_0\!\to\!0$, gives the Efimov trimers. Second, the term proportional to $\sqrt{n_0}$ in the Hamiltonian mixes the states with different particle numbers; in particular it couples the Efimov states with part of the state describing the single-excitation dressing of the impurity, resulting in a level crossing between the two states close to a scattering length $a\simeq a_{-}$. Interestingly, the energy of the polaron is significantly lowered as a result of three-body correlations even for scattering lengths where no three-body bound state exists, as shown in Fig.~\ref{fig:polaronEfimov} (a). Moreover, the energy of the attractive polaron is lower, but follows the bare Efimov trimer, suggesting that the presence of the condensate stabilizes the Efimov trimer.

\begin{figure}
    \centering
    \includegraphics[scale=0.4]{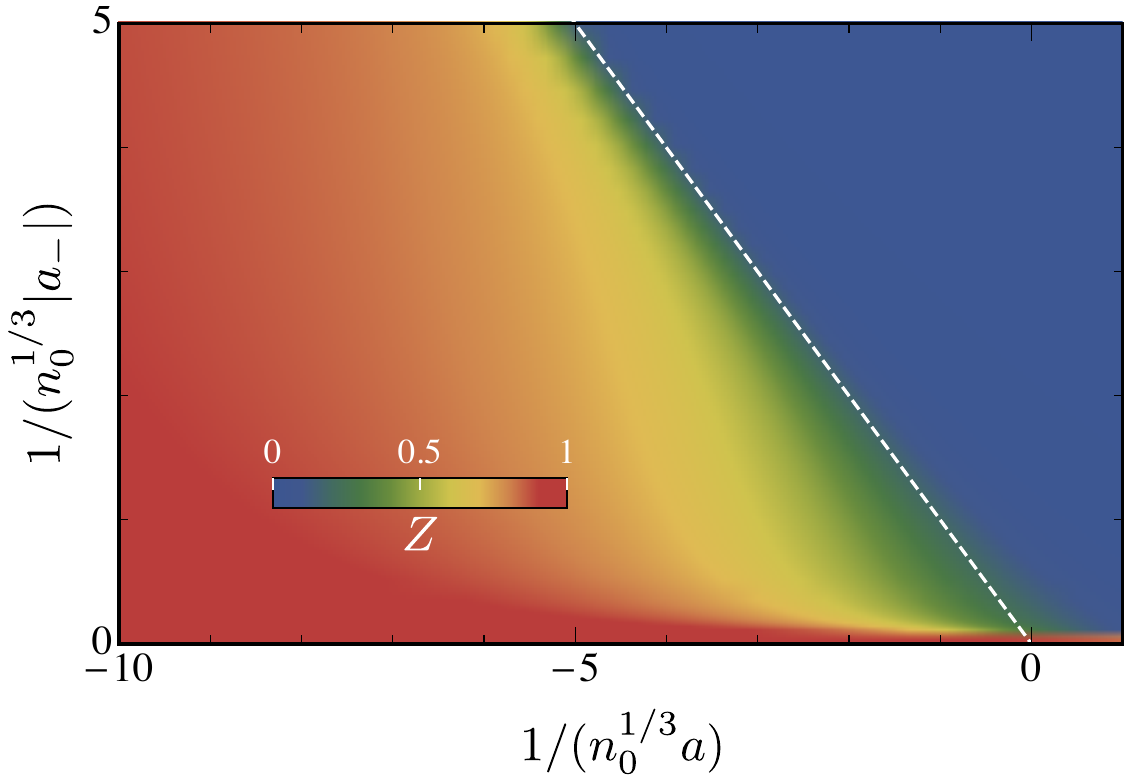}
    \caption{\textbf{Polaron quasiparticle residue} as a function of $1/(n_0^{1/3} |a_{-}|)$ and $1/(n_0^{1/3}a)$. For $n^{1/3}_0 |a_{-}| \ll 1$, the polaron remains a well-defined quasiparticle across unitarity. The crossover region is located around $a \sim a_{-}$, with its width decreasing as $1/(n^{1/3}_0 |a_{-}|)$ increases. Figure is reprinted from Ref.~\cite{Levinsen2015}.}
    \label{fig:polaronEfimovZ}
\end{figure}

The effect of Efimov correlations also causes a significant decrease in $Z$, as depicted in Fig.~\ref{fig:polaronEfimov} (b). Compared to a variational ansatz which excludes $\ket{\psi_2}$ (dashed blue line in Fig.~\ref{fig:polaronEfimov}). The dependence of $Z$ on $1/(n^{1/3}_0 |a_{-}|)$ and $1/(n^{1/3}_0 a$ is further illustrated in Fig.~\ref{fig:polaronEfimovZ}. At $1/(n^{1/3}_0 a_{-}) \ll 1$, the polaron remains a well-defined quasiparticle across unitarity into the repulsive side. In this regime, polaronic dressing contribution is much stronger than the bare Efimov energy, thus three-body effects can be neglected. On the other hand, for $1/(n^{1/3}_0|a_{-}|)\gtrsim1$, the sharp reduction of $Z$ occurs for $a\sim a_{-}$ as Efimov correlations build up. Furthermore, the crossover region around $a\sim a_{-}$ narrows by increasing $1/(n^{1/3}_0 |a_{-}|)$, as a result of scaling of the coupling strength between the trimer and the dressed impurity by $a/m_{\mathrm{red}}$, which decreases close to the crossover region $a \sim a_{-}$. The $a/m_{\mathrm{red}}$ scaling of polaron coupling to lower Efimov states further justifies neglecting the role of higher order Efimov states: the latter states emerge at scattering lengths $|a|<|a_{-}|$, making the coupling to these states smaller. 

\begin{figure}
    \centering
    \includegraphics[scale=0.4]{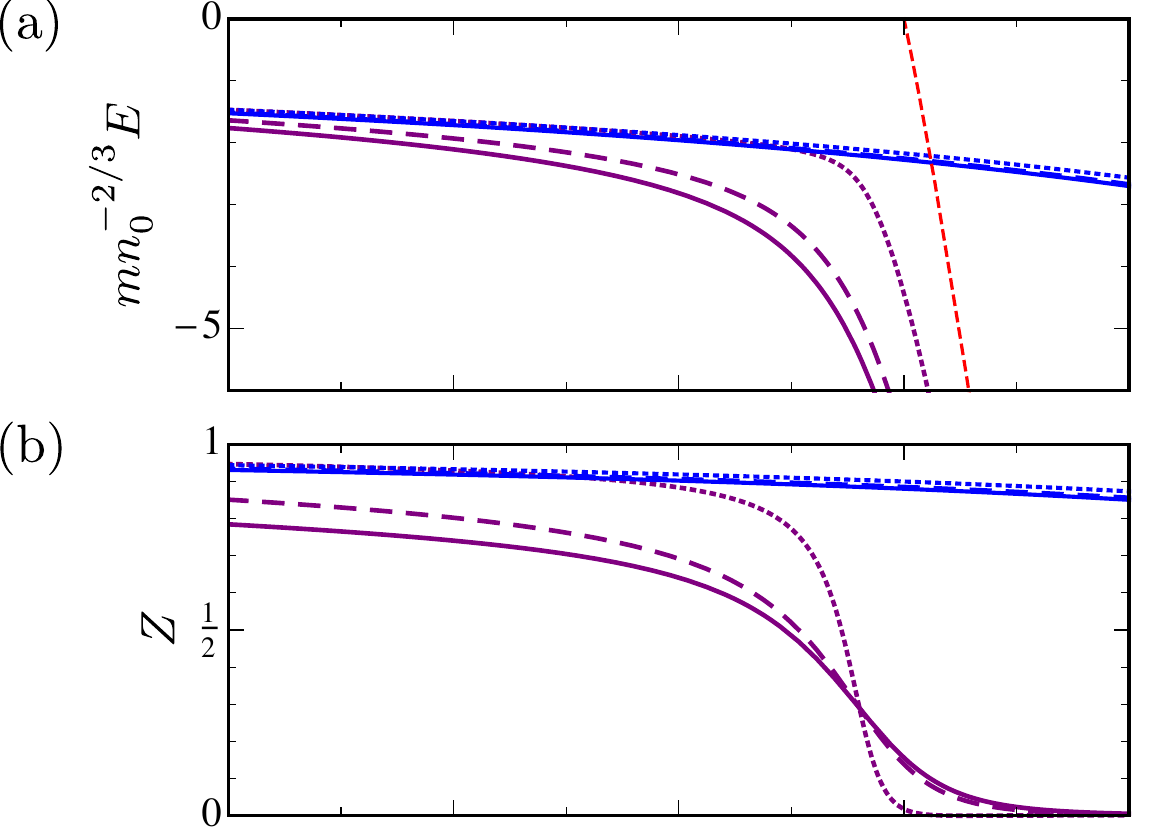}
    \caption{\textbf{Bose polaron properties for different three-body parameter values $a_-$.} Polaron energy $E$ (a), quasiparticle residue $Z$ (b) for $1/(n^{1/3}_0 a_{-}) = - 1/4$ for $a_{\mathrm{BB}}/|a_{-}|$ equal to $0.01$ (purple solid line), $0.1$ (purple dashed line), and $0.5$ (purple dotted line). Stronger boson-boson repulsion -- included within the Bogoliubov approximation -- results in suppression of Efimov correlations and the crossover becomes sharper. The effect of Efimov physics on the energy and quasiparticle residue is sizable away from the crossover region. The blue lines correspond to variational states that neglect effects due to three-body correlations and effective range. Figure is reprinted from Ref.~\cite{Levinsen2015}.} 
    \label{fig:polaronEfimovGamma}
\end{figure}

To understand the role of boson-boson repulsion within the Bogoliubov approximation, the energy and quasiparticle residue of the polaron is depicted in Fig.~\ref{fig:polaronEfimovGamma} (a)-(b) for $a_{\mathrm{BB}}/a=0.01,\,0.1$ and $0.5$. In both energy and quasiparticle residue, the crossover region shrinks for increasing $a_{\mathrm{BB}}$, indicating that the contribution of Efimov correlations is suppressed due to boson-boson repulsion.

\subsection{Universality of an impurity in a BEC}
\label{sec:universalityPolaronEfimov}
In a unitary Bose gas, Efimov trimers and higher-body bound states can form, whose ground state properties are typically non-universal and depend on microscopic length scales. As a consequence, the behavior of a unitary Bose gas can inherit this non-universality. In this regard, the impurity-boson system is suitable to study in which regimes few-body universality carries over to the many-body context.  

\begin{figure}
    \centering
    \includegraphics[scale=0.6]{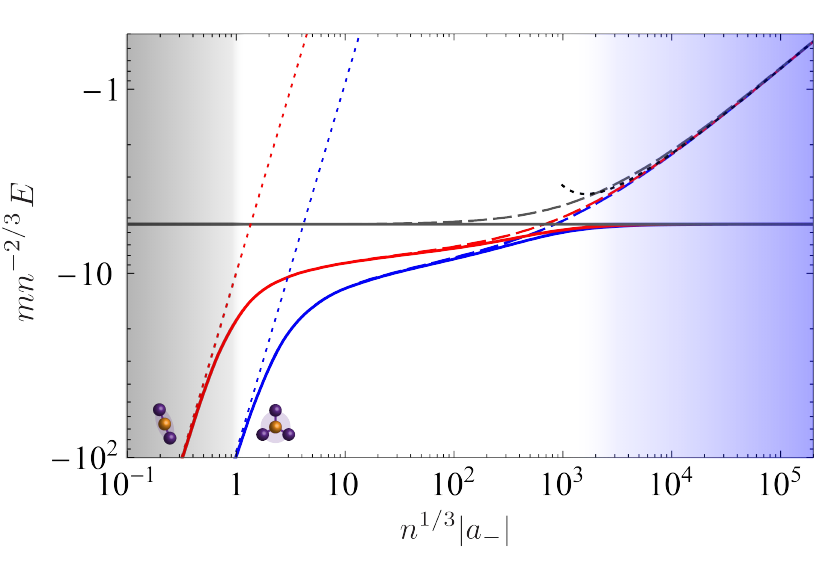}
    \caption{\textbf{Universality of Bose polarons.} Polaron energy at unitarity including up to one (grey lines), two (red lines) and three (blue lines) Bogoliubov excitations in the variational ansatz, obtained with the $r_0$ model (dashed lines) and the $\Lambda$ model (solid lines). In the low density regime $r_{\mathrm{eff}} \ll n^{-1/3}_0 \ll |a_{-}|$, the variational energies approach the ones for vacuum Efimov states, $E=-\eta/\big(m|a_{-}|^2\big)$, with $\eta=9.91$ for trimer (dotted red line) and $\eta=92.7$ for tetramer (dotted blue line). At high densities $n^{-1/3}_0 \lesssim |r_{\mathrm{eff}}|, \, \Lambda^{-1}$, the polaron energy becomes model-dependent. Thus, the results predicted by different models deviate, with the $r_0$ model approaching the high-density perturbative result. Figure is reprinted from Ref.~\cite{Yoshida2018PRX}.} 
    \label{fig:univ-polaronEfimovE}
\end{figure}

The universality of the impurity-boson system from a few-body perspective was studied by Yoshida et al.~\cite{Yoshida2018PRX} using the same model as Eq.~\eqref{eq:BosePolaronTowChannelModel} but with an extension of the variational state in Eq.~\eqref{eq:varState} to $\ket{\psi}=\ket{\psi_0}+\ket{\psi_1}+\ket{\psi_2}+\ket{\psi_3}$, where
\begin{equation}
\begin{split}
    \ket{\psi_3} & = \bigg( \frac{1}{6} \sum_{\vc{k}_1,\vc{k}_2,\vc{k}_3} \, \alpha_{\vc{k}_1\vc{k}_2\vc{k}_3} \, \hat{c}^{\dagger}_{-\vc{k}_1-\vc{k}_2-\vc{k}_3} \, \hat{\beta}^{\dagger}_{\vc{k}_1} \hat{\beta}^{\dagger}_{\vc{k}_2} \hat{\beta}^{\dagger}_{\vc{k}_3} \\ & + \frac{1}{2} \sum_{\vc{k}_1,\vc{k}_2} \, \gamma_{\vc{k}_1\vc{k}_2} \, \hat{d}^{\dagger}_{-\vc{k}_1-\vc{k}_2} \, \hat{\beta}^{\dagger}_{\vc{k}_1} \hat{\beta}^{\dagger}_{\vc{k}_2} \, \bigg) \ket{\emptyset} \, ,
\end{split}
\end{equation}
accounts for impurity interaction with three Bogoliubov excitations, and can capture polaron coupling to tetramers. The authors first investigated the few-body physics of an impurity interacting with two and three bosons, and then included the few-body physics in the many-body context of Bose polarons by means of a variational ansatz that recovers the three- and four-body equations in the limit of vanishing density. In the low density regime $r \ll n^{-1/3} \ll |a_{-}|$ where $r$ characterizes the range of interactions, Efimov physics can have sizeable effects on the Bose polaron, whereas in the high density regime $n^{1/3} r \gg 1$, polaron properties are non-universal and depend on density, see Fig.~\ref{fig:univ-polaronEfimovE}. 

To establish independence of the few-body physics with respect to short range details of the model, the authors of Ref.~\cite{Yoshida2018PRX} employ two models: (1) a model which includes effective range $r_{\mathrm{eff}}$, and relates the three-body parameters to two-body scattering properties (\emph{$r_0$ model}). (2) A model that takes $r_{\mathrm{eff}} \to 0$ but sets a high energy momentum cut-off $\Lambda$ on the momenta involved in scattering processes which leads to an effective three-body repulsion (\emph{$\Lambda$ model}). The interaction range $r$, thus, can be $|r_{\mathrm{eff}}|$ and $\Lambda^{-1}$ in the $r_0$ and $\Lambda$ model, respectively.

The relevant few-body states of the system include two Efimov trimers as well as two tetramers tied to the ground state trimer. The Efimov scale factor for the impurity-boson system, $\lambda_0\simeq1986.1$, is much larger compared to the identical boson case. The large separation between the typical length scales of Efimov physics and short range details, given by $|a_{-}|\simeq2467|r_{\mathrm{eff}}|$, as well as the large Efimov scale factor result in the universality of the few-body physics. Table I in Ref.~\cite{Yoshida2018PRX} presents the dimensionless ratios of few-body quantities for both the $r_{0}$ and $\Lambda$ models. The dimensionless ratios for both models are close to their universal values predicted by the Efimov theory. Especially, the ratio of the ground state tetramer to trimer energy at unitarity $E^{(1)}_{4\mathrm{B},-}/E_{-}$ is found to be $9.35$ and $9.43$ for $r_{0}$ and $\Lambda$ models, respectively, which is close to the value $9.7$ obtained by diffusion quantum Monte Carlo. This agreement indicates that there is no additional length scale required to set the ratio $E^{(1)}_{4\mathrm{B},-}/E_{-}$, which further supports the universality of the few-body physics.

\begin{figure}
    \centering
    \includegraphics[scale=0.6]{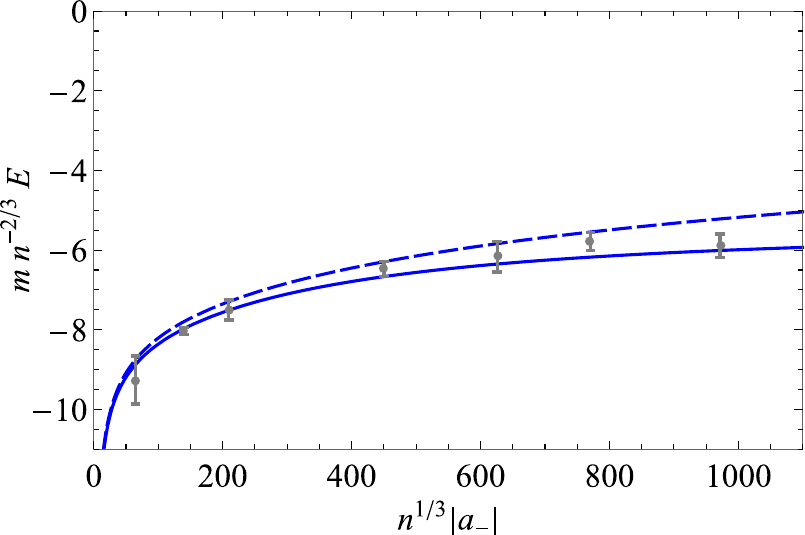}
    \caption{\textbf{Comparison of polaron energy at unitarity} obtained from the $r_0$- and $\Lambda-$ models obtained using the three-excitation ansatz. Grey dots denote Monte Carlo results by Peña Ardila et al.~\cite{Ardila2016}. The collapse of data on curves indicate the model-independence and universality of the polaron energy. The difference between the $r_0$ and $\Lambda$ models is due to $n_0^{1/3}|r_{\mathrm{eff}}|$ corrections. Figure is reprinted from Ref.~\cite{Yoshida2018PRX}.}
    \label{fig:univ-polaronEfimovQMC}
\end{figure}

In addition to the universality of few-body physics, the ground state energy in the many-body limit was found in Ref.~\cite{Yoshida2018PRX} to be universal over a wide range of densities. Interestingly, the Monte Carlo results for the polaron energy as a function of $n^{1/3}|a_{-}|$ collapse on the curves obtained from the three-excitation ansatz for both the $r_0-$ and $\Lambda-$ models, see Fig.~\ref{fig:univ-polaronEfimovQMC}. The agreement between different models suggests the universal dependence of the polaron energy on $n^{1/3}|a_{-}|$ regardless of microscopic details. 

In the low density limit, the polaron energy obeys the universal form $E=-\eta /(m|a_{-}|^2)$, where $\eta$ is a universal constant associated to each few-body state ($\eta=92.7$ for tetramer and $\eta=9.91$ for trimer for equal impurity and boson masses). As evident in Fig.~\ref{fig:univ-polaronEfimovE}, the polaron energy evaluated with variational wavefunctions including up to two (red lines) and three (blue lines) Bogoliubov excitations is model-independent for a wide range of densities, and approaches the respective Efimov state in the zero density limit. The model independence extends to densities as large as $n^{1/3} |a_{-}| \sim 10^2$. In the high density regime where $|r_{\mathrm{eff}}|,\, \Lambda^{-1} \gtrsim n^{-1/3}$, the polaron energy depends on the microscopic details, with the $r_{0}$ model approaching the perturbative result for high densities (dashed black line), and the $\Lambda$-model resembling the behavior of a single-excitation variational ansatz. 

In the high density regime $n^{1/3} r\!\gg\!1$, the inter-particle spacing becomes less than the interaction ranges, and the microscopic details of the models start to matter. Nevertheless, the polaron energies obtained using the $r_{0}$ model approach the perturbation theory result (dotted black line in Fig.~\ref{fig:univ-polaronEfimovE})\cite{Gurarie2007}. To the lowest order in the coupling constant $g_{\mathrm{IB}}$ and assuming weak boson-boson interactions ($n a^{3}_{\mathrm{BB}} \ll a_{\mathrm{BB}}/|r_{\mathrm{eff}}| \ll 1$), the ground state polaron energy at unitarity is bounded from below and is given by
\begin{equation}\label{eq:EpolHighDens}
E \simeq -\frac{1}{m} \sqrt{\frac{8 \pi n}{|r_{\mathrm{eff}}|}} + \frac{1}{m} \sqrt{\frac{3}{7}} \bigg( \frac{8 \pi n}{|r_{\mathrm{eff}}|^5} \bigg)^{1/4} \, .
\end{equation}
The finiteness of polaron energy results from the fact that scattering of a single boson into the closed channel molecule prevents other bosons from interacting with the impurity, leading to an effective boson-boson repulsion which restricts the boson density localized around the impurity. This important effect is further discussed later in more detail.

In addition to polaron energy, other quasiparticle properties were also characterized in Ref.~\cite{Yoshida2018PRX}. The residue $Z$ vanishes in the low density limit, as the polaron approaches few-body states with vanishing overlap with the non-interacting state. In the high density limit, the residue starts to depend on the model and the number of excitations in the variational ansatz, making $Z$ less universal than the energy. In fact, $Z$ depends on an infrared cut-off length scale $\ell_{\rm IR}$ according to
\begin{equation}\label{eq:ZhighDens}
    Z^{-1} = 2 - \sqrt{\frac{3}{7}} \frac{2}{\big( 8 \pi n |r_{\mathrm{eff}}|^3 \big)^{1/4}} + \frac{\ell_{\rm IR}}{|r_{\mathrm{eff}}|} \, .
\end{equation}
For a BEC, the low-energy length scale $\ell_{\rm IR}$ is proportional to $\xi$. Thus, the residue vanishes as $\xi \to \infty$ (non-interacting bosons), a behavior that requires an infinite number of excitations in the variational ansatz to account for the impurity dressing. 

Similar to the residue, the behavior of the effective mass is non-universal for high densities, while it approaches the mass of few-body bound states at low densities. For the contact $C$, the results of few-excitation variational states at unitarity approach the perturbative result at high densities for the $r_0$ model. At low densities, the contact shows less dependence on the model and approaches the universal few-body values, with $C|a_{-}|\simeq171.4\,(\simeq172.2)$ for the trimer and $C|a_{-}|\simeq535.5\,(\simeq542.6)$ for the tetramer within the $r_0$ model ($\Lambda$ model). 

\subsection{Static impurity in a BEC: few- and many-body physics}
\label{sec:MutiBodyRes}
Similar to their mobile impurity counterparts, static impurities in BECs are also experimentally relevant. They can be realized by immersing impurity atoms much heavier than bosons or confining the impurities using species- or state-dependent optical traps. Such static impurity-boson systems exhibit interesting few- and many-body physics that can, in some cases, be markedly different from mobile impurity systems. 

\begin{figure}
    \centering
    \includegraphics[scale=0.32]{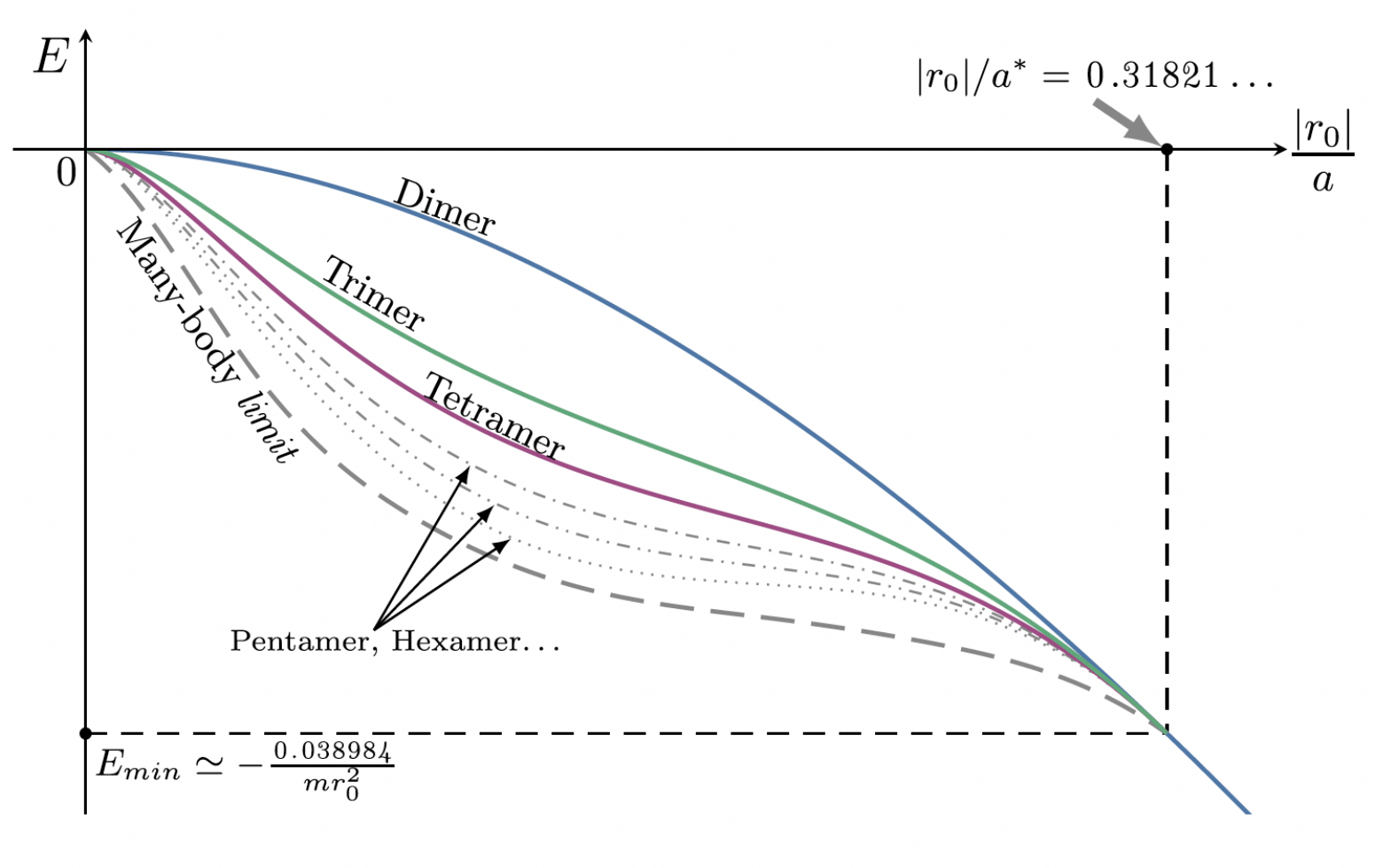}
    \caption{\textbf{Schematic energy spectrum of the multi-body bound states} of the $(N+1)$-body system consisting of a static impurity coupled to $N$ non-interacting bosons. The $N$-body bound states appear as $1/a\to 0^{+}$, and dissolve into the atom-dimer continuum as $a$ approaches $a^{*}=|r_{\mathrm{eff]}}|/0.31821\cdots$, where $r_{\mathrm{eff}}$ is the impurity-boson effective range. The binding energies increase with $N$, while the many-body limit $N\to\infty$ is well-defined. Figure is reprinted from Ref.~\cite{Shi2018Multibody}.}
    \label{fig:MBR_sch}
\end{figure}

The few-body states of an $(N+1)$-body system consisting of a static impurity coupled to $N$ non-interacting bosons were characterized Yoshida et al.~\cite{Yoshida2018PRA,Yoshida2018PRX} and Shi et al.~\cite{Shi2018Multibody} by introducing a simpler variant of the two-channel model, 
\begin{equation}\label{eq:bosonicAndersonModel}
\begin{split}
    \hat{H} & = \sum_{\vc{k}} \, \epsilon_{\vc{k}} \hat{b}^{\dagger}_{\vc{k}} \hat{b}_{\vc{k}} + \nu_0 \, \hat{d}^{\dagger} \hat{d} + g_{\mathrm{IB}} \sum_{\vc{k}} \, \Big ( \hat{d}^{\dagger} \, \hat{b}_{\vc{k}} + h.c. \Big ) \\ & + \frac{U}{2} \, \hat{d}^{\dagger}\hat{d}^{\dagger}\hat{d}\hat{d} \, , 
\end{split}
\end{equation}
that resembles the Anderson impurity model and thus is called \emph{the (bosonic) Anderson model}. The Anderson model maps to the two-channel model with Hamiltonian
\begin{equation}
    \hat{H}_{\rm 2ch}  = \sum_{\vc{k}} \, \epsilon_{\vc{k}} \hat{b}^{\dagger}_{\vc{k}} \hat{b}_{\vc{k}} + \nu_0 \, \hat{d}^{\dagger} \hat{d} + g_{\mathrm{IB}} \sum_{\vc{k}} \, \Big ( \hat{d}^{\dagger} \hat{c} \, \hat{b}_{\vc{k}} + h.c. \Big )
\end{equation}
in the limit $U \to \infty$, where $U$ is an on-site boson-boson repulsion that prevents double occupancy of the localized closed-channel dimer. Within this bosonic Anderson model, the system demonstrates the emergence of $N$-body bound states for $1/a\to0^{+}$ and $1/a\to1/(a^{*}+0^{+})$, where $a^{*}=|r_{\mathrm{eff}}|/0.31821\cdots$ and $r_{\rm eff}$ is the effective range. The two critical scattering lengths $a=+\infty$, respectively, $a=a^{*}$, correspond to multi-body resonances, where all the $N$-body bound states simultaneously appear in the spectrum, respectively, dissolve into the atom-dimer continuum, see Fig.~\ref{fig:MBR_sch}.

In the vicinity of the $a=+\infty$ critical point, the $N$-boson state has energy
\begin{equation}\label{eq:unitarityNbosonMBR}
    E_{N+1} = \bigg ( -N + N(N-1)\frac{\pi}{\mathrm{ln}\big(a/r\big)} \bigg ) \, E_{\mathrm{B}} \, ,
\end{equation}
where $E_{\mathrm{B}}$ is the two-body binding energy, and $r$ is a length scale that depends on the detailed mechanism underlying the effective boson-boson repulsion. In the limit $a \to a^{*}$, the $(N+2)$-body system consisting of the impurity and $N+1$ bosons has the ground state energy 
\begin{equation}\label{eq:astrEnpl2}
    E_{N+2} \simeq \, - E_{\mathrm{B}} + \bigg ( - N + N(N-1) \frac{\pi}{\mathrm{ln}(a_{\mathrm{ad}}/r)} \bigg ) \, E^{(\mathrm{ad})}_{\mathrm{B}}\, ,
\end{equation}
where $a_{\mathrm{ad}}$ and $E^{(\mathrm{ad})}_{\mathrm{B}}$ are the atom-dimer scattering length and bound state energy, respectively. Interestingly, there is a correspondence between $a^{*}$ and the unitarity critical points: the physics of the $(N+2)$-body system is formally equivalent to the physics of $(N+1)$ bosons scattering off the impurity-boson dimer. In fact, the atom-dimer scattering length $a_{\mathrm{ad}}$ diverges at $a=a^{*}$, indicating that $a^{*}$ corresponds to the unitarity limit of the atom-dimer scattering.

\begin{figure}
    \centering
    \includegraphics[scale=0.21]{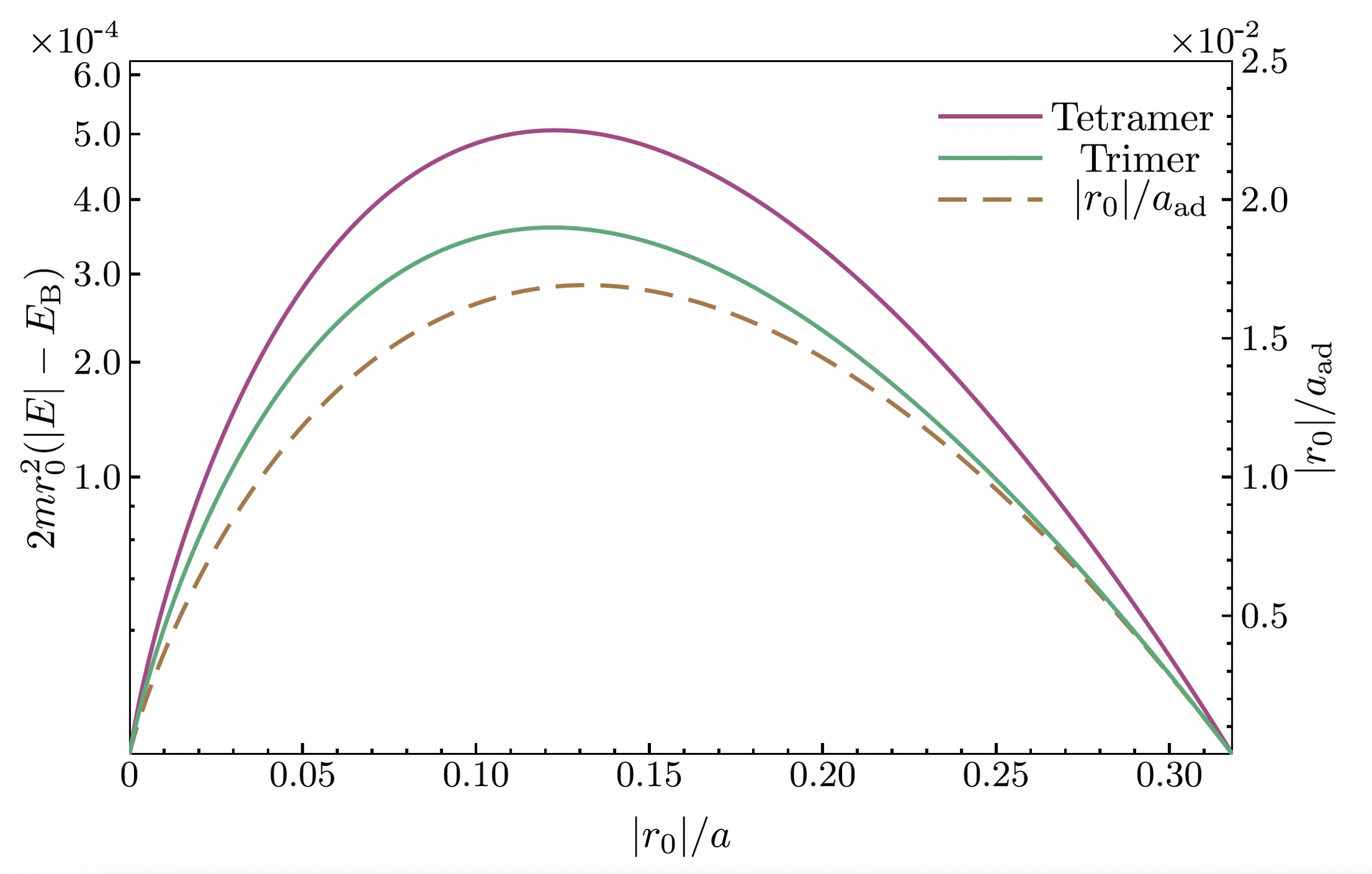}
    \caption{\textbf{Trimer and tetramer binding energy} and atom-dimer scattering length for $0<1/a<1/a^{*}$. The trimer energy is close to the universal value $1/\big(2ma^{2}_{\mathrm{ad}}\big)$, reflecting the approximate universality of the trimer state. The atom-dimer scattering length diverges at $a=a^{*}$. Figure is reprinted from Ref.~\cite{Shi2018Multibody}.}
    \label{fig:MBR_1}
\end{figure}

\begin{figure}
    \centering
    \includegraphics[scale=0.31]{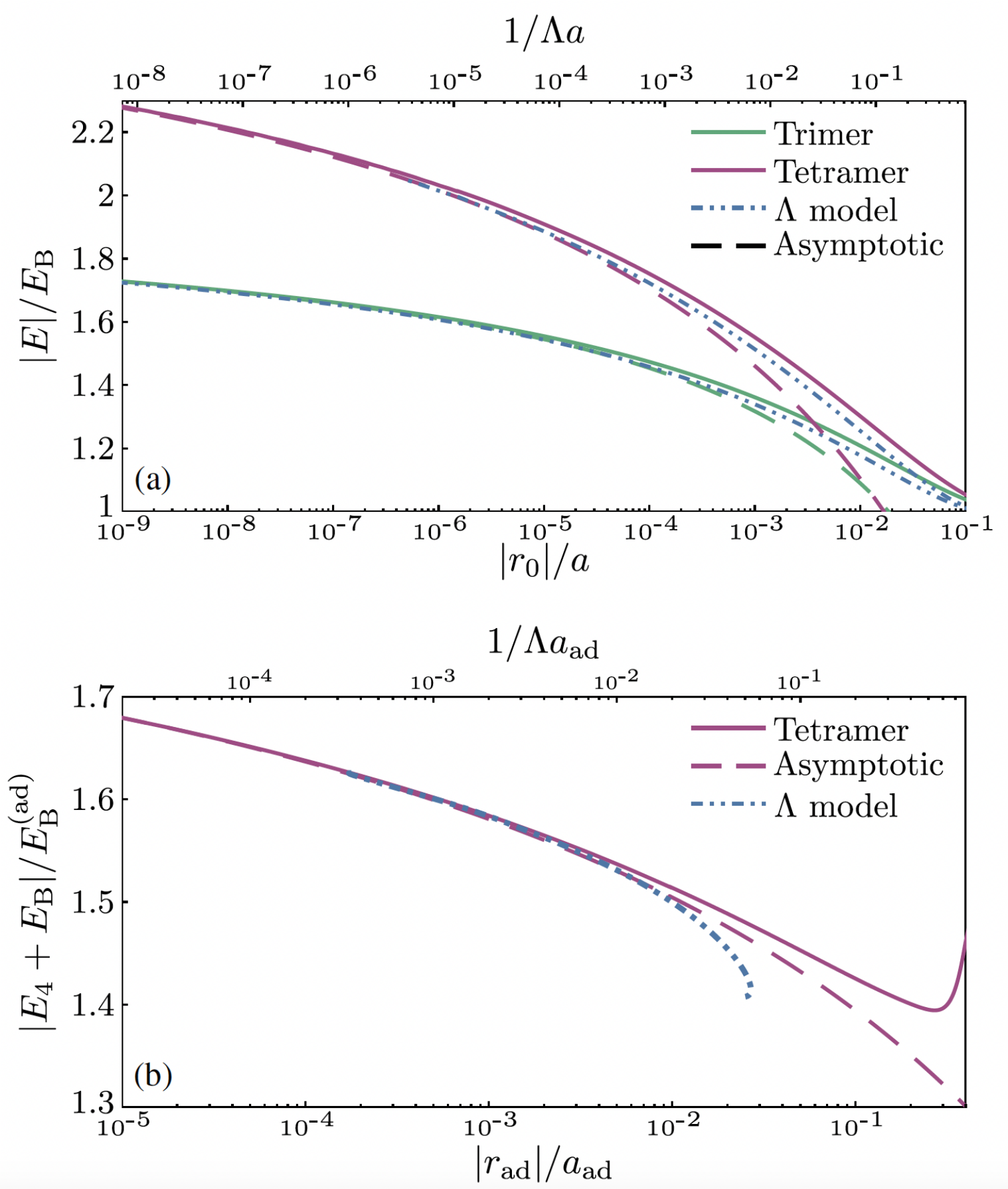}
    \caption{\textbf{Asymptotic behavior of trimer and tetramer energies} near (a) $a=+\infty$ and (b) $a=a^{*}$, obtained from the $r_0$ model (solid lines), the $\Lambda$ model (dashed-dotted lines), and the asymptotic forms in Eqs.~\eqref{eq:unitarityNbosonMBR}, \eqref{eq:astrEnpl2}. Figure is reprinted from Ref.~\cite{Shi2018Multibody}.}
    \label{fig:MBR_2}
\end{figure}

Although it is conjectured in Ref.~\cite{Shi2018Multibody} that all $N$-body bound states exist for $0<1/a<1/a^{*}$, the energies of trimer and tetramer states were obtained using an exact analytical method. In Fig.~\ref{fig:MBR_1}, the binding energies of the trimer and tetramer are depicted. Interestingly, the trimer energy closely follows the universal relation $1/\big(2m a^{2}_{\mathrm{ad}}\big)$, signalling the universality of the trimer energy. To test the universality of the binding energies, the trimer and tetramer energies are depicted in Fig.~\ref{fig:MBR_2} for $a\to +\infty$ and $a\to a^{*}$, where the predictions of the $r_0$ model agrees well with the $\Lambda$ model, and reproduce the analytical results in Eqs.~\eqref{eq:unitarityNbosonMBR} and \eqref{eq:astrEnpl2}.

The disappearance of $N$-body bound states at $a^{*}$ further implies the boundedness of the ground state energy in the thermodynamic limit. This is the consequence of the fact that the contact $C \propto \partial E_{N+1}/\partial(-1/a)$ is positive. The lower bound is given by the dimer energy at $a^{*}$, i.e. $E_{N+1} \geq - 0.038984/ m r^2_{\mathrm{eff}}$ for all $N$ and $a\leq a^{*}$. The existence of a well-defined ground state in the thermodynamic limit pertains to the fact that the scattering of a single boson into the closed-channel dimer prohibits other bosons from scattering off the impurity.

In the many-body limit at finite boson densities, the size of multi-boson bound states becomes larger than the inter-particle density at a critical boson number $N_V$, and higher-body bound states will break apart. An estimate of $N_V$ is obtained by noting that the size of the $(N+1)$-body bound state $l_{N+1} \sim (m|E_{N+1}-E_{N}|)^{-1/2}$, thus $m|E_{N_V+1}-E_{N_V}|\! \sim\!(N_V/V)^{2/3}$. 

\subsection{Quantum blockade of the localized impurity from inter-boson interactions}
\label{sec:qblock}
The Anderson model considered in Refs.~\cite{Shi2018Multibody,Yoshida2018PRA} to characterize the few-body states of the $(N+1)$-system neglects boson-boson interactions. Nevertheless, boson repulsion can significantly affect the behavior of the system. Indeed, in a non-interacting Bose gas, the ground state is realized when all bosons occupy the lowest-energy single-particle state of the system. Thus, for impurity-boson potentials supporting a bound state, the resulting non-interacting ground state energy does not have a well-defined thermodynamic limit, signaling the essential role of boson repulsion to stabilize the system. 

The influence of boson repulsion on the $(N+1)$-body physics described in Sec.~\ref{sec:MutiBodyRes} was studied by Levinsen et al.~\cite{Levinsen22} using quantum Monte Carlo (QMC). On the few-body level, the QMC results reveal that the trimer and tetramer states disappear for $a<a^{*}$, where $a^{*}\simeq10.0\,a_{\mathrm{BB}}$ is a critical scattering length. Furthermore, the QMC energies for both the trimer and tetramer states agree with the predictions of the Anderson model, when the effective range of the Anderson model $r_{\mathrm{eff}}\simeq-3\,a_{\mathrm{BB}}$, to match the $a^{*}$ obtained from QMC. The agreement of the Anderson model with QMC further indicates the dominant effect of impurity blocking by a single boson. 

\begin{figure}
    \centering
    \includegraphics[scale=0.27]{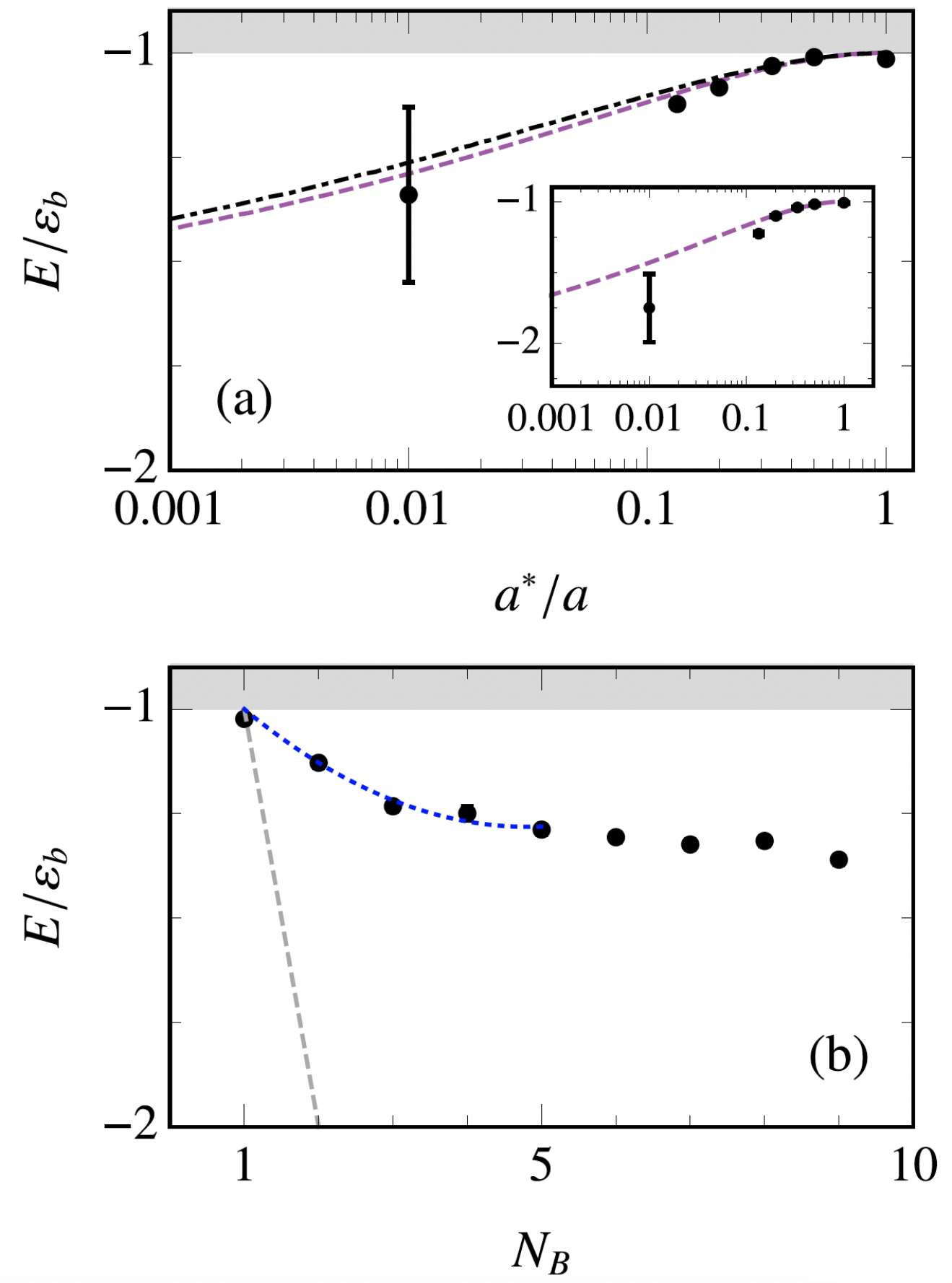}
    \caption{\textbf{Quantum blockade of a localized impurity.} (a) Trimer binding energy of an impurity bound to two bosons with attractive contact interaction (black dashed-dotted line), non-interacting bosons within the bosonic Anderson model of Refs.~\cite{Shi2018Multibody,Yoshida2018PRA} (purple dashed line), and QMC data (black dots). The inset shows the comparison of tetramer binding enegy obtained from QMC and the bosonic Anderson model. The few-body states cease to exist for $a>a^{*}$, where $a^{*}\simeq|r_{\mathrm{eff}}|/0.31821\cdots$ is the critical scattering length of multi-body resonances (see text). (b) The boson number dependence of the ground state energy obtained from QMC for $a/a_{\mathrm{B}}=75$. For $N_{\mathrm{BB}} \leq 5$, the energy agrees well with Eq.~\eqref{eq:FewBodyEnergy} (blue dashed line). Figure is reprinted from Ref.~\cite{Levinsen22}.}
    \label{fig:quantumBlockingFewBody}
\end{figure}

Due to the single-boson blocking of the impurity, the rest of the bosons scatter from the effective potential of the impurity-boson dimer, which has a longer range $\sim a$. The QMC data provides evidence for this picture, as the $N$-body bound state energies for $N\leq 5$ satisfy the relation 
\begin{equation}\label{eq:FewBodyEnergy}
E_{N}= - \varepsilon_{\mathrm{B}} - (N-1) \, \varepsilon_{\mathrm{T}} + U/2 \, (N-1)(N-2) \, ,
\end{equation}
in the limit $a \gg a_{\mathrm{BB}}$ and $\varepsilon_{\mathrm{B}} \ll \varepsilon_{\mathrm{T}}$, see Fig.~\ref{fig:quantumBlockingFewBody}. In Eq.~\eqref{eq:FewBodyEnergy}, $\varepsilon_{\mathrm{T}}$ is the trimer binding energy and $U \sim a_{\mathrm{BB}}/m a^{3}$ is an effective boson repulsion strength. The boson number dependence of energy in Eq.~\eqref{eq:FewBodyEnergy} indicates that $N-1$ bosons occupy the boson-dimer bound state, further supporting the single-boson blocking picture.

In the many-body limit at finite boson density, the polaron state takes the general form
\begin{equation}\label{eq:polState}
\begin{split}
    \ket{\Psi} = \Big ( \alpha_0 & + \sum_{\vc{k}}\,\alpha_{\vc{k}}\,\hat{b}^{\dagger}_{\vc{k}} \\ & + \frac{1}{2} \sum_{\vc{k}_1,\vc{k}_2}\,\alpha_{\vc{k}_1\vc{k}_2}\,\hat{b}^{\dagger}_{\vc{k}_1} \hat{b}^{\dagger}_{\vc{k}_2} + \cdots \Big ) \ket{\emptyset}\, .
\end{split}
\end{equation}
Requiring that states of the form in Eq.~\eqref{eq:polState} be eigenstates of the Anderson model Hamiltonian yields the polaron ground state energy to leading order in $n^{1/3}\,a_{\mathrm{B}}\ll1$,
\begin{equation}\label{eq:EpolLeadOrder}
    E = n \Bigg [ \frac{m}{2 \pi a} + \sum_{\vc{k}} \, \bigg( \frac{1}{\epsilon_{\vc{k}}+G_{\vc{k}}} - \frac{1}{\epsilon_{\vc{k}}}\bigg) \Bigg ]^{-1} \, .
\end{equation}
In Eq.~\eqref{eq:EpolLeadOrder}, the effect of spatial boson-boson correlations on the polaron energy is captured by the function $G_{\vc{k}}$,
\begin{equation}\label{eq:Gk}
    G_{\vc{k}}=g \sqrt{n} \, \bigg( \sum_{\vc{k}'} \, \alpha_{\vc{k}\vc{k}'}/\alpha_{\vc{k}} - \sum_{\vc{k}'} \, \alpha_{\vc{k}'}/\alpha_0 \bigg) \, .
\end{equation}
Thus, in the extremely dilute regime ($n^{1/3}\,a_{\mathrm{B}}\ll1$), boson-boson correlations in the polaron state can significantly affect the polaron energy. In particular, it signifies the necessity for applying a method that takes into account the two-boson correlations exactly, such as a truncated basis variational ansatz. 

\begin{figure}
    \centering
    \includegraphics[scale=0.18]{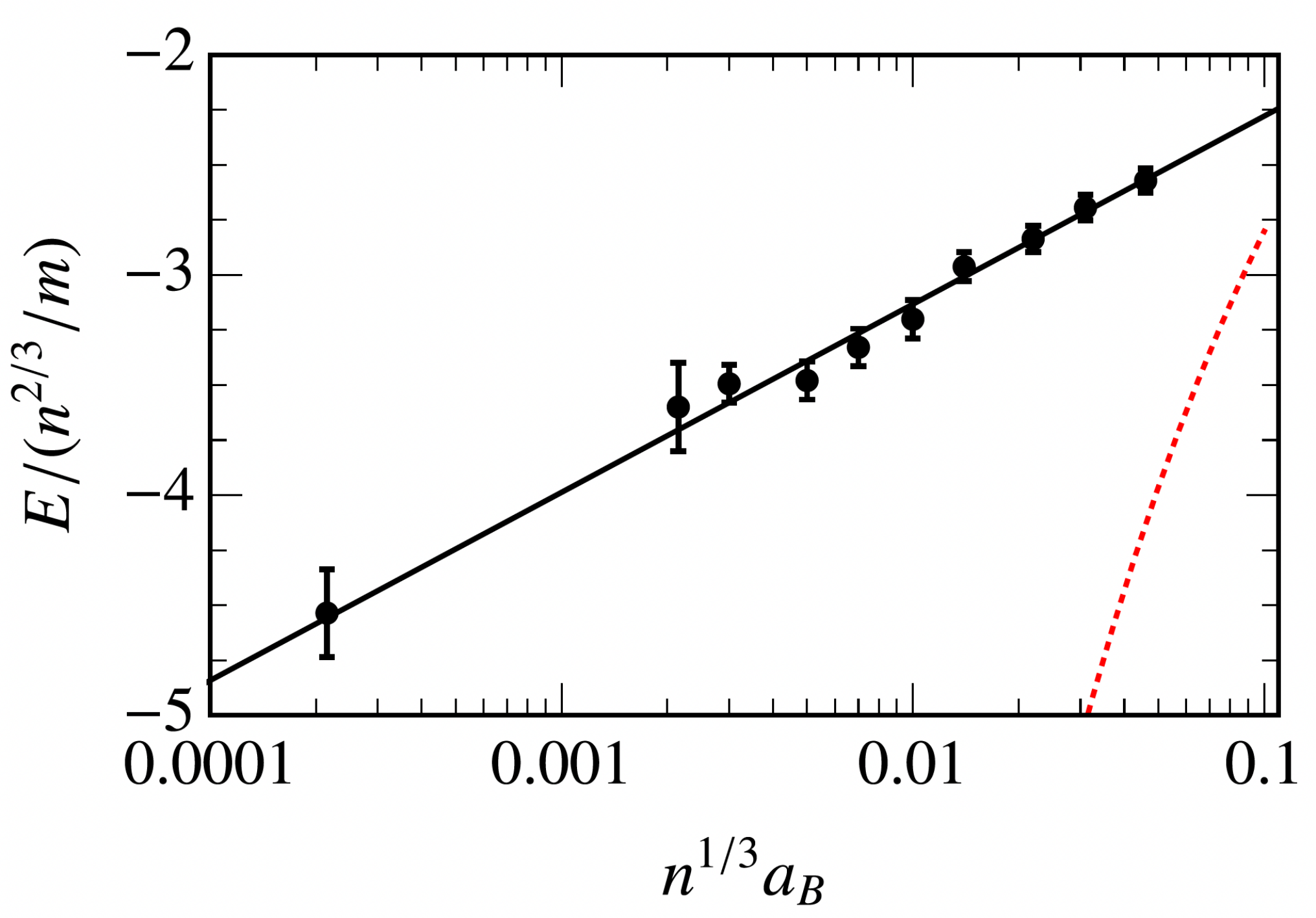}
    \caption{\textbf{Energy of a localized impurity at unitarity.} The QMC data (dots) shows a logarithmic dependence of the polaron energy on $a_{\mathrm{BB}}$ (denoted $a_{\mathrm{B}}$ in the figure) of the form $E^{\mathrm{QMC}}\simeq 0.37\,\mathrm{ln}\big(0.022 \, n^{1/3}a_{\mathrm{BB}}\big) n^{2/3}/m$. The red dashed line represents the polaron energy assuming a coherent state ansatz of Bogoliubov excitations. Figure is reprinted from Ref.~\cite{Levinsen22}.}
    \label{fig:quantumBlockingU}
\end{figure}

The effect of boson-boson interactions is especially notable at unitarity, where the polaron energy takes the universal form
\begin{equation}
    E=-f\big(n^{1/3}\,a_{\mathrm{BB}}\big)\,n^{2/3}/m \, ,
\end{equation}
The QMC data for polaron energy in the extreme dilute regime is compatible with a log-dependence of the form $E=0.37\,\mathrm{ln}\big(0.022\,n^{1/3}\,a_{\mathrm{BB}}\big)\,n^{2/3}/m$, see Fig.~\ref{fig:quantumBlockingU}.

\subsection{Interplay of polaronic dressing and Efimov physics: Light impurities}
\label{sec:polaronEfimov}
The Efimov effect also leads to the formation of large Efimov clusters involving many bosons bound via impurity-induced attraction. Such Efimov clusters form the ground state of the impurity-boson system for interaction strengths beyond what is necessary to form the largest cluster. In these settings, polaron states exist as metastable states which can decay to large Efimov clusters. 

A suitable class of variational states to capture the physics of Efimov clusters are Gaussian states of the form introduced in Sec.~\ref{sec:NonGaussianState}. The advantage of Gaussian states is their ability to account for correlations arising from impurity-induced inter-boson interactions and an arbitrary number of excitations in the polaron cloud while remaining computationally feasible. More specifically, Gaussian states capture Efimov correlations since the expectation value of the impurity-mediated interaction in the LLP Hamiltonian acquires non-zero values that can reduce the energy upon correlating the bosons. Furthermore, the generic Gaussian state is a superposition of even particle number states, thus containing a fluctuating number of particles with a mean excitation number $\langle \hat{N} \rangle$ that can be fixed by introducing a chemical potential term $\mu_{N}\hat{N}$ in the Hamiltonian and tuning $\mu_{N}$ to fix $\langle \hat{N} \rangle$ \cite{ChristianenPRA_2022}. Thus, Gaussian states can include many-body dressing effects underlying polaron physics. 

\begin{figure}
    \centering
    \includegraphics[scale=0.35]{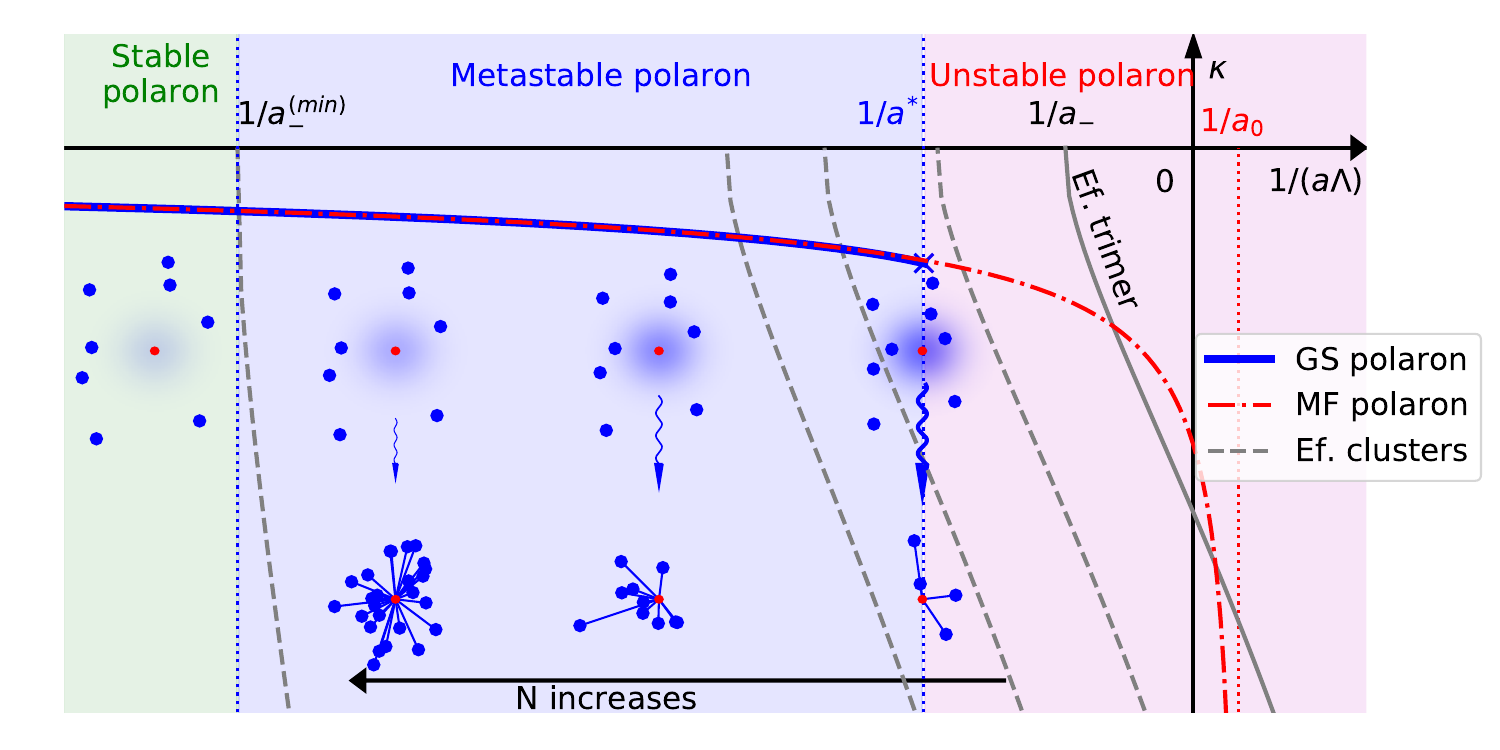}
    \caption{\textbf{Many-body spectrum of the beyond Fr\"ohlich model} (schematic) obtained from a Gaussian state variational scheme. The characteristic wave vector $\kappa=-\sqrt{M|E|}/\Lambda$ of the polaron is a function of $1/(a\Lambda)$, where $\Lambda$ is a UV momentum cut-off. The first Efimov cluster emerges at $a^{(\mathrm{min})}_{-}$. For $|a|<|a^{(\mathrm{min})}_{-}|$, the polaron is a stable state, while for $|a|>|a^{(\mathrm{min})}_{-}|$ it is metastable and can decay to the low-lying Efimov clusters. At $a=a^{*}$, the polaron ceases to exist as a metastable state. Figure reprinted from Ref.~\cite{ChristianenPRA_2022}.}
    \label{fig:GaussEfimov_intro}
\end{figure}

In Refs.~\cite{ChristianenPRL_2022,ChristianenPRA_2022}, Christianen et al. studied the interplay of the Efimov effect and polaron physics in systems comprising light impurities where the Efimov effect is more prominent. For the experimentally relevant case of $^{6}$Li impurities immersed in $^{133}$Cs condensates with impurity boson mass ratio $m_{\rm I}/m_{\rm B}=6/133$, they demonstrated that the build-up of Efimov correlations can lead to instability of the polaron, see Fig.~\ref{fig:GaussEfimov_intro}. The latter can also be interpreted as a many-body shift of the Efimov resonance. To model the impurity-BEC system close to a broad Feshbach resonance, Refs.~\cite{ChristianenPRL_2022,ChristianenPRA_2022} considered a mobile impurity interacting via a single-channel potential with Bogoliubov excitations of a weakly interacting Bose gas as in Sec.~\ref{secExtdBogoFroh}, including beyond Fr\"ohlich terms. They used Gaussian variational wavefunctions to describe the polaronic states, and obtained the variational ground state energies fixing the mean excitation number $\langle N_{\mathrm{ex}}\rangle$ for scattering lengths $a<0$. 

Christianen et al.~\cite{ChristianenPRL_2022,ChristianenPRA_2022} found that many-body bound states always exist for scattering lengths $|a|>|a^{(\mathrm{min})}_{-}|$, where $a^{(\mathrm{min})}_{-}$ is the smallest scattering length for which an Efimov cluster can exist. Thus, for $|a|<|a^{(\mathrm{min})}_{-}|$, the attractive polaron is the minumum-energy stable state, while for $|a|>|a^{(\mathrm{min})}_{-}|$, it appears as a metastable state corresponding to a local minimum of the energy functional. However, further Gaussian states with $\langle \hat{N}\rangle$ different than the polaron can exist with energies arbitrarily lower than the polaron energy, representing large Efimov clusters. For increasing $|a|$, the mean excitation number of clusters with the same energy as the polaron state decreases, and eventually, at a critical scattering length $a^{*}$, the polaron becomes unstable, see Fig.~\ref{fig:GaussEfimov_intro}.

\begin{figure}
    \centering
    \includegraphics[scale=0.55]{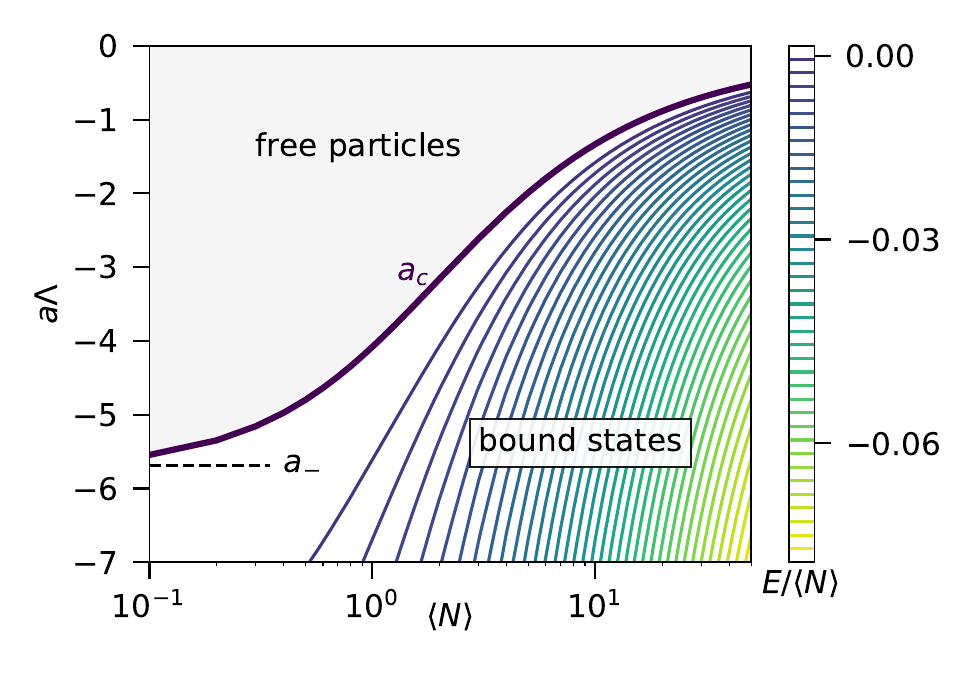}
    \caption{\textbf{Contour plot of variational energy} per excitation $E/\langle \hat{N}\rangle$, as a function of $a\Lambda$ and $\langle \hat{N}\rangle$, in the zero condensate density limit. The critical scattering length at which bound states emerge ($E<0$) is denoted by $a_{c}$. The increase in $|E|/\langle \hat{N}\rangle$ by increasing $\langle \hat{N}\rangle$ signifies the cooperative binding effect. For $\langle \hat{N}\rangle \ll 1$, $a_{c}$ approaches the few-body limit $a_{-}$, where the first Efimov trimer emerges from the continuum. Figure is reprinted from Ref.~\cite{ChristianenPRA_2022}.}
    \label{fig:coopBind}
\end{figure}

To understand the polaron instability, it is instructive to address the few-body physics within the framework of Gaussian state theory. In the zero-condensate density limit, Gaussian cluster states exist with negative energies for all scattering lengths beyond a certain critical value $a_c(\langle \hat{N}\rangle)$, corresponding to multi-body bound states, see Fig.~\ref{fig:coopBind}. For $\langle \hat{N}\rangle \ll 1$, $a_c$ approaches the Efimov three body parameter, as the main contribution to the binding energy comes from the Efimov trimer. For constant $a\Lambda$, bound states exist for all $\langle \hat{N}\rangle$ larger than $\langle \hat{N}\rangle$ at $a=a_c$. Interestingly, the binding energy per particle $E/\langle \hat{N}\rangle$ increases with $\langle \hat{N}\rangle$, which can be understood as a \textit{cooperative binding effect}. Christianen et al. also revealed signatures of the cooperative binding effect in the ratio of the boson-impurity binding energy to the interaction energy, $E^{(\mathrm{bos})}_{\mathrm{kin}}/E_{\mathrm{int}}$ and $E^{(\mathrm{imp})}_{\mathrm{kin}}/E_{\mathrm{int}}$, respectively~\cite{ChristianenPRA_2022}.

\begin{figure}
    \centering
    \includegraphics[scale=0.54]{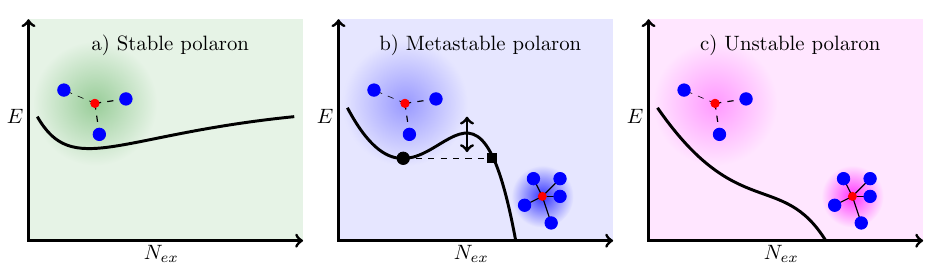}
    \caption{\textbf{Gaussian state energy landscape} (qualitative behavior), as a function of $N_{\mathrm{ex}}=\langle \hat{N}\rangle$, for different scenarios depicted in Fig.~\ref{fig:GaussEfimov_intro}. (a) For $|a|\!<\!|a^{(\mathrm{min})}|$, the polaron is the stable ground state, associated with the minimum of the energy landscape. (b) For $|a^{(\mathrm{min})}|\!<\!|a|\!<\!|a^{*}|$, the polaron corresponds to a local minumum, separated by an energy gap from states with increasing excitation number and decreasing energy, representative of Efimov clusters. (c) For $|a^{*}|<|a|$, the local minimum disappears, corresponding to the polaron instability. Figure is reprinted from Ref.~\cite{ChristianenPRA_2022}.}
    \label{fig:energyLand}
\end{figure}

In the presence of a finite condensate density, the energy landscape over the Gaussian states changes qualitatively, and a local minimum appears corresponding to the attractive polaron, see Fig.~\ref{fig:energyLand}. Thus, the attractive polaron appears as a metastable state, with a finite potential barrier preventing its decay to lower energy Efimov states. Christianen et al.~\cite{ChristianenPRL_2022} further noticed that the mean excitation number $\langle N_{\mathrm{ex}}\rangle$ in the polaron cloud increases for higher condensate densities due to an increase in scattering rate of bosons from the condensate to the polaron cloud captured by the Fr\"ohlich term. On the other hand, the critical $\langle N_{\mathrm{ex}}\rangle$ required to form an Efimov cluster decreases for larger $|a|$, as stronger attraction facilitates multi-body bound state formation. Eventually, at a critical scattering length $a^{*}(n_0)$, the polaron becomes unstable and disappears as a local minimum in the energy landscape (see Fig.~\ref{fig:energyLand} (c)). Since the energy landscape for $|a^{*}|<|a|$ qualitatively resembles the few-body case where the energy of Gaussian states monotonically decrease with $\langle N_{\mathrm{ex}}\rangle$, $a^{*}$ is interpreted as a many-body shifted Efimov resonance. 

In Refs.~\cite{ChristianenPRL_2022,ChristianenPRA_2022}, deeply bound few-body clusters do not appear as metastable solutions, since the model admits clusters with arbitrary large binding energy upon increasing the particle number. However, the same authors investigate an improved model in Ref.~\cite{christianen2024phase} by including respulsive inter-boson interactions. Indeed, the energy penalty for binding more bosons increases with the cluster size. Thus, the competition between impurity attraction and boson repulsion leads to stabilization of clusters. Interestingly, including boson repulsion terms in the Hamiltonian together with the Gaussian variational ansatz reveals the intricate interplay between polaron physics and cluster formation, and leads to a rich phase diagram for attractive polaron depending on inter-boson repulsion, BEC density, and impurity-boson mass ratio (see, e.g. Figs.~7 and 8 in Ref.~\cite{christianen2024phase}).

For weak enough attraction, the polaron is a stable state in the form of an impurity dressed by few excitations of the BEC. By increasing the attraction, few-body impurity-boson clusters in the form of $(N+1)$-body bound states appear in the few-body spectrum of the system, crossing the polaron line as their binding energy increases. Nevertheless, the polaron remains a metastable state, protected from quantum fluctuations against decay to cluster states by an energy barrier. This energy barrier vanishes at a critical impurity-boson attraction, leading to polaronic instability. The critical scattering length at which polaron becomes unstable, as well as the window of polaron metastability depends on several factors, including inter-boson repulsion, BEC density and impurity-boson mass ratio. 

\begin{figure*}
    \centering
    \includegraphics[scale=0.5]{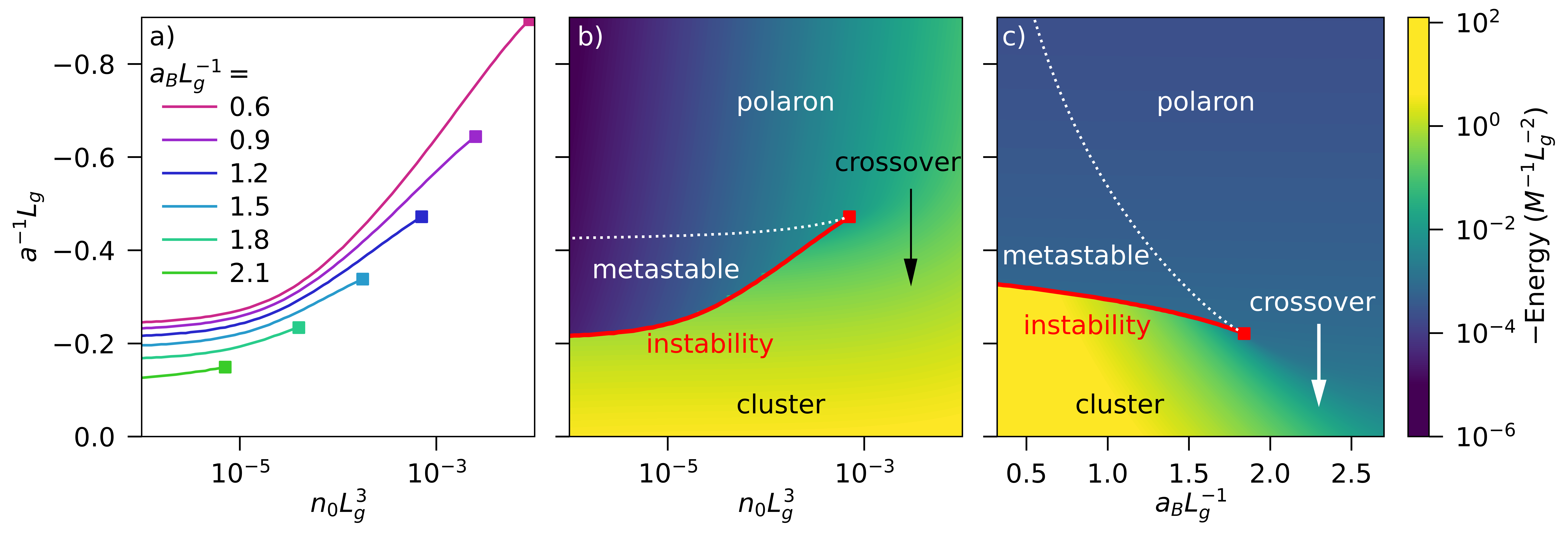}
    \caption{\textbf{Polaron instability.} a) Density dependence of polaronic instability critical scattering length for various inter-boson interactions. At small densities, the critical scattering length approaches the trimer formation threshold, which shifts to larger values for stronger inter-boson interactions. Increasing the density lowers the critical scattering length, as weaker attraction is needed for the instability. At a critical density, the instability turns into a crossover. b) Attractive polaron energy in terms of density and scattering length for fixed $a_{\rm{BB}}\equiv a_{\rm B}=1.2\,L_g$, where $L_g$ denotes the range of the Gaussian impurity-boson potential (see Eq.~(10) in \cite{christianen2024phase}). The red line indicates the critical scattering length of instability. The white dotted line indicates the scattering length where the cluster energy becomes lower than the polaron, making the polaron a metastable state. c) Polaron energy as a function of $a_{\rm{BB}}$ at fixed density $n_0 \,L^{3}_g=10^{-4.5}$. The cluster-polaron crossing boundary varies more strongly by interboson scattering length. Figure reprinted from Ref.~\cite{christianen2024phase}}
    \label{fig:instability-crossover}
\end{figure*}

In Fig.~\ref{fig:instability-crossover}, the change in ground state energy is investigated as a function of BEC density and inter-boson repulsion. At vanishing densities, the critical scattering length of polaronic instability approaches the trimer formation threshold, which shifts to larger values for stronger inter-boson interactions as stronger attraction is required to destabilize the polaron (see Fig.~\ref{fig:instability-crossover} (a,b)). Interestingly, increasing the density lowers the critical scattering length, since density increase sharply reduces the polaron energy, thus reducing the polaron-cluster energy barrier. At a critical density, the instability terminates and polaron smoothly crosses over to the cluster. On the other hand, stronger interboson interaction has a more drastic effect on increasing the cluster energy, causing the metastability to occur at smaller scattering lengths and the instability to occur for stronger impurity-boson attraction, see Fig.~\ref{fig:instability-crossover} (c).

\subsection{Few-body bound states with heavy impurities and inter-boson repulsion} 
The polaronic instability described in the last section has at its heart the impurity-mediated inter-boson attraction and Efimov physics, and is predicted in Ref.~\cite{christianen2024phase} to occur in systems with impurity-boson mass ratio $m_{\rm I}/m_{\rm B} \!<\! 1$. When the impurity is much heavier than the boson, $m_{\rm I}\!>\!m_{\rm B}$, the influence of Efimov physics on the polaron state is highly suppressed. Nevertheless, the intricate physics of many-body bound state formation can still survive \cite{Drescher2020,Grusdt2016}. Various aspects of these many-body bound states have been investigated for neutral heavy impurities \cite{Shi2018Multibody,Yoshida2018PRA,Levinsen22} see Secs.~\ref{sec:MutiBodyRes} and \ref{sec:qblock}, as well as ionic \cite{Astrakharchik2021,Christensen2021} and Rydberg impurities \cite{Schmidt2016,Camargo2018}. 

The problem of a heavy (or even static) impurity strongly coupled to an ideal BEC is peculiar in several regards. At coupling strengths sufficient for impurity-boson dimer formation, there is an energy gain equal to the dimer binding energy for each boson forming a dimer with the impurity. Thus, impurity-boson dimer formation continues until the BEC collapses into a gigantic molecule, with all bosons occupying the same dimer state, leading to an unbounded ground state energy. The cause of this pathological behaviour is the complete absence of inter-boson repulsion in the model of an ideal BEC, which prevents the indefinite cascaded collapse. This problem was first noticed in Ref.~\cite{Grusdt2016} and explored in other works \cite{Drescher2020,Chen2018,Levinsen22,Mostaan2023,Massignan2021,Yegovtsev2022}, predicting different ground state properties controlled by several length scales. Such involvement of several length scales, the necessity to account for many interacting particles and the strong coupling nature of the physics poses substantial challenges for theoretical descriptions. Fortunately, several promising theoretical schemes have recently been developed that provide plausible descriptions for certain regimes \cite{Drescher2020,Chen2018,Shi2018Multibody,Yoshida2018PRA,Levinsen22,Mostaan2023,Massignan2021,Yegovtsev2022,ChristianenPRA_2022,christianen2024phase}. In the following, we discuss the theoretical approach proposed by Mostaan et al.~\cite{Mostaan2023}.

The strong coupling, heavy impurity problem is amenable to an effective modelling in the regime where the binding energy of the highest energy impurity-boson many-body bound state is much larger than the typical excitation energy of Bogoliubov quasiparticles. This condition necessitates a large energy separation between the impurity-boson two-body bound state and impurity-boson scattering states. In this situation, single particle Bogoliubov modes can be grouped into a set $\mathcal{S}=\{\hat{b}_{\alpha}\}^{|\mathcal{S}|}_{\alpha=1}$ of strongly interacting modes with annihilation operators $\hat{b}_{\alpha}$ and a set $\mathcal{W}=\{\hat{\beta}_{k}\}^{|\mathcal{W}|}_{k=1}$ where $\mathcal{W}$ constitutes a set of weakly interacting modes. Here, the strongly interacting modes are defined by modes whose interactions according to the full Hamiltonian Eq.~\eqref{eqHBogoFull} should be treated non-perturbatively, whereas the weakly interacting modes interact perturbatively with one another and with the strongly interacting modes; both will be specified further below. This condition can be stated more formally by expressing the full Hamiltonian by
\begin{equation}\label{eq:HSW}
    \hat{\mathcal{H}}=\hat{\mathcal{H}}_{\mathcal{S}}[\hat{b}^{\dagger}_{\alpha},\hat{b}_{\alpha}] + \hat{\mathcal{H}}_{\mathcal{S}\mathcal{W}}[\hat{b}^{\dagger}_{\alpha},\hat{b}_{\alpha};\hat{\beta}^{\dagger}_{k},\hat{\beta}_{k}] + \hat{H}_{\mathcal{W}}[\hat{\beta}^{\dagger}_{k},\hat{\beta}_{k}] \, .
\end{equation}
In Eq.~\eqref{eq:HSW}, $\hat{\mathcal{H}}_{\mathcal{S}}$ and $\hat{\mathcal{H}}_{\mathcal{W}}$ denote the Hamiltonian of strongly interacting and weakly interacting modes, respectively, and $\hat{\mathcal{H}}_{\mathcal{S}\mathcal{W}}$ describe their interaction. The explicit form of $\hat{\mathcal{H}}_{\mathcal{S}},\,\hat{\mathcal{H}}_{\mathcal{S}\mathcal{W}}$ and $\hat{\mathcal{H}}_{\mathcal{W}}$ are given in Ref.~\cite{Mostaan2023}. The main utility of such mode separation noted in Ref.~\cite{Mostaan2023} is that the strong coupling Hamiltonian Eq.~\eqref{eqHBogoFull} can be re-expressed in terms of a continuum of modes whose interactions can be treated perturbatively, and are weakly coupled to the strongly interacting modes. Although mode separation is not unique (there might be many sets $\mathcal{S}$ and $\mathcal{W}$ satisfying the aforementioned properties), the predicted values of physical observables do not depend on the particular choice of mode separation and are only affected by the convergence properties of the perturbation series. 

To apply the above mentioned ideas to the strong coupling Bose polaron problem, a natural choice for mode separation is obtained by diagonalizing the quadratic Hamiltonian $\hat{\mathcal{H}}_{2\mathrm{ph}}$ into bound state modes $\hat{b}_{\alpha}$ with negative energies $-\varepsilon_{\alpha}$ $(\,[\hat{\mathcal{H}}_{2\mathrm{ph}},\hat{b}^{\dagger}_{\alpha}]=-\varepsilon_{\alpha}\,\hat{b}^{\dagger}_{\alpha}\,)$ and scattering state modes with positive energy $\varepsilon_{\vc{k}}$ $(\,[\hat{\mathcal{H}}_{2\mathrm{ph}},\hat{\beta}^{\dagger}_{\vc{k}}]=\varepsilon_{\vc{k}}\,\hat{\beta}^{\dagger}_{\vc{k}}\,)$\,. In case where only a single boson-impurity bound state exists, the annihilation operator for bosonic field fluctuations $\delta\hat{\phi}(\vc{x})$ can be decomposed as 
\begin{equation}\label{eq:phibbeta}
\begin{split}
\delta\hat{\phi}(\vc{x}) & = \Big( u_{\mathrm{B}}(\vc{x})\,\hat{b}+\sum_{\vc{k}}\,u_{\vc{k}}(\vc{x})\,\hat{\beta}_{\vc{k}} \Big) \\ & + \Big( v_{\mathrm{B}}(\vc{x})\,\hat{b}^{\dagger}+\sum_{\vc{k}}\,v_{\vc{k}}(\vc{x})\,\hat{\beta}^{\dagger}_{\vc{k}} \Big) \, ,
\end{split}
\end{equation}
where $(u_{\mathrm{B}}(\mathbf{x}),v_{\mathrm{B}}(\mathbf{x}))$ and $(u_{\vc{k}}(\vc{x}),v_{\vc{k}}(\vc{x}))$ are the real space form of Bogoliubov factors associated to $\hat{b}$ and $\hat{\beta}_{\vc{k}}$, respectively. In practice this is achieved by expanding around the repulsive polaron saddle-point solution which is associated with one bound mode $\hat{b}$, see Sec.~\ref{subsubsecMFsaddlePointAndBoundStates}.

Using the decomposition in Eq.~\eqref{eq:phibbeta},  the full Hamiltonian takes the form in Eq.~\eqref{eq:HSW}. Here, $\hat{\mathcal{H}}_{\mathcal{S}}[\hat{b}^{\dagger},\hat{b}]$ describes a single interacting bosonic mode, $\hat{\mathcal{H}}_{\mathcal{W}}[\hat{\beta}^{\dagger}_{\vc{k}},\hat{\beta}_{\vc{k}}]$ is the weakly interacting Hamiltonian of Bogoliubov modes, and $\hat{\mathcal{H}}_{\mathcal{S}\mathcal{W}}[\hat{b}^{\dagger},\hat{b};\hat{\beta}^{\dagger}_{\vc{k}},\hat{\beta}_{\vc{k}}]$ is the perturbative coupling between the bound state and scattering state modes. The fact that $\hat{\mathcal{H}}_{\mathcal{S}\mathcal{W}}$ is perturbative pertains to the smallness of the inter-boson interaction matrix elements between scattering states and the bound state, compared to the many-body bound state energy separation due to the suppression of initial-final state overlap in the strong coupling regime. Furthermore, for $\hat{\mathcal{H}}_{\mathcal{W}}$ to be a weakly interacting Hamiltonian, it is important for the Lee-Low-Pines term to have a vanishingly small contribution to the physics. To satisfy this condition, the impurity mass needs to be sufficiently larger than the boson mass, $m_{\rm I} > m_{\rm B}$. 

Concerning the quantum state of the impurity-boson system, the stable states of $\hat{\mathcal{H}}$ with energy $E$ have the general form $\ket{\Psi}=\sum_{i}\,c_i\ket{\psi_i}_{\mathrm{B}}\otimes\ket{i}_{\mathrm{sc}}$ where $\ket{\psi_i}_{\mathrm{B}}$ is the $i$'th eigenstate of $\hat{\mathcal{H}}_{\mathcal{S}}$ in the form of a multi-body resonance that belongs to the Fock space of $\hat{b}$ ($\mathcal{F}_{\mathrm{B}}$), and the states $\ket{i}_{\mathrm{sc}}$ are normalized but in general non-orthogonal states belonging to the Fock space of scattering-state Bogoliubov modes ($\mathcal{F}_{\mathrm{sc}}$). To zeroth order in terms of $\hat{\mathcal{H}}_{\mathcal{S}\mathcal{W}}$, the eigenstates of $\hat{\mathcal{H}}$ are of the form $\ket{\psi_i}_{\mathrm{B}}\otimes\ket{\chi_j}_{\mathrm{sc}}$, where $\ket{\chi_j}_{\mathrm{sc}}$ is an eigenstate of $\hat{\mathcal{H}}_{\mathcal{W}}$. Of particular interest is the effect of $\hat{\mathcal{H}}_{\mathcal{S}\mathcal{W}}$ on $\ket{\psi_i}_{\mathrm{B}}\otimes\ket{\emptyset}_{\mathrm{sc}}$ where $\ket{\emptyset}$ is the vacuum of $\hat{\mathcal{H}}_{\mathcal{W}}$ excitations. As $\hat{\mathcal{H}}_{\mathcal{S}\mathcal{W}}$ is turned on, the largest component of each eigenstate $\ket{\Psi}$ still comes from a single $\ket{\psi_i}_{\mathrm{B}}\otimes\ket{i}_{\mathrm{sc}}$, as the effect of $\hat{\mathcal{H}}_{\mathcal{S}\mathcal{W}}$ is perturbative. This allows to separate the component with the largest weight $\ket{\Psi}=c_i\,\ket{\psi_i}_{\mathrm{B}}\otimes\ket{i}_{\mathrm{sc}}+\sum_{j\neq i}\,c_j\ket{\psi_j}_{\mathrm{B}}\otimes\ket{j}_{\mathrm{sc}}$ since the components $\ket{\psi_j}_{\mathrm{B}}\otimes\ket{j}_{\mathrm{sc}}$ with $j\neq i$ are far separated in energy from $\ket{\psi_i}_{\mathrm{B}}\otimes\ket{i}_{\mathrm{sc}}$, and one can neglect them compared to $\ket{\psi_i}_{\mathrm{B}}\otimes\ket{i}_{\mathrm{sc}}$. In this manner, one would expect that the state $\ket{\psi_i}_{\mathrm{B}}\otimes\ket{i}_{\mathrm{sc}}$ to be a plausible approximation of the true quantum state, and to be useful as a basis for a variational approach to approximate the eigenstates of $\hat{\mathcal{H}}$. Such variational approach is based on states of the form 
\begin{equation}
    \ket{\Psi_{\mathrm{var}}} = \ket{\Psi_{\mathrm{B}}}_{\mathrm{B}}\otimes\ket{\Psi_{\mathrm{sc}}}_{\mathrm{sc}} \, ,
\end{equation}
where $\ket{\Psi_{\mathrm{B}}}_{\mathrm{B}}$ and $\ket{\Psi_{\mathrm{sc}}}_{\mathrm{sc}}$ belong to the Fock spaces $\mathcal{F}_{\mathrm{B}}$ and $\mathcal{F}_{\mathrm{sc}}$, respectively. In particular, $\ket{\Psi_{\mathrm{B}}}_{\mathrm{B}}=\sum^{\infty}_{n=0} \, \varphi_n \ket{n}_{\mathrm{B}}$, where $\ket{n}_{\mathrm{B}}$ is a Fock state of $\hat{b}$, with the amplitudes $\varphi_n$ a set of variational parameters. 

\begin{figure}
    \centering
    \includegraphics[scale=0.6]{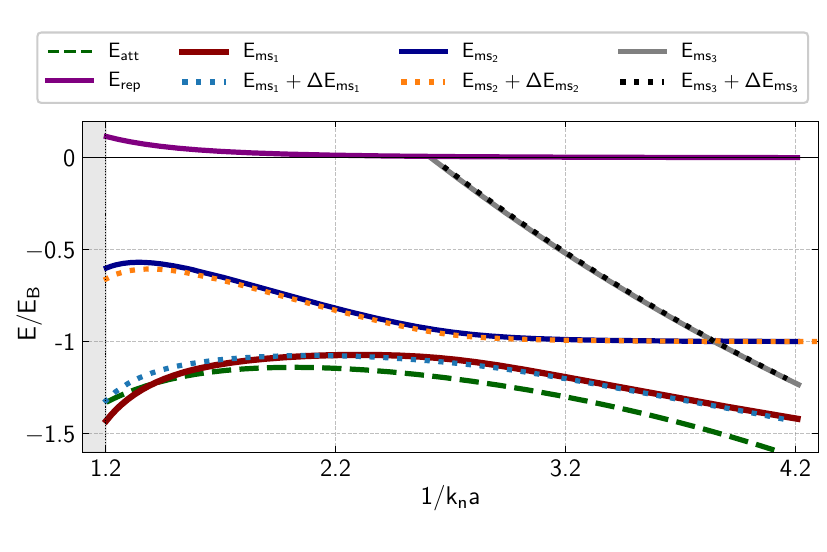}
    \caption{\textbf{Many-body bound states energies} in units of impurity-boson dimer binding energy obtained from variational states of the form in Eq.~\eqref{eq:varstate}, for parameters corresponding to the JILA experiment. The attractive (dashed green line) and repulsive (solid purple line) polaron branches are obtained using a coherent state variational ansatz of GPE type. In addition, many-body bound states $ms_i,\,i=2,3$ appear between the attractive and the repulsive branch; the lowest-lying variational state $ms_1$ was conjectured to correspond to the attractive polaron. The solid lines are energies $E_{ms_i}$ obtained by setting the coherent state part, describing corrections to the BEC deformation, equal to $\ket{\emptyset}$, while the dotted lines with energies $E_{ms_i}+\Delta E_{ms_i}$ are evaluated including the coherent states in Eq.~\eqref{eq:varstate}. Figure is reprinted from Ref.~\cite{Mostaan2023}.}
    \label{fig:nonGauss_E_withCS}
\end{figure}

\begin{figure}
    \centering
    \includegraphics[scale=0.98]{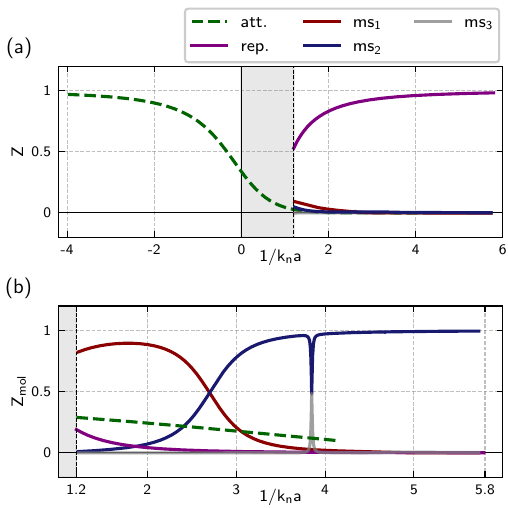}
    \caption{\textbf{Quasiparticle overlaps of many-body bound states.} (a) Quasiparticle residues $Z$ of the many-body bound states labeled $ms_1$ to $ms_3$, compared to $Z$ of the repulsive and attractive polaron. The residue of all states except the repulsive polaron are vanishingly small on the repulsive side, and approach zero in the limit of non-interacting bosons. (b) Molecular Z-factor of the states in (a). The significant magnitude and the non-monotonic behavior of the molecular weight associated to states $ms_1$ and $ms_2$ can be regarded as a smoking gun for their existence, as this behavior is substantially different from the coherent state prediction for the attractive polaron. The sharp spikes in $Z_{\mathrm{mol}}$ for $ms_1$ and $ms_3$ is due to a level crossing. Figure is reprinted from Ref.~\cite{Mostaan2023}.}
    \label{fig:nonGauss_Z}
\end{figure}

To establish a variational structure for $\ket{\Psi_{\mathrm{sc}}}_{\mathrm{sc}}$, we note that $\ket{\Psi_{\mathrm{sc}}}_{\mathrm{sc}}$ has to be an eigenstate of the effective Hamiltonian $\hat{\mathcal{H}}_{\mathrm{eff},\mathrm{sc}}=\hat{\mathcal{H}}_{\mathcal{W}} + \bra{\Psi_{\mathrm{B}}} \hat{\mathcal{H}}_{\mathcal{S}\mathcal{W}} \ket{\Psi_{\mathrm{B}}}_{\mathrm{B}}$. Terms in $\hat{\mathcal{H}}_{\mathrm{eff},\mathrm{sc}}$ linear in $\hat{\beta}_{\vc{k}}$ and $\hat{\beta}^{\dagger}_{\vc{k}}$ can be removed by a coherent state displacement $\beta_{\vc{k}}$ of the field operators, and the resulting Hamiltonian is in normal ordered form, with no terms solely containing $\hat{\beta}^{\dagger}_{\vc{k}}$. Thus, $\hat{\mathcal{H}}_{\mathrm{eff},\mathrm{sc}}$ has the same vacuum state as $\hat{\mathcal{H}}_{\mathcal{W}}$, and one can approximate the eigenstates adiabatically connected to $\ket{\psi_i}_{\mathrm{B}}\otimes\ket{\emptyset}$ by states of the form 
\begin{equation}\label{eq:varstate}
    \ket{\Psi_{\mathrm{var}}} = \ket{\Psi_{\mathrm{B}}}_{\mathrm{B}}\,\otimes \prod_{\vc{k}} \ket{\beta_{\vc{k}}} \, .
\end{equation}
The variational parameters $\varphi_n$ and $\beta_{\vc{k}}$ then have to be determined self-consistently by minimizing $\bra{\Psi_{\mathrm{var}}}\hat{\mathcal{H}}\ket{\Psi_{\mathrm{var}}}$ subject to the condition $\braket{\Psi_{\mathrm{var}}}{\Psi_{\mathrm{var}}}=1$. Note that the large separation of energy scales between the dynamics of states belonging to $\mathcal{F}_{\mathrm{B}}$ and $\mathcal{F}_{\mathrm{sc}}$ also provides a justification for the separable structure of the quantum state in Eq.~\eqref{eq:varstate}, an assumption underlying similar approximations such as the Born-Oppenheimer approximation.

In Ref.~\cite{Mostaan2023} the above framework was applied to the JILA experiment setting described in Ref.~\cite{Hu2016} to investigate the qualitative features of the discrete (i.e. few-body) part of the spectrum. In Fig.~\ref{fig:nonGauss_E_withCS}, the resonance energies for different multi-body bound states are shown, and compared to the energy of the repulsive polaron saddle-point (top curve in the plot) around which the expansion was performed. To reveal the effect of corrections to the condensate deformation from the multi-body bound states, variational energies corresponding to states in Eq.~\eqref{eq:varstate} are compared to the energies of $\ket{\psi_i}\otimes\ket{\emptyset}$ for all $\psi_i$ states with energy below the repulsive polaron. This comparison demonstrates that the coherent state amplitudes $\beta_{\vc{k}}$ are much smaller than the coherent state distortion around the uniform condensate corresponding to the repulsive polaron, and substituting the coherent state part by $\ket{\emptyset}$ has a minute effect on the calculated energies. 

To elucidate the nature of the different few-body bound states, Mostaan et al.~\cite{Mostaan2023} calculated the quasiparticle residue $Z$, describing how well such states can be observed in impurity spectroscopy experiments. As depicted in Fig.~\ref{fig:nonGauss_Z}, on the repulsive side $Z$ is dominated by the repulsive polaron, with vanishing contributions from the other metastable states. Notably, the molecular quasiparticle residue, defined by the overlap of an impurity-boson molecular initial state with the interacting states as $Z_{\mathrm{mol}}(i)=|\langle {\rm mol} | \Psi_i \rangle|^2$, provide much stronger signatures for the existence of the many body bound states. The drop of $Z_{\mathrm{mol}}$ at larger values of $1/(k_n a)$ directly reveals a hybridization of one- and two-boson bound states, see Fig.~\ref{fig:nonGauss_Z}. $Z_{\mathrm{mol}}(i)$ can be directly measured in ejection RF spectroscopy after adiabatic preparation of the few-body bound state $\Psi_i$.

\section{Quantum dynamics and non-equilibrium polaron physics}
\label{secQuantDynamics}
Experimentally it is often more natural to study out-of-equilibrium settings than the polaron ground states, for different reasons. The ground state is challenging to prepare and has a limited life-time due to the atomic losses associated with the Feshbach resonance used to realize and tune impurity-boson interactions. On the other hand, dynamical probes provide new insights, either realizing qualitatively new phenomena or giving direct access to key properties that characterize the considered polaron ground or excited state. 

In this section we will review which dynamical probes have been considered so far, and how they allow to unravel the underlying mechanisms of polaron formation. We will start by discussing dynamical protocols which work closest to the equilibrium polaron (ground) state, i.e. with low excitation energies, and proceed to more dramatic quenches corresponding to higher excitation energies. 

We will mention the employed theoretical methods along the way, since they are mostly time-dependent extensions of the equilibrium techniques reviewed in Sec.~\ref{secEffHamiltonians}. Generally speaking, variational approaches can be extended to study dynamical processes using time-dependent variational parameters, making them the main workhorse in the study of polaron dynamics. Except in a few analytically tractable cases, no numerically exact method exists until today that can capture arbitrary polaron dynamics; quantum Monte Carlo approaches, which provide the most accurate description of equilibrium states, are plagued by the notorious sign problem when studying out-of-equilibrium dynamics. In lattice systems, time-dependent variants of matrix-product / tensor-product states have recently gained significant attention \cite{Paeckel2019a}.

\subsection{Polaron oscillations and effective mass}
\label{subSecPolaronOscillations}
A key characteristic of any quasiparticle is its renormalized mass, which provides a direct means to evaluate how strongly renormalized the interacting system is. One of the most natural ways to measure the effective polaron mass is through the renormalization of non-equilibrium dipole oscillations in a harmonic trapping potential. There are two main mechanisms which have been studied that lead to renormalized dipole oscillations: (i) through renormalization of the trapping potential by impurity-boson interactions, or (ii) through renormalization of the impurity kinetic energy, i.e. directly via mass renormalization. Both effects need to be taken into account, and can be used to extract the renormalized mass of the polaron.

\begin{figure}
	\centering
	\includegraphics[width=\linewidth]{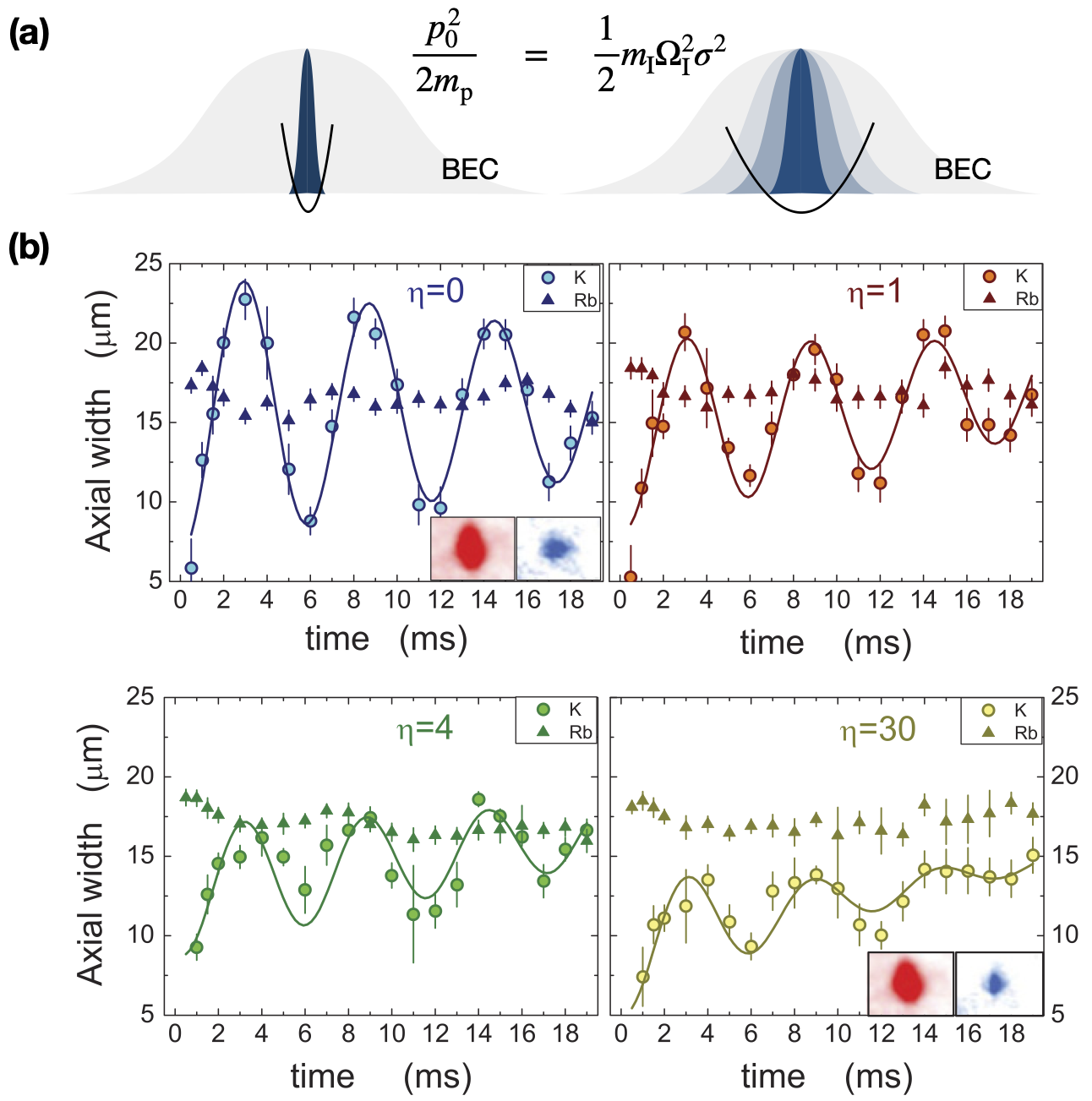}
	\caption{\textbf{Breathing oscillations of trapped polarons} in $d=1$-dimensional tubes were used by Catani et al.~\cite{Catani2012} to measure the renormalized polaron mass $m_{\rm p}$. (a) An initially relatively tightly trapped polaron cloud (blue) is suddenly released into a much shallower trap (frequency $\Omega_{\rm I})$, in which it expands. Thereby the initial average kinetic energy $p_0^2/2m_{\rm p}$, due to the tight localization, is transformed into potential energy, reaching its maximum value $m_{\rm I} \Omega_{\rm I}^2 \sigma^2/2$, with $\sigma$ the maximum width of the expanding cloud. (b) The measured width $\sigma(t)$ of the oscillating cloud of the minority (impurity $~^{41} {\rm K}$ atoms) and majority (atomic $~^{87} {\rm Rb}$ BEC) is shown for different values of the dimensionless impurity-boson coupling strength $\gamma$. Part (b) is re-printed from Ref.~\cite{Catani2012}.}
	\label{figCataniExp}
\end{figure}

\subsubsection{Experiments.}
The first measurement of the effective Bose polaron mass was performed in quasi one-dimensional systems by Catani et al.~\cite{Catani2012} in Florence. This experiment was reviewed already in Sec.~\ref{subsecFlorenceExp}; now we discuss in more detail the dynamical sequence realized, and how it can be related to the effective polaron mass. The experimental sequence is illustrated in Fig.~\ref{figCataniExp} (a): It starts from a $~^{41} {\rm K}$ impurity atom which is relatively tightly localized by an optical dipole potential, and interacts with the surrounding quasi 1D $~^{87} {\rm Rb}$ Bose gas. This leads to the formation of a Bose polaron, which is localized on a length scale $\sigma(0) = \sqrt{\langle x^2(0) \rangle}$ independent of the interaction strength. Next, the dipole trap is suddenly released and the polaron has an average kinetic energy $E_{\rm kin} = p_0^2/(2 m_{\rm p})$, where $p_0 \simeq 1/\sigma(0)$ is the initial momentum scale. While the latter is independent of the impurity-boson interaction strength $\eta$, the effective polaron mass $m_{\rm p}$ depends on $\eta$. Notably, the kinetic energy is renormalized, depending on $m_{\rm p}$ instead of the bare impurity mass.

The experiment analyzed the dynamics of the post-quench impurity distribution, $\sqrt{\langle x^2(t) \rangle}$, see Fig.~\ref{figCataniExp} (b). Both the impurity and the bosons remain subject to weak harmonic trapping potentials, $(1/2) m_{\alpha} \Omega_{\alpha}^2 x^2(t)$, with $\alpha={\rm I,B}$ denoting impurity (I) and boson (B) cases. While the extent of the surrounding Bose gas remains essentially constant, the impurity atoms undergo breathing oscillations. For sufficiently weak interactions, $\eta \lesssim 4$, the maximum extent $\sigma$ of the impurity cloud lies outside of the Bose gas or reaches its edge. At this turning point, the impurity atoms no longer interact with any surrounding bosons and their energy is entirely given by the potential energy, $(1/2) m_{\rm I} \Omega_{\rm I}^2 \sigma^2$. Neglecting dissipative effects during the first oscillation period, the initial kinetic energy of the polaron and the potential energy at the first maximum can be equated \cite{Catani2012}, yielding
\begin{equation}
    \frac{p_0^2}{2 m_{\rm p}(\eta)} = \frac{1}{2} m_{\rm I} \Omega_{\rm I}^2 \sigma^2(\eta)
\end{equation}
with interaction ($\eta$-) dependencies explicitly indicated. Comparing to the maximum amplitude $\sigma(\eta=0)$ in the non-interacting case thus yields 
\begin{equation}
    \sqrt{\frac{m_{\rm I}}{m_{\rm p}(\eta)}} = \frac{\sigma(\eta)}{\sigma(\eta=0)}.
\end{equation}
This relation was used by Catani et al. to extract the renormalized polaron mass -- we will review their results and comparisons with theoretical models below in Sec.~\ref{SubsubsecCatani1D}.

The experiment also found that the frequency of the observed breathing oscillations is essentially unchanged, within error bars, when changing the interaction strength \cite{Catani2012}. This behavior is strikingly different from what would be expected in a homogeneous Bose gas. In the latter case, when the impurity only feels the external trapping potential, the renormalized mass directly affects the trapping frequency $\Omega_{\rm p}$ experienced by the polaron, as can be seen by writing:
\begin{equation}
    \frac{1}{2} m_{\rm I} \Omega_{\rm I}^2 x^2 = \frac{1}{2} m_{\rm p} \underbrace{\left( \sqrt{\frac{m_{\rm I}}{m_{\rm p}}} \Omega_{\rm I} \right)^2}_{=\Omega_{\rm p}^2} x^2.
\end{equation}
Hence, in a homogeneous Bose gas, one expects a renormalized frequency of dipole oscillations:
\begin{equation}
    \Omega_{\rm p} = \sqrt{\frac{m_{\rm I}}{m_{\rm p}}} \Omega_{\rm I}.
    \label{eqOmegaRenorm}
\end{equation}
This relation could be used in future experiments, if a sufficiently homogeneous Bose gas can be realized, e.g. in a box-shaped trap~\cite{Etrych24}.

The experimental observation by Catani et al.~\cite{Catani2012} of almost constant frequency, as a function of the interaction strength, points towards effects resulting from the inhomogeneous trapping potential experienced by the polaron. Through impurity-boson interactions, the inhomogeneous density $n(x)$ of the Bose gas leads to a modified effective trapping potential. Ignoring quantum fluctuations, the latter contains an additional contribution $g_{\rm IB} n(x)$. Such renormalization of the trapping potential has been directly observed and analyzed quantitatively in a different set of experiments with weakly interacting mixtures, by Ferrier-Barbut et al.~\cite{FerrierBarbut2014}. These authors worked in a regime where additional mass-renormalization effects can be neglected. 

In related experiments by Palzer et al.~\cite{Palzer2009} in Cambridge (UK) and by Fukuhara et al.~\cite{Fukuhara2013} in Munich, spin-flip impurities were created and subsequently released inside quasi one-dimensional Bose gases. The former worked in continuum, while the latter included a longitudinal lattice. In both cases the subsequent time-evolution revealed strongly renormalized propagation velocities of the impurity atoms due to interaction effects with the bath. These are particularly pronounced in the presence of a lattice \cite{Fukuhara2013}. In both experiments, the use of in-situ microscopy further allowed to visualize the dynamics of the polaronic dressing cloud forming dynamically around the impurity.

\subsubsection{Theoretical analysis.}
The dynamical sequence studied experimentally by Catani et al.~\cite{Catani2012} was subsequently analyzed in great detail theoretically. Johnson et al.~\cite{Johnson2011} developed an effective theory, based on a time-dependent Gross-Pitaevskii equation coupling the impurity wavefunction $\chi(x,t)$ to the mean-field condensate wavefunction $\varphi(x,t)$. They showed that the essential qualitative features observed by Catani et al.~\cite{Catani2012} can be captured by an analytical treatment based on the Thomas-Fermi approximation and the use of Gaussian variational states. Their theory also included pronounced self-trapping effects, signatures of which were not found by Catani et al.~\cite{Catani2012}; this was attributed to possible finite-temperature effects by Johnson et al.~\cite{Johnson2011}.

Further theoretical analysis of the dynamical protocol by Catani et al.~\cite{Catani2012} was performed by Grusdt et al.~\cite{Grusdt2017RG1D} using time-dependent mean-field calculations based on the beyond-Fr\"ohlich Hamiltonian, Eq.~\eqref{eqH2ph}. Effects of the trapping potentials were included within the local density approximation (LDA). In the idealized case of a homogeneous Bose gas, full numerical simulations of the dynamics confirmed that polaronic mass renormalization can be accurately extracted from the renormalized frequency of dipole oscillations, through Eq.~\eqref{eqOmegaRenorm}; i.e. non-adiabatic effects beyond the quasiparticle picture do not lead to significant corrections, at least on the level of the employed beyond-Fr\"ohlich Hamiltonian which neglects phonon non-linearities. In contrast, the renormalized oscillation amplitude, employed by Catani et al., was shown to be insufficient in the case of a homogeneous Bose gas for extracting the renormalized polaron mass at intermediate interaction strengths, owing to non-adiabatic effects which can be associate with phonon emission / dissipation. 

The authors of \cite{Grusdt2017RG1D} further studied the case of an inhomogeneous Bose gas, assuming an additional adiabatic impurity-boson interaction potential which is treated within LDA. While quantitative comparisons to the experiment by Catani et al.~\cite{Catani2012} are challenging, the theoretical analysis revealed strong possible effects of the inhomogeneous impurity-boson interaction energy. It is speculated that the latter effect plays a role for understanding the weak /  non-detectable influence of impurity-boson interaction strength on the frequency of the observed breathing oscillations.

\subsection{Radio-frequency and Ramsey spectroscopy}
\label{secRFandDynamics}
As already indicated in Secs.~\ref{secKeyConcepts}, \ref{subsecExpAarhus}, \ref{subsecExpJILA}, radio-frequency (RF) spectroscopy constitutes one of the most powerful probes of polaron physics, giving direct access to the entire excitation spectrum. The most natural implementation of RF spectroscopy works in the frequency domain, where a weak perturbation is modulated at frequency $\omega$ and the linear response of the system is recorded, see e.g. \cite{Jorgensen2016,Hu2016}. An alternative, but equivalent, incarnation works directly in the time domain, where a variant of a Ramsey interferometer is realized dynamically. Here we review the connection of the time domain and frequency domain approaches (Sec.~\ref{SubSecRFasDyn}), and summarize existing experiments based on Ramsey spectroscopy (Sec.~\ref{SubSecRamseyInterfero}). 

\subsubsection{RF spectroscopy as a dynamical probe.}
\label{SubSecRFasDyn}
The setup required to perform RF spectroscopy starts with a host Bose gas with an impurity immersed in it. The latter is assumed to have two internal spin states, which we denote as $\ket{1}$ and $\ket{2}$. Initially the impurity atoms are prepared in the $\ket{1}$ state. In the \emph{direct} (ejection) RF protocol \cite{Koschorreck2012}, the initial $\ket{1}$ state is assumed to be interacting with the host Bose gas, and it is subsequently driven into the non-interacting $\ket{2}$ final state by a weak time-dependent perturbation
\begin{equation}
    \H_{\rm RF}(t) = \Omega_{\rm RF} e^{i \omega t} \ket{1} \bra{2} + \hc .
\end{equation}

\begin{figure}
	\centering
	\includegraphics[width=0.8\linewidth]{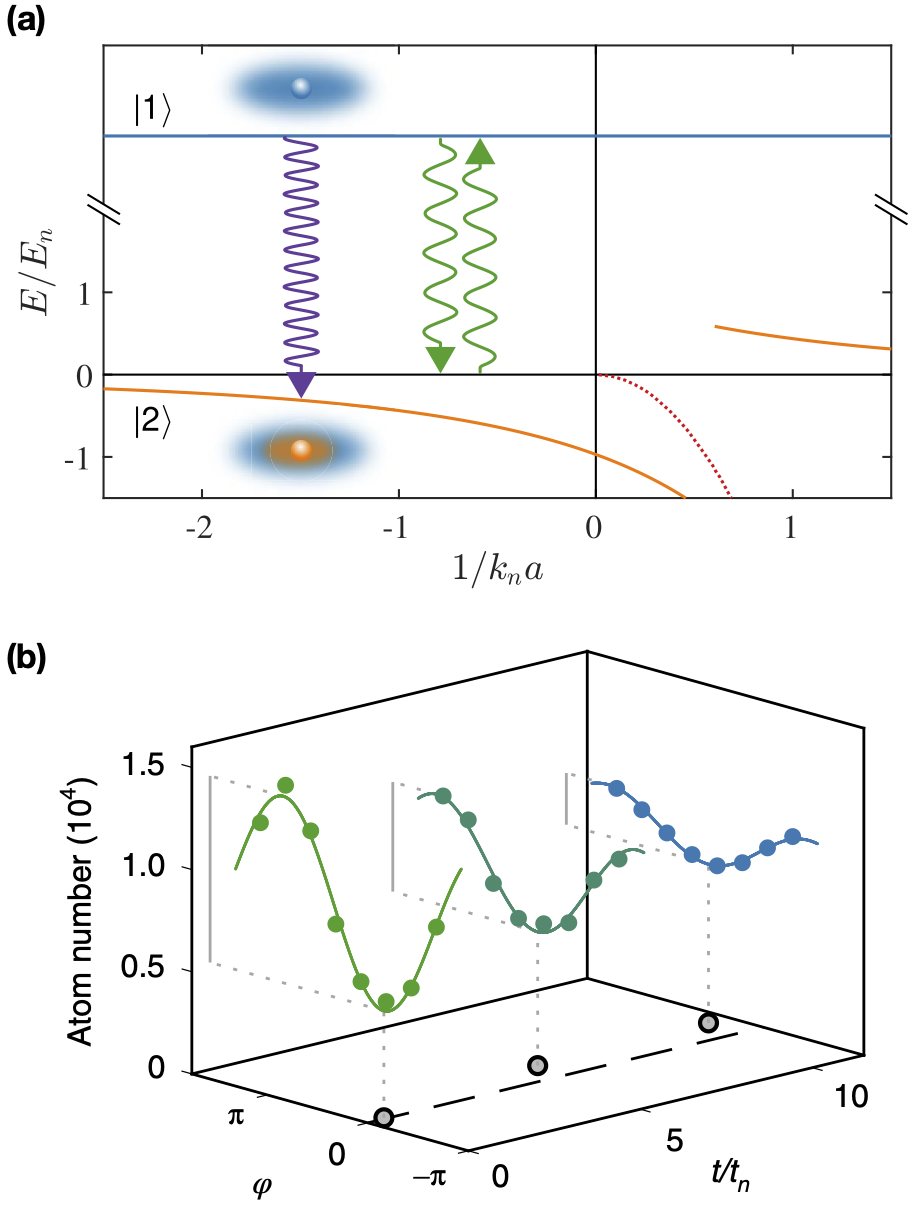}
	\caption{\textbf{Dynamical probes of the polaron spectrum} (a) can be performed using a weak radio-frequency pulse (purple) or by creating a coherent superposition of interacting and non-interacting states using a sequence of short Ramsey pulses (green arrows). (b) By varying the relative phase $\varphi$ of the Ramsey pulses and the evolution time $t$ in between, a subsequent measurement of the atom number allows to extract the phase and amplitude of the spectrum $C(t) \in \mathbb{C}$ in the time-domain. Figures in (a) and (b) are re-printed from \cite{Skou2022} and \cite{Skou2021}, respectively.}
	\label{figRFRamsey1}
\end{figure}

Vice-versa, in the \emph{inverse} (injection) RF protocol, the initially non-interacting $\ket{1}$ state is driven to the interacting $\ket{2}$ final states, see Fig.~\ref{figRFRamsey1} (a). In the following we will focus on the inverse RF protocol for concreteness. It is more often employed in practice, see e.g.~\cite{Jorgensen2016,Hu2016}, since it provides access to the interacting eigenstates and does not require preparation of the interacting initial polaron state with its finite life-time. See e.g. Ref.~\cite{Liu2020b} for a discussion of the relation between injection and ejection spectroscopy.

Assuming a weak drive, i.e. sufficiently small $\Omega_{\rm RF}$, the spin-flip probability can be calculated using Fermi's golden rule. The resulting spin-flip rate in the linear-response regime, which direclty constitutes the experimentally accessible RF signal, becomes
\begin{equation}
    I(\omega) \propto \sum_n | \bra{n,2} \H_{\rm RF} \ket{\psi_{\rm in},1} |^2 \delta(\omega - E_{n,2} + E_{{\rm in},1}),
    \label{eqIwFermiGolden}
\end{equation}
where $\ket{\psi_{\rm in},1}$ is the initial non-interacting eigenstate with energy $E_{{\rm in},1}$ and $\ket{n,2}$ denote the interacting final eigenstates with energies $E_{n,2}$ which are summed over. If the system has translational invariance and starts from an initial eigenstate $\ket{\psi_{\rm in},1,\vec{p}}$ at total momentum $\vec{p}$, the momentum-resolved spectrum $I(\vec{p},\omega)$ can be similarly obtained. 

Eq.~\eqref{eqIwFermiGolden} can be equivalently expressed as
\begin{equation}
    I(\omega) = {\rm Re} \frac{1}{\pi} \int_0^\infty dt ~ e^{i \omega t} C(t),
    \label{eqIwTime}
\end{equation}
where $C(t)$ is the spectral function in the time domain,
\begin{equation}
    C(t) = e^{i E_{{\rm in},1}t} \bra{\psi_{\rm in},2} e^{- i \H t} \ket{\psi_{\rm in},2}
\label{eqAtDef}
\end{equation}
Here $\H$ denotes the complete system Hamiltonian without the drive, and $\ket{\psi_{\rm in},2}$ is the initial state with its spin flipped from $1$ to $2$; see e.g.~\cite{Shashi2014RF}. I.e., from the time-dependent overlap $C(t)$ the spectrum can be calculated by a Fourier transform.

The time-dependent overlap $C(t)$ describes a dynamical sequence, where the spin of the initial state is flipped at time zero, through application of $\H_{\rm RF}$. The so-obtained interacting state is then time-evolved for time $t$ with the full interacting Hamiltonian $\H$, and the overlap with the original non-interacting state is calculated after fully reversing the spin-flip through the second application of $\H_{\rm RF}$.

This description of the spectrum in the time-domain highlights how the ability to calculate time-dependent polaron wavefunctions goes hand-in-hand with the ability to predict the polaron spectral function $I(\omega)$. For Bose polarons in a BEC, this one-to-one correspondence was first utilized in numerical calculations by Shashi et al.~\cite{Shashi2014RF}. Using the Bogoliubov-Fr\"ohlich Hamiltonian, they made a time-dependent Lee-Low-Pines mean-field ansatz and obtained the spectral function; a refined path-integral formulation, tailored to saddle-point calculations of $A(t)$ was also proposed in that work. 

A second insight from the formulation, Eq.~\eqref{eqIwTime}, in the time domain is the direct accessibility of $A(t)$ in a time-dependent quench experiment. The idea is to realize exactly the dynamical sequence described in Eq.~\eqref{eqAtDef}, i.e. suddenly switch on interactions and perform the subsequent time evolution. In order to measure the final overlap with the initial state, an interferometric sequence can be employed: This leads to the Ramsey protocol discussed next.

\begin{figure*}
	\centering
	\includegraphics[width=\linewidth]{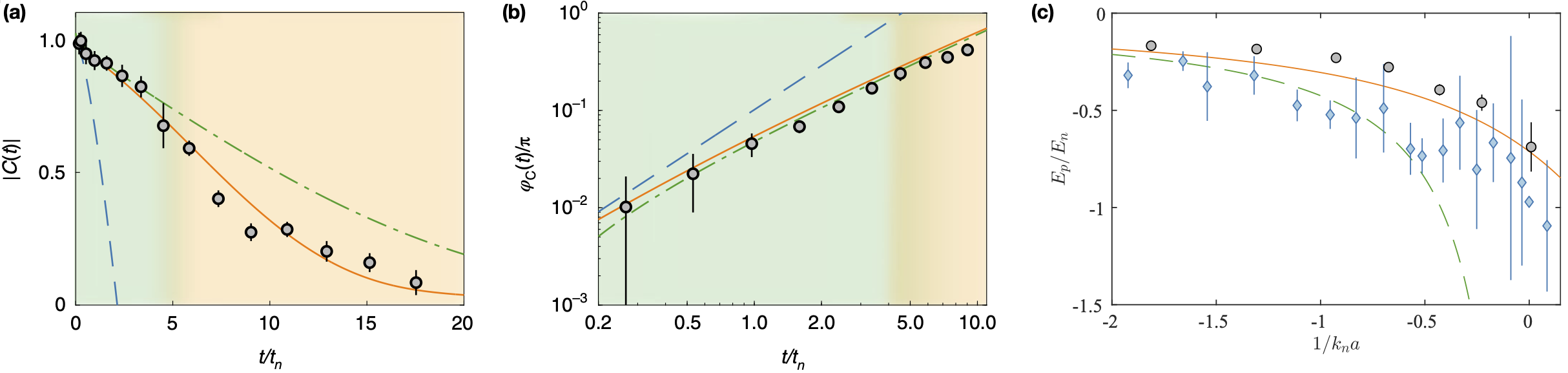}
	\caption{\textbf{Extracting the polaron ground state properties from Ramsey interferometry.} The measurements of the time-dependent interferometric amplitude $|C(t)|$ (a) and phase $\varphi_{\rm C}(t)$ (b) allow to extract the quasiparticle energy (c) and weight (not shown here). In (c) the results of this interferometric method (gray dots) are compared to results from direct RF spectroscopy (blue dots). In all plots, comparisons with theoretical predictions are provided (lines), allowing to distinguish qualitatively distinct dynamical regimes as indicated by the shaded background colors in (a) and (b). Figures in (a), (b) [(c)] are re-printed from \cite{Skou2021} [from \cite{Skou2022}].}
	\label{figRFRamsey2}
\end{figure*}

\subsubsection{Ramsey interferometry and quasiparticle decoherence.}
\label{SubSecRamseyInterfero}
To measure the time-dependent overlap, $C(t)$ in Eq.~\eqref{eqAtDef}, an interferometric sequence can be employed, as theoretically proposed by Knap et al.~\cite{Knap2012,Knap2013a}. Starting from the non-interacting $\ket{1}$ impurity state, a Ramsey-$\pi/2$ pulse is applied to create a coherent superposition $\ket{1}+\ket{2}$. While the non-interacting state only picks up a phase, the interacting state $\ket{2}$ evolves with the interacting Hamiltonian, $\ket{2} \to e^{-i \H t} \ket{2}$, creating entanglement between the impurity atom and bath atoms. Applying a control phase $\varphi$ to $\ket{2}$ and a second $\pi/2$ Ramsey pulse at time $t$, the population in the $\ket{2}$ state at time $t$ becomes
\begin{equation}
    N_2(\varphi,t) \propto |C(t)| \cos(\varphi - \varphi_C(t)),
\end{equation}
see Fig.~\ref{figRFRamsey1} (b), where the time-dependent overlap (also known as the Loschmidt-echo) is parametrized as
\begin{equation}
    C(t) = |C(t)| e^{i \varphi_C(t)}.
\end{equation}

Experimentally, Ramsey sequences of this type have been successfully recorded. In an early experiment in Kaiserslautern, Schmidt et al. \cite{SchmidtF2018} measured the amplitude $|C(t)|$ for a $~^{133}{\rm Cs}$ impurity immersed in a $~^{87}{\rm Rb}$ BEC. From their measurement they extracted a characteristic decoherence time scale on which $|C(t)|$ decays, owing to the developing system-bath entanglement of the interacting impurity state. A closely related experiment was performed by Cetina et al.~\cite{Cetina2015} using an impurity atom interacting with a Fermi sea.

In 2016, Cetina et al.~\cite{Cetina2015} performed another set of experiments in a fermionic system, and demonstrated explicitly that the full time-dependent overlap $C(t)$ can be measured up to long times. They extracted both amplitude and phase information, and confirmed that the spectral function $I(\omega)$ can be accurately reconstructed from measurements performed entirely in the time domain. 

In the context of Bose polarons the Ramsey technique was experimentally pioneered  by the Aarhus group, see Fig.~\ref{figRFRamsey1}. In Skou et al.~\cite{Skou2021} the group measured amplitude and phase of $C(t)$, as shown in Fig.~\ref{figRFRamsey2} (a) and (b). They identified regimes with different universal quantum dynamics, and studied their dependence on the impurity-boson interaction strength. While short-time dynamics are governed by universal two-body physics~\cite{parishJesper2016quantum}, many-body effects manifest themselves in the long-time dynamics. 

It was further pointed out by Parish et al.~\cite{parishJesper2016quantum}, Cetina et al.~\cite{Cetina2015} and Skou et al.~\cite{Skou2021}, that the phase evolution at long times can be used directly to measure the ground state (quasiparticle) energy through the relation
\begin{equation}
    E_0 = - {\rm lim}_{t \to \infty} \hbar \frac{d}{dt} \varphi_C(t),
\end{equation}
provided that the quasiparticle weight is non-vanishing, $Z\neq0$. The extracted energies, based on a refined analysis by the same group \cite{Skou2022}, are shown in Fig.~\ref{figRFRamsey2} (c). The interferometrically obtained energy estimates are significantly less noisy and match the theoretical models more closely than peak positions extracted directly from the spectrum $I(\omega)$ taken in the frequency domain. This technique was subsequently applied by Etrych et al.~\cite{Etrych24} in a homogeneous system with a box-shaped trapping potential, yielding some of the most accurate experimental results for the Bose polaron energy to date and establishing that Ramsey interferometry and RF spectroscopy yield essentially consistent results.

In subsequent works, the dynamics of interaction-induced impurity decoherence was studied theoretically by Nielsen et al.~\cite{Nielsen2019} using the Master-equation approach, on the level of an effective Fr\"ohlich Hamiltonian. In this work the authors predicted a further suppression of coherence upon increasing the initial impurity momentum. Ardila~\cite{Ardila2021} later included beyond-Fr\"ohlich terms and performed a time-dependent Lee-Low-Pines mean-field analysis of the time-dependent overlap $C(t)$. This work revealed additional experimental imperfections which need to be included in order to perform quantitative comparisons of theory and experiment; specifically: inhomogeneous broadening, inelastic losses and magnetic field fluctuations.

\subsection{Subsonic and supersonic polarons}
The dynamical probes reviewed so far only involved polarons at small (or zero) momentum $|\vec{k}|<k_c$, i.e. with velocities well below the speed of sound $c$ of the host BEC.  These subsonic polarons are stable quasiparticles, as manifested by a delta-function peak in the momentum resolved zero-temperature spectrum,

\begin{equation}
    I(\omega,\vec{k}) = Z_{\vec{k}} \delta(\omega-E_{\rm p}(\vec{k})) + I_{\rm incoh}(\omega,\vec{k}),
\end{equation}

Here $Z_{\vec{k}}$ is the momentum-dependent quasiparticle residue, $E_{\rm p}(\vec{k})$ is the renormalized polaron dispersion relation and $I_{\rm incoh}(\omega,\vec{k})$ is the incoherent contribution to the spectrum, with $I_{\rm incoh}(\omega,\vec{k})=0$ for $\omega < E_{\rm p}$.

When the velocity of the polaron reaches the speed of sound it enters the \emph{supersonic} regime, where spontaneous emission of phonon excitations is possible. Here, energy and momentum conservation can be simultaneously satisfied, and the resulting phonon radiation can be viewed as Cherenkov emission. This effect gained significant attention as a direct probe of the superfluidity of weakly interacting BECs. For simplicity, external potentials were considered which move through the condensate at a constant speed, see e.g. Astrakharchik and Pitaevskii \cite{Astrakharchik2004}. This corresponds to an infinitely massive impurity, highlighting the direct relation to the polaron problem. 

\subsubsection{Quasiparticle breakdown.}
The momentum dependence of the Bose polaron was first studied by Shashi et al.~\cite{Shashi2014RF} using the Bogoliubov-Fr\"ohlich Hamiltonian. Using the self-consistent Lee-Low-Pines mean-field theory at non-zero total momentum $\vec{K}_{\rm tot}$, they found an expression for the coherent amplitudes in the ground state,
\begin{equation}
    \alpha_{\vec{k}}^{\rm MF} = - \frac{V_{\vec{k}}}{\omega_{\vec{k}} + \frac{\vec{k}^2}{2 M} - \frac{\vec{k}}{M} \cdot \left( \vec{K}_{\rm tot} - \sum_{\vec{k}'} \vec{k}' |\alpha_{\vec{k}'}|^2 \right)},
    \label{eqAlphaKMFfiniteK}
\end{equation}
which diverge if the denominator has zeros. Taylor-expanding the denominator and using the mean-field equation for the renormalized polaron mass $m_{\rm p}$ Shashi et al.~\cite{Shashi2014RF} showed that such divergences are present if the polaron velocity
\begin{equation}
    v_{\rm p} = |\vec{K}_{\rm tot}| / m_{\rm p} > c
\end{equation}
exceeds the speed of sound.

Remarkably, this condition exactly reproduces Landau's criterion for superfluidity, even though quantum fluctuations of the mobile impurity are included. Other theoretical approaches, including even advanced methods such as the self-consistent T-matrix \cite{Rath2013}, do not capture this behavior. 

When the polaron velocity exceeds the speed of sound, the phonon ground state according to Eq.~\eqref{eqAlphaKMFfiniteK} has a divergent phonon number. This leads to a complete suppression of the quasiparticle weight,
\begin{equation}
    Z_{\vec{K}_{\rm tot}} = 0, \quad \text{for} \quad |\vec{K}_{\rm tot}| > k_c = m_{\rm p} c,
\end{equation}
Through the renormalized mass $m_{\rm p}$, this quasiparticle breakdown depends strongly on the impurity-boson interaction strength \cite{Shashi2014RF}.

\begin{figure}
	\centering
	\includegraphics[width=\linewidth]{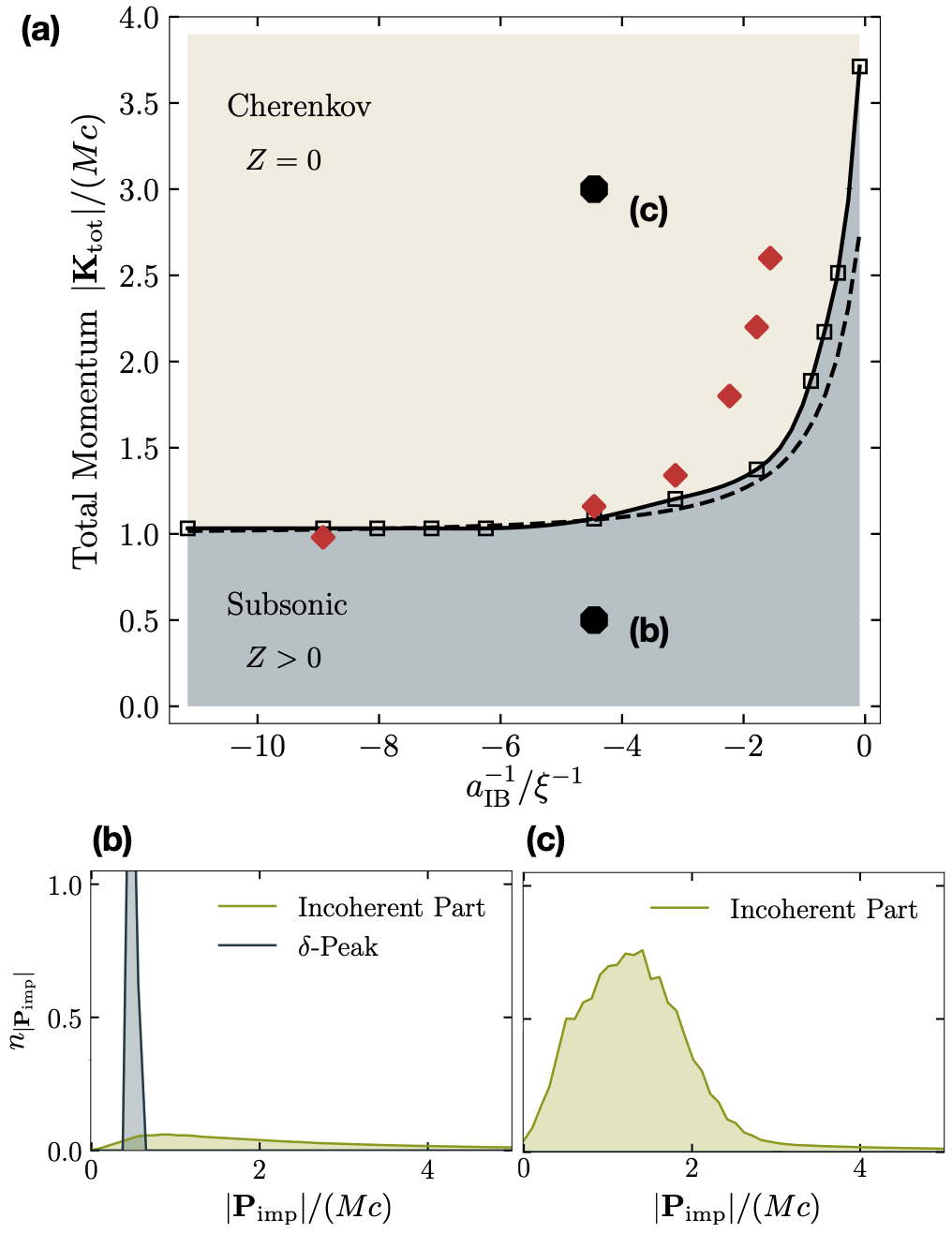}
	\caption{\textbf{Quasiparticle breakdown at the subsonic-to-supersonic phase transition}, which can be controlled by the interaction strength (a). The subsonic regime is realized for small total momentum $|\vec{K}_{\rm tot}|$, where the spectral function exhibits the characteristic delta-function peak at the ground state polaron energy, corresponding to a non-vanishing quasiparticle residue $Z>0$ (b). Different symbols correspond to different methods used to determine the critical momentum. At an interaction-dependent critical total momentum, the velocity of the impurity reaches the speed of sound and phonons start to accumulate indefinitely in the ground state. This leads to quasiparticle breakdown, signaled by the disappearance of the coherent delta-function peak in the spectral function (c), corresponding to $Z=0$. This figure was adapted from \cite{Seetharam2021arXiv}, see also \cite{Seetharam2021}.}
	\label{figSeetharam}
\end{figure}

Qualitatively identical results were found by Seetharam et al.~\cite{Seetharam2021,Seetharam2021arXiv} for models including two-phonon terms, i.e. beyond the Fr\"ohlich Hamiltonian. Their results are shown in Fig.~\ref{figSeetharam} (a). For strong interactions, an extended regime of subsonic polarons with $Z_{\vec{k}} > 0$ has been identified. As another characteristic signature of the subsonic-to-supersonic transition, Seetharam et al.~\cite{Seetharam2021arXiv} demonstrated that the full distribution function $n_{P} = \langle \sum_{|\vec{P}_{\rm imp}|=P} \ket{\vec{P}_{\rm imp}} \bra{\vec{P}_{\rm imp}}\rangle$ of the bare impurity momentum $\vec{P}_{\rm imp}$ shows a delta-function peak on top of an incoherent background in the subsonic regime, see Fig.~\ref{figSeetharam} (b). In contrast, in the supersonic regime the delta-function disappears and a more pronounced incoherent feature is predicted, as shown in Fig.~\ref{figSeetharam} (c).

Seetharam et al.~\cite{Seetharam2021} also studied how the Loschmidt-echo $|C(t)|$, defined in Eq.~\eqref{eqAtDef} above, decays when the total momentum is in the supersonic regime. Notably, they found a power-law decay $C(t) \simeq t^{- \gamma}$, highlighting the involved non-Markovian dynamics. This is in contrast to the exponential decay of $C(t)$ predicted in the same setting using a Markovian master equation approach \cite{Nielsen2019}.

\subsubsection{Hydrodynamics and Boltzmann equation.}
While the description of the phonon bath as Markovian is questionable at very low temperatures, this assumption can be justified at more elevated temperatures. Treating impurity-boson interactions perturbatively, Lausch et al.~\cite{Lausch2018} derived a finite-temperature master equation governing the dynamics of the impurity density matrix. In the low-temperature limit and from one-phonon (Fr\"ohlich) terms they obtain a similar result as Nielsen et al.~\cite{Nielsen2019}; however Lausch et al.~\cite{Lausch2018} also included two-phonon terms (beyond Fr\"ohlich) and thermal fluctuations. Next, from the master equation, they derived a Boltzmann equation describing the dynamics of a given distribution of impurity momenta. 

\begin{figure}
	\centering
	\includegraphics[width=\linewidth]{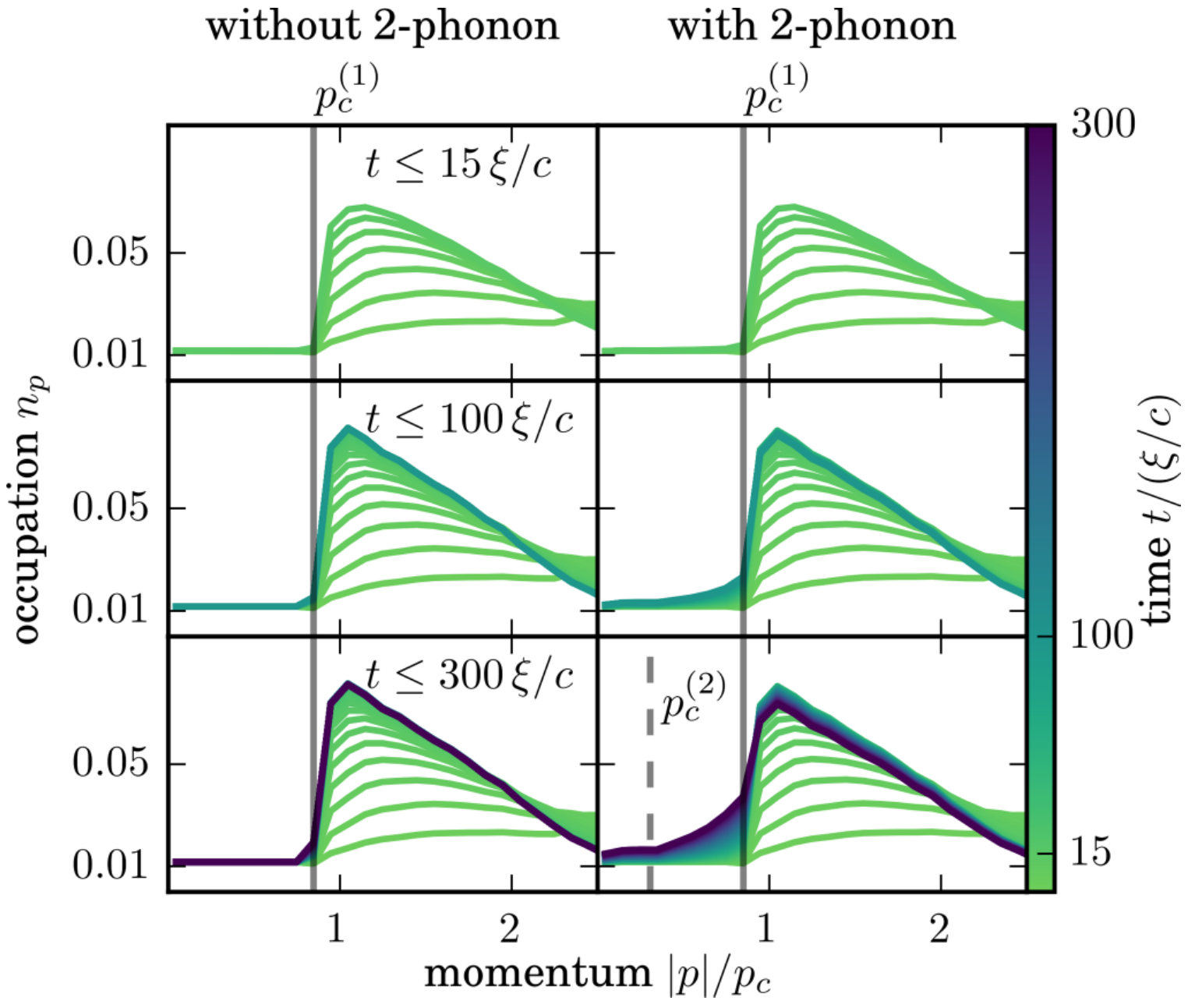}
	\caption{\textbf{Two-step cooling dynamics} of an impurity in a BEC can be identified in the polaron momentum distribution $n_p$ shown here as a function of time. An initially hot polaron scatters phonons and is sympathetically cooled down by the BEC. Above the critical momentum $|p|>p_c= M c$ determined by the speed of sound in the BEC, a pre-thermal stationary state is quickly established through one-phonon processes (left column). Two-phonon processes can also equilibrate lower momenta, but on a much longer time scale (right column). This figure was re-printed from \cite{Lausch2018}.}
	\label{figLausch}
\end{figure}

Using the so-obtained hydrodynamic description, Lausch et al.~\cite{Lausch2018} investigated the cooling dynamics of an initially hot gas of heavy impurity atoms in a three-dimensional BEC. The latter was assumed to be at a temperature of $0.1 T_c$, below the critical condensation temperature $T_c$. At short times they demonstrated that a quasi-steady state is quickly reached due to one-phonon scattering processes. Remarkably, this state is found to be very non-thermal for impurities heavier than the surrounding host bosons, $M > m_{\rm B}$, see Fig.~\ref{figLausch} (left panel), with only a small population of low-energy (subsonic) polaronic states. This observation was explained by the strongly constrained nature of one-phonon scattering events by energy and momentum conservation -- the same effect that ultimately leads to the superfluidity of the BEC.

As shown in Fig.~\ref{figLausch} (right panel), the inclusion of two-phonon scattering terms eventually leads to thermalization of the impurity cloud with the surrounding BEC. However owing to the significantly reduced matrix elements of the two-phonon scattering events, due to the absence of Bose enhancement by the BEC in these terms, 
\begin{equation}
    \frac{\Gamma_{\rm 2 ph}}{\Gamma_{\rm 1ph}} \sim \frac{1}{2 \pi \xi^3 n_0},
\end{equation}
the final equilibration takes a order of magnitude longer than the initial establishment of the pre-thermal quasi-steady state.  

In a subsequent work, Lausch et al.~\cite{Lausch2018a} studied similar cooling dynamics using their hydrodynamic approach in lower dimensional systems. In this case they found that thermalization is dominated by two-phonon processes, owing to infrared divergences in the corresponding scattering rates. This physics is related to the accumulation of low-energy thermal phonon excitations which also underlies the Mermin-Wagner-Hohenberg theorem.

\subsection{Polaron motion and transport}
Another topic of great interest involves the motion of polarons, or interacting impurity atoms, themselves. This includes problems where an initially moving impurity is inserted and can be slowed down by its interaction with the surrounding Bose gas, or even reflected; or situations in which an external force, or field, is applied, and polaron transport is studied. Such situations are of great experimental relevance due to their conceptually simple nature. On the other hand, complete theoretical descriptions are challenging; this makes for an ideal test case for developing and benchmarking new theoretical approaches, however.

\subsubsection{Bloch oscillations without a lattice in 1D.}
Interaction effects are most strongly pronounced in low-dimensional quantum systems, due to the restricted available phase space. A particularly striking effect was observed by Meinert et al.~\cite{Meinert2017}, considering an impurity atom accelerated inside a strongly interacting one-dimensional Bose gas. 

\begin{figure}
	\centering
	\includegraphics[width=0.9\linewidth]{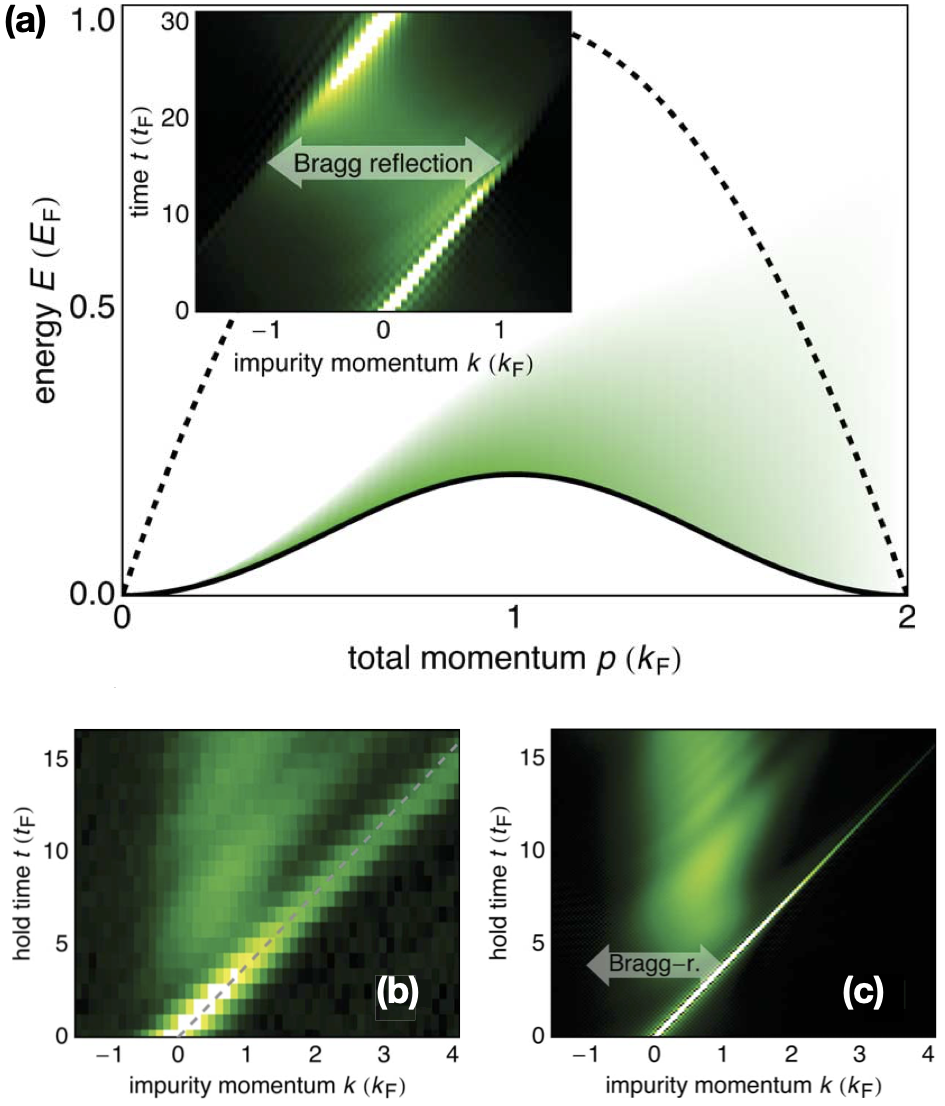}
	\caption{\textbf{Bloch oscillations without an underlying lattice} can be realized by mobile impurities moving through a one-dimensional Bose gas. The latter effectively fermionizes in one dimensions, with Fermi momentum $k_{\rm F}$, and exhibits a many-body energy spectrum with approximate $2k_F$ periodicity (a). This realizes the Bragg-reflections $\Delta k = \pm 2 k_{\rm F}$ of an accelerating impurity which parallels Bloch oscillations in a lattice. The effect can be observed in time-of-flight measurements of the impurity \cite{Meinert2017}: (b) experiment and (c) theory. This figure was re-printed from \cite{Meinert2017}.}
	\label{figPolaronBOMeinert}
\end{figure}

The team realized an ensemble of one-dimensional Bose gases, in a hyperfine state of bosonic Ceasium and initialized a small concentration of impurities (on average one per tube) in a different hyperfine state by applying a short resonant radio-frequency pulse. The smaller magnetic moment of the impurities causes them to experience a weaker magnetic trapping potential, and they start to accelerate along the tubes due to the gravitational force. The gas parameter $\gamma \gg 1$ of the host bosons was chosen to be large, which effectively fermionizes them, and leads to the periodic excitation spectrum shown in Fig.~\ref{figPolaronBOMeinert}. Its period is given by twice the corresponding Fermi momentum, $k_{\rm F} = \pi n_{1D}$, which in turn is determined by the density $n_{1D}$ of the Bose gas.

In the subsequent dynamics, Meinert et al.~\cite{Meinert2017} studied the impurity density distribution by the time-of-flight method. First, they observe an increase of the impurity momentum with time; however, around the time when the momentum reaches $2 k_{\rm F}$, a pronounced back-scattering to momenta around zero is observed. This sequence repeats and leads to damped periodic oscillations, see Fig.~\ref{figPolaronBOMeinert} (b) (experiment) and (c) (numerical simulations).

The observed oscillations are reminiscent of Bloch oscillations, a hallmark of coherent dynamics in systems with a broken translational symmetry. In contrast, the system studied by Meinert et al.~\cite{Meinert2017} does not break any symmetries and remains translationally invariant. Following earlier theoretical predictions by Gangardt and co-workers \cite{Gangardt2009,Schecter2012a}, the observed phenomenon can nevertheless be viewed as a closely related effect: The strong interactions among the bosons lead to short-range crystalline correlations, effectively providing a periodic structure probed by the impurity. In a related fashion, the kinematically allowed scattering processes are constrained due to the periodicity with $2 k_{\rm F}$ of the low-energy excitations in the Bose gas, which leads to the emergence of an effective Brillouin zone structure. 

The experimental demonstration of this effect by Meinert et al.~\cite{Meinert2017} provides one of the most striking manifestations of the special role of interactions in one dimensions. In fact, there exists a large body of theoretical work on related problems involving impurity atoms moving through an interacting one-dimensional gas. For example, Knap et al.~\cite{Knap2014a} studied persistent oscillations, so-called quantum flutter, after injecting a moving impurity atom in a one-dimensional chain; Schecter et al.~\cite{Schecter2012} studied peculiarities of the superfluid critical velocity owing to the special properties of one-dimensional systems; Zvonarev et al.~\cite{Zvonarev2007} studied beyond-Luttinger liquid effects in quantum dynamics of spin-flip excitations in a ferromagnetic Bose gas; Schecter et al.~\cite{Schecter2012a} provided a hydrodynamic theoretical description of the predicted Bloch-like oscillations in strongly interacting Bose gases and analyzed the impurity's mobility and dynamical phase diagram.

\subsubsection{Polaron trajectories in higher dimensions}
\begin{figure}
	\centering
	\includegraphics[width=0.85\linewidth]{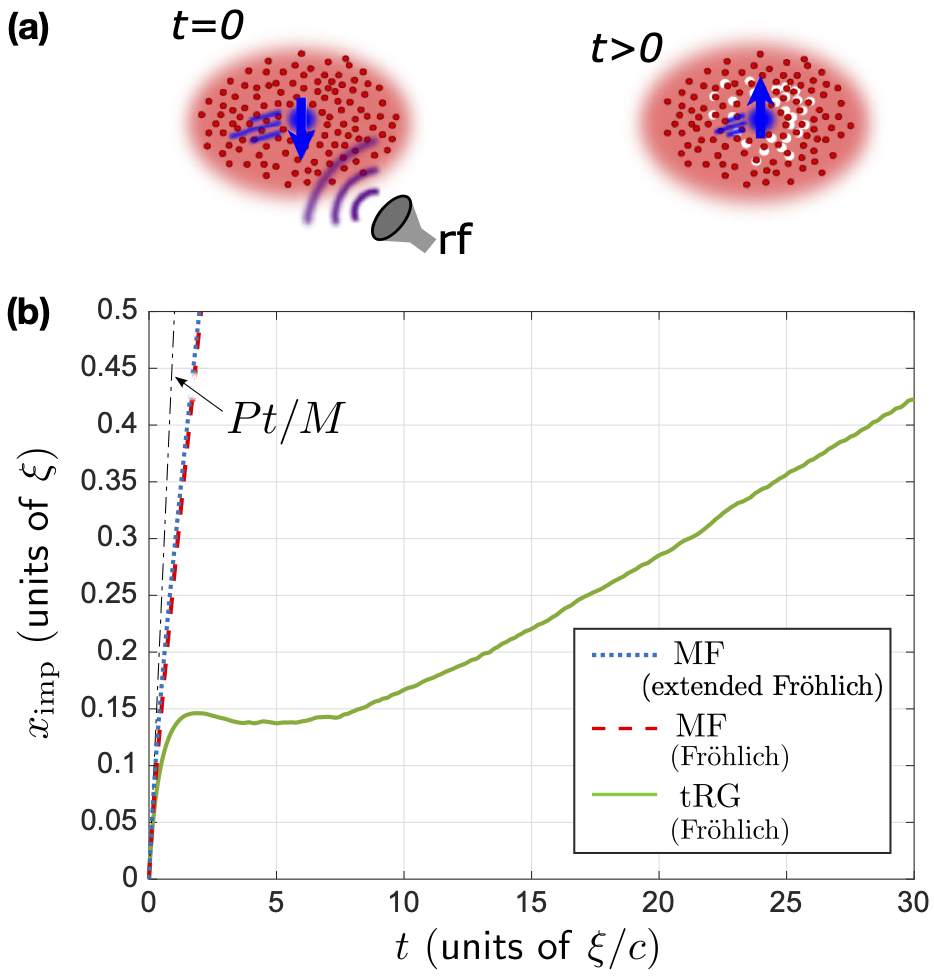}
	\caption{\textbf{Slow-down of a mobile impurity} following a sudden interaction quench into the strong-coupling regime, which could be realized experimentally by a sudden radio-frequency (rf) pulse (a). Theoretical predictions for the time-dependent ($t$) impurity position $x_{\rm imp}(t)$ indicate the possibility of strong non-perturbative effects (b): Time-dependent RG calculations within the Bogoliubov-Fr\"ohlich model predict a dramatic slow-down of the dynamically dressed impurity (green line), in a regime where mean-field predictions of the Fr\"ohlich and extended Fr\"ohlich models consistently predict a less pronounced slow-down. This figure was adapted from \cite{Grusdt2018dRG}.}
	\label{figTrajectoriesDRG}
\end{figure}

While interaction effects are less pronounced in higher dimensions, the problem of an impurity atom injected into a BEC with a non-zero initial velocity $v_{\rm I}$ and starting to interact with the BEC remains interesting to study the emergence of friction. Experimentally this setting can be relatively easily realized by first accelerating a non-interacting impurity and transferring it into an interacting hyperfine state using a fast radio-frequency pulse, see Fig.~\ref{figTrajectoriesDRG} (a). 

The ensuing polaron trajectories were studied theoretically in three dimensions by Grusdt et al.~\cite{Grusdt2018dRG}, using a time-dependent variant of the renormalization group method and time-dependent Lee-Low-Pines mean-field wavefunctions. In the subsonic regime and for weak interactions, they predict a gradual slow-down of the impurity which can be explained by the dynamical formation of the polaronic dressing cloud and the simultaneous increase of the renormalized polaron mass. In a related setting, Drescher et al.~\cite{Drescher2019} used time-dependent Lee-Low-Pines mean-field theory to predict oscillating impurity trajectories on the repulsive side of the Feshbach resonance, which they related to an impurity-boson bound state occupied by multiple phonons. Similar oscillations associated with bound state formation have also been found in other quantities~\cite{Shchadilova2016,Drescher2021} but have more recently been argued to be tied to the absence of phonon-phonon interactions~\cite{Mostaan2023}.

In the strong coupling regime but with the initial impurity velocity still subsonic, Grusdt et al.~\cite{Grusdt2018dRG} predict the possibility of strongly renormalized impurity dynamics resulting from beyond mean-field effects, see e.g. Fig.~\ref{figTrajectoriesDRG} (b). This suggests that the polaronic dressing cloud in the dynamical setting can become more pronounced than expected for equilibrium polarons. Their studies of beyond mean-field effects so far were limited to the effective Fr\"ohlich model, however, and further investigations - both in theory and experiment - of this setting would be desirable.  

Another popular approach to describing impurity dynamics in a Bose polaron context is based on the use of the (quantum) Brownian motion formalism. There, the phonon excitations of the BEC are modeled as providing a bath, which generates noise and thus affects the impurity motion. Bonart et al.~\cite{Bonart2012} considered a one-dimensional system, modelling the experiment by Cantani et al.~\cite{Catani2012}, and demonstrated that phonons have a super-Ohmic character with low-frequency spectral density $S(\omega) \sim \omega^3$. They used their approach to calculate relaxation dynamics after a projective measurement and to describe the breathing oscillations observed by Catani et al.~\cite{Catani2012}. A generalization of this approach to higher dimensions was developed by Lampo et al.~\cite{Lampo2017}, who derived a quantum analogue of a stochastic Langevin equation describing the impurity dynamics. They identified significant memory effects in the bath and predicted that the latter give rise to super-diffusive behavior of an untrapped impurity atom.

\subsubsection{Polaron transport: continuum and lattice.}
A topic of particular relevance due to its close connection to solid state systems \cite{alexandrov2009advances}, is the question of transport coefficients that several authors have addressed. 

Punk and Sachdev~\cite{Punk2013} used field-theory methods to calculate the diffusion constant $D$ of a single impurity atom as a function of temperature, in the quantum critical regime above a super-fluid to Mott transition in two dimensions. The latter can be directly measured experimentally from the mean-square displacement,
\begin{equation}
    \langle \hat{\vec{x}}^2 \rangle = 4 D t,
\end{equation}
as a function of time $t$, which is accessible in modern quantum gas microscopy setups. Punk and Sachdev~\cite{Punk2013} showed that although the quantum criticality of the bulk is governed by a single time scale $\tau_R \sim \hbar / (k_B T)$, the resulting impurity time scale depends on couplings which are formally irrelevant at the critical point, namely the impurity inverse mass $1/M$ and the associated energy scale $M c^2$, where $c$ is the velocity of bulk bosonic excitations at the critical point. Using a hydrodynamic description, Punk and Sachdev~\cite{Punk2013} showed that the impurity diffusion constant in the quantum critical regime is given by the scaling form
\begin{equation}
    D \simeq \frac{c^6}{g_{\rm IB}} \Phi \bigg(\frac{T}{m c^2},\frac{\Delta}{m c^2} \bigg),
\end{equation}
where $\Phi(\tilde{T},\tilde{\Delta})$ is a scaling function which depends on the unit-less temperature $\tilde{T}$ and Mott gap $\tilde{\Delta}$. 

Further effort has been invested in studying transport of polarons in optical lattices. Bruderer et al.~\cite{Bruderer2008} formulated a strong-coupling theory, based on the Lang-Firsov transformation, and derived a generalized Master equation. They solved the latter in one dimension and found a cross-over as a function of temperature from coherent to incoherent polaron motion at elevated temperatures. Next they added a potential gradient and calculated the voltage-current relation at different temperatures, assuming a Markovian bath. They found a characteristic Esaki-Tsu type relation \cite{Esaki1970} describing Ohmic transport in the weak-field limit and negative-differential conductance, $dI/dV < 0$, in the limit of a strong potential gradient. 

In a follow-up work, Johnson et al.~\cite{Johnson2011} performed time-evolving block decimation (TEBD) calculations -- a time-dependent variant of DMRG -- to study Bose polaron transport in one-dimensional lattices quasi-exactly. They pointed out, and established numerically, that impurity transport can be used as a non-destructive probe of the excitation spectrum of the Bose gas. They also analyzed deviations of the voltage-current relation from the Esaki-Tsu relation in the Mott insulating regime, and argued that the latter reveal information about the host Bose gas. 

Grusdt et al.~\cite{Grusdt2014BO} revisited the problem of Bose polaron transport along one dimension, assuming however that the impurity is interacting with a $d$-dimensional host Bose gas. They observed strongly renormalized Bloch oscillations by interactions, which can be used as for direct measurements of the renormalized polaron dispersion relation,  using the Lee-Low-Pines mean-field calculations. Moreover, they extracted the voltage-current relation and demonstrated a sub-Ohmic dependence for small fields if the host Bose gas has dimensionality $d>1$. In this case, significant deviations from the standard Esaki-Tsu relation are predicted.  

In a further development, Yin et al.~\cite{Yin2015PRA} studied multi-band effects for impurities in a lattice which is immersed in a host BEC. They used a strong coupling (Lang-Firsov) approach and predicted a new interband self-trapping effect as well as strongly renormalized interband relaxation dynamics caused by polaronic effects.

\section{Bose polarons in low dimensions}
\label{secLowDimPolarons}
Up to this point in the review we have not devoted much time to discussions of the role of dimensionality of the system: On one hand, many qualitative effects including polaronic dressing and mass enhancement are not qualitatively altered when changing spatial dimension. On the other hand, a number of striking effects have been identified over the years that depend extremely sensitively on dimensionality: For example, the Efimov effect relies on an emergent scale-invariance which can only be realized in $d=3$ spatial dimensions. 

In this chapter, we will focus on low dimensional polaron problems, specifically in $d=1$ and $d=2$. As will be discussed, this leads to peculiarities in the polaronic phase diagrams, and especially in 1D systems provides access to a number of exactly solvable limits providing extremely valuable benchmarks for approximate analytical methods of all kinds. Experimentally, the ability to use intense laser light to confine ultracold atoms to lower-dimensional manifolds has allowed thorough investigations of $d=1,2$ dimensional polaron problems.

\begin{figure*}
	\centering
	\includegraphics[width=\linewidth]{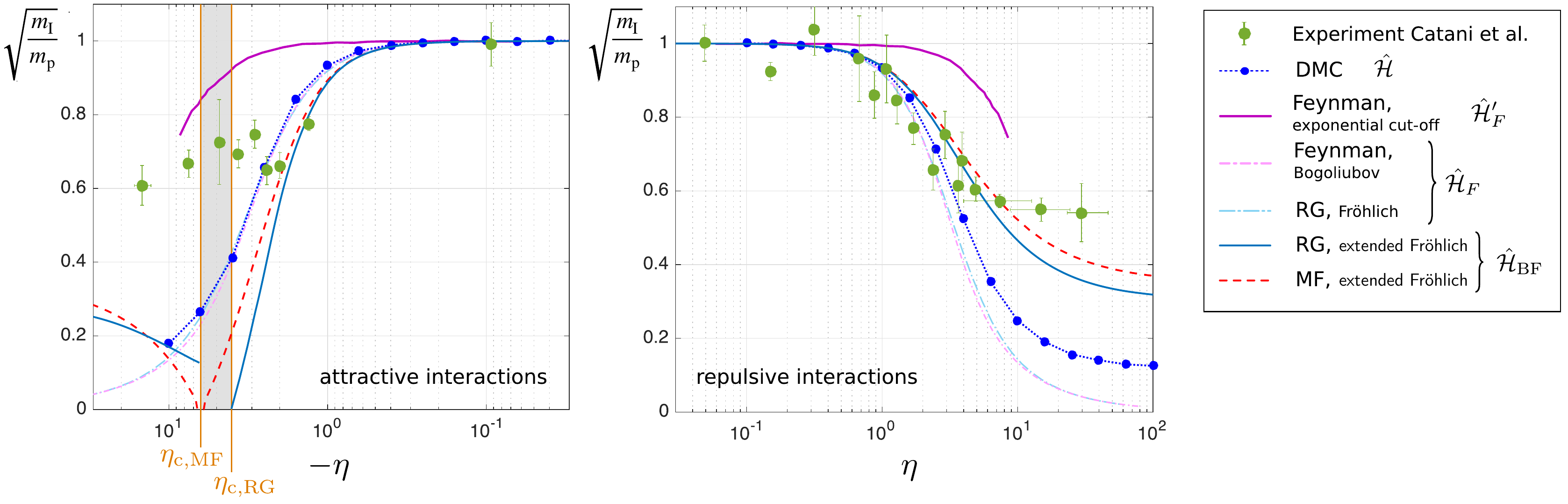}
	\caption{\textbf{Renormalized polaron mass} $m_{\rm p}$ in $d=1$-dimensional tubes measured by Catani et al.~\cite{Catani2012} is compared to different theoretical predictions. In the experiment \cite{Catani2012} the ratio $\sqrt{m_{\rm I}/m_{\rm p}}$ is directly related to the amplitude renormalization $\sigma/\sigma_0$ of the breathing oscillations described in Fig.~\ref{figCataniExp}, with $\sigma_0$ the amplitude of non-interacting impurities, $\eta=0$. The original experimental data was compared to the Feynman path-integral prediction based on a Fr\"ohlich model with exponential UV cut-off (solid purple line) \cite{Catani2012} which can be improved by working with the Bogoliubov phonon dispersion \cite{Grusdt2017RG1D}. The most accurate theoretical prediction should be from the diffusion Monte Carlo (DMC) approach (blue dots). In the strong coupling regime, systematic discrepancies remain, likely originating from effects beyond the simplified model of breathing oscillations assuming adiabatic polarons moving in a homogeneous BEC. The figure is adapted from \cite{Grusdt2017RG1D}, experimental data from \cite{Catani2012}.}
	\label{figCataniThy}
\end{figure*}

\subsection{ Bose polaron in  1D}
\label{Sec:1DBosePolaron}
Both in theory and experiments, strongly interacting one-dimensional quantum gases have been intensely studied, see e.g. \cite{Cazalilla2011} for a review or the classic textbook by Giamarchi \cite{Giamarchi2003}. In the context of Bose polarons, a particularly important role is played by the Lieb-Lininger gas, i.e. a system of $N$ one-dimensional bosons with coordinates $x_1,...,x_N$ interacting through contact interactions of strength $g_{\rm BB}$. Its Hamiltonian,
\begin{equation}
    \H = - \frac{1}{2m_{\rm B}} \sum_{n=1}^N \frac{\partial^2}{\partial x_n^2} + g_{\rm BB} \sum_{n<m} \delta(x_n-x_m),
    \label{eqHB1d}
\end{equation}
is the first-quantized version of Eq.~\eqref{eqHB} for $d=1$. 

As shown by Lieb and Lininger in 1963 \cite{Lieb1963a}, Eq.~\eqref{eqHB1d} is exactly solvable by Bethe ansatz, rendering the properties of the host Bose gas analytically accessible for arbitrary interaction strengths. In particular, Lieb and Lininger showed that the model is characterized by just one dimensionless number, the so-called gas parameter:
\begin{equation}
    \gamma = \frac{m_{\rm B} g_{\rm BB}}{n_0} = \frac{2}{n_0 |a_{\rm BB}|}.
\end{equation}
Here $n_0=N/L$ is the density of the Bose gas and in the second step we introduced the one-dimensional s-wave scattering length
\begin{equation}
    a_{\rm BB} = - 2 / (m_{\rm B} g_{\rm BB}).
\end{equation}

When $\gamma \lesssim 2$ is sufficiently small, Lieb and Lininger further showed that many properties of the Bose gas, including the dispersion relation of phonon excitations, are accurately described by Bogoliubov theory. This provides an excellent starting point to study Bogoliubov-Fr\"ohlich polarons \cite{Mathey2004,Tempere2009,Catani2012} upon introducing an additional impurity particle into the Lieb-Lininger gas. On the other hand, the breakdown of Bogoliubov-Fr\"ohlich theory in a strongly interacting host gas can be explored when $\gamma > 2$ \cite{Grusdt2017RG1D}.

In such a setting in 1D, the impurity-boson interaction is conveniently characterized by a dimensionless parameter,
\begin{equation}
    \eta = g_{\rm IB} / g_{\rm BB}.
    \label{eqDefEta}
\end{equation}
Here, as in the following, it is assumed that impurities have contact interactions with the host bosons,
\begin{equation}
    \H_{\rm IB} = g_{\rm IB} \sum_{n=1}^N \delta(r_{\rm I} - x_n),
\end{equation}
where $r_{\rm I}$ denotes the one-dimensional coordinate of the impurity.

\subsubsection{Florence experiment: Catani et al.~\cite{Catani2012}.}
\label{SubsubsecCatani1D}
Now we turn to a discussion of the 1D polaron physics observed in the seminal experiment by Catani et al.~\cite{Catani2012}. We have already reviewed the experimental setup and the dynamical sequence it realizes in the previous chapters. Here we will focus on the measured effective polaron mass $m_{\rm p}$ and its theoretical interpretation.

In Fig.~\ref{figCataniThy} we show the experimentally extracted mass ratio $\sqrt{m_{\rm I} / m_{\rm p}}$ as a function of the dimensionless impurity-boson interaction $\eta$, see Eq.~\eqref{eqDefEta}, separately for attractive (left) and repulsive (right) interactions. On both sides, significant mass renormalization is observed as the interaction strength increases, reaching values between $m_{\rm p}/m_{\rm I} = 2 ... 4$ at the strongest couplings. While systematic errors in relating the observed dynamical amplitude suppression to the effective mass likely play an increasingly important role for stronger interactions, the experiment reveals a strikingly strong influence of impurity-boson interactions on the underlying polaronic state.

Already in the original paper by Catani et al.~\cite{Catani2012} a direct quantitative comparison with theoretical models was performed. Starting from a Luttinger liquid description of the host Bose gas, the authors derived an effective Fr\"ohlich Hamiltonian $\mathcal{H}_{\rm F}'$ similar to the Bogoliubov-Fr\"ohlich model in Eqs.~\eqref{eqHBogo0}, \eqref{eqHIBFroh}. The main difference to the latter was the assumption of a purely linear phonon dispersion, $\omega_k = c k$, and the use of an exponential cut-off in the impurity-phonon scattering amplitude $V_k$, i.e. 
\begin{equation}
    V_k  = \frac{g_{\rm IB}}{2 \pi} \sqrt{K |k|} e^{- |k|/(2 k_c)}
\end{equation}
where $K$ is the Luttinger parameter of the host Bose gas and $k_c$ a phenomenological cut-off parameter. The momentum cut-off $k_c$ is set by the healing length of the Bose gas, $k_c^{-1} \sim \xi$, below which the field-theoretical (bosonization) description breaks down.

Surprisingly, the mass renormalization predicted by the field-theoretic Fr\"ohlich Hamiltonian $\H_{\rm F}'$ is significantly weaker than observed experimentally, see solid magenta line in Fig.~\ref{figCataniThy}. In particular, the disagreement starts already at weak couplings, where both the employed Feynman path-integral method as well as the underlying effective Fr\"ohlich Hamiltonian ignoring two-phonon terms should provide an accurate description. Nevertheless, within the effective model an effective interaction $\tilde{g}_{\rm IB}$ about a factor of $3.15$ stronger than measured needs to be used to explain the experimental results.

This discrepancy was later clarified by Grusdt et al.~\cite{Grusdt2017RG1D} by working with the Bogoliubov-Fr\"ohlich Hamiltonian. Since $\gamma \approx 0.44 \ll 2$ in the experiment, the use of Bogoliubov approximation is justified in this regime. Within the Bogoliubov-Fr\"ohlich model the momentum cut-off is still defined around $1/\xi$, but in contrast to the field-theoretical model $V_{k \to \infty}$ remains finite in the UV limit, see Eq.~\eqref{eqVkDef}. Instead the momentum cut-off is realized by the cross-over of the Bogoliubov phonon-dispersion from $\omega_k \simeq ck$ for $k \lesssim 1/\xi$ to $\omega_k \simeq k^2 / (2m_{\rm B})$ for $k \gtrsim 1/\xi$. As a result, the dressing with high-energy phonons at energies $\omega > c /\xi$ contributes significantly more to the renormalized polaron mass. This leads to the excellent agreement of the experiment with theoretical predictions based on the Bogoliubov-Fr\"ohlich model at weak couplings $|\eta| \lesssim 2$ shown in Fig.~\ref{figCataniThy} (dash-dotted lines).

At stronger couplings, beyond-Fr\"ohlich effects are expected to play a role, which should lead to an asymmetry between attractive and repulsive interactions $\pm \eta$. The measured polaron mass indeed starts to become asymmetric under changes of the sign of interactions for $|\eta| \gtrsim 3$. However, even the most accurate diffusion Monte Carlo simulations (blue dots in Fig.~\ref{figCataniThy}), which capture numerically exactly the ground state properties of the impurity in a Lieb-Lininger gas, do not yield quantitative agreement with the experiment at strong couplings. This suggests systematic problems in directly relating the dynamical signal by Catani et al.~\cite{Catani2012} to the mass ratio $\sqrt{m_{\rm I} / m_{\rm p}}$. Indeed, Catani et al.~\cite{Catani2012} already point out for the entire parameter range that special care has to be taken in applying this one-to-one relation. 

To date, the experiment by Catani et al.~\cite{Catani2012} remains the only study of Bose polarons reporting significant mass renormalization. The successful comparison with theoretical models should provide additional motivation to carry out similar experiments in higher-dimensional systems in the future.

\subsubsection{Exact solutions.}
The first exact solution of a 1D polaron problem was derived long before any concrete experiments on those systems could even be envisioned. In 1964 McGuire \cite{McGuire1965} used Bethe ansatz wavefunctions to solve exactly the problem of a single down-spin impurity immersed in a sea of up-spin fermions. He had to assume identical masses $m_{\rm I} = m_{\rm B}$ of the impurity and host particles, but allowed a tunable delta-function inter-particle potential of strength $g_{\rm IB} \geq 0$. In the context of Bose polarons, McGuire's result is relevant since his non-interacting host fermions can be mapped directly to hard-core host bosons \cite{Girardeau1960}. I.e. McGuire's solution, which includes an analytic expression for the polaron energy $E_{\rm p}$, describes an equal-mass impurity inside a Tonks-Girardeau gas of host bosons with a large gas parameter $\gamma \gg 1$. 

In a subsequent paper, McGuire extended his analysis to include attractive impurity-boson interactions $g_{\rm IB} < 0$ in the otherwise identical setting \cite{McGuire1966}. This case is even more interesting, since McGuire found that the ground state involves binding of the impurity to a host particle. Nevertheless, the physical properties of the ground state are analytical continuations of the results obtained in the repulsive case. 

Exact Bethe ansatz solutions of the Bose polaron problem are not limited to the Tonks-Girardeau limit of hard-core bosons or single impurities. Already in 1967 Gaudin \cite{Gaudin1967} and Yang \cite{Yang1967} extended McGuire's solution based on a single flipped spin in a Fermi sea of free up-spins to mixtures of $N_\uparrow$ ($N_\downarrow$) fermions interacting with contact interactions $g_{\rm IB}$. Following the advent of cold atom quantum gas experiments, Li et al.~\cite{Li2003} further extended this analysis to bosonic mixtures of $SU(2)$ (iso-) spins with contact interactions. 

\begin{figure}
	\centering
	\includegraphics[width=0.85\linewidth]{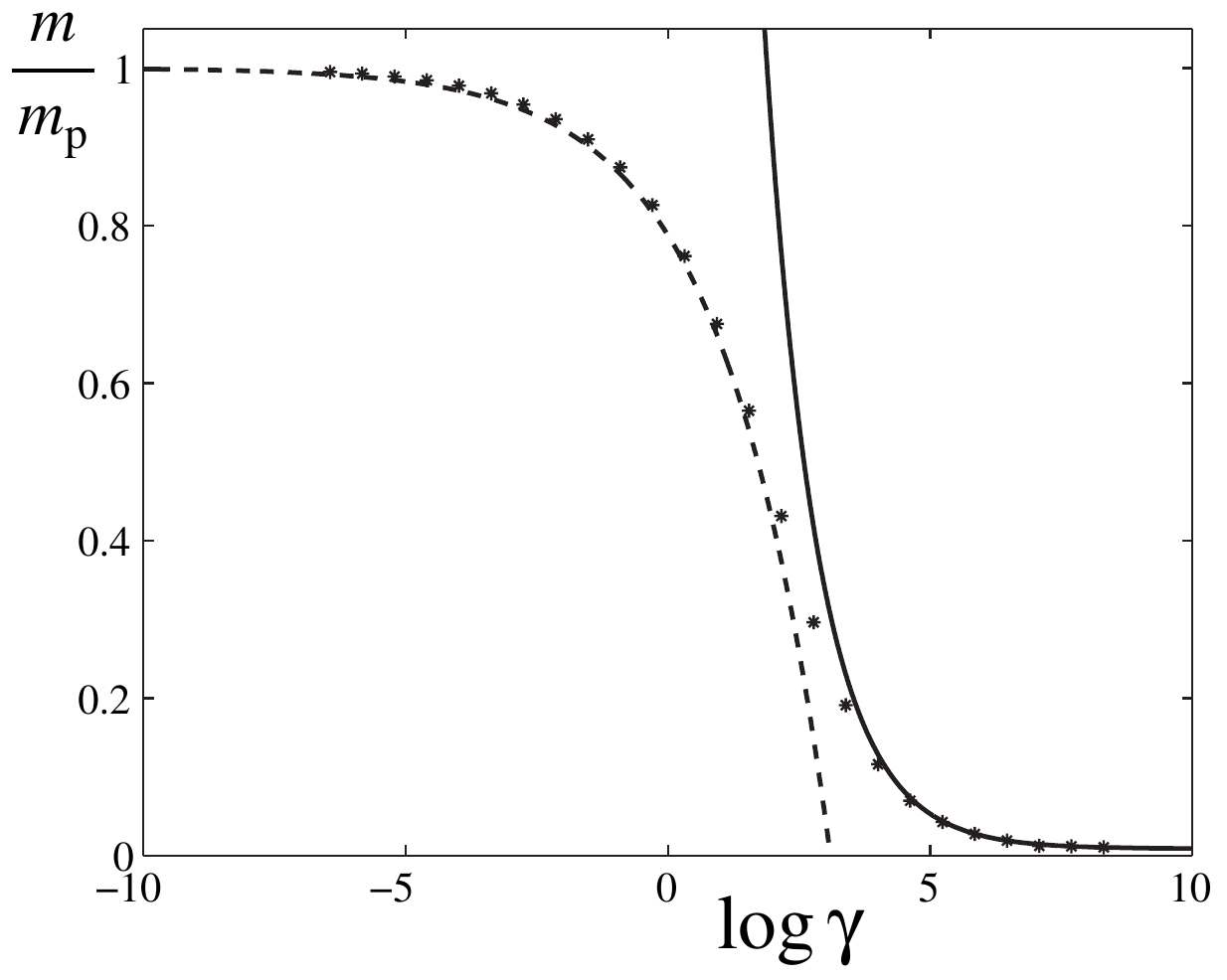}
	\caption{\textbf{Exactly solvable Bose polarons via Bethe ansatz} are realized in $d=1$ dimension for equal-mass $m_{\rm I}=m_{\rm B}=m$ and equal interaction $\gamma = \eta$ mixtures \cite{Gaudin1967,Yang1967}. Here Fuchs et al.~\cite{Fuchs2005} calculated the inverse renormalized polaron mass $m_{\rm p}$ exactly using Bethe ansatz for a system with total $N=111$ particles (stars), and compared to weak- and strong coupling asymptotic expressions respectively. The figure is adapted from \cite{Fuchs2005}.}
	\label{figFuchsBethe}
\end{figure}

In this setting the case of a single spin-flip excitation, taking the role of an individual mobile impurity, in a Lieb-Lininger gas of majority bosonic spins was analyzed by Fuchs et al.~\cite{Fuchs2005}, again using the Bethe ansatz. In the language of Bose polarons, their solution corresponds to the case of an equal-mass ($m_{\rm I}=m_{\rm B}$) impurity immersed in a Lieb-Lininger gas of arbitrary gas parameter $\gamma$, controlled by the host gas density $n_0$, and assuming equal-strength contact interactions $g_{\rm IB}=g_{\rm BB}$. Under these conditions, Fuchs et al. calculated the renormalized polaron mass $m_{\rm p}$ of the impurity as a function of the host gas parameter $\gamma$, which we show in Fig.~\ref{figFuchsBethe}. The exact Bethe-ansatz result interpolates smoothly between weak and strong coupling solutions by Bogoliubov-Fr\"ohlich theory (dashed line) and perturbative expansion in $1/\gamma$ (solid line), respectively. The crossover from one to the other regime takes place around $\gamma \approx 2$. Series expansions of the Bethe ansatz result to higher orders in $\gamma$ and $1/\gamma$ were derived by Ristivojevic~\cite{Ristivojevic2021}.

A particularly noteworthy result by Fuchs et al.~\cite{Fuchs2005} concerns the Tonks-Girardeau limit $g_{\rm IB}=g_{\rm BB} \to \infty$ of impenetrable particles. In this case, the Bethe ansatz equations can be solved exactly and Fuchs et al.~\cite{Fuchs2005} showed that 
\begin{equation}
    m_{\rm p} = N m_{\rm B} \to \infty, \qquad \text{as}~~g_{\rm IB} \to \infty.
\end{equation}
I.e. the impurity is self-localized in the 1D Bose gas. This result can be understood intuitively, since the hard-core interactions make it entirely impossible for host bosons to tunnel across the impurity. Hence any motion of the impurity would involve motion of the entire surrounding Bose gas, which endows the emergent excitation with a total mass of order $N m_{\rm B}$. This is a consequence of the highly constrained nature of one-dimensional systems, and the result is not expected to remain valid in higher dimensions $d>1$.

\begin{figure}
	\centering
	\includegraphics[width=\linewidth]{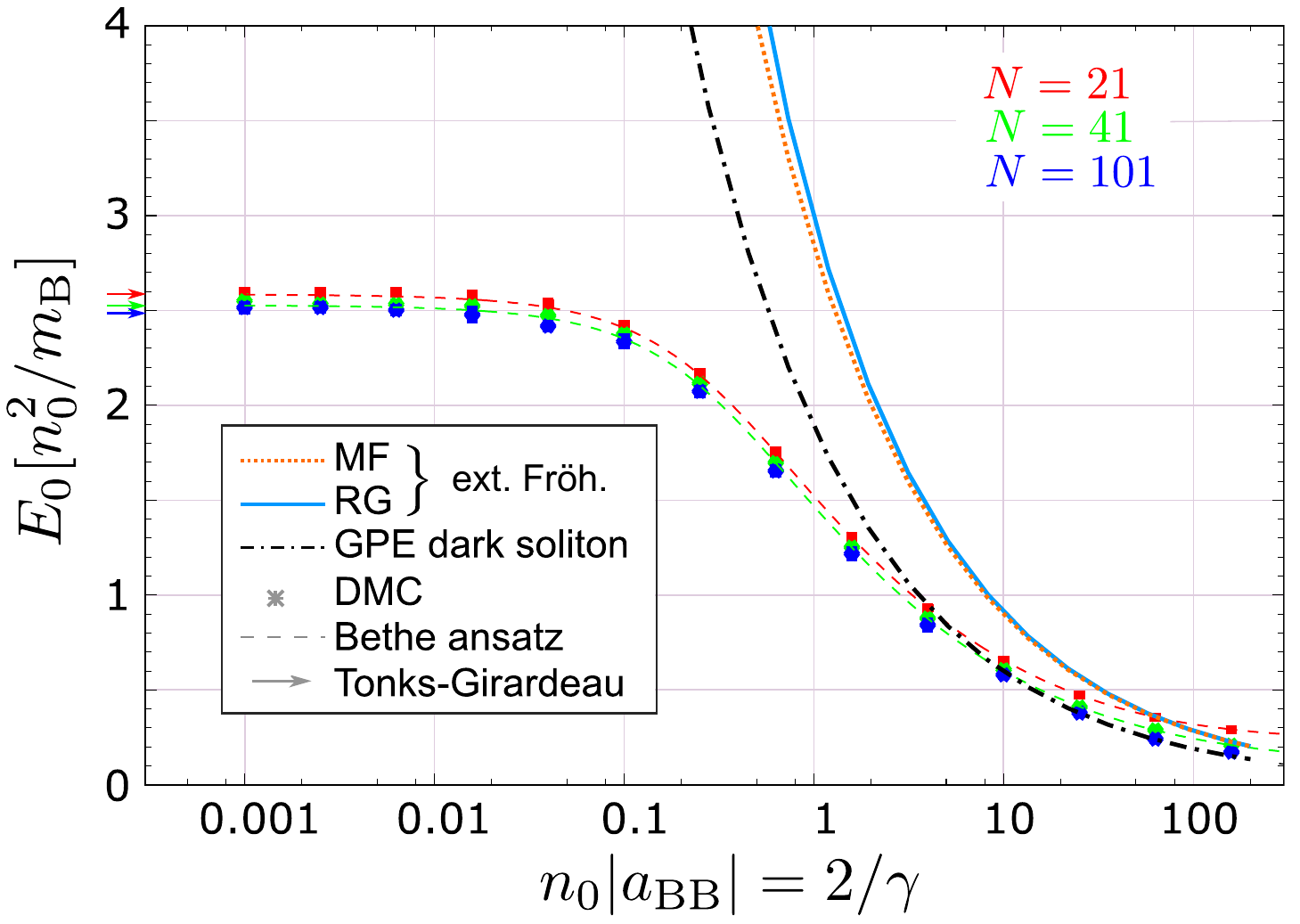}
	\caption{\textbf{Another exactly solvable limit of the Bose polaron} in $d=1$ dimension corresponds to an impenetrable $\eta=\infty$ and localized impurity with $m_{\rm I} \to \infty$. The corresponding ground state polaron energy $E_0$ is calculated here as a function of the inverse Bose-Bose interaction gas parameter $\gamma$. The exact results from Bethe ansatz are compared with numerically exact diffusion Monte Carlo (DMC), as a benchmark of the latter, and with various approximate methods working in the weak-coupling (small $\gamma$) regime: LLP mean-field theory (MF) and RG rely on the extended Bogoliubov-Fr\"ohlich Hamiltonian (i.e. including two-phonon terms); the GPE dark soliton uses the full microscopic Hamiltonian. The figure is re-printed from \cite{Grusdt2017RG1D}.}
	\label{figMinfgIBinf}
\end{figure}

Yet another exactly solvable limit of the Bose 1D Bose polaron is obtained by assuming an impenetrable, $g_{\rm IB} \to \infty$, and localized, $m_{\rm I} \to \infty$, impurity. In this case, Bethe ansatz can be directly used to describe the surrounding Lieb-Lininger gas of arbitrary gas parameter $\gamma$ and mass $m_{\rm B}$. Such calculations were performed in Ref.~\cite{Grusdt2017RG1D} to obtain the polaron energy, and compared to predictions by various approximate methods, see Fig.~\ref{figMinfgIBinf}. Only deep in the weak-coupling regime agreement is obtained with models which do not include phonon-phonon interactions. The latter are included in the GPE approach, in which the impenetrable localized impurity corresponds to a stationary dark-soliton solution. None of these methods are able to capture the saturation of the polaron energy at strong couplings, to its value in the Tonks-Girardeau gas. 

The weak-coupling calculations (MF and RG) in Fig.~\ref{figMinfgIBinf} themselves rely on approximate solutions of the extended Bogoliubov-Fr\"ohlich Hamiltonian. This simplified model can been solved exactly, however, in the case of a static impurity, $m_{\rm I} \to \infty$. The solution, provided by Kain and Ling \cite{Kain2018}, relies on the use of multi-mode Gaussian squeezed-coherent states to fully diagonalize the quadratic extended Bogoliubov-Fr\"ohlich Hamiltonian.

\subsubsection{Phase-diagram of the 1D Bose polaron.}
\begin{figure}
	\centering
	\includegraphics[width=\linewidth]{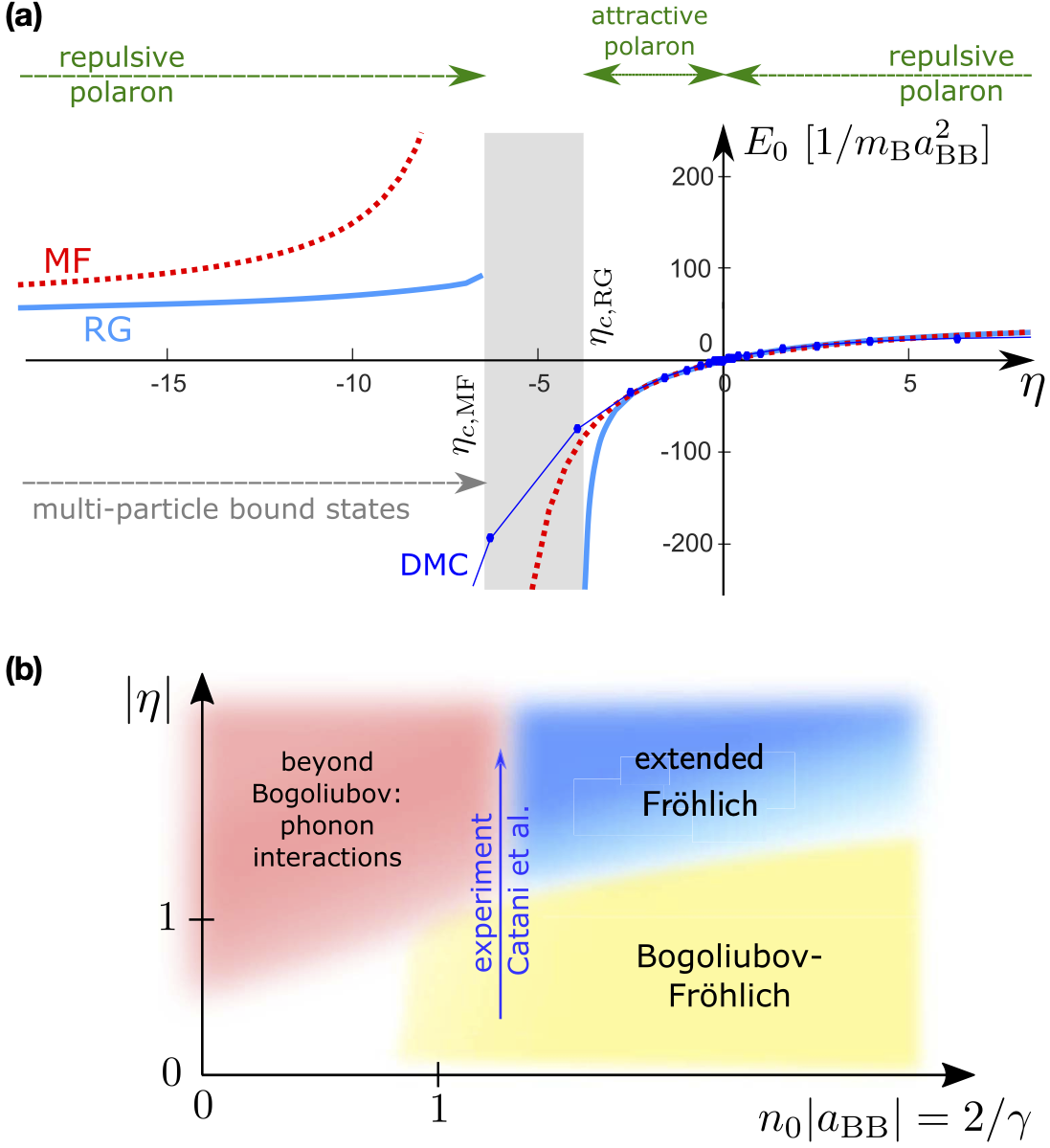}
	\caption{\textbf{Structure of the Bose polaron in $d=1$ dimension.} (a) The ground state polaron energy $E_0$ is calculated for repulsive ($\eta>0$) and attractive ($\eta < 0$) impurity-boson interactions. For weak attraction, corresponding to small negative $\eta$, the attractive polaron is realized. For repulsive $\eta > 0$, the repulsive polaron is found, which continuously turns into a meta-stable repulsive polaron state when $\eta=+\infty \to \eta = -\infty$. In this regime of strong negative $\eta$, the overall ground state is dominated by boson-boson interactions $\gamma$. (b) Regimes of validity of the different effective model Hamiltonians. The figure is adapted from \cite{Grusdt2017RG1D}.}
	\label{fig1DphaseDiag}
\end{figure}

Next we discuss the phase diagram of Bose polarons in 1D, i.e. how the polaron branches depend on the impurity-boson coupling $\eta = g_{\rm IB} / g_{\rm BB}$. To this end we focus on the weak coupling regime, $\gamma < 2$, where the extended Bogoliubov-Fr\"ohlich Hamiltonian provides an adequate starting point. The phase diagram of this model has been mapped out using RG and mean-field saddle-point analysis by Grusdt et al.~\cite{Grusdt2017RG1D}, as shown here in Fig.~\ref{fig1DphaseDiag} (a).

On the repulsive side, $g_{\rm IB}>0$ (i.e. $\eta>0$), the ground state is constituted by the repulsive polaron. On the attractive side, the physics is significantly more complex: For weak attraction, the ground state corresponds to an attractive polaron. In contrast to the two-particle limit, where infinitesimal attraction is sufficient to introduce a bound state at the bottom of the spectrum, an extended regime is found in the many-body setting between $\eta_c < \eta < 0$, where the ground state corresponds to a Bose polaron instead of a bound (molecular) state. Within mean-field theory, the Feshbach resonance is effectively shifted to $\eta_{c,{\rm MF}} < 0$ due to interaction effects; for smaller $\eta < \eta_{c,{\rm MF}}$, the mean-field ground state corresponds to a multi-particle bound state with several bosons bound to the impurity \cite{Mostaan2023,Grusdt2017RG1D}. Between $\eta_{c,{\rm MF}} < \eta < 0$, mean-field theory yields a spectrum unbounded from below and associated with a dynamically unstable mode. This is an artifact of neglecting phonon-phonon interactions \cite{Mostaan2023}, as confirmed by numerically exact diffusion Monte-Carlo (DMC) calculations \cite{Grusdt2017RG1D}, blue dots in Fig.~\ref{fig1DphaseDiag} (a). The latter find a well-defined ground state with polaron character on the attractive side. In the most strongly coupled regime around $\eta_{c,{\rm MF}}$ the RG analysis of the model breaks down completely, indicating a strong-coupling regime in the model without phonon non-linearities.

For even stronger attractive interactions, $\eta < \eta_{c,{\rm MF}}$, saddle-point and RG analysis suggests the existence of a repulsive polaron branch at positive energies. This state is meta-stable, existing in the many-body continuum above the molecular ground state, and is hence challenging to address in exact numerics. This is similar to the repulsive polaron branch in higher dimensions. Notably, the asymptotic repulsive polaron states reached on opposite ends, for $\eta \to \pm \infty$, coincide. I.e. the repulsive polaron on the repulsive side $\eta > 0$ develops a more and more pronounced dressing cloud, reaches $\eta = + \infty$, and continuously evolves from $\eta=-\infty$ to $\eta_{\rm c,{\rm MF}} < 0$ where it ceases to exist as a well-defined resonance. This behavior is similar to the super-Tonks Girardeau gas realized when a 1D Bose gas continuously evolves from the repulsive to the attractive side across a 1D Feshbach resonance \cite{Astrakharchik2005}.

The general structure of the ground state in extended Bogoliubov-Fr\"ohlich Hamiltonians is common to all dimensions, and is confirmed qualitatively by its exact solution in the limit of infinite impurity mass due to Kain and Ling \cite{Kain2018}. However, as alluded to in several places above, phonon-phonon interactions neglected in the extended Fr\"ohlich model can become relevant. They cannot be neglected when either $\eta \approx \eta_c$ is around the renormalized Feshbach resonance, where the ground state involves multiple bosons; and they always play a role when the surrounding Bose gas is strongly interacting, $\gamma \gtrsim 2$. The regimes of validity of the various simplified models are sketched in Fig.~\ref{fig1DphaseDiag} (b), following Ref.~\cite{Grusdt2017RG1D}.

\begin{figure*}
	\centering
	\includegraphics[width=\linewidth]{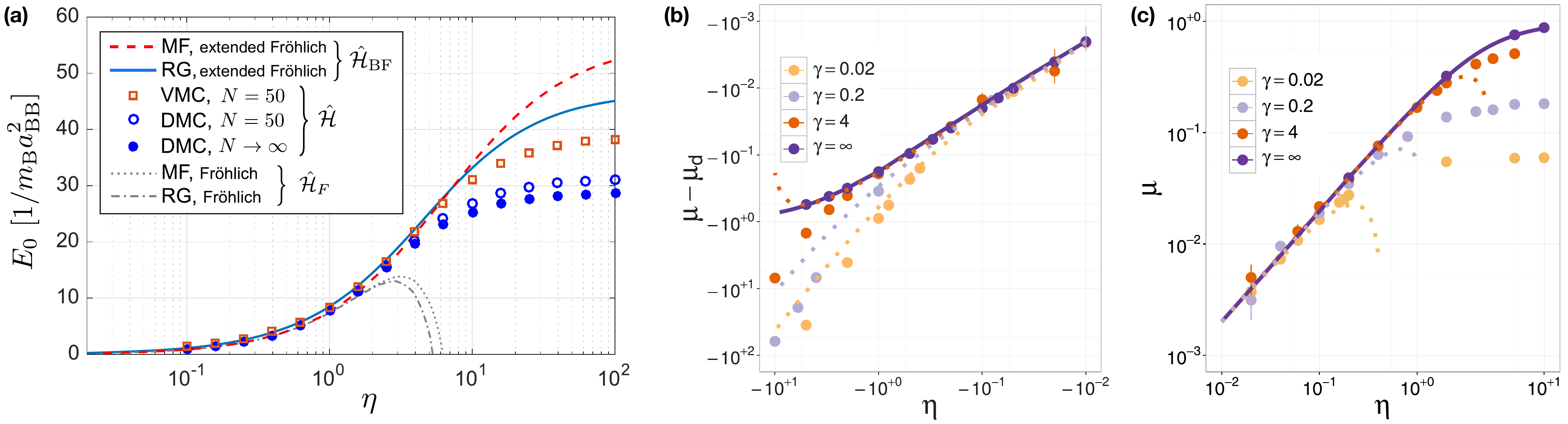}
	\caption{\textbf{Ground state polaron energy} $E_0$ calculated from different numerical methods. (a) For repulsive impurity-boson interactions $\eta > 0$ different theoretical methods are compared: numerically exact diffusion Monte Carlo (DMC), approximate variational Monte Carlo (VMC) predictions based on the full model, and RG and LLP mean-field (MF) predictions based on the (extended) Bogoliubov-Fr\"ohlich Hamiltonian. (b) and (c) show DMC predictions for the polaron energy by Parisi and Giorgini, separately for attractive and repulsive interactions and for different values of the gas parameter $\gamma$ as indicated in the legend. Note the double-logarithmic scale. They assumed equal masses, $m_{\rm B}=m_{\rm I}$, and plot the binding energy in units of $\varepsilon_F \equiv \hbar^2 (\pi n_0)^2 / (2 m_{\rm B})$; in (b) the contribution from the two-particle dimer $\mu_d=-\eta^2/(2 \pi^2)$ is subtracted. The solid line for $\gamma = \infty$ corresponds to the exact result by McGuire \cite{McGuire1965,McGuire1966}; dotted lines correspond to low-order perturbative results. The figure is adapted from \cite{Grusdt2017RG1D} (a) and \cite{Parisi2017} (b) and (c).}
	\label{fig1DenergiesAstrakharchikParisi}
\end{figure*}

\subsubsection{Infrared divergencies and orthogonality catastrophe.}
One of the most intriguing properties of one-dimensional many-body systems is the break-down of the quasiparticle picture owing to highly divergent phase-space integrals. I.e. low-energy excitations accumulate in the infrared (IR) regime and lead to excessive dressing and vanishing quasiparticle residues $Z \to 0$. Well-known manifestations of this phenomenon include the break-down of Fermi-liquids in 1D \cite{Haldane1981,Giamarchi2003} and Anderson's orthogonality catastrophe \cite{Anderson1967}: In the latter, an infinitesimal $s$-wave potential of strength $\varepsilon$ inside a Fermi-liquid leads to a completely orthogonal ground state $\bra{\Psi(0)} \Psi(\varepsilon) \rangle = 0$ for any $\varepsilon \neq 0$. The involved $s$-wave scattering effectively renders this problem one-dimensional in this case.

Similar IR divergences have been predicted to occur in 1D Bose polarons by Grusdt et al.~\cite{Grusdt2017RG1D}. They showed that the number of low-energy IR phonons in the polaron cloud diverges logarithmically with the IR momentum cut-off $\lambda_{\rm IR}$,
\begin{equation}
    N_{\rm ph} \simeq - \log(\lambda_{\rm IR}) \to \infty,
\end{equation}
which in turn causes the quasiparticle weight to vanish accordingly,
\begin{equation}
    Z \simeq e^{- N_{\rm ph}} \simeq \lambda_{\rm IR} \to 0.
\end{equation}
This is a direct manifestation of Anderson's orthogonality catastrophe \cite{Anderson1967}, and can be directly revealed in the associated dynamics or impurity spectral function as shown by Kantian et al.~\cite{Kantian2014}.

An even more dramatic consequence of the accumulation of IR phonons is the formal divergence of the mean-field ground state energy in 1D. Grusdt et al.~\cite{Grusdt2017RG1D} pointed out that the Bogoliubov-Fr\"ohlich Hamiltonian includes a logarithmically IR divergent constant contributing 
\begin{equation}
    \sim - g_{\rm IB} \log(\lambda_{\rm IR}) \to {\rm sgn}(g_{\rm IB}) \times \infty
\end{equation}
to the energy. Remarkably, by performing an RG analysis, they could show that this IR divergence is exactly canceled when treating quantum fluctuations beyond mean-field approximation within the extended Bogoliubov-Fr\"ohlich model -- a great success of the RG analysis. In particular it is not necessary to include phonon-phonon interactions in order to regularize these IR divergences.  

\subsubsection{Numerical studies of 1D Bose polarons.}
Finally we turn to a review of existing numerical studies of Bose polarons in one dimension. Initial work motivated by the experiments of Catani et al.~\cite{Catani2012} focused on the weak coupling Fr\"ohlich regime and employed Feynman's variational path-integral approach to calculate the polaron energy and its renormalized mass \cite{Catani2012,Casteels2012}. A systematic perturbative treatment of the full Hamiltonian was provided at weak ($\gamma \ll 1$) and strong ($\gamma \gg 1$) coupling of the host Lieb-Lininger gas by Pastukhov \cite{Pastukhov2017}. They used free bosons (fermions) respectively as a starting point for the perturbative analysis. Moreover, motivated by the lattice Bose polaron experiment of Fukuhara et al.~\cite{Fukuhara2013}, numerically precise DMRG simulations were performed by Dutta et al.~\cite{Dutta2013}.

The continuum system (no lattice) that we focus on in this review, the most accurate treatment of 1D Bose polarons to date is constituted by diffusion Monte Carlo (DMC) calculations performed by Parisi and Giorgini~\cite{Parisi2017} and by Astrakharchik in Ref.~\cite{Grusdt2017RG1D}. In Fig.~\ref{fig1DenergiesAstrakharchikParisi} (a) we show DMC results from \cite{Grusdt2017RG1D} on the repulsive side, $\eta > 0$, for parameters as in the experiment of Catani et al.~\cite{Catani2012} (dark blue data points). These results are compared to closely related variational Monte Carlo (VMC) simulations (orange squares) of the full Hamiltonian based on a Jastrow-factor trial state. The latter correctly predicts the qualitative shape of the ground state energy $E_0(\eta)$, although quantitative deviations are clearly visible. Notably the qualitative dependence on $\eta$ is also captured correctly by LLP mean-field and RG calculations based on the extended Bogoliubov-Fr\"ohlich Hamiltonian which neglects phonon-phonon interactions. Even though $\gamma \approx 0.44$ is well below one, i.e. in the weak coupling regime, strong impurity-boson coupling $\eta \gg 1$ implies that phonon-phonon interactions play a significant role. In contrast, predictions based only on the weak-coupling Bogoliubov-Fr\"ohlich Hamiltonian (gray lines in Fig.~\ref{fig1DenergiesAstrakharchikParisi} (a)) predict completely different qualitative shape.

In Fig.~\ref{fig1DenergiesAstrakharchikParisi} (b) and (c), DMC results by Parisi and Giorgini~\cite{Parisi2017} for the polaron binding energy are shown in a mass-balanced gas, i.e. for $m_{\rm I}=m_{\rm B}$, and for different gas parameters $\gamma$. The weak- and strong coupling results, $\gamma \ll 1$ and $\gamma \gg 1$, show qualitatively similar behavior, and in all cases perturbative analysis (dotted lines) breaks down when $\eta$ becomes sizable. Note that in (b) energies are measured relative to the two-particle dimer energy $\mu_d$, and the obtained negative energies indicate that the dimer is dressed by further excitations. In their extensive analysis, Parisi and Giorgini also calculated the renormalized polaron mass $m_{\rm p}$, the contact parameter $C$, and considered the case of a localized impurity, $m_{\rm I} = \infty$, see Ref.~\cite{Parisi2017}. Furthermore, both Refs.~\cite{Grusdt2017RG1D,Parisi2017} calculated density profiles of the Bose gas relative to the mobile impurity (not plotted here).

Beyond DMC simulations, various semi-analytical approaches have been developed and applied to 1D Bose polarons. In contrast to full-blown Monte Carlo calculations, which provide by far the most reliable predictions for ground state properties, most of these methods have to advantage of being applicable to non-equilibrium settings or finite temperatures. This makes them versatile tools for the exploration of Bose polarons. Notable works include semi-analytical few-body calculations based on a GPE-type descriptions of the Bose gas, by Dehkharghani et al.~\cite{Dehkharghani2015}, Volosniev et al.~\cite{Volosniev2017} and Jager et al.~\cite{Jager2020}, which all provide accurate results for the polaron energies when benchmarked through comparison to DMC simulations \cite{Grusdt2017RG1D,Parisi2017}; In another notable theoretical work, Brauneis et al.~\cite{Brauneis2021} used the Hamiltonian-based Wegner flow-equation approach \cite{Kehrein2006} to predict very accurately various Bose polaron properties in one dimension; a powerful path-integral approach was developed and benchmarked on DMC results by Panochko and Pastukhov \cite{Panochko2019}. Another mean-field study by Pastukhov \cite{Pastukhov2019} described the effect of more exotic three-body interactions on Bose polarons in one dimension.

\subsection{Bose polaron in  2D}
The field of condensed matter physics has seen a surge of interest in two-dimensional atomic systems, particularly in spin-imbalanced Fermi gases. These systems are closely linked to phenomena such as Stoner's itinerant ferromagnetism~\cite{stoner33} and the celebrated Fulde-Ferrell-Larkin-Ovchinnikov (FFLO) state~\cite{FFLO1,FFLO2}, boosting the motivation for the study of Fermi polarons in 2D~\cite{SchmidtEnss2012,ngampruetikorn2012repulsive}.
In 2012, the first 2D Fermi polarons were created via momentum-resolved photoemission spectroscopy~\cite{Koschorreck2012}. The spectral function of these polarons showed the presence of two branches in their polaron energy: attractive and repulsive polaron branches, similar to the phenomenology of Bose polarons that we will return to in a moment. The attractive Fermi polaron features dressing with fermions that are effectively attracted by the impurity; however, as interactions become stronger, a discontinuous transition from the polaron to a dressed molecule appears due to the two-body bound-state inherent in $d=2$ dimensions for attractive microscopic interactions~\cite{Massignan2014}. Properties of the compound object, such as the effective mass, increase dramatically above the polaron-to-molecule transition due to the enhanced effect of quantum fluctuations in low dimensions.

More recently, Fermi polarons were also observed in other platforms such as atomically thin semiconductors, in particular, transition metal dichalcogenides (TMD) monolayers such as Molybdenum diselenide $\mathrm{MoSe_2}$ and disulfide $\mathrm{MoS_2}$ or Tungsten diselenide $\mathrm{WSe_2}$ and disulfide $\mathrm{WS_2}$, increasing thus the interest in polaron physics in semiconductor heterostructures~\cite{Sidler2017}. Boosting nonlinear effects in these materials can be achieved through polaron polariton-mediated interactions~\cite{Tan20} or excitonic impurities interacting with a $d=2$ dimensional electron gas~\cite{Muir2022}. The resulting polaronic quasiparticles can interact over longer distances than bare polaritons or excitons, making them potentially interesting for quantum simulation and communication applications.

Investigating Bose polarons in $d=2$ presents greater challenges compared to their fermionic counterpart, especially in the strongly interacting regime. This is primarily due to the increased complexity inherent to their theoretical description and the absence of Pauli pressure reducing the effects of interactions among host particles. Nevertheless, a strong motivation stems from the fact that induced interactions mediated by bosonic baths tend to be stronger. For instance, the realization of Bose polaron-polaritons in $d=2$ constitutes a promising setting to explore such strong nonlinearities (as we discuss in Sec.~\ref{Sec:Light-matter interactions and polarons}). This section of our review provides an overview of the most important theoretical predictions concerning Bose polarons in quantum gases in $d=2$, a setting that largely remains to be explored experimentally.

Experimental realizations of Bose polarons in two dimensions are feasible with current technologies using ultracold quantum gases, for instance, by utilizing pancake-shaped optical trapping potentials. The strong confinement along one direction (let us suppose $z$) to a few micrometer wide region restricts the motion of particles into the plane ($x$ and $y$), where much weaker harmonic potentials can be used to trap the particles on much larger length scales. The corresponding trapping frequencies thus satisfy $(\omega_{x},\omega_{y})\ll\omega_{z}$. To study Bose polarons at low temperatures, it is important to remain in the superfluid regime by cooling the host Bose gas to temperatures below the Berezinskii-Kosterlitz-Thouless (BKT) critical temperature~\cite{Bloch2008,Hadzibabic2006}. The atomic densities in two dimensions typically range from 10 to 100 atoms per $\mathrm{(\mu m)^{2}}$~\cite{Ville18}. The accurate value of the scattering length in two dimensions can be calculated by combining the harmonic confinement length scale, $l_{z}=\sqrt{\hbar/m_{\mathrm {B}}\omega_{z}}$, and the three-dimensional scattering length, thus $ a_{\mathrm{BB}}^{(d=2)}=1.863l_{z}\exp\left[-\left(\pi/2\right)l_{z}/a_{\mathrm{BB}}^{(d=3)}\right]$~\cite{Bloch2008}. For typical values of $30<l_{z}/a_{\mathrm{BB}}^{(d=3)}<50$, the gas parameters in $d=2$ are significantly smaller compared to the ones in $d=3$, with the typical range $10^{-40}<na_{\mathrm{BB}}^{2}\lesssim10^{-20}$; here $a_{\mathrm{BB}}=a_{\mathrm{BB}}^{(d=2)}$ is the pure boson-boson $s-$wave scattering length in two dimensions.

\subsubsection{Hamiltonian and system.} 
A homogeneous Bose gas in $d=2$ dimensions interacting with a single impurity -- both confined in a box of linear size $L$ -- and with an average density of the bosonic bath of $n = N/L^{2}$ is described by the following Hamiltonian in first quantization,
\begin{equation}
\begin{array}{c}
\mathcal{\hat{H}}=-\frac{1}{2m_{\mathrm{B}}}\sum_{i=1}^{N}\nabla_{i}^{2}+\sum_{i<j}V_{\mathrm{BB}}\left(r_{ij}\right)\\-\frac{1}{2m_{\mathrm{I}}}\nabla_{{\rm I}}^{2}+\sum_{i=1}^{N}V_{\mathrm{IB}}\left(r_{i{\rm I}}\right).
\end{array}
    \label{eq:H2D}
\end{equation}
The first line accounts for the $d=2$ Bose gas of $N$ bosons, where $r_{ij}=|\vec{r}_i-\vec{r}_j|$ is the inter-particle distance. The second line describes the kinetic energy of the impurity and its interaction with the bosons, where $r_{i{\rm I}}=|\vec{r}_i-\vec{r}_{\rm I}|$. The boson-boson and impurity-boson interactions are proportional to the coupling strengths $g_{\mathrm{BB}}$ and $g$, respectively.  

Contrary to the $d=3$ case, in the low-energy limit the coupling strength in $d=2$ acquires a dependence on the scattering length and the density. The exact ground-state energy of the Hamiltonian can be found using QMC by using a finite range potential that reproduces the desirable scattering lengths. Alternatively, the Hamiltonian in Eq.~\eqref{eq:H2D} can be conveniently written in momentum space as discussed in  Sec.~\ref{secEffHamiltonians} to facilitate the use of variational or diagrammatic approaches. In this case, the impurity-boson coupling strength $g$ beyond the mean field needs to be renormalized as the scattering matrix includes an ultraviolet (UV) divergence for $\mathbf{q} \rightarrow \infty$; see Sec.~\ref{subsecUVdiv}. In order to properly regularize $g$, we rely on the fact that for attractive interactions, a bound state with binding energy $\epsilon_{b}=-1/(2m_{\mathrm{red}}a^{2})$ always exists, satisfying
\begin{equation}
    \frac{1}{g} = -\left[\frac{m}{4 \pi} \ln \left(\frac{m\epsilon_b +\Lambda^2}{m\epsilon_b}\right)\right],
\end{equation}
with $\Lambda$ an ultraviolet momentum cutoff related to the inverse of the impurity-boson potential range. In the limit  $\mathbf{q} \rightarrow 0$ the integral energy is well-behaved due to the presence of the bound-state energy scale. Considering the limit of a zero-range potential, with $\ensuremath{\Lambda^{2}}/m\gg\epsilon_{b}$, the UV cutoff needs to be replaced by another scale: the inter-particle distance of the host bose gas, $\ensuremath{\Lambda^{2}}\propto n \propto k_{\rm F}^{2}$~\cite{Pastukhov2018_2D}. This relation leads to the dependence of the coupling strength on the density $g=-\frac{4\pi}{m}\frac{1}{\ln\left(na^{2}\right)}.$

\begin{figure}
\includegraphics[width=\linewidth]{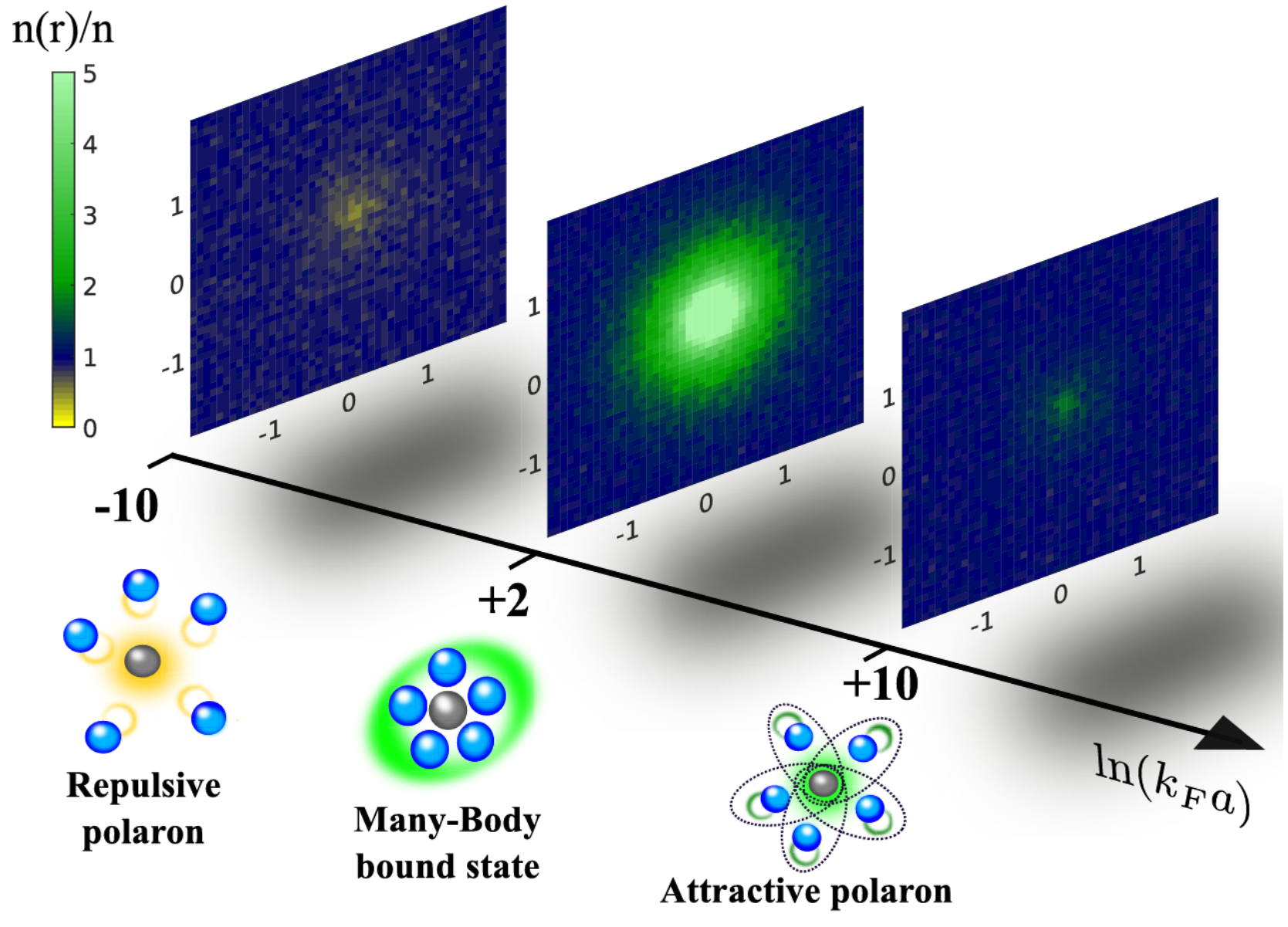}
	\caption{\textbf{Attractive and repulsive Bose polaron branches} in $d=2$ dimensions appear for weak interactions, $|\ln \left(k_{\mathrm{F}} a\right)| \ll 1$. In the strong-coupling regime in between, around where $\ln \left(k_{\mathrm{F}} a\right)$ changes sign, many-body bound states are found that involve up to a few tens of particles. The color plots show the host boson density $n(r)/n$ around the impurity in the center at $r=0$, normalized by the unperturbed host density $n$; distances in the insets are in units of the healing length $\xi=\hbar/\sqrt{2m_{\mathrm{B}}g_{\mathrm{BB}}n}$. The impurity-boson correlations $n(r)/n$ were calculated using QMC by Ardila et al~\cite{Ardila2020}, assuming $m_{\mathrm{B}}=m_{\mathrm{I}}$ and a gas parameter $na_{\mathrm{BB}}=1\times10^{-40}$. The figure is re-printed from~\cite{Ardila2020}.}
	\label{fig:Density2D}
\end{figure}

\subsubsection{Fröhlich and perturbative regime.}
In the weakly interacting regime, i.e. when $\left|\ln(k_{\mathrm{F}}a)\right|\gg1$ and the inter-particle distance is much smaller than the typical size of the bound-state, the Bose polaron Hamiltonian can be recast in the form of a Fr\"ohlich-type Hamiltonian; see Sec.~\ref{secEffHamiltonians}. The corresponding ground state was calculated perturbatively in $g$ by Pastukhov~\cite{Pastukhov2018_2D}; following proper ultraviolet regularization, the polaron energy up to second-order in $g$ and for zero total momentum, $\vec{P}=0$, reads 
\begin{equation}
   E_0 = E_{\mathbf{P}=0}^{(2)} + \mathcal{O}(g^4), 
\end{equation}
with
\begin{equation}
\small
\frac{E_{\mathbf{P}=0}^{(2)}}{E_{n}}=\frac{1}{|\ln\left(1/na^{2}\right)|}\left(1-\frac{\ln\left|\ln\left[a_{\mathrm{BB}}^{2}n\right] \right|+2\gamma+\ln\pi+1}{\left|\ln\left[a^{2}n\right]\right|}\right).
\label{eq:FrohlichEnergy2D}
\end{equation}
Here $E_{n}=\hbar^{2}k_{\mathrm{F}}^{2}/2m_{\mathrm{B}}$ and $\gamma=0.577$ is the Euler constant; in Eq.~\eqref{eq:FrohlichEnergy2D} we moreover assumed that $m_{\rm B}=m_{\rm I}$, but a closed analytical expression for general mass ratios $m_{\rm I}/m_{\rm B}$ is also provided in Ref.~\cite{Pastukhov2018_2D}. 

Similar to $d=1$ and $d=3$, for finite momentum, one can expand the polaron energy in powers of $\mathbf{P}^2$ and extract the effective mass~\cite{Pastukhov2018_2D,Ardila2020,Nakano2023},
\begin{equation}
\frac{m_\mathrm{I}}{m^{*}}=1-\frac{m_{\rm B}}{2 m_{\rm I}}\frac{\left|\ln\left(na_{\mathrm{BB}}^{2}\right) \right|}{\ln^{2}\left(na^{2}\right)}.
\label{eq:FrohlichMass2D}
\end{equation}
The leading order result for the quasiparticle residue is~\cite{Pastukhov2018_2D}
\begin{equation}
Z^{-1}=1+\frac{\left|\ln\left(na_{\mathrm{BB}}^{2}\right)\right|}{\ln^{2}\left(na^{2}\right)} \frac{m_{\rm B} + m_{\rm I}}{2 m_{\rm I}}.
    \label{eq:FrohlichResidue2D}
\end{equation}

\begin{figure*}
	\centering
\includegraphics[width=\linewidth]{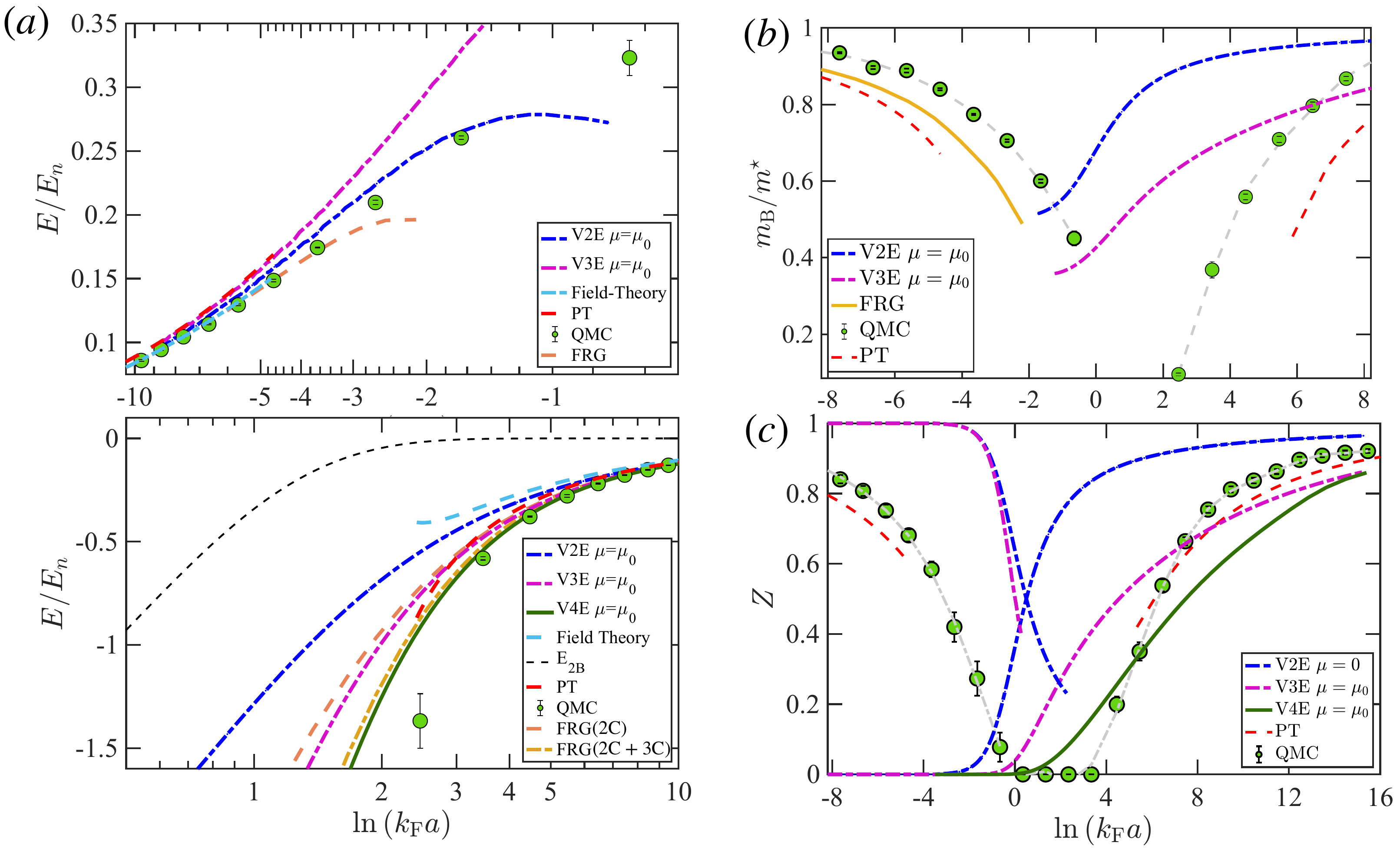}
	\caption{\textbf{Quasiparticle properties of the Bose Polaron in $d=2$.} (a) Repulsive and attractive polaron branches with energy $E/E_n$ are shown as a function of the impurity-boson coupling strength $\ln\left(k_{\mathrm{F}}a\right)$. Numerically exact QMC results~\cite{Ardila2020} are compared to a variational approach containing up to two (V2E), three (V3E) and four-body correlations (V4E) (only for the attractive branch)~\cite{Nakano2023} for a weakly interacting Bose gas with $\mu_0=0.136n/m_{\mathrm{B}}$; to perturbation theory ($\mathrm{PT}$); and finally to functional renormalization group ($\mathrm{FRG}$)~\cite{Isaule21} calculations. In particular, in the $\mathrm{FRG}$ scheme for the attractive branch, there is a distinction between including two $\mathrm{FRG (2C)}$ and three-particle correlations $\mathrm{FRG (2C + 3C)}$ respectively. In addition, for the attractive branch also, the energy of the vacuum-bound state $\epsilon_{B}=-1/2m_{\mathrm{red}}a^{2}$ is shown for comparison. (b)  Effective mass and (c) quasiparticle residue as a function of the impurity-boson coupling strength $\ln\left(k_{\mathrm{F}}a\right)$ using some of the same methods for which the energy is shown in (a). Here $E_{n}=\hbar^{2}k_{\mathrm{F}}^{2}/2m_{\mathrm{B}}$ and $k_{\mathrm{F}}=\sqrt{4\pi n}$.}
	\label{fig:PolaronEnergy2D}
\end{figure*}

\subsubsection{Polaron spectrum.}
The Bose polaron in two dimensions has rich physics beyond the Fr\"ohlich paradigm, which is challenging to describe theoretically due to the important role of quantum fluctuations on one hand and the lack of analytically tractable limits on the other hand. These properties were extremely useful in the previously discussed cases of $d=3$ and $d=1$ dimensions, respectively.  Furthermore, in $d=2$ an impurity-boson bound state is always present, even at weak coupling, in stark contrast to the $d=3$ case, where the formation of the boson-impurity dimer takes place once the resonance is crossed, i.e. when $a^{(d=3)}>0$. Nevertheless, the ground state of the 2D Bose polaron exhibits a sharp crossover from a polaronic to a many-body bound state characterized by two-body molecular correlations as the coupling $\ln \left(k_{\mathrm{F}} a\right)$ is increased; here $k_{\mathrm{F}}=\sqrt{4\pi n}$. This can be observed in the local density of host atoms around the impurity, computed directly from the pair correlation function from QMC by Ardila et al.~\cite{Ardila2020} for different coupling strengths, see Fig.~\ref{fig:Density2D}. The size of the many-body bound-state formed in the strong coupling regime also increases with the coupling strength $\ln\left(k_{\mathrm{F}}a\right)$.

Recent \textit{ab initio} quantum Monte Carlo calculations by Ardila et al.~\cite{Ardila2020} confirmed the emergence of two branches for the $d=2$ Bose polaron, in close analogy to its fermionic counterpart~\cite{Koschorreck2012} and as expected from the perturbative Fr\"ohlich model. In Fig.~\ref{fig:PolaronEnergy2D} (a) the polaron energy $E$ is plotted as a function of the coupling strength $\ln(k_{\mathrm{F}}a)$, computed using different methods. In the plots the energy scale is set by $E_{n}=\hbar^{2}k_{\mathrm{F}}^{2}/2m_{\mathrm{B}}$. The polaron energy exhibits two branches: an attractive branch with $E/E_n<0$ for $\ln(k_{\mathrm{F}}a)>0$  and a repulsive branch for $\ln(k_{\mathrm{F}}a)<0$ with $E/E_n>0$. In the weakly interacting regime, $\left|\ln(k_{\mathrm{F}}a)\right|\gg1$, path integral methods based on one-loop expansion at low temperatures by Pastukhov~\cite{Pastukhov2018_2D} predict 
analytical expressions for the polaron properties, recovering the exact results from a simple many-body perturbation theory approach, see Eq.~(\ref{eq:FrohlichEnergy2D}).

In addition, QMC results valid for all coupling strengths and containing all possible correlations in the system are compared against recent functional renormalization group theory approaches ($\mathrm{FRG}$) by Isaule et al.~\cite{Isaule21}, including two and three-particle correlations. A remarkable agreement is found between the two methods when higher-order correlations are included in the $\mathrm{FRG}$ scheme — akin to the $d=3$ case (see Sec.~\ref{secStrongCplgPolaron}). On the same line of analytical approaches that include few-body effects explicitly, a recent variational ansatz by Nakano et al.~\cite{Nakano2023} uses a two-channel model (see Eq.~\eqref{eq:BosePolaronTowChannelModel}), that can incorporate up to four-body correlations and builds upon the Bogoliubov approximation. The simplest case corresponds to the inclusion of two-body correlations (impurity and one boson) where no difference appears between an ideal gas $\mu=0$ and the interacting Bose gas $\mu=\mu_0$. In the weakly interacting regime, this simple ansatz agrees quite well with respect to other methods such as perturbation theory, FRG, and QMC; even the inclusion of higher-order correlations is imperceptible. As one approaches the strongly interacting regime on the attractive side, the role of few-body excitations or correlations, as well as the interacting nature of the bath, becomes relevant. FRG and the variational results progressively approach the QMC as more correlations are taken into account. In fact, by including three body correlations (1 impurity + 2 bosons) in the FRG and four body excitations (1 impurity + 3 bosons) in the variational ansatz, the results agree with QMC for strong coupling. This trend is analogous to the behavior observed in the $d=3$, where the variational ansatz wavefunctions progressively approach the QMC results as more correlations are included. 

Recently, non-self-consistent T-matrix approximation (NSCT) was also used to predict the polaron properties for the $d=2$ system by C\'ardenas-Castillo et al.~\cite{Cardenas23}, however akin to the $d=3$ case, this formalism is equivalent to the low-order variational ansatz~\cite{Nakano2023}, namely, a trial wavefunction truncated up to two-body correlations. The diagrammatic approach needs the inclusion of higher-order correlations and bound-state physics, both essential components to describe the strongly coupled polaron in $d=2$.  

The variational approach by Nakano et al.~\cite{Nakano2023} can include several correlations as the FRG and QMC, yet there is still an open question on whether the Bogoliubov approximation remains reliable for very strong interactions $\ln\left(k_{\mathrm{F}}a\right)\sim0$. Finally, for the repulsive branch, the variational ansatz wavefunctions capture the QMC results by truncating the wavefunction up to two-body correlations. Higher correlations are probably sensible because near to $\ln\left(k_{\mathrm{F}}a\right)\sim-1$, the polaron has a shorter lifetime, and the quasiparticle may broaden due to its decay into a lower-lying continuum of states.

\subsubsection{Effective mass and quasiparticle residue.}
In Fig.~\ref{fig:PolaronEnergy2D} (b-c), we summarize the theoretical predictions for the effective mass and quasiparticle residue of the 2D Bose polaron, as a function of the coupling strength $\ln(k_{\mathrm{F}}a)$, using QMC, perturbation theory (see for instance Eq.~{\eqref{eq:FrohlichMass2D} and Eq.~{\eqref{eq:FrohlichResidue2D}}) and variational approaches~\cite{Nakano2023}. For very weak interactions, $\left|\ln(k_{\mathrm{F}}a)\right|\gg1$, QMC results for the effective mass agree with the perturbative result for both the attractive and repulsive branches, as expected.

For the intermediate coupling and following the attractive side $1\lesssim\ln\left(k_{\mathrm{F}}a\right)\lesssim 10$, perturbation theory starts to deviate and eventually breaks down as it includes neither bound-state physics nor higher-order correlations. In fact,  QMC predicts a sharp increase of the quasiparticle's effective mass in the strongly interacting regime. NSCT~\cite{Cardenas23} and variational ansatzes, including up to three Bogoliubov excitations~\cite{Nakano2023}, have been plotted in Fig.~\ref{fig:PolaronEnergy2D}(b). The effective mass using three-body excitations in the variational ansatz approaches the QMC and perturbation theory predictions in the weakly interacting regime, yet it deviates considerably from QMC in the strongly interacting regime. For example, for $\ln\left(k_{\mathrm{F}}a\right)\sim2$ the effective mass is roughly $m^{\star}\sim1.6m_{\mathrm{B}}$ in contrast to QMC which predicts $m^{\star}\sim10 m_{\mathrm{B}}$. A plausible explanation for the differences is the lack of four-body correlations that may be needed to converge towards the QMC predictions, as observed for the energy and the residue, see Fig.~\ref{fig:PolaronEnergy2D} (a) and (c). $\mathrm{FRG}$ instead follows the QMC trend very closely, and small deviations are observed in the strong coupling regime, which is consistent with the energy results in Fig.~\ref{fig:PolaronEnergy2D}(a). Apart from QMC and perturbation theory calculations, results are not reported for the effective mass in $d=2$ yet.

Finally we focus on the quasiparticle residue $Z$ plotted in Fig.~\ref{fig:PolaronEnergy2D}(c). For the attractive polaron, one particular feature of the quasiparticle residue is that QMC predicts a rapid decrease and for $0< \ln(k_{\mathrm{F}}a)\lesssim3$ the quasiparticle picture breaks down, as signaled by $Z\sim 0$ consistent with zero. This QMC prediction should be contrasted to the approximate NSCT, which predicts sizable values for the residue ($Z\approx 0.7$) and the effective mass ($m^{\star}/m=1.1$), supporting the quasiparticle picture. However, as explained before, NSCT is restricted to only one Bogoliubov excitation and hence underestimates the surrounding atomic cloud forming around the impurity. 

This picture is further corroborated by comparing to the variational approach by Nakano et al.~\cite{Nakano2023} including variable number of Bogoliubov excitations. While including two-body correlations is equivalent to NSCT, the inclusion of three-body correlations -- $\mathrm{V3E}$ in Fig.~\ref{fig:PolaronEnergy2D} (c) -- leads to considerable corrections in the predicted value of $Z$. Considering first an ideal case, $\mu=0$, and two-body correlations $\mathrm{V2E}$, no agreement is found with either perturbation theory or QMC, highlighting the importance of both bath interactions and correlations as discussed for the energy. Indeed, by including three-body correlations, $\mathrm{V3E}$ follows perturbation theory and QMC closely; at strong coupling, the inclusion of four-body correlations is important to obtain results approaching the QMC values. Interestingly, the variational ansatz predicts a critical coupling strength for which the quasiparticle residue vanishes, thus reproducing a key qualitative feature of the QMC results. While the quantitative value of the critical coupling strength differs from the QMC result, it approaches the latter upon including larger number of excitations.

The regime between $0< \ln(k_{\mathrm{F}}a)\lesssim3$ where the quasiparticle picture breaks down remains poorly understood. It is believed that the formation of many-body bound states involving multiple excitations causes the observed breakdown.  An open question concerns microscopic physics of this regime and whether the polaron leads to an orthogonality catastrophe~\cite{Knap2012,Mistakidis2019} or rather self-localization~\cite{sacha2006self}. 

Finally, the quasiparticle residue for the repulsive polaron is overestimated by the variational ansatz of Nakano et al.~\cite{Nakano2023}, where a change of the number of excitations has a much weaker effect on the quasiparticle residue than on the attractive side.

\section{Interacting polarons and induced interactions.}
\label{secInducedInteractions}


\subsection{Induced and mediated interactions between polarons.}
The study of many-impurity systems is of great interest in the field of polarons from various perspectives. While the single-impurity problem disregards any properties inherent to the quantum statistics of the impurities, statistics play a crucial role as the number of impurities increases. In solid-state systems, electrons, i.e. fermions, in a lattice form polarons and may bind together to form Cooper pairs~\cite{Cooper56}. The formation of these bound states requires a weak attractive interaction, which is mediated by phonons in conventional superconductors. Cooper pairing condensation underlies the explanation of conventional superconductivity within the BCS theory ~\cite{BCSTheory_RMP}. Related binding mechanisms originating from quasiparticle-mediated interactions, such as para-magnon exchange~\cite{Scalapino1986,Bruegger2006,Tacon2011}, constitute a popular ingredient for understanding unconventional (e.g., high-temperature) superconductivity~\cite{Bednorz1986,RMPKresin,Zhang2023Bipolaron}. In particular, high-temperature superconductors exhibit a very rich phase diagram as a function of the temperature and doping~\cite{keimer2018, paglione2010,stewart2017,fradkin2015}; therefore, understanding how impurities and polarons interact can provide insights into the nature of the pseudo-gap phase and the microscopic pairing mechanism governing these systems.

Ultracold atomic quantum gases and their ability to precisely control the ratio of atom numbers of different atomic species open up a direct pathway to study the physics of quantum mixtures, all the way from the impurity limit to the number-balanced case. Regardless of statistics, the highly imbalanced regime can be described by the physics of individual impurities interacting with the environment, as demonstrated in the experiments reviewed above, e.g.~\cite{Jorgensen2016,Hu2016,Yan2020,Cambridgde2023,Ardila2019}. In fact, if the impurity gas is very dilute, possible interactions between quasiparticles are negligible and difficult to track. Following the pioneering work by Baym and Pethick~\cite{LandauFermi-LiquidTheory} on quasiparticle interactions in Helium mixtures, recent works have started to extend beyond the Landau paradigm of quasiparticle interaction~\cite{Bruunprx2018,Fujii2022,Drescher2023}. 

For higher impurity densities where the impurity statistics becomes relevant, the situation turns out to be more intriguing, especially for bosonic impurities, where even the sign of the medium-induced inter-impurity interactions is subject to
debate. While arguments based on the Landau’s Fermi liquid theory predict the sign of the effective quasiparticle interaction to be attractive (repulsive) for bosonic (fermionic) impurities~\cite{Bruunprx2018,ZhenhuaYu2012}, phase
space filling arguments demonstrate that bosonic impurities interact repulsively (attractively) on the attractive (repulsive)
polaron branch under consideration~\cite{tan2020interacting,Tan2023,Muir2022}. Interestingly, a recent theoretical work has pointed out a new type of medium-induced interaction between bosonic impurities, which is argued to be the leading order for weak impurity-medium interctions~\cite{levinsen2024medium}. In this context, it is argued that the exchange of Bogoliubov modes can enhance the repulsion between the bosonic impurities, while exchange-induced attraction is a higher order effect in the impurity-medium coupling strength. These counterintuitive results further illustrate the richness of the medium-induced interaction effects in binary mixtures.

Polarons can interact with each other, even in the absence of direct interaction between bare impurities, by exchanging low-energy excitations of the surrounding medium, such as phonons associated with density modulations of the host crystal. One of the first proposals to study Landau's effective interaction between Bose polarons, formed with either bosonic or fermionic impurities, relied on diagrammatic methods, employed first by Camacho-Guardian et al.~\cite{Bruunprx2018}. Induced interactions between quasiparticles are attractive for bosonic impurities, while for fermionic impurities their sign may depend on the energy transfer between the two polarons and the mass ratio $m_{\mathrm{I}}/m_{\mathrm{B}}$. 

As a natural starting point to discuss induced interactions among impurities in a Bose gas, we consider a system of well-defined polaronic quasiparticles immersed in a BEC. The total energy of the system, up to second order in the momentum-space densities of the impurities $n_{\mathbf{p}\sigma}$, can be written as 
\begin{equation}
E=E_{0}+\sum_{\mathbf{p},\sigma}\epsilon_{\mathbf{p}\sigma}^{0}n_{\mathbf{p}\sigma}+\frac{1}{2}\sum_{\mathbf{p},\sigma,\mathbf{p}^{\prime},\sigma^{\prime}}f_{\mathbf{p}\sigma,\mathbf{p}^{\prime}\sigma^{\prime}}n_{\mathbf{p}\sigma}n_{\mathbf{p}^{\prime}\sigma^{\prime}}.
    \label{eq:EnergyInd}
\end{equation}
Here $E_0$ is the energy of the system without any impurities, the second term describes the energy of free polaronic quasiparticles and the third term accounts for their interaction, characterized by the Landau parameters $f_{\mathbf{p}\sigma,\mathbf{p}^{\prime}\sigma^{\prime}}$. 

In the weakly interacting regime, the retarded Landau effective quasiparticle interaction between two polarons with momentum $\mathbf{p_{1}}$ and $\mathbf{p_{2}}$ was shown by Camacho-Guardian et al.~\cite{Bruunprx2018} to follow,
\begin{equation}
f(\mathbf{p_{1}},\mathbf{p_{2}},\omega)=\pm g^{2}n_{0}\frac{p^{2}}{m_{\mathrm{B}}(\omega^{2}-\omega_{\mathrm{\mathbf{p}}}^{2})},
\label{eq:f_ind}
\end{equation}
where $+(-)$ corresponds to bosonic (fermionic) impurities. In this expression, $g$ is the coupling of the impurity with the bosonic medium with homogeneous density $n_0$, and the remaining term arises from the BEC density-density correlation function; $\vec{p}=\vec{p_1}-\vec{p_2}$ is the momentum exchange and $\omega_{\vec{p}}$ denotes the Bogoliubov dispersion of the collective excitations of the BEC mediating the interaction. In the  static limit, $\omega \to 0 $, the previous equation corresponds to the well-known Yukawa form $(d=3)$ in real space,
\begin{equation}
  f(r)=-\frac{g^{2}n_{0}m_{\mathrm{B}}}{\pi}\frac{\exp\left(\sqrt{2}r/\xi\right)}{r}, 
   \label{eq:f_r}
\end{equation}
where $\xi$ is the healing length of the BEC and $r$ denotes the distance between the two impurities. The minus sign comes from the fact that the induced interaction $V_{\text {ind}}\left(\mathbf{p}-\mathbf{p}^{\prime}, \omega_{\mathbf{p}^{\prime}}-\omega_{\mathbf{p}}\right)$ is always attractive. In general the Landau quasiparticle relates to the induced interaction via $f\left(\mathbf{p}, \mathbf{p}^{\prime}\right)= \pm V_{\text {ind }}\left(\mathbf{p}-\mathbf{p}^{\prime}, \omega_{\mathbf{p}^{\prime}}-\omega_{\mathbf{p}}\right)$~\cite{paredes2024}.

In contrast, in the strongly interacting regime, calculating the induced interactions between two impurities becomes more challenging because it requires the inclusion of several high-order diagrams and therefore the extension of the standard $\mathrm{T}-$ matrix ladder approximation is needed as proposed by Camacho-Guardian et al.~\cite{Bruunprx2018}.
In the low impurity density limit, the Landau quasiparticle interaction $f(\mathbf{p’},\mathbf{p})$ in general depends on the impurity's statistics, the impurity-boson mass ratio $m_{\mathrm{B}}/m_{\mathrm{I}}$, the coupling strength $g$ and the impurity's momentum transfer. Camacho-Guardian et. al~\cite{Bruunprx2018} considered the case of momentum transfer between a polaron from $\mathbf{p=0}$ to a finite momentum $\mathbf{p}$. The resulting Landau quasiparticle interaction presents a different behavior as a function of the momentum. In the weakly interacting regime and for $m_{\mathrm{I}}/m_{\mathrm{B}}=1$, which corresponds to the experimental realization in the Aarhus experiment~\cite{Jorgensen2016}, $f(\mathbf{p’},\mathbf{p})$ is independent of the momentum,  $f(0,\mathbf{p})=-g^{2}/g_{\mathrm{BB}}$. In contrast, in the strongly interacting regime $\left(k_{n}a\right)^{-1}\gg1$ the interaction enhances and behaves non-monotonously as a function of the momentum transfer: it is attractive for small momentum and decreases monotonously until reaching a minimum at the critical value $\left|\mathbf{p}_{c}\right|=m_{\mathrm{I}}c$; for momenta larger than $\mathbf{p}_{c}$ the interaction increases, turning repulsive for  $\mathbf{\left|p\right|} \gg \left|\mathbf{p}_{c}\right|$.

The previous paragraph discussed the Landau quasiparticle interaction for the equal mass case, $m_{\mathrm{B}}=m_{\mathrm{I}}$. However, most relevant experimental realizations involve the imbalanced case. Camacho-Guardian et al.~\cite {Bruunprx2018} also discussed these cases. For example, for heavy impurities as the $^{41}\mathrm{K}-^{6}\mathrm{Li}$ mixture in the Innsbruck experiments~\cite{Fritsche2021,RianneLous17,Baroni2024}, where $m_{\mathrm{I}}/m_{\mathrm{B}}=40/7$, the Landau quasiparticle interaction is repulsive (fermionic impurities) for small momentum and decreases monotonously with it, until it becomes almost independent of the impurity-boson coupling strength $g$. On the contrary, for light fermionic impurities, as in the case of the JILA experiment~\cite{Hu2016}, where  $m_{\mathrm{I}}/m_{\mathrm{B}}=40/87$, the induced interaction is repulsive for small momentum and increases up to a critical momentum where it diverges, meaning that the impurity momentum is resonant with the Bogoliubov sound mode $\omega_{p}$, namely, $p^{2}/2m_{\mathrm{I}}=\omega_{p}$. For $p>k_{n}$, the interaction changes sign and turns attractive, vanishing for $p \gg  k_{n}$.  In summary, the most suitable candidate for observing strong induced interaction is a system consisting of lighter impurities.

\begin{figure}[htb]
\centering
\includegraphics[width=9cm]{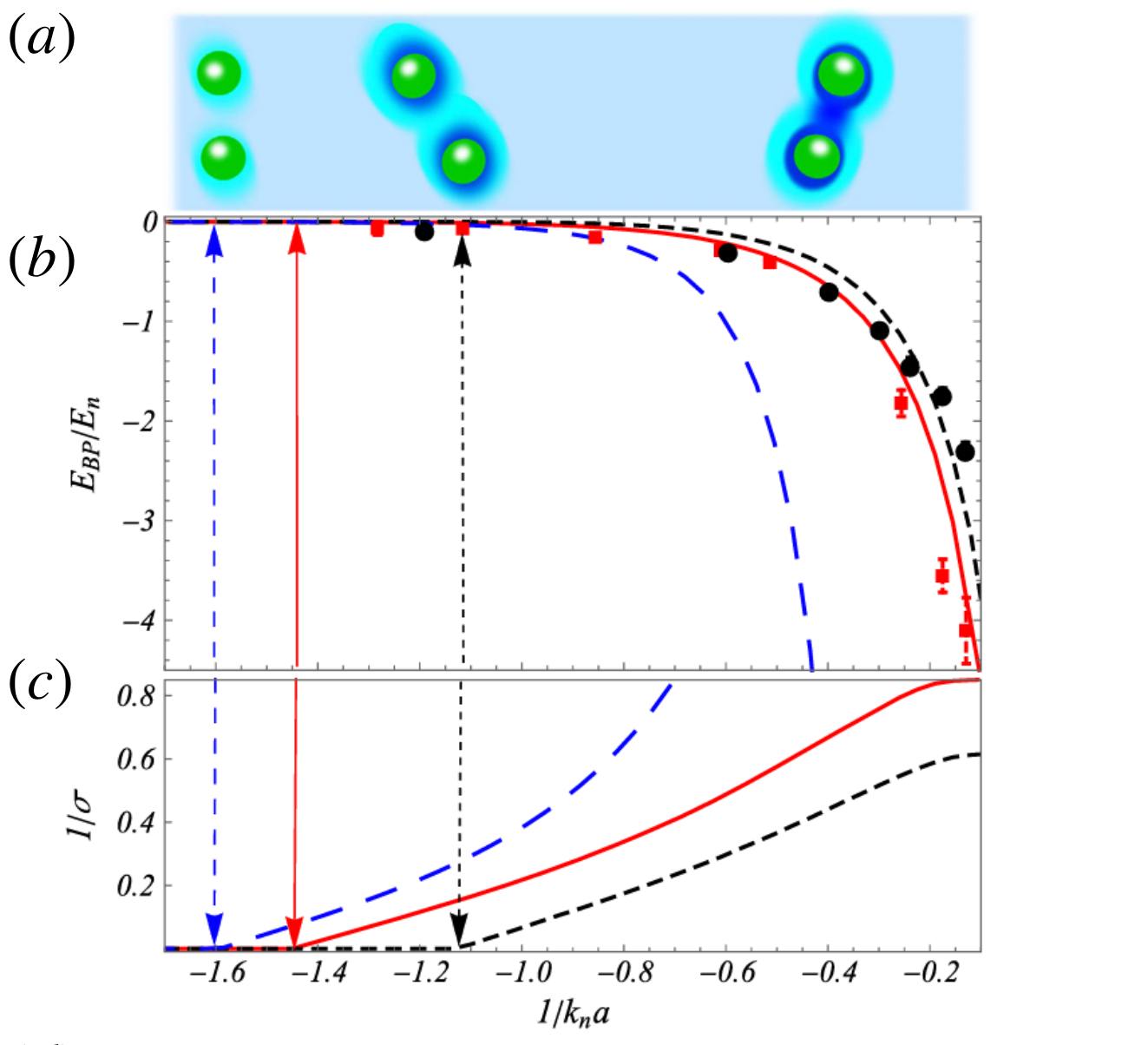}
	\caption{\textbf{Bipolaron formation of impurities in a BEC.} (a) The cartoon shows Bose polarons forming a bipolaron as a consequence of a mediated interaction. (b) Binding energy $E_{BP}$ of the bipolaron as a function of the impurity-boson interaction strength for two gas parameters $na_{\mathrm{BB}}^{3}$ with $m_{\mathrm{I}}=m_\mathrm{B}$. The red solid and black dashed lines are solutions of the Schr\"odinger equation for two polarons interacting via an instantaneous interaction using diagrammatic approaches, whereas the symbols depict quantum Monte-Carlo results~\cite{Camacho2018bip}. The two gas parameters are $n_\mathrm{B}a_\mathrm{BB}^{3}=10^{-6}$ (red symbols and red line) and $n_\mathrm{B}a_\mathrm{BB}^{3}=10^{-5}$ (black symbols and black line) respectively. The blue long-dashed line is the ground state energy of the Yukawa interaction Eq.~\eqref{eq:f_ind} for $n_\mathrm{B}a_\mathrm{BB}^{3}=10^{-6}$. (c) The corresponding inverse size $\sigma^{-1}=\xi_\mathrm{B}/\sqrt{\langle r^2\rangle}$ of the bipolaron wavefunction. Vertical arrows denote the critical strength to form a bound state between polarons. Figure taken from~\cite{Camacho2018bip}.}
	\label{Bipolaron3D}
\end{figure}

\subsection{Bipolarons}
\label{subsecBipolaron}
Polarons can either repel or attract each other. In the latter case, the mediated interaction can bind two polarons, forming a bipolaronic bound state. A system of several impurities interacting with a BEC can form polarons locally. Even without direct interaction between the impurities, these quasiparticles remain essentially non-interacting unless a critical impurity-boson coupling strength is reached; at this point, polarons can interact strongly via enhanced density fluctuations, boosting the mediated interaction. The critical value for bipolaron formation depends on the strength of the impurity-bath coupling, parametrized by the coupling $(k_na)^{-1}$, and on the compressibility of the bath, which depends on both its density and the boson-boson interaction strength.

Bipolarons in ultracold quantum gases were first predicted using path integral methods within the Fröhlich framework by Casteels et al.~\cite{CasteelsBipol}. Subsequently, Camacho-Guardian et al.~\cite{Camacho2018bip} predicted the formation of bipolaronic states by using non-perturbative field theory approaches going beyond the ladder approximation along with QMC methods. As discussed before, in the weakly interacting regime, the induced interaction is of Yukawa-type, and it is very weak: polarons can be regarded as isolated quasiparticles. Approaching the strongly interacting regime, there is a critical impurity-boson coupling strength for which the polarons attract each other sufficiently strongly to form a two-body bound state, a bipolaron, see Fig.~\ref{Bipolaron3D}(a). In addition, Jager et al. ~\cite{Jager2022} showed that the induced interaction is also enhanced as the bath becomes more compressible by using traditional GPE approaches. An analog situation occurs in a spinor condensate as in Bighin et al.~\cite{Bighin2022_2}, where the induced attractive interaction is resonantly enhanced when the boson-boson scattering length is tuned as the soft spin mode of the Bose mixture can induce resonantly enhanced pairing.

 Camacho-Guardian et al.~\cite{Camacho2018bip} computed the polaron binding energy defined as $E_{BP}=E(2,N)-2E(1,N)+E(0,N)$, where $E(N_{\rm I},N)$ is the ground-state energy of the system with $N_{\rm I}$ impurities and $N$ bosons. A bipolaron bound-state is formed when $E_{\mathrm{BP}}<0$, see  Fig.~\ref{Bipolaron3D}(b) and its energy increases as the binding (impurity-boson) gets stronger, and as a consequence, the size of the polaronic bound-state $1/\sigma=\xi_{B}/\sqrt{\left\langle r^{2}\right\rangle}$, decreases as depicted in  Fig.~\ref{Bipolaron3D}(c) as one gets to the strongly interacting regime. In addition, the limit $1/\sigma =0$ signals the critical coupling strength where bipolarons are formed; the bipolaron radius diverges when the polarons unbind.

From a methodological point of view, Camacho-Guardian et al.~\cite{Camacho2018bip} solved the Bethe-Salpeter equation (BSE)~\cite{BetheSalpeter} that describes the scattering between polarons. Solving this equation is a very demanding task even within the ladder approximation because the explicit form of the equation for two polarons depends on the impurity's Green function and the Landau induced interaction. However, in the static limit, an effective Schrödinger equation can be recovered, which is more tractable. This theoretical approach agrees with QMC predictions also computed in~\cite{Camacho2018bip}, see Fig.~\ref{Bipolaron3D} (b), where the bipolaron energy is computed with both methods. 

Besides the bath compressibility and the impurity-boson coupling strength, 
the impurity-boson mass ratio $\mathrm{m_{\rm I}}/\mathrm{m_{\rm B}}$ plays an important role in bipolaron formation. This can be understood by noting that induced interactions can give rise to few-body Efimov physics at very short distances, see Sec.~\ref{secFewBodyEfimov}. In the regime where the impurity-boson mass ratio differs significantly from one, a different mediated interaction potential appears: an Efimov-induced potential. The interplay between polaronic bound states and Efimov trimers arising in the strongly interacting regime was investigated by Naidon~\cite{Pascal2018}. In their work, it was shown that while virtual particles mediate the Yukawa-induced interaction, the Efimov-induced interaction is mediated by real particles. In the former case, the excitations are phonon-like, while in the latter the excitations are high energy free-particle like. Specifically, two particles are bound together due to the presence of a third one. For heavy bosons and light impurities, the Efimov effect predicts a ground-state configuration forming a strong trimer consisting of two bosons and one impurity (see Sec. ~\ref{secFewBodyEfimov}). 

\begin{figure*}[htbp]
\centering
	\centering
\includegraphics[width=17cm]{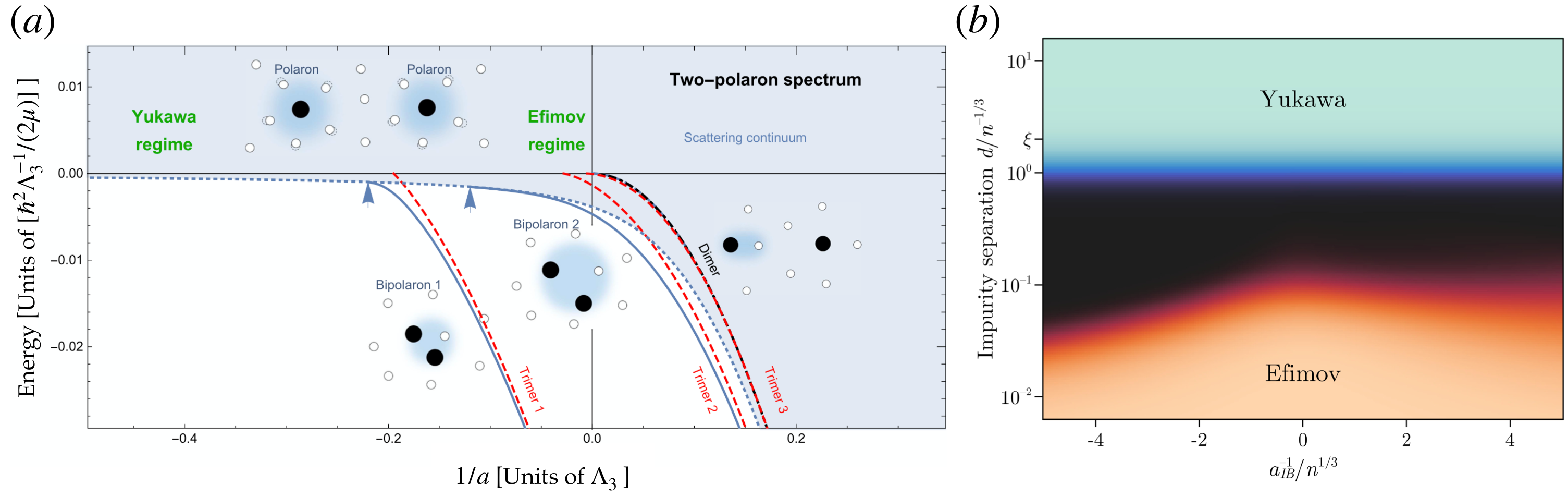}
	\caption{\textbf{Energy spectrum of two impurities in a BEC characterized by an Efimov-Yukawa scaling crossover.} (a) Energy spectrum of two light polarons with bare impurity mass $m_{\mathrm{I}}$ interacting with a condensate of bosons of mass $m_{\mathrm{B}}$, with $m_{\mathrm{\rm B}}/m_{\mathrm{\rm I}}=19$, as a function of inverse scattering length $1/a$ between the impurities and the bosons; the condensate gas parameter is  $na_{\mathrm{B}}^3=0.0017$. The shaded area represents the scattering continuum of the two attractive polarons and its threshold is shown as a dotted curve. The solid curves correspond to the bound states between two impurities (bipolarons). The arrows indicate the critical coupling strength threshold. For reference, the black dot-dashed curve shows the boson-impurity dimer energy in a vacuum, and the red dashed curves correspond to the boson-impurity-impurity trimer energies in vacuum. (b) Induced interaction scaling for two heavy polarons of mass $m_{\mathrm{I}}$ interacting with a condensate of bosons of mass $m_{\mathrm{B}}$, with $m_{\mathrm{B}}/m_{\mathrm{I}}\rightarrow 0$ as a function of the impurity-boson coupling strength $a_{\mathrm{IB}}^{-1}/n^{1/3}$ and the impurity separation $d/n^{-1/3}$. The gas parameter is $na_{\mathrm{BB}}^{3}=10^{-6}$. For impurity separation smaller than the interparticle distance, $d\ll n^{-1/3}$ there exists a three-body bound-state; for distances comparable with interparticle distance $d\sim n^{-1/3}$ the induced interaction is of Yukawa type, Eq.~\eqref{eq:f_ind} and for large distances $d\gg n^{-1/3}$, the interaction scales as $\propto 1/r^7$~\cite{Fujii2022}. Figures taken from (a)~\cite{Pascal2018} and (b)~\cite{Drescher2023} respectively.}\label{BipolaronPascal}
\end{figure*}

Naidon~\cite{Pascal2018} utilized a variational wavefunction, truncated to a single excitation, to minimize the total energy of a system consisting of two impurities and a BEC. This work predicted the formation of bipolarons and a few exotic few-body states. The system is characterized by an ultra-dilute gas with a gas parameter $na_{\mathrm{BB}}^{3} = 0.0017$. The specific case of heavy impurities, such as $^{133}\mathrm{Cs}$ immersed in a $^{7}\mathrm{Li}$ BEC, was considered. The energy spectrum obtained from the variational calculation was plotted as a function of the inverse impurity-boson coupling strength, $1/a$, expressed in units of a typical cutoff, which is usually set as the inverse of the typical potential range, see Fig.~\ref{BipolaronPascal}(a). In contrast to the case of equal masses studied by Camacho-Guardian et al.~\cite{Camacho2018bip}, two bipolaronic branches with negative energy emerge for $1/a<0$, whereas a Yukawa regime and an Efimov regime appear in the scattering continuum with positive energies. As the coupling strength increases from negative to positive, one of the bipolaronic branches (Bipolaron 1) approaches the ground trimer state of two impurities and one boson in vacuum (Trimer 1). Its binding energy corresponds to the energy of two independent polarons in the weakly interacting regime, namely $E\sim 2E(1, N)$. In this theoretical model by Naidon, the induced interaction ranges from a Yukawa interaction Eq.~\eqref{eq:f_ind} as predicted in reference \cite{Camacho2018bip}, to an Efimov interaction with two polarons attracting each other mediated by a third particle, as illustrated in Fig.~\ref{BipolaronPascal}(a).

Additionally, in Fig.~\ref{BipolaronPascal}(a), the energy of the excited trimer state of two impurities and one boson in vacuum (Trimer 2) is plotted together with the corresponding bipolaron energy (Bipolaron 2). The attractive collective interaction due to the bosons in the neighborhood of the impurities enhances the formation of the bipolaron with respect to the vacuum, and the effect is more evident close to the resonance $1/a_\mathrm{IB}\sim 0$, where there is a sizeable difference between the bipolaron branch and the vacuum trimer for positive scattering lengths. The critical value of forming the trimer is lower in the presence of the condensate with respect to the vacuum; in other words, at weaker impurity-boson interaction strength, the trimer is formed by the presence of the condensate, whereas in the vacuum, they require a larger coupling strength. For larger positive scattering lengths, the two-polaron scattering threshold is expected to be $2 E_d$, namely, each impurity strongly binds a nearby boson. In contrast, the energy in Fig.~\ref{BipolaronPascal} (a) converges to the single dimer energy $E_{d}$. The reason for the apparent contradiction lies in the fact that Naidon's method is restricted only to one bosonic excitation, and therefore, processes beyond binding the impurity with two or more excitations are not included.

Induced interactions between impurities and a few-body composite state were also explored by Drescher et al.~\cite{Drescher2023}; they used a non-local Gross-Pitaevskii equation formalism. The three-body (impurity and two heavy impurities) energy approaches the trimer binding energy, exhibiting an Efimov scaling $1/r^2$ for short distances, while for large distances, a Yukawa potential is recovered. At intermediate distances (on the order of the coherence length of the condensate), the Efimov-Yukawa crossover is sketched and is almost independent of the impurity-boson coupling strength, see Fig.~\ref{BipolaronPascal}(b). Drescher et al.'s~\cite{Drescher2023} findings differ from Naidon's prediction~\cite{Pascal2018} since the Yukawa potential was exclusively emerging in weak coupling and Efimov for strong coupling. Note that the Yukawa-Efimov crossover is absent in the work by Camacho et al.~\cite{Camacho2018bip} as Efimov physics is irrelevant for the equal mass case due to the difference in length scales between the typical scattering length and the three-body parameter (see Sec.~\ref{secFewBodyEfimov}).

\subsection{Induced interaction beyond Yukawa}
The general behavior of the induced interaction between impurities in a superfluid was studied by Fujii et al.~\cite{Fujii2022}. They showed by using effective field theory that for two heavy impurities pinned in a superfluid of healing length $\xi$, the Yukawa-induced interaction is smeared out at large distances ($d\gg \xi$) by a Casimir - Polder scaling, $r^{-7}$~\cite{Casimir48}. The origin of this interaction arises from the exchange of two phonons, in contrast to the Yukawa potential for  ``short" distances $d\sim n^{-1/3}$ that stems from one phonon exchange~\cite{pethick_smith_2008,Pascal2018,Camacho2018bip, Bruunprx2018}. Fujii et al.~\cite{Fujii2022} found additional contributions arising from thermal fluctuations: a Coulomb contribution scaling as $\sim T^{6}/r$ appearing for distances $\xi<r\ll c/T$, where $c$ is the speed of sound of the condensate. Instead, for $r>c/T$, the induced interaction has the non-relativistic contribution $T/r^6$. The situation differs for polaronic systems that exhibit significant finite-range effects, such as atom-ion hybrid systems~\cite{Astrakharchik2021,Christensen2021}; in this case the Yukawa potential holds for short distances, whereas for large distances, a Casimir force scales as $1/r^4$~\cite{Ding2022,Astrakharchik2023} (see also Sec.~\ref{Sec:SubSecAtomIon}).

\subsection{Induced interactions in low dimensions}
Nonlinear effects arising from the interaction between polarons are also boosted by the dimensionality. In particular, in lower dimensions, quantum fluctuations are enhanced with respect to the $d=3$ case, rendering the bipolaron more strongly bound and facilitating its formation. In $d=2$, polarons and bipolarons have been investigated in bidimensional lattices using a variational ansatz within the ladder approximation by Ding and et al.~\cite{ShanshanDing} and recently in solid-state set-ups such as excitons polaritons emerging in monolayer transition metal dichalcogenides (TMD)~\cite{Tan2022}. However, direct measurements of induced interactions between Bose polarons in pure atomic quantum gases remain to be reported. On the other hand, $d=1$  offers a very appealing platform for studying induced interactions.

In homogeneous systems in $d=1$, Recati et al.~\cite{Recati2005} and Klein et al.~\cite{Klein05} predicted that the induced interaction displays a Yukawa-type behavior with the characteristic $d=1$ exponential scaling, $V_{\mathrm{ind}}(r)\sim\exp(-r/\xi)$ in the weakly interacting regime. This perturbative result is valid for distances smaller than the coherence length $\xi$ of the one-dimensional quasicondensate. Using a GPE formalism including quantum fluctuations, Reichert et al.~\cite{Reichert_2019} found a general functional form for the induced interaction for arbitrary distances. For short distances $r<\xi$, the interaction asymptotically recovers the result by Recati et al.~\cite{Recati2005}. On the other hand, for large distances $r>\xi$, the general formula asymptotically approaches a Casimir-type $1/r^3$ interaction. This power-law trend was also recovered by Schecter et al.~\cite{Schecter2014} using an effective low-energy theory for mobile impurities. Recent calculations by Will et al.\cite{Will2021} used both a sophisticated mean-field approach that accounts for density deformations along with ab initio QMC 
techniques~\cite{Grusdt2017} for arbitrary coupling. At weak coupling and for short distances, Will et al.\cite{Will2021} also recover the aforementioned Yukawa-type induced potential. For strong coupling, a universal linear scaling is found, consistently within both the analytical and numerical approaches. Contrary to the Casimir $1/r^3$ prediction at large distances derived in~\cite{Reichert_2019,Schecter2014}, Will et al.~\cite{Will2021} found an exponentially decaying form for the induced potential, and for arbitrary coupling strengths.

Recent calculations by Dehkharghani et al.~\cite{Dehkharghani2018} predicted the formation of bipolaronic bound states by studying the induced impurity-impurity interaction in a harmonically trapped system, spanning from weak to strong interactions. For very large impurity-boson interactions ($a_{\mathrm{IB}}\rightarrow\infty$), the bosons tend to cluster locally around the impurities, an effect explained as resulting from an attractive, effective potential between the impurities. Additional comparisons with the homogeneous case were made by using a ring as a boundary condition, and no $1/r^3$ Casimir-induced interaction~\cite{Reichert_2019,Schecter2014} was found, although the perturbative Yukawa potential was still recovered for weak coupling. Trapped systems were further investigated using Bogoliubov theory~\cite{Klein05} and variational multilayer multiconfiguration time-dependent Hartree methods~\cite{Mistakidis2020Induced,Mistakidis2019Correlated,Mistakidis2020Many}.

\subsection{Experimental signature of induced interactions}
Induced interactions arising in a system of several impurities were explored by computing the optical absorption in electron/hole systems by Tempere and Devreese~\cite{Tempere2001}. Energy shifts did arise due to the polaron-photon scattering, as well as a reduction in the optical absorption of the quasiparticle formed by the plasmon-phonon interaction. Theoretical predictions of this formalism agree with measurements in Cuprate Oxides~\cite{Lupi1992,Lupi1999}, where the electron-phonon interaction is believed to be well-described by the solid-state Fr\"ohlich Hamiltonian~\cite{alexandrov2009advances}. Closely inspired by this technique, the same methodology was applied to a system of impurities interacting with a Bose-Einstein condensate using the Fr\"ohlich Hamiltonian~\cite{CasteelsMany2011} and recent calculations by Loon et al.~\cite{Loon2018} include beyond-Fr\"ohlich terms, providing a better accuracy in the strongly interacting regime for atomic systems. This latter work employed the procedure pioneered in solid-state systems~\cite{Tempere2001}, where the static structure factor of the impurities serves as the key observable for characterizing the gas of polarons. A shift in the ground-state properties appears due to the presence of impurities, which is more pronounced in the case of fermionic impurities with respect to their bosonic counterpart. However, as the resonance is approached, the theoretical approach by Loon et al.~\cite{Loon2018} becomes less accurate as the approach relies on the Bogoliubov approximation, and ab initio methods provide a more suitable description of the many-polaron physics in the strongly interacting regime~\cite{Ardila2022Atoms}.

\begin{figure}
\centering
	\centering
\includegraphics[width=7.5cm]{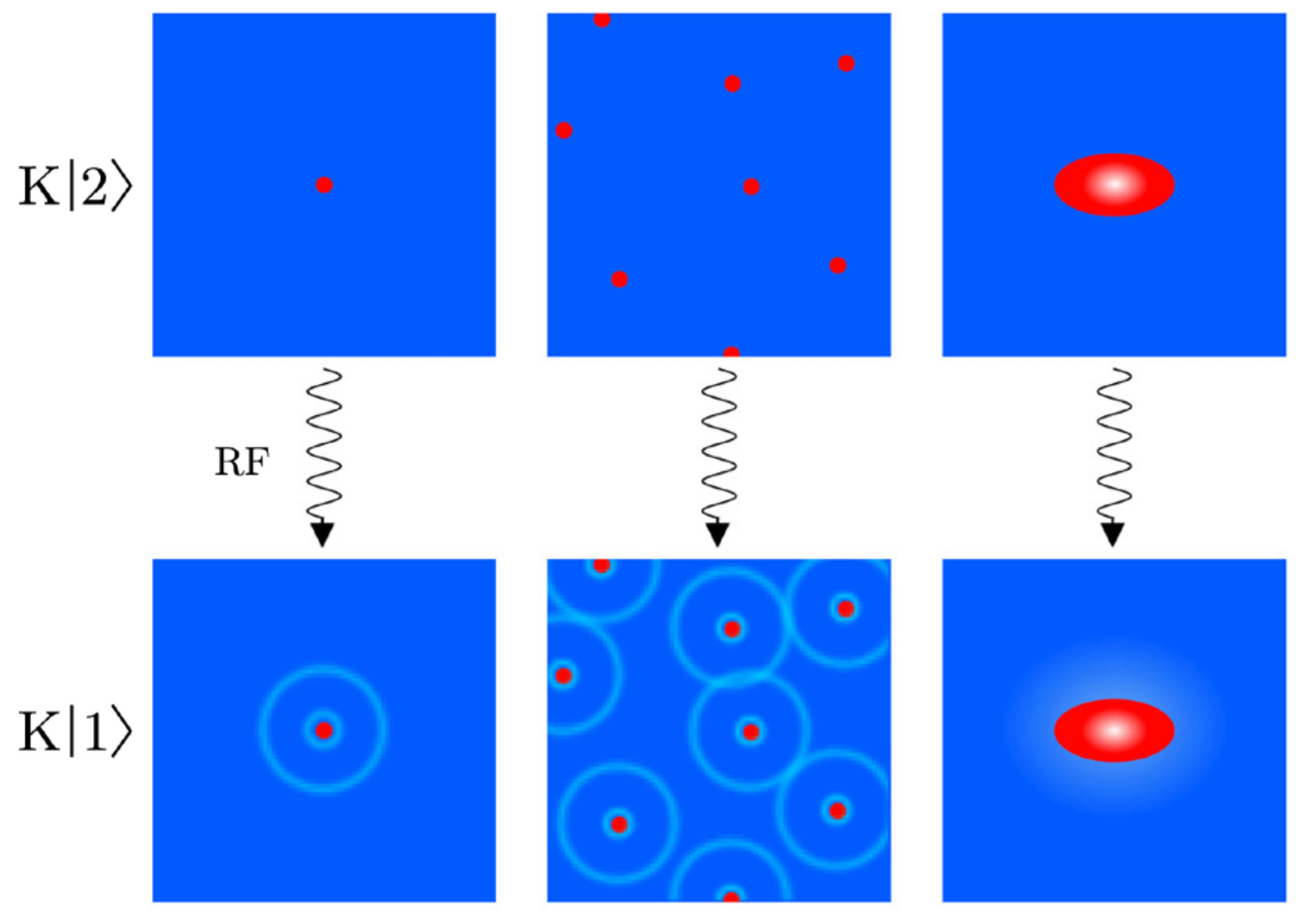}
	\caption{\textbf{Illustration of the Fermi-Bose mixture $^{41}\mathrm{K}$ and $^{6}\mathrm{Li}$ for three different impurity density}. The upper (lower) row shows the noninteracting $\mathrm{K}\left|2\right\rangle $ (interacting $\mathrm{K}\left|1\right\rangle $) impurities immersed in a Fermi sea, which is represented by the blue background. Injection RF is used to drive the system from noninteracting to a state where the impurity strongly interacts with the bath.
 From left to right, the three columns indicate the increase in the bosonic density from a single impurity (fermi polaron) to high densities and finally to a mixed phase containing a large BEC component. Figure taken from~\cite{Fritsche2021}.} \label{ExperimentInducedInteraction2}
\end{figure}

One of the first experimental signatures of induced interactions was obtained in terms of thermodynamic quantities~\cite{ZhenhuaYu2012}. They were observed 
in strongly imbalanced Bose-Fermi mixtures of $^{41}\mathrm{K}$ and $^{6}\mathrm{Li}$ by Fritsche et al.~\cite{Fritsche2021}. By using injection radio frequency spectroscopy, both Fermi and Bose polarons were observed depending on whether the majority component hosting the impurities was a Fermi or Bose gas, see Fig.~\ref{ExperimentInducedInteraction2}. 
Starting from a low concentration of bosonic impurities, the system resembles the single impurity dressed by low energy particle-hole excitations of the Fermi sea: a Fermi polaron is formed. While increasing the number of bosonic impurities, an attractive Landau induced interaction between the Fermi polarons arises~\cite{Bruunprx2018}. Increasing further the bosonic density, for a critical density, the thermal $\mathrm{K}$ impurities undergo a Bose-Einstein condensate transition. The trap density of the BEC is much larger than the one of the fermions, and one ends up with a situation in which the fermionic $\mathrm{Li}$ atoms play the role of impurities immersed in a BEC of $\mathrm{K}$ recovering the Bose polaron case. Thus, the bosons swapped roles: they became the bath, and the fermions play the role of impurities.

Later experiments quantified the polaron-polaron interaction exclusively for Fermi polarons. Baroni et al.~\cite{Baroni2024} directly observed polaron-polaron interactions and the induced interaction depends on the quantum statistics of the impurities. In order to extract the induced interaction, several measurements of the polaron energy as a function of impurity concentration were performed for a fixed coupling strength. Following the model in Eq.~\eqref{eq:EnergyInd}, the induced interaction can be extracted as the slope of the energy-concentration plot. In fact, in the experiment~\cite{Baroni2024} the chemical composition of the gas can be easily changed from a Fermi-Fermi $^{6}\mathrm{Li}-^{40}\mathrm{K}$  mixture to a Fermi- Bose $^{6}\mathrm{Li}-^{41}\mathrm{K}$ one, thereby changing the impurity’s statistics at (almost) fixed impurity/bath mass ratio. Contrary to the Fermi polaron problem~\cite{Fritsche2021}, where induced interactions where fully characterized along a wide range of parameters, induced interactions between Bose polarons have not been observed yet in pure atomic systems. However, very recently, signatures of Bose polaron interactions  were observed in a cavity-
coupled monolayer semiconductors~\cite{Tan2023} (see also Sec.~\ref{Sec:Light-matter interactions and polarons}).

\section{Finite temperature Bose polaron}
\label{secFiniteT}
The possibility of Bose-Einstein condensation in bosonic systems at finite temperatures can have notable consequences for polaron formation. Below the condensation temperature, the mobile impurity interacts with both the thermally excited and the condensed bosons. The resulting interplay between the two interaction processes can lead to rich physics and needs to be properly understood in order to theoretically model and analyze experiments which always work at non-zero temperature. 

\begin{figure}
    \centering
    \includegraphics[scale=0.7]{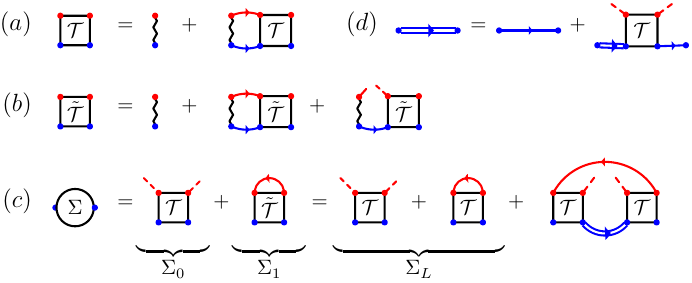}
    \caption{ \textbf{Diagrammatics involved in obtaining polaron properties at finite temperature}, as derived by Guenther et al.~\cite{Guenther2018}. (a) Impurity-boson T-matrix Dyson equation within the ladder approximation. The bare impurity (Bogoliubov) propagator is denoted by blue (red) lines. (b) Extended T-matrix, including impurity interaction with Bogoliubov excitations and condensed bosons (dashed red lines). (c) Self energy evaluated within the extended ladder approximation. The double wavy line corresponds to the condensate-dressed impurity propagator, satisfying the Dyson equation in (d). $\Sigma_{L}$ is the ladder self energy. Wavy lines stand for the vacuum T-matrix $\mathcal{T}_{v}$. Figure reprinted from Ref.~\cite{Guenther2018}.}
    \label{fig:finT_diag}
\end{figure}

Finite temperature effects in Bose polaron formation was studied in the non-interacting case by Drescher et al.~\cite{drescher2024bosonic} using an extension of the functional determinant approach to BECs at finite temperature, which circumvents the grand canonical catastrophe. The exact result for a heavy impurity in an ideal BEC is surprisingly that the Bose polaron peak becomes narrower with increasing temperature at strong coupling. In the weak coupling~\cite{levinsen2017finite} and the strong coupling~\cite{Guenther2018} regimes, Bose polaron formation at finite temperature was studied using diagrammatic techniques. In Ref.~\cite{Guenther2018}, Guenther et al. employed a diagrammatic formalism based on a ladder approximation of the impurity-boson T-matrix. To account for impurity scattering off the thermally excited as well as the condensed bosons, the T-matrix must include the bare interaction vertices corresponding to both scattering processes, yielding the diagrams depicted in Fig.~\ref{fig:finT_diag}. Thus, besides the standard ladder self-energy $\Sigma_{L}$ from impurity scattering with Bogoliubov excitations, the resulting self-energy $\Sigma$ includes an additional contribution, see Fig.~\ref{fig:finT_diag} (c). The spectrum obtained within this ``extended" ladder approximation by Guenther et al.~\cite{Guenther2018} demonstrates features markedly different from the zero temperature case, as illustrated in Fig.~\ref{fig:finiteTSpec}. At $T=0$, the attractive polaron consists of a single quasiparticle peak that splits into two equally weighted peaks for $0<T\ll T_c$. Increasing the temperature leads to a red-shift and broadening of the lower energy branch, while the higher energy branch blue-shifts and simultaneously loses its weight until it disappears close to $T_c$.

\begin{figure}
    \centering
    \includegraphics[scale=0.35]{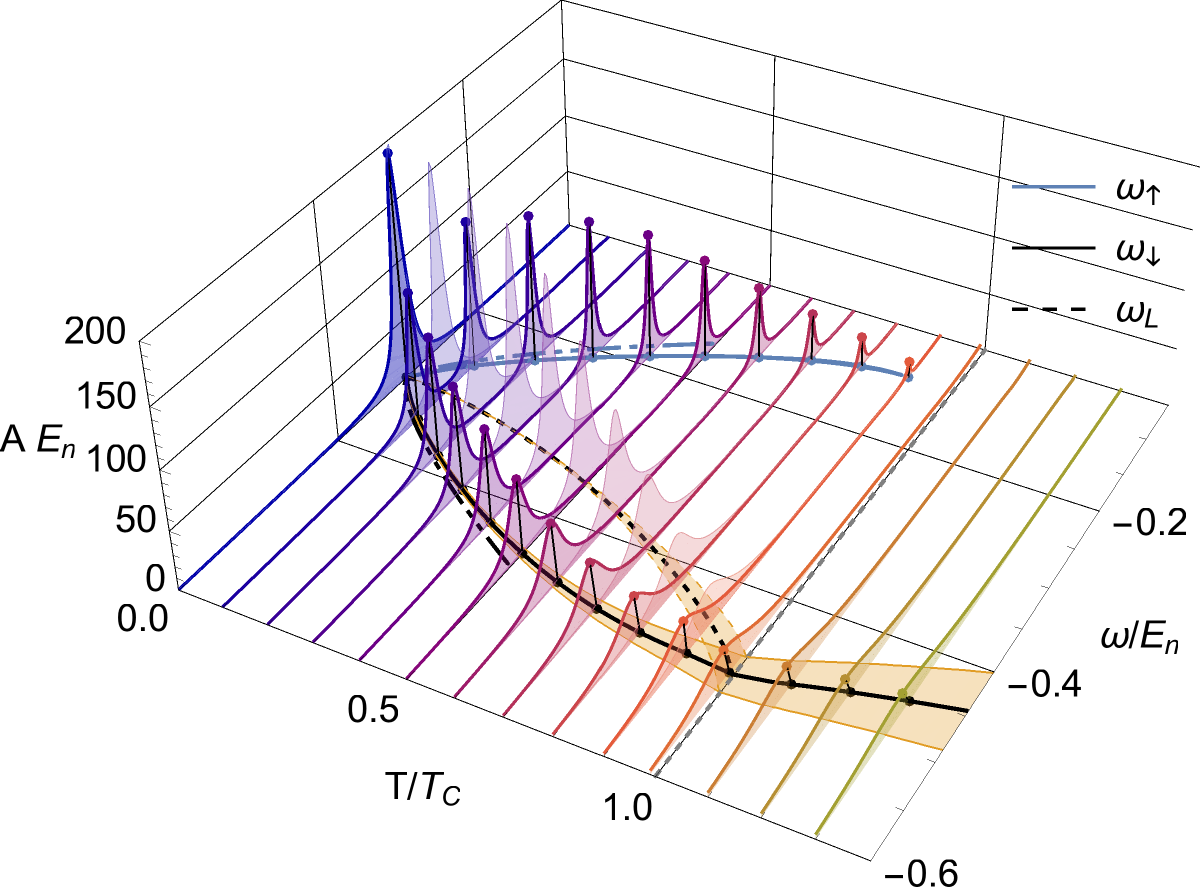}
    \caption{\textbf{Polaron spectrum as a function of temperature}, calculated by Guenther et al.~\cite{Guenther2018} using the extended T-matrix formalism. At $T=0$, the attractive polaron is the only well-defined quasiparticle peak. For $0<T<T_c$, two polaron branches appear in the spectrum with equal weights for $T\ll T_c$. The lower energy branch ($\omega_{\downarrow}$) red-shifts while broadening and decreasing its weight, whereas the higher-energy branch ($\omega_{\uparrow}$) blue-shifts and loses weight until it disappears for $T$ near $T_c$. The lighter peak corresponds to the polaron energy within the standard ladder approximation, with the mean peak position denoted by a dashed line. Figure reprinted from Ref.~\cite{Guenther2018}.} 
    \label{fig:finiteTSpec}
\end{figure}

To understand why the polaron peak splits in two at non-zero temperature, the authors of Ref.~\cite{Guenther2018} examined the pole condition of the dressed impurity Green's function, which is obtained from the solution to $\omega=\mathrm{Re}[\Sigma(\omega)]$. The extended T-matrix $\tilde{\mathcal{T}}(\vc{p},\omega)$ has a pole that is close to the weak coupling energy $n_0(\mathcal{T}_{v}-\mathcal{T}_{\mathrm{B}})$ for $\vc{p}\to0$, where $\mathcal{T}_{v(\mathrm{B})}$ is the vacuum T-matrix of the impurity-boson (boson-boson) scattering. This pole corresponds to the resonant interaction of a single Bogoliubov particle excited out of the condensate by the impurity and eventually scattered back into the condensate. This process gives the $\Sigma_{0}$ contribution to the self-energy. At $T=0$, the pole related to $\Sigma_{0}$ is the only quasiparticle energy. However, at finite temperatures, the occupation number of low energy Bogoliubov excitations is infrared divergent, which gives a significant contribution to the self-energy close to the pole of $\tilde{\mathcal{T}}$ with a sign change across the pole. Hence, the resonant shape of $\mathrm{Re}[\Sigma(\omega)]$, which causes the emergence of two quasiparticles, comes from the impurity interaction with a diverging number of thermal Bogoliubov excitations. This interaction process hybridizes with the impurity-condensate interaction, leading to the predicted polaron splitting. Furthermore, the red-shift of the lower polaron branch is due to the larger number of Bogoliubov excitations available for the impurity to scatter, compared to $T=0$ where only few excitations are available. The behavior of the self energy is shown in Fig.~\ref{figSelfEnergyFiniteT}.

\begin{figure}
    \centering
    \includegraphics[scale=0.5]{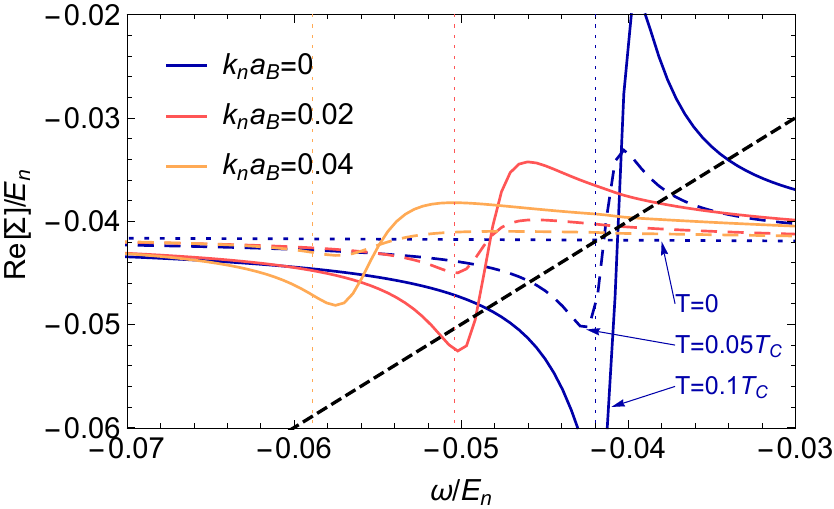}
    \caption{\textbf{Dependence of the self energy features on temperature}, calculated by Guenther et al.~\cite{Guenther2018} using the extended T-matrix formalism. The resonant shape of $\mathrm{Re}[\Sigma(\omega)]$ around $\omega\sim n_0 \mathcal{T}_{v}$ gives rise to three quasiparticle solutions, determined by the intersection of $\mathrm{Re}[\Sigma(\omega)]$ with the black dashed line. The solution around $n_0 \mathcal{T}_{v}$ has a large imaginary part and does not correspond to a well-defined quasiparticle peak. Figure reprinted from Ref.~\cite{Guenther2018}.}
    \label{figSelfEnergyFiniteT}
\end{figure}

The finite temperature structure of the polaron spectrum discussed in Ref.~\cite{Guenther2018} was further investigated by Field et al.~\cite{Field2020} by means of a finite temperature variational method. In summary, the variational ansatz amounts to approximating the Heisenberg-picture zero-momentum impurity creation operator $\hat{c}_{0}^{\dagger}(t)$ by the variational operator $\vc{\hat{c}_{0}^{\dagger}}(t)$, and finding stationary solutions for $\vc{\hat{c}_{0}^{\dagger}}(t)$. The polaron model considered in Ref.~\cite{Field2020} is the two-channel model similar to Eq.~\ref{eq:BosePolaronTowChannelModel} described in Sec.~\ref{secFewBodyEfimov} (see also Eq.~(5) in \cite{Field2020}). A suitable variational ansatz for $\vc{\hat{c}_{0}^{\dagger}}(0)$ including up to $N$ particle-like (boson) and $M$ hole-like excitations, is of the form
\begin{equation}\label{eq:varCcompact}
\begin{split}
\vc{\hat{c}_{0}^{\dagger}}(0) & = \sum^{N}_{n=0}\sum^{M}_{m=0}\,\sum_{\substack{\vc{k}_1,\cdots,\vc{k}_n}}\sum_{\substack{\vc{q}_1,\cdots,\vc{q}_m}}
\\&\alpha^{\vc{q}_1,\cdots,\vc{q}_m}_{\vc{k}_1,\cdots,\vc{k}_n}\,\hat{c}^{\dagger}_{\vc{Q}-\vc{K}}\,
\hat{b}^{\dagger}_{\vc{k}_1}\cdots\hat{b}^{\dagger}_{\vc{k}_n}
\hat{b}_{\vc{q}_1}\cdots\hat{b}_{\vc{q}_m}\\
& + \sum^{1}_{n=0}\sum^{2}_{m=0}\,\sum_{\substack{\vc{k}_1,\cdots,\vc{k}_n}}\sum_{\substack{\vc{q}_1,\cdots,\vc{q}_m}}
\\&\gamma^{\vc{q}_1,\cdots,\vc{q}_m}_{\vc{k}_1,\cdots,\vc{k}_n}\,\hat{d}^{\dagger}_{\vc{Q}-\vc{K}}\,
\hat{b}^{\dagger}_{\vc{k}_1}\cdots\hat{b}^{\dagger}_{\vc{k}_n}
\hat{b}_{\vc{q}_1}\cdots\hat{b}_{\vc{q}_m} \, ,
\end{split}
\end{equation}
with \(\vc{Q}=\sum^{m}_{i=1}\vc{q}_i\,\), \(\vc{K}=\sum^{n}_{i=1}\vc{k}_i\,\); moreover, \(\alpha^{\vc{q}_1,\cdots,\vc{q}_m}_{\vc{k}_1,\cdots,\vc{k}_n}\) (\(\gamma^{\vc{q}_1,\cdots,\vc{q}_m}_{\vc{k}_1,\cdots,\vc{k}_n}\)) are coefficients of terms describing the correlation of the impurity $\hat{c}$ (dimer $\hat{d}$) with \(n\) particle-like and \(m\) hole-like excitations out of the Bose gas. The Hamiltonian dynamics then couples the states corresponding to different terms in Eq.~\eqref{eq:varCcompact}, as illustrated in Fig.~\ref{fig:scat_proc}. 

\begin{figure}
    \centering
    \includegraphics[scale=0.25]{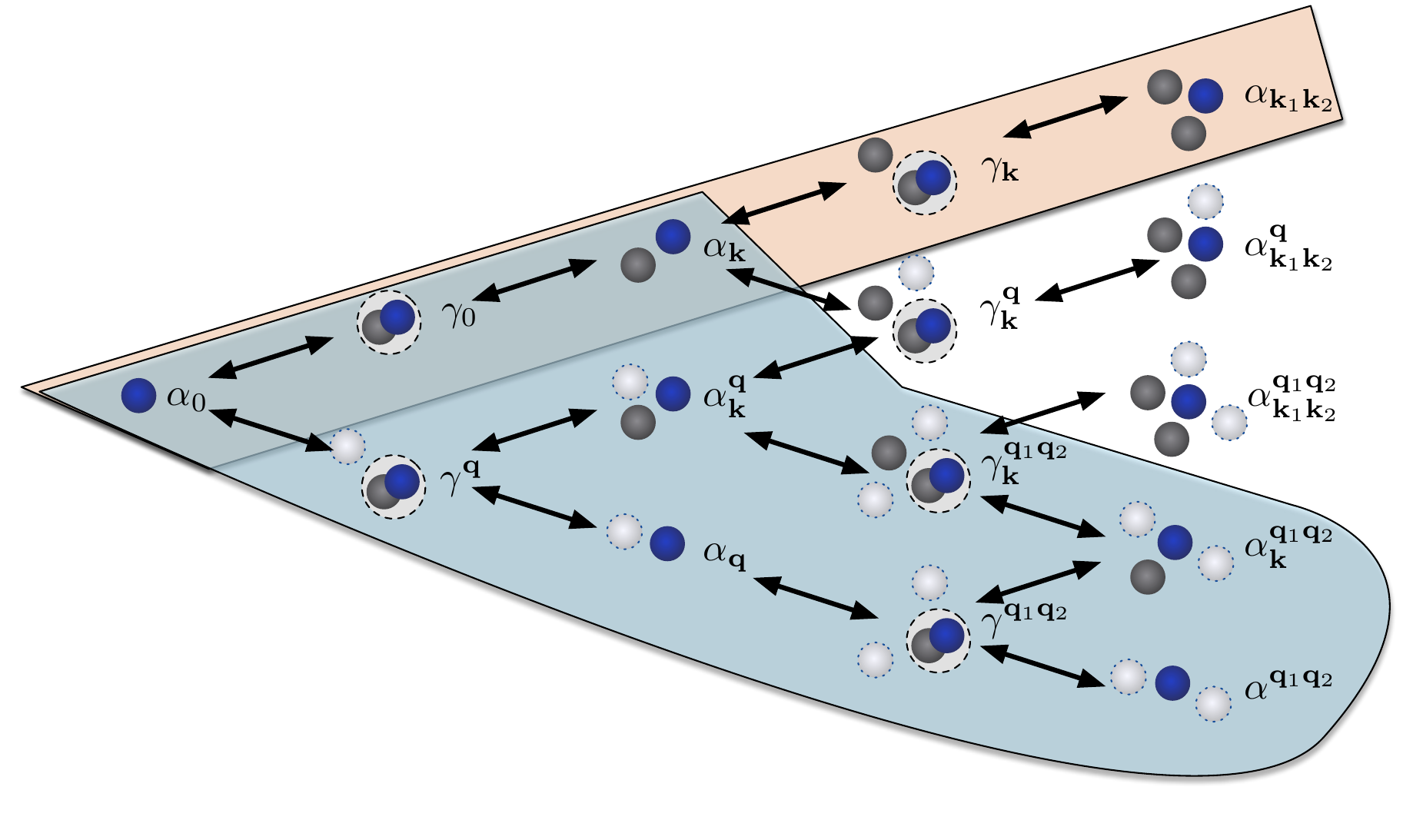}
    \caption{\textbf{Main scattering processes underlying polaron spectral features within the few-body variational ansatz} by Field et al.~\cite{Field2020}. Illustration of different scattering processes contained in the variational ansatz in Eq.~\eqref{eq:varCcompact}. The impurity, bosons and hole-like excitations are depicted by dark blue, dark grey and white hallow dots, respectively. The shaded orange region highlightes processes responsible for Efimov physics. The shaded blue region describes processes involving one-particle terms with several numbers of hole-type excitations that result in quasiparticle splitting.  Figure reprinted from Ref.~\cite{Field2020}.}
    \label{fig:scat_proc}
\end{figure}

\begin{figure}
    \centering
    \includegraphics[scale=0.4]{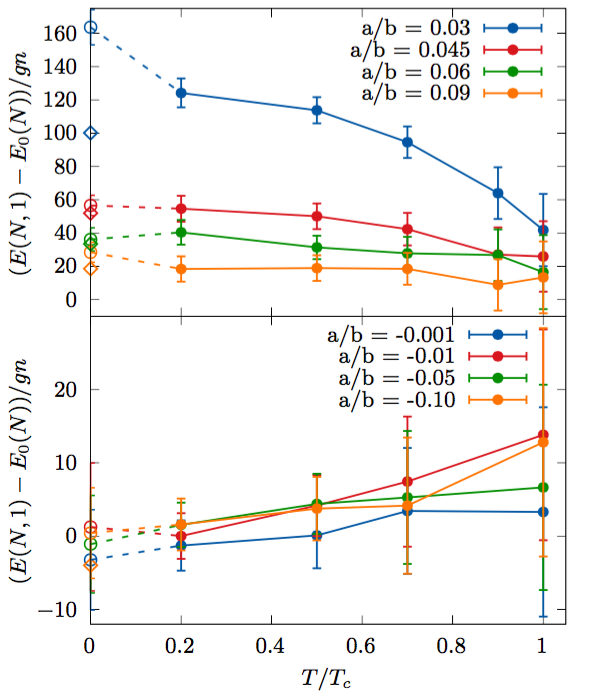}
    \caption{\textbf{Temperature dependence of the repulsive and attractive polaron resonances obtained with path integral Monte Carlo} by Pascual et al.~\cite{pascual2021}, for different values of inverse impurity-boson scattering length $b\equiv a_{\rm IB}$ in terms of inter-boson scattering length $a=a_{\rm BB}$. By increasing the temperature, the location of the repulsive polaron resonance (top panel) depends less on the scattering length, and decreases in energy, whereas the attractive polaron energy (bottom panel) increases and merges with the repulsive polaron as the critical temperature $T_c$ is reached. Figure is reprinted from Ref.~\cite{pascual2021}.}
    \label{fig:finT_E}
\end{figure}

By means of a two-body variational ansatz of the form in Eq.~\eqref{eq:varCcompact} with $N\!=\!1, \, M\!=\!1$, Field et al.~\cite{Field2020} computed the impurity spectral function at finite temperature, as shown in Fig.~\ref{fig:2B_spec_T}. For $T<T_c$, the attractive polaron splits into two peaks, as predicted earlier in Ref.~\cite{Guenther2018}. Moreover, another peak around zero energy emerges for $T \lesssim T_c$, with increasing weight as $T$ approaches $T_c$. This peak contains many eigenstates with small overlap with the non-interacting state, with the dominant contribution from the $\alpha_{\vc{k}}$ terms in Eq.~\eqref{eq:varCcompact}. Such terms represent the scattering of bosons out of the condensate into the low-momentum states of the thermal cloud, which are enhanced due to the singular nature of Bose distribution function at small momenta. 

\begin{figure*}
    \centering
    \includegraphics{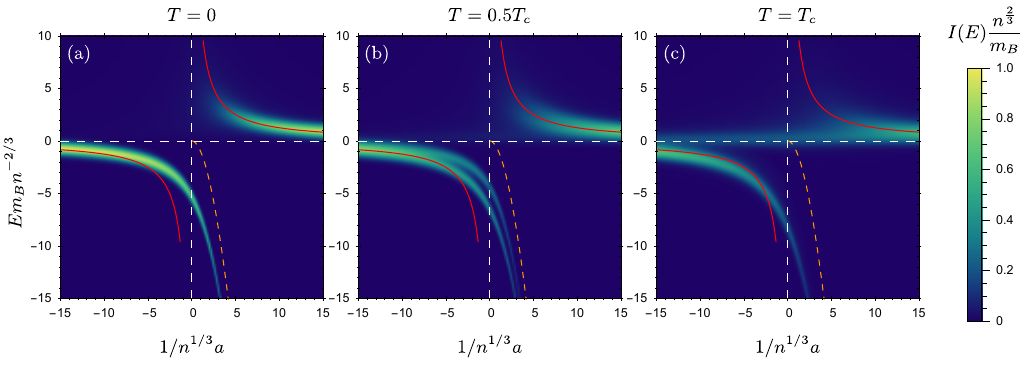}
    \caption{\textbf{Finite-temperature impurity spectral function across the Feshbach resonance} obtained from the ansatz Eq.~\eqref{eq:varCcompact} by Field et al.~\cite{Field2020} for $N=M=1$. Parameters are $m_{I}=m_{B}$, at (a) $T=0$, (b) $T=0.5 \,T_c$ and (c) $T=T_c$. At $T>0$, the attractive polaron splits into two branches. The splitting increases with $T$ until the upper attractive branch disappears at $T=T_c$. The solid red lines are the mean-field result, and the dashed orange line is the dimer energy. Figure is reprinted from Ref.~\cite{Field2020}.}
    \label{fig:2B_spec_T}
\end{figure*}

An interesting finding in Ref.~\cite{Field2020} is the correspondence of the number of quasiparticle branches and the number of impurity-excited hole-type excitations out of the thermal cloud. At $T=0$, the impurity resonantly interacts with bosons it excites out of the BEC, whereas at low temperatures $T\ll T_c$ and $T\ll |E_{\mathrm{att}}|$ with $E_{\mathrm{att}}$ the attractive polaron energy, thermal excitations dress the $T=0$ impurity propagator. Within a diagrammatic approximation, the number of poles of the dressed impurity Green's function is one larger than the number of hole propagators included in the diagrammatics. The energy correction to each branch scales as $|E_{\mathrm{att}}|\sqrt{n_{\mathrm{ex}}/n_0}$, where $n_{\mathrm{ex}}$ is the thermal excitation density. Thus the attractive polaron is expected to form a broad peak with a peak width $\Gamma \propto |E_{\mathrm{att}}|\sqrt{n_{\mathrm{ex}}/n_0}$. Since $n_{\mathrm{ex}}$ scales as $(T/T_c)^{3/2}$ for an ideal gas, $\Gamma \propto (T/T_c)^{3/4}$. 

The finite temperature properties of Bose polarons were further studied using path integral Monte Carlo (PIMC) by Pascual et al.~\cite{pascual2021}, using the standard Hamiltonian of $N$ bosons interacting through a two-body potential $V_B$, and a mobile impurity interacting with the bosons via a two-body potential $V_{I}$. Figure~\ref{fig:finT_E} depicts the temperature dependence of the attractive and repulsive polaron resonances, obtained for attractive and repulsive impurity-boson potentials, respectively. The decrease of the repulsive polaron energy with temperature indicates that the dominant contribution to the repulsive polaron energy comes from the impurity interacting with the condensate. The attractive polaron energy shows a monotonous increase with temperature, while it does not capture the initial energy decrease observed in the experiments up to impurity-boson scattering length $T/T_c=0.2$ \cite{Yan2020}. 

In addition to the polaron energy, the influence of the impurity on several properties of the Bose bath, such as the condensate and superfluid density, as well as the boson-boson correlation function, is characterized in Ref.~\cite{pascual2021} as a function of temperature. To this end, a finite density of impurities is considered. At finite temperatures, increase in kinetic energy overcomes the repulsive interboson interaction, resulting in increased boson bunching. The addition of a repulsive impurity leads to a local compression of the Bose gas, which further increases the correlations. Furthermore, the inclusion of the impurity leads to excitation out of the condensate and the non-condensed component of the superfluid. This effect is quantified in Ref.~\cite{pascual2021} by evaluating the ratio of the condensate (superfluid) density in the absence of the impurity $\rho_{s,B}$ ($n_{0,B}$) to the one in the presence of the impurity $\rho_{s,I}$ ($n_{0,I}$). the results indicate that at low temperatures, a larger fraction of the excitations come from the condensate, whereas increasing the temperature leads to more excitations out of the non-condensed component of the superfluid relative to the condensed fraction.

Further aspects of finite-temperature effects on Bose polaron physics were studied by Dzsotjan et al.~\cite{dzsotjan2020dynamical}, who developed a finite-temperature time-dependent extension of the coherent state variational ansatz for the beyond Fr\"ohlich model introduced in Sec.~\ref{secExtdBogoFroh}. The good agreement these authors achieved between the predictions of the thermal coherent state theory and the corrected experimental data -- see Ref.~\cite{Ardila2019} -- from the Aarhus experiment~\cite{Jorgensen2016} demonstrate that inclusion of finite temperature corrections is crucial to explain the discrepancy between the zero-temperature coherent state theory results and the experimental data when $1/k_n a >0$. 

The thermal coherent state theory by Dzsotjan et al.~\cite{dzsotjan2020dynamical} works as follows. The main quantity of interest to obtain the finite temperature polaron spectrum is the \textit{dynamical overlap} $S(t)$ defined by
\begin{equation} 
    S(t) = \mathrm{Tr}\big\{ e^{i \hat{H}_0 t} e^{-i \hat{H} t} \, \hat{\rho} \big\} \, ,
    \label{eqStDzsotjan}
\end{equation}
where $\hat{\rho}$ is the initial density matrix of Bogoliubov phonons,  and $\hat{H}_0$ ($\hat{H}$) is the impurity-bath Hamiltonian in the absence (presence) of impurity-boson interaction. The Fourier transform of $S(t)$ yields the injection absorption spectrum via $A(\omega)=2\,\mathrm{Re}\int^{\infty}_{0}\,d t\,e^{i\omega t}S(t)$. For the calculation of $S(t)$ the trace over thermal phonon states in Eq.~\eqref{eqStDzsotjan} is represented as an integral over coherent phonon initial states, each of which is variationally time-evolved. The resulting density-averaged spectrum $\bar{A}(\omega)$ is shown in Fig.~\ref{fig:dzsotjan_spec}. The mean-peak value of the thermal spectrum on the repulsive side evaluated at $T=160\,\mathrm{nK}$ agrees well with the corrected experimental data of Ref.~\cite{Ardila2019}, in contrast to the zero-temperature coherent state theory predictions. On the attractive side, the onset of the polaron branch matches the zero-temperature results of Ref.~\cite{Ardila2019} evaluated using Quantum Monte Carlo. On the other hand, there is no indication of finite-temperature quasiparticle splitting discussed earlier.

The thermal coherent state theory further demonstrates an exponentially decaying asymptotic behavior for $S(t)$ as $|S(t \to \infty)|=Z(T)\,e^{-\gamma(T)t}$, where $Z(T)$ and $\gamma(T)$ are temperature-dependent quasiparticle residue and decay rate, respectively. Interestingly, away from unitarity, both $Z(T)$ and $\gamma(T)$ show power-law scaling in terms of $T/T_c$, while close to unitarity, $\gamma(T)$ behaves linearly in $T/T_c$, indicating quantum critical behavior, also discussed in Ref.~\cite{Yan2020}. 

\begin{figure}
    \centering
    \includegraphics[scale=0.3]{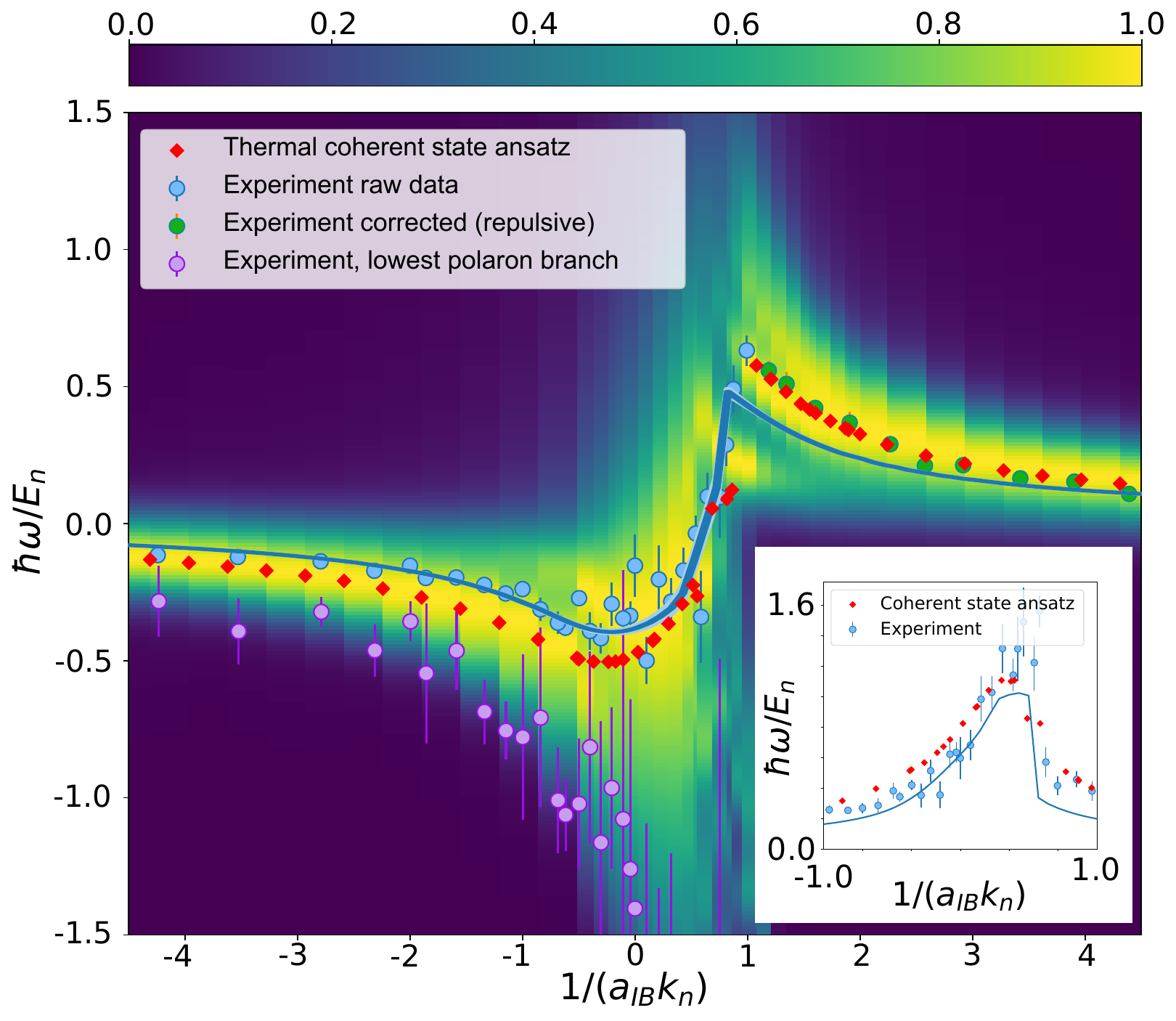}
    \caption{\textbf{Finite-temperature polaron spectrum} obtained using the thermal coherent state theory in Ref.~\cite{dzsotjan2020dynamical} at $T=160\,\mathrm{nK}$ corresponding to the Aarhus experiment~\cite{Jorgensen2016}. The mean peak position and half width half max (HWHM) (inset) evaluated by the thermal coherent state theory (red dots) shows good agreement with the corrected experimental data (green dots) on the repulsive side, compared to the zero-temperature coherent state theory predictions (solid blue line). Figure is reprinted from Ref.~\cite{dzsotjan2020dynamical}.}
    \label{fig:dzsotjan_spec}
\end{figure}

\begin{figure}
    \centering
    \includegraphics[scale=0.21]{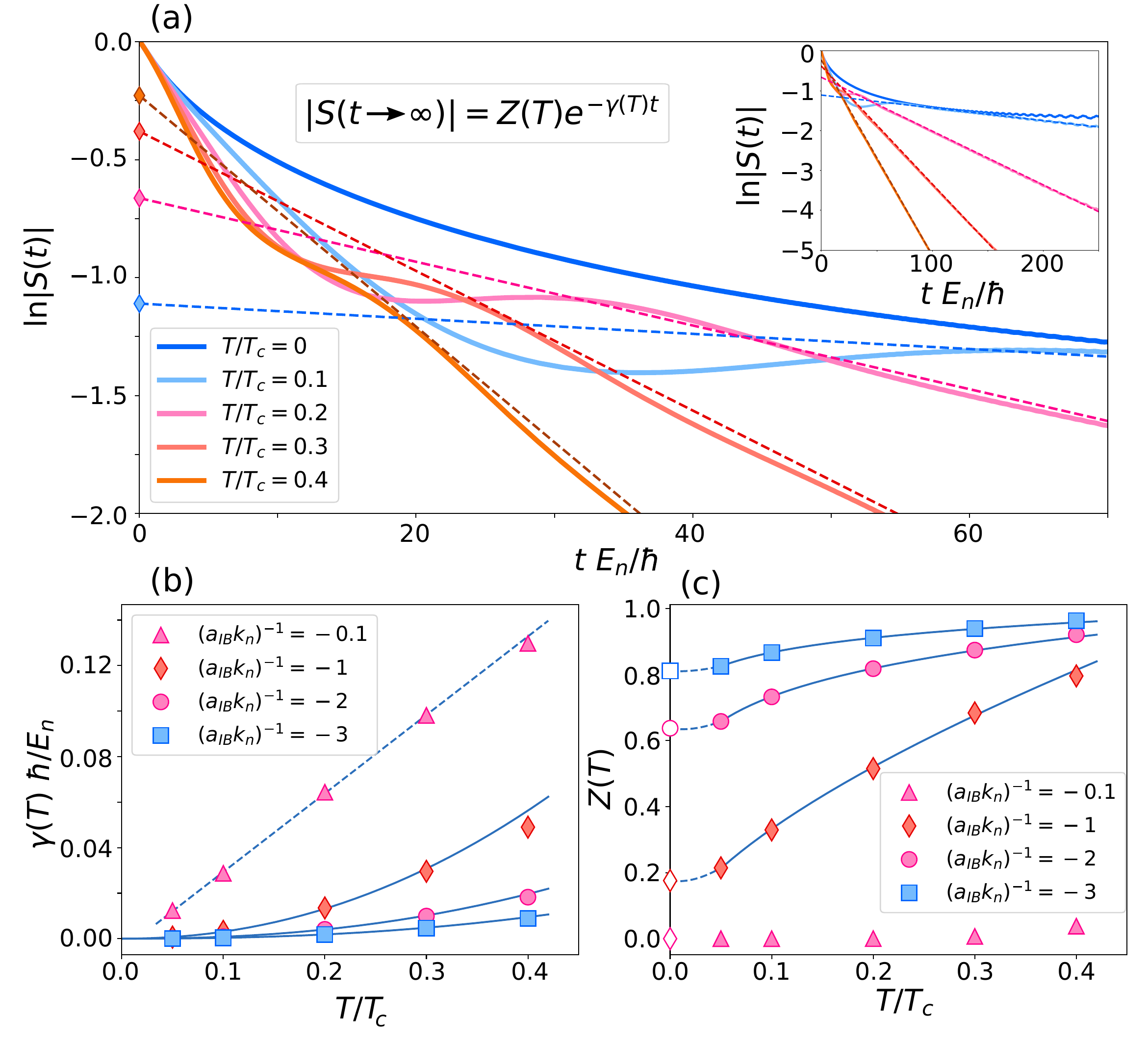}
    \caption{\textbf{Time-dependent overlap and quasiparticle properties at finite temperature.} (a) Asymptotic behavior of $S(t)$ shows an exponential decay $Z(T)e^{-\gamma(T)t}$, with temperature dependent quasiparticle weight $Z(T)$ and decay rate $\gamma(T)$. Temperature dependence of the decay rate $\gamma(T)$ (b) and the quasiparticle weight $Z(T)$ for various values of $1/k_n\,a$ on the attractive side. The decay rate shows a power-law temperature dependence away from unitarity, while it behaves linear in temperature close to the unitarity. Figure is reprinted from Ref.~\cite{dzsotjan2020dynamical}.}
    \label{fig:dzsotjan_S}
\end{figure}

\section{Bose polarons in other settings}
\label{secOtherSettings}
Although the focus of this review is on the traditional problem of a mobile quantum impurity interacting with a continuous Bose gas, as realized in experiments with ultracold atomic mixtures, many of the basic principles explored here have much wider applicability. As we discussed in the introduction, the field of Bose polarons itself was motivated by drawing connections to polarons realized in solids, when itinerant electrons are dressed by collective phonon excitations of the host crystal. But many other exciting settings exist, which resemble Bose polarons more or less closely. In the following, we provide short overviews of a selection of such systems, emphasizing however that the list is far from complete. 

\subsection{Atom-ion hybrid platforms}
\label{Sec:SubSecAtomIon}
Bose-Einstein condensates (BEC) have emerged as an alternative platform for investigating ions in superfluid media, due to their pristine nature and high controllability. Both experimental and theoretical progress have advanced in recent years thanks to the extensive research on the neutral case side. From the experimental point of view, ions can be created within a degenerate quantum gas by exciting the outermost electron of an ordinary atom to a state with a very large principal quantum number. 

Experiments in Stuttgart by Kleinbach et al.~\cite{Kleinbach2018} and in Houston by Camargo et al.~\cite{Camargo2018} utilized trapped BECs of Rubidium and Strontium atoms, respectively. One can study Rydberg polarons or ionic polarons depending on the density and trapping of the condensate, as well as the principal quantum number of the outermost electron of the impurity Rydberg atom. In the former, the size of the Rydberg orbit is comparable to the typical size of the condensate as demonstrated in the experiment by Camargo et al.~\cite{Camargo2018}; thus, the electron-atom and ion-atom interaction are both relevant. In the second case, the typical size of the condensate is much smaller than the Rydberg orbit and consequently, one can naively disregard the electron’s degree of freedom and focus exclusively on interaction of the ionic core with the BEC atoms, as demonstrated in the experiment by Kleinbach et al.~\cite{Kleinbach2018}.

The full electron $(\mathrm{e})$ and atom-ion $(\mathrm{ai})$ interaction can be decomposed into $
V(\mathbf{r})=V_{\mathrm{e}}(\mathbf{r})+V_{\mathrm{ai}}(\mathbf{r})$ with 
\begin{equation}
V_{\mathrm{e}}(\mathbf{r})=\frac{2\pi\hbar^{2}}{m_{e}}a_{s}(\mathbf{k})|\Psi(\mathbf{r})|^{2}+\frac{6\pi\hbar^{2}}{m_{e}}a_{p}^{3}(\mathbf{k})|\vec{\nabla}\Psi(\mathbf{r})|^{2}
\label{eq:Ve}
\end{equation}
and 
\begin{equation}
V_{\mathrm{ai}}(\mathbf{r})=-\frac{C_{4}}{r^{4}},
\label{eq:Vai}
\end{equation}
where $V_{\mathrm{e}}(\mathbf{r})$  accounts for an electron-atom interaction and $\Psi(\mathbf{r})$ is the wavefunction of the electron in the Rydberg state. Moreover, $a_{s/p}$ are the momentum-dependent $s$-wave and $p$-wave scattering lengths. Instead, $V_{\mathrm{ai}}(\mathbf{r})$ is the asymptotic form of the atom—ion interaction that arises from the electric dipole induced in a neutral atom as it gets closer to the ion~\cite{Tomza2019}.

 In the experiments by Camargo et al.~\cite{Camargo2018}, the Rydberg polarons were created by exciting an electron to a highly excited Rydberg state $n=49,60$ and $n=72$ starting from a Strontium BEC. A macroscopic occupation of the molecular bound state is observed via spectroscopic measurements. Rydberg polarons exist as a series of excited states of bound states and scattering states as shown in the broader spectral function. This contrasts with polarons having short-ranged interactions where the ground state is the attractive many-body polaron branch. From a theoretical point of view, the Rydberg polaron problem cannot be described with the standard Fr\"ohlich paradigm and multi-scattering events need to be included~\cite{Schmidt2018Ryd}, akin to the strongly coupled Bose polaron with short-ranged interactions. High-energy processes between the impurity and the bosons are essential in both cases. Interestingly, the experiment and theory show an excellent agreement even though the latter does not account for interactions between the condensate atoms due to the large differences in energy scales between the typical energy range of Rydberg aggregates and the chemical potential of the bosons.

As mentioned earlier, when the Rydberg orbit is very large, one can exploit the atom-ion interactions. In this case, the interaction potential simplifies to a $1/r^{4}$ interaction as described in Eq.~\eqref{eq:Vai}. Pioneering research conducted by Côté et al.~\cite{Cote2002} demonstrated that the polarization potential can confine several condensed atoms, giving rise to a mesoscopic molecular ionic state formed by weakly bound atoms. In addition, at absolute zero temperature ($T=0$), the primary scattering processes are inelastic. In particular, a boson is captured by the ionic potential and releases kinetic energy by emitting excitations (phonons) into the surrounding condensate. From the perspective of a single impurity, it was shown by  Massignan et al.~\cite{MassignanIon} that a single heavy ionic impurity immersed can locally trap several atoms. In the low-density
regime, the number of trapped atoms is directly proportional to the ratio of ion-boson and boson-boson scattering lengths. In the high-density regime, the local density excess is computed by using a Gross–Pitaevskii approach and several solutions are correlated with the number of bound states.

In the experiment by Kleinbach et al.~\cite{Kleinbach2018}, the atoms participating in the scattering with the ions are confined in a length scale of approximately $R^{\star}=\sqrt{2m_{\mathrm{red}}C_{4}}/\hbar\approx 5000 a_0$, allowing one to neglect the contributions arising from the electron-atom interaction. In other words, the Rydberg state is prepared so that its size is much larger than $R^{\star}$. Despite these efforts, the achieved temperatures in the system are higher than the typical s-wave scattering limit, which for Rubidium is expected to be on the order of $E^{\star}/k_{\mathrm{B}}=\frac{\hbar^{2}}{2m_{\mathrm{red}}(R^{\star})^{2}}=0.08\mu\mathrm{K}$. Unlike to neutral cases, electric fields can be used easily to control the transport of ions and study the collisional dynamics. The charge transport of ions in the condensate shows that the ion moves diffusely through the BEC due to the large atom-ion cross section and the high local density that causes fast atom-ion collision processes as measured by Dieterle et al.~\cite{Dieterle2021}.

One of the milestones of the current experiments based on atom-ion hybrid platforms is achieving the regime of full quantum degeneracy, where only a few partial waves in atom-ion collisions contribute to scattering physics. Yet, cooling down the ion and the atomic ensemble to the degenerate regime presents two main challenges: First, the low energies associated with the low partial waves in the atom-ion collision need very low temperatures. Reaching that limit needs energies lower than the sub$-\mu \mathrm{K}$ (in the case of the heavy ion)  much smaller than the typical temperatures that Doppler cooling reaches. The second reason is the intrinsic motion of the ion, known as micromotion, which arises from the oscillating radio frequency quadrupole potential in a typical Paul trap. Ion micromotion is one of the biggest problems in the realization of degenerate mixtures due to the excess kinetic energy of the ion with respect to the host gas. Placing the ion at the minimum of the RF potential is a suitable solution; however, the complex dynamics and the short-range nature of the atom-ion potential prevent the ion from remaining at the center of the trap, making it more likely to move. One way to circumvent this issue is to use a very heavy ionic impurity. For example, Fig.~\ref{fig:AtomIonMixture} depicts the expected phase diagram for an ionic highly imbalanced $^{171}\mathrm{Yb}^{+}-^{6}\mathrm{Li}$ mixture; for this case, the mass ratio is almost $m_{\mathrm{Yb}}/m_{\mathrm{Li}}=30$. For comparison the energy scales required to achieve degeneracy are also indicated in the figure.

\begin{figure}
    \centering
    \includegraphics[scale=0.25]{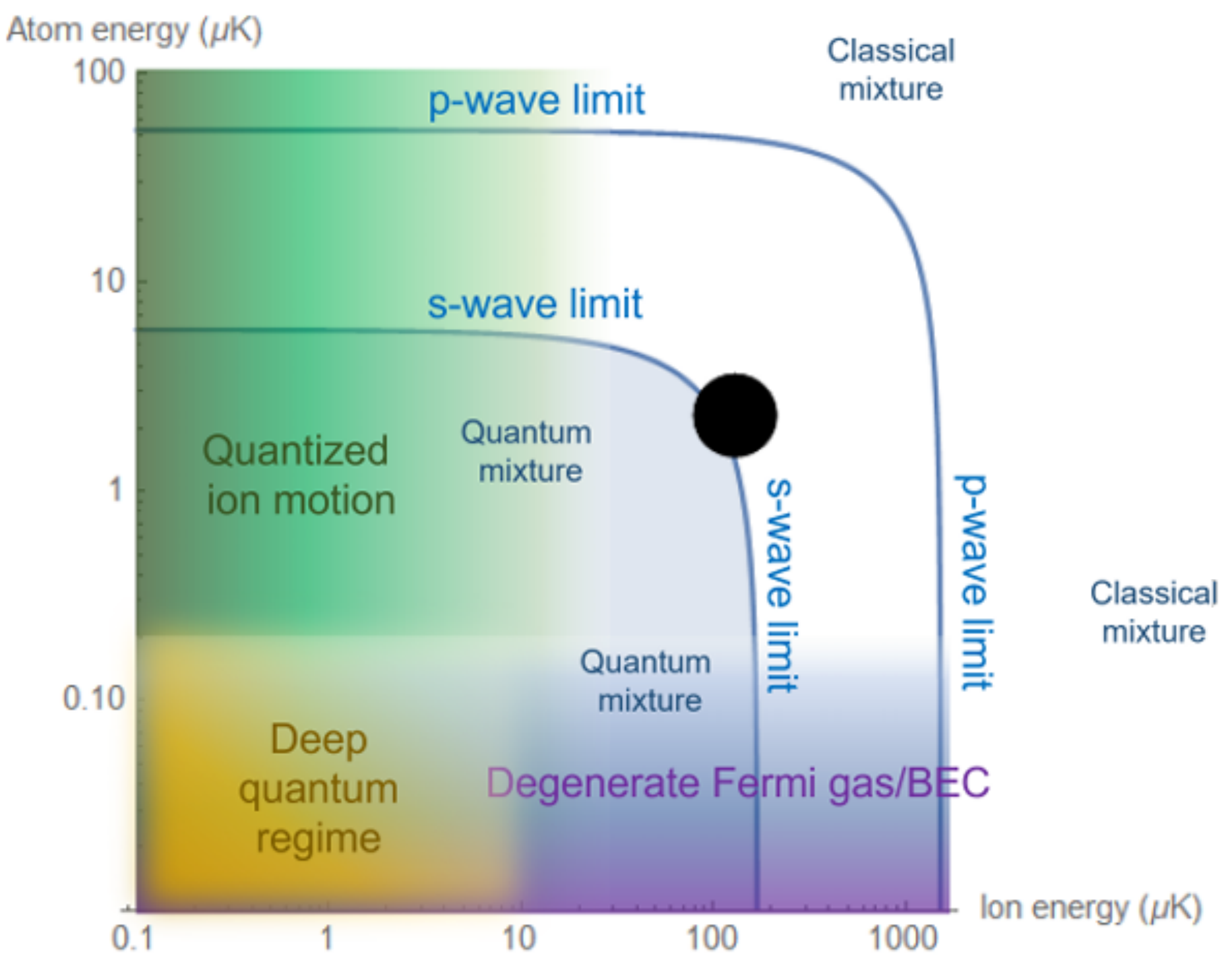}
    \caption{\textbf{Typical phase diagram of an atom-ion hybrid system.} Shown are the expected energy scales for atoms and ions, going from the classical towards the quantum degenerate atom-ion mixture. The particular example of a $^{171}\mathrm{Yb}^{+}-^{6}\mathrm{Li}$ mixture is presented. Figure taken from Lous et al.~~\cite{Lous2022}.} 
    \label{fig:AtomIonMixture}
\end{figure}

Besides the significant progress on atom-ion mixtures, the possibility of creating Bose polarons with ultracold atom-ion hybrid system is appealing because, on the one hand, the precise control of atom-ion interaction can be achieved akin to the neutral case using Feshbach resonances~\cite{Weckesser2021}. On the other hand, temporal and spatial correlations of polarons can be detected with high resolution using modern quantum microscopes~\cite{Veit2021}. Combining both of these techniques, polarons featuring long-range interactions can unveil unknown phenomena in regimes inaccessible in solid-state systems~\cite{Bissbort2013}.

A renormalized version of the atom-ion potential, capturing the asymptotic long-range $1/r^4$~\cite{kry15}  can be written as 
\begin{equation}
   V_{\mathrm{ai}}(\mathbf{r}) = -C_4\left(\frac{r^{2}-c^{2}}{r^{2}+c^{2}}\right)\frac{1}{(r^{2}+b^{2})^{2}} ,
   \label{eq:vai}
\end{equation}
where $b$ determines the depth of potential, and $c$  sets the finite short-range repulsive contribution. The parameterization of the potential strength is determined by $C_4$, which is equal to $\alpha e^2 / 2$ and relies on the atomic static polarizability $\alpha$, with $e$ representing the elementary electrostatic charge. The natural length scale is the range of the potential $r_{\star}$, defined as $\left(2m_{\mathrm{red}} C_4 /\hbar^2\right)^{1/2}$, where $m_{\mathrm{red}}^{-1}=m_{\mathrm{atom}}^{-1}+m_{\rm ion}^{-1}$ denotes the reduced mass of the atom-ion system.  Instead, the energy scale $E^{\star}$ is directly proportional to the height of the centrifugal barrier for any partial wave.

One of the earliest theoretical works on charged polarons was conducted by Casteel et al.~\cite{Casteels2011Ion}. They computed the polaron properties of an ion immersed in a condensate using an all-coupling path integral approach with the Fröhlich Hamiltonian. More recently, by using the renormalized atom-ion potential Eq.~(\ref{eq:vai}), the ground-state properties of the charged ionic Bose polarons were computed using quantum Monte-Carlo approaches by Astrakharchik et al.~\cite{Astrakharchik2021}, where a transition from a polaron to a many-body bound state was found. The latter is characterized by a sizeable deformation of the bath around the ion. Christensen et al.\cite{Christensen2021} utilized diagrammatic approaches supported by a coherent variational ansatz and predicted also the formation of bound states: molecular ions.  Alternatively, methods based on modified mean-field approaches~\cite{Schmidt2022,olivas2024} are relevant as they take into account the large density deformation around the ion. Finally, the induced interactions between ionic polarons have been studied in~\cite{Ding2022, Astrakharchik2023}. Due to the considerable density deformation of the bath, the strength of the induced interaction between ions is stronger than in the neutral case~\cite{Astrakharchik2023}, and clear experimental protocols are devised for its detection~\cite{Ding2022}.  In addition,  recent Fixed-node diffusion Monte Carlo (FNDMC)  calculations also shown a strong dependence of the Fermi polaron properties.~\cite{Pessoa2021, Pessoa2024}.

\subsection{Light-matter interactions and polarons}
\label{Sec:Light-matter interactions and polarons}
Bose polarons can be explored in settings where the mobile impurity is coupled to a light field, realizing a polariton. Through the matter component of the polariton, one naturally obtains strong interactions of the impurity with the surrounding Bose gas, and by changing the strength of the light-matter coupling tunable model parameters can be realized. These advantages were pointed out by Grusdt and Fleischhauer~\cite{Grusdt2016PRL}, where a concrete experimental setup was also proposed. They considered an ultracold quasi-2D BEC, trapped inside a high-finesse optical cavity which provides a strong transverse confinement for a light mode. The latter is coherently coupled in a three-level configuration realizing electromagnetically-induced transparency (EIT), with long-lived coherent dark-state polaritons \cite{Fleischhauer2000} involving a meta-stable hyperfine state of the BEC. 

On the one hand, such optical setups provide direct experimental access to the impurity spectral function through absorption spectroscopy. Another main advantage of this platform~\cite{Grusdt2016PRL} is its ability to realize very light impurities, well in the strong-coupling regime of an effective Fr\"ohlich model description: The transverse photon mass $m_{\rm ph}$ can be orders of magnitude smaller than typical atomic masses and the polariton mass $m_{\rm I}$ is determined by the tunable EIT mixing angle $\theta$ through 
\begin{equation}
 m_{\rm I}^{-1} = m_{\rm ph}^{-1} \cos^2 \theta + m_{\rm B}^{-1} \sin^2 \theta.   
\end{equation}
This paves the way for studies of ultra-light impurities near a Feshbach resonance, the physics of which remains largely unexplored. 

A related setup, with polariton impurities coupled to atomic BECs in an EIT $\Lambda$ configuration, was analyzed by Camacho-Guardian et al.~\cite{CamachoGuardian2020}. They went beyond the Fr\"ohlich paradigm and demonstrated a breakdown of the quasiparticle picture in the vicinity of an inter-atomic Feshbach resonance. In the strongly interacting regime, the polaron-polariton is replaced by a more complex state of interacting matter and light fields. Specifically, scattering to photons at energies deviating from the two-photon resonance underlying EIT lead to light-induced damping of the optically induced excitation that the authors modeled in detail. In a subsequent study by Knakkergaard Nielsen et al.~\cite{Nielsen2020}, the same authors demonstrated an even more dramatic effect of the light field, enabling a superfluid flow of polaron-polaritons above Landau's critical velocity. This is enabled by a suppression of spontaneous Cherenkov-type emission of phonons through a light shift that renders impurity-phonon collisions inelastic. 

Going a step further, the host Bose gas can also be replaced by atom-light constituents: Several works reviewed below considered a coherent exciton-polariton condensate in a semiconductor~\cite{Carusotto2013} as a starting point. Making use of a valley-degree of freedom, two types of excitons can be independently addressed using laser light of opposite circular polarization $\sigma_\pm$. Levinsen et al.~\cite{Levinsen2019} proposed to realize an exciton-polariton BEC in one valley by a strong $\sigma_+$ drive, which interacts with a $\sigma_-$ impurity exciton that is weakly excited by a $\sigma_-$ probe pulse. Placing the monolayer semiconductor into a microcavity which confines the light fields along the transverse direction thus yields an effective two-dimensional Bose polaron problem. Levinsen et al.~\cite{Levinsen2019}, and later Bastarrachea-Magnani et al.~\cite{BastarracheaMagnani2019}, modelled this setting by a beyond-Fr\"ohlich Hamiltonian of the form, 
\begin{multline}
    \H = \sum_{\vec{k}} \left( \omega_{{\rm X}\vec{k}} \hat{b}^\dagger_{\vec{k}} \hat{b}_{\vec{k}} + \omega_{{\rm C}\vec{k}} \hat{c}^\dagger_{\vec{k}} \hat{c}_{\vec{k}} + \frac{\Omega_R}{2} \left( \hat{b}^\dagger_{\vec{k}} \hat{c}_{\vec{k}} +  \hc \right)\right) \\
    + \sum_{\vec{k}} \left( \omega_{{\rm LP}\vec{k}} - \omega_{{\rm LP}\vec{0}} \right)  \hat{L}^\dagger_{\vec{k}} \hat{L}_{\vec{k}} + \sum_{\vec{k},\vec{k'}, \vec{q}} g_{\vec{k}\vec{k}'} \hat{L}^\dagger_{\vec{k}} \hat{b}^\dagger_{\vec{q}-\vec{k}} \hat{b}_{\vec{q}-\vec{k}'} \hat{L}_{\vec{k}'}\\
    + \sqrt{n} \sum_{\vec{k}, \vec{q}} g_{\vec{k}\vec{0}} \hat{b}^\dagger_{\vec{q}-\vec{k}} \hat{b}_{\vec{q}} \left( \hat{L}_{\vec{k}}^\dagger + \hat{L}_{-\vec{k}} \right).
    \label{eqHexcPolPol}
\end{multline} 
Here $\hat{b}_{\vec{k}}$ and $\hat{c}_{\vec{k}}$ denote the probe exciton and cavity (i.e. photon) modes, with transverse dispersion relations $\omega_{{\rm X}\vec{k}}$ and $\omega_{{\rm C}\vec{k}}$; the operator $\hat{L}_{\vec{k}}$ annihilates a lower polariton (LP) at momentum $\vec{k}$, around a condensate of LP at $\vec{k}=\vec{0}$ with density $\propto n$; the LP dispersion is $\omega_{{\rm LP} \vec{k}}$, the photon-exciton coupling is $\Omega_{R}$. Finally, $g_{\vec{k}\vec{k}'}$ describes interactions between the majority LPs and the minority exciton, which can be tuned through a biexciton Feshbach resonance. Levinsen et al.~\cite{Levinsen2019} furthermore suggested the possibility of observing a triexciton resonance.

Both theoretical works~\cite{Levinsen2019,BastarracheaMagnani2019} calculated the Bose polaron spectrum, using a Chevy-type ansatz revealing genuine many-body correlations in this system~\cite{Levinsen2019} and using a T-matrix approach~\cite{BastarracheaMagnani2019}. Bastarrachea-Magnani et al.~\cite{BastarracheaMagnani2019} furthermore applied their formalism to model earlier pump-probe experiments in an exciton-polariton setup~\cite{Takemura2017,NavadehToupchi2019}. Although the latter did not tune across a biexciton resonance, direct signatures of strong impurity-boson interactions could be observed. The experimental results agree reasonably well with the theoretical description~\cite{Levinsen2019,BastarracheaMagnani2019} even though the latter ignores interactions among majority excitons. Another experimentally relevant effect in relation to exciton-polaritons is the dressing of exciton-polaritons with density fluctuations of an optically dark excitonic medium~\cite{choo2024polaronic}. Such an incoherent bath of dark excitons are often created during the formation of exciton-polaritons, and can interact with the polaritons through their excitonic component, leading to the appearance of coherent polaron-like states, studied in Choo et al.~\cite{choo2024polaronic}.

Experimentally, the first direct observation of repulsive and attractive Bose polaron branches across a tunable biexciton Feshbach resonance was achieved in a cavity-coupled ${\rm MoSe}_2$ monolayer by Tan et al.~\cite{Tan2023}. Their setup is sketched in Fig.~\ref{figTan2023} (a), where the monolayer semiconductor ${\rm MoSe}_2$ is sandwiched between hexagonal boron nitride (hBN) layers and a graphene gate below, all placed inside a distributed Bragg reflector (DBR) realizing the optical microcavity. The detuning of the cavity can be used to tune across a biexciton Feshbach resonance between exciton polaritons of opposite chirality that was reported earlier in this system~\cite{Takemura2014}. The experiment starts by resonantly exciting a bath of majority $\sigma_+$ polaritons, followed by a measurement of the transmission spectrum of the minority $\sigma_-$ polaritons. The normalized transmission spectrum as a function of detuning and measured with respect to the energy of the non-interacting polariton is shown here in Fig.~\ref{figTan2023} (b). It reveals both an attractive and repulsive polaron branch, and was modelled theoretically by the authors building upon the theoretical model from Eq.~\eqref{eqHexcPolPol}. 

\begin{figure}
	\centering
	\includegraphics[width=0.7\linewidth]{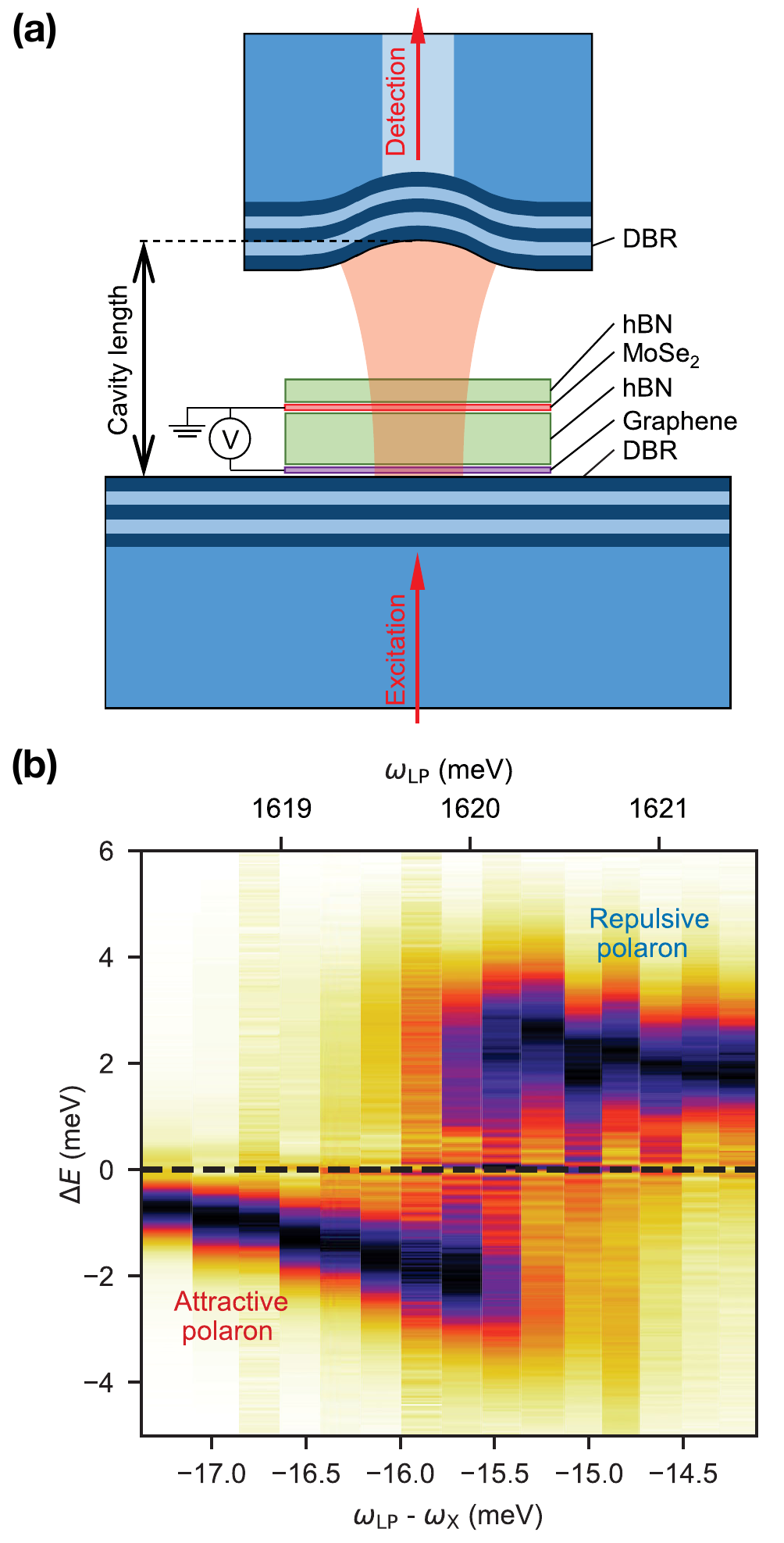}
	\caption{\textbf{Bose polarons realized by exciton-polaritons} in a cavity-coupled monolayer semi-conductor by Tan et al.~\cite{Tan2023}. (a) The setup, consisting of a two-dimensional monolayer ${\rm MoSe}_2$ inside an optical microcavity. (b) By changing the detuning of the cavity, i.e. the polariton energy $\omega_{\rm LP} - \omega_{\rm X}$, attractive and repulsive polaron branches are identified in the transmission spectrum of minority exciton polaritons. Figure adapted from Ref.~\cite{Tan2023}.}
	\label{figTan2023}
\end{figure}

By measuring how the polaronic energy shift depends on the density of impurity and bath exciton polaritons, Tan et al.~\cite{Tan2023} furthermore extracted the strength of bath-induced interactions, demonstrating a remarkable tunability of the latter via variations of the bath density. These experiments are in line with recent theoretical analysis of mediated interactions between exciton-polaritons by Camacho-Guardian et al.~\cite{CamachoGuardian2021} and pave the way towards non-linear optical materials with strong and tunable photon-photon interactions. 

These experimental advances pave the way towards more detailed studies of Bose polarons in light-matter systems. For example, Vashisht et al.~\cite{Vashisht2022} have proposed to study polaron dynamics, the effective mass and drag forces acting on the exciton-polariton impurities in these settings. Within a Bogoliubov-Fr\"ohlich approach but including a dissipative bath term, they predicted different dynamical regimes and further suggested to use exciton-polariton impurities as probes of superfluidity in quantum fluids of light. As another extension, Cardenas-Castillo et al.~\cite{CardenasCastillo2023} recently suggested to consider electrons in a lightly filled conduction band as impurities which interact with an exciton-polariton BEC, yielding another route towards Bose polaron formation. Recently, Bose polarons have also been observed when photo-excited direct excitons are created as impurities inside a degenerate Bose gas of indirect excitons in an electron-hole bilayer realized in GaAs heterostructures~\cite{Szwed24}.

\subsection{Angulons}
The polaron concept was successfully extended to rotors featuring an angular momentum, or spin, that is conserved in isolation. When placed inside a many-body environment, interactions can lead to an exchange of angular momentum between these quantized rotors and the bath, similar to the exchange of linear momentum in the traditional Bose polaron problem. The microscopic theoretical description of this setting was pioneered by Schmidt and Lemeshko~\cite{Schmidt2015}, who argued that interactions cause dressing of the quantum rotor with the surrounding bosonic quantum field and lead to the formation of a new quasiparticle - termed the \emph{angulon}. In analogy with the effective mass of Bose polarons, the rotational constant $B$ of the rotor is renormalized, $B \to B^*$, upon angulon formation. Likewise, the ground state energy is modified by a rotational Lamb shift and the impurity rotational spectrum features a many-body induced fine structure~\cite{Schmidt2015}.

To describe the emergent many-body physics of angulons, Schmidt and Lemeshko~\cite{Schmidt2015} proposed a Fr\"ohlich-type angulon Hamiltonian. It describes how the bare quantum rotor, with Hamiltonian $\H_0 = B \hat{\vec{J}}^2$ and angular momentum $\hat{\vec{J}}$, interacts with an irrotational bosonic bath:
\begin{multline}
    \H_{\rm ang} = B \hat{\vec{J}}^2 + \sum_{k\lambda\mu} \omega_k \bd_{k\lambda\mu} \b_{k\lambda\mu} + \\
    + \sum_{k\lambda\mu} U_\lambda(k) \left[ Y^*_{\lambda\mu}(\hat{\theta},\hat{\phi}) \bd_{k\lambda\mu} + \hc \right].
\end{multline}
Here $k=|\vec{k}|$ denotes the length of momentum $\vec{k}$, and $\lambda$ ($\mu$) is the quantum number of angular momentum (its projection on the $z$ axis) of the bosonic excitations, $\b_{k\lambda\mu} = k (2\pi)^{-3/2} \int d\Omega_k \b_{\vec{k}} (-i)^\lambda Y_{\lambda\mu}(\Omega_k)$. Moreover, $\omega_k$ is the dispersion relation of the bosons and their interactions with the rotor are characterized by $U_\lambda(k)$. Finally, $\hat{\theta}$ and $\hat{\phi}$ are the angles defining the rotor's orientation, conjugate to its angular momentum, and $Y_{\lambda\mu}$ is a spherical harmonic. 

The angulon problem was proposed to be realized experimentally for molecules, or molecular ions, immersed in superfluids. As a first example, this includes quantum rotors in weakly interacting BECs~\cite{Schmidt2015,Midya2016}, in close analogy with the Bose polarons reviewed here. Another notable platform is constituted by molecules immersed in superfluid Helium~\cite{Lemeshko2017}: They have been studied for more than 25 years, and provide one of the most compelling cases for the angulon concept. As shown by Lemeshko~\cite{Lemeshko2017}, angulon theory is able to explain to a large degree the experimentally observed renormalization of the rotational constant $B^*/B$ for a variety of molecules immersed in superfluid Helium. The largest deviations between theory and experiment were found in the theoretically most challenging regime of intermediate couplings, where advances in the theoretical modelling have recently been reported by Liu et al.~\cite{Liu2023}.

Another remarkable case where angulon theory has be successfully applied corresponds to the problem of electron-to-lattice angular momentum transfer. Mentink et al.~\cite{Mentink2019} modelled the interplay of electron-phonon coupling in a solid with spin and orbital angular momentum of the underlying electrons. Their study captures ultrafast demagnetization dynamics on the shortest timescales and paves the way towards a deeper understanding of this effect and its associated time scales. 

As in the field of traditional Bose polarons, the impurity-nature of the angulon problem allows for significant theoretical modelling. Since the initial proposal of the angulon concept, many key steps have been taking. While we cannot provide an in-depth review of these advances, highlights we would like to mention include progress in generalizing the LLP canonical transformation to quantum rotors~\cite{Schmidt2016PRX} - making the conservation of total angular momentum apparent; progress in variational approaches - ranging from Chevy-type wavefunctions to coherent states~\cite{Zeng2023} and combinations thereof~\cite{XiangLi2019}; progress in diagrammatic approaches - including path-integral formalism~\cite{Bighin2017} and numerically exact diagrammatic Monte Carlo simulations~\cite{Bighin2018}. 

Applying these theoretical approaches, various effects resembling the phenomenology of Bose polarons have been revealed. Notable examples include angular self-localization studied by Li et al.~\cite{Li2017}, where the spherical symmetry is broken and the angulon ground state features a discontinuity in the first derivative of the ground-state energy. In another study by Li et al.~\cite{Li2020} it was shown that interactions between two quantum rotors can be mediated by the bosonic bath, and bound states of the latter -- so-called bi-angulons -- can form.

\subsection{Dipolar polarons and mixtures} 
Ultracold atoms with high magnetic or electric dipole moments have recently attracted much attention. In particular,  dipolar interactions between atoms are long-range and anisotropic in space. They have been shown to facilitate the realization of new states of the matter, such as quantum
droplets and the long-sought supersolid states. Both were realized in the laboratory and explored widely from a theoretical point of view, see Refs.~\cite{Lahaye_2009,Chomaz_2023} for recent reviews on this topic.

In a fully polarized gas where all atoms are aligned with the external magnetic field, the dipole-dipole interaction (DDI) between two atoms with a dipole moment $\mu$ takes the form, 
\begin{equation}
V(\mathbf{r})=g\delta(\mathbf{r})+\frac{\mu_{0}\mu^{2}}{4\pi\left|\mathbf{r}\right|^{3}}\left(1-3\cos^{2}\theta\right),
\label{eq:DDI}
\end{equation}
where $\vec{r}$ is the distance between the two dipoles, $\theta$ is the angle between $\vec{r}$ and the applied field and $\mu_0$ is the vacuum permeability. The first term is the typical short-ranged term characterized by a coupling strength $g$. In $d=3$, the dipole-dipole interaction (DDI) is long-ranged and anisotropic. Namely, when dipoles are side-by-side, the interaction is effectively repulsive, while in a head-to-tail configuration, the effective interaction is attractive. The strength of the DDI is parametrized by $\epsilon=\frac{\mu_{0}\mu^{2}}{3g}$.

Bose-Einstein condensation of binary dipolar mixtures motivates the further exploration of impurity and polaron physics in uncharted regimes. Ultracold dipolar mixtures in this context are achieved employing partially dipolar combinations of magnetic and no magnetic species, for instance, $\mathrm{Er-Li}$~\cite{Schafer2022}, $\mathrm{Yb-Er}$~\cite{Schafer2023}, as well as dual species with large dipole moments as is the case of $\mathrm{Er-Dy}$~\cite{Mingwu2012, Trautmann18, Durastante20}. In contrast to the neutral, isotropic scenario or the long-range atoms-ion hybrid platform, the DDI does not require a large difference in length scales to manifest exotic physics. Instead, a delicate equilibrium between mean-field energy (consisting of the sum of short and long-range interactions) and quantum fluctuations introduce novel characteristics to both single and binary dipolar condensates. 

A mobile impurity in a dipolar mixture can be modelled by an effective Fr\"ohlich Hamiltonian,
\begin{eqnarray}
\hat{\mathcal{H}} &=& \sum_{\mathbf{k}} \omega_{\mathbf{k}} \hat{b}_{\mathbf{k}}^{\dagger} \hat{b}_{\mathbf{k}} + \frac{\hat{\mathbf{p}}^2}{2 m} + n V_{12}(\mathbf{k}=0) \nonumber \\
&& + \frac{\sqrt{n}}{\sqrt{V}} \sum_{\mathbf{k}} V_{12}(\mathbf{k}) W_{\mathbf{k}} e^{-i \mathbf{k} \cdot \hat{\mathbf{r}}} \left(\hat{b}_{\mathbf{k}}+\hat{b}_{-\mathbf{k}}^{\dagger}\right),
\label{eqHdipFroh}
\end{eqnarray}
as described in Eq.~\eqref{eqHFroh} in Chapter~\ref{secEffHamiltonians}. Here $V_{12}(\mathbf{k})$ is the Fourier transform of the dipolar interaction potential Eq.~\eqref{eq:DDI}, and the Bogoliubov spectrum of the dipolar host BEC becomes $\omega_{\mathbf{k}}=\sqrt{\epsilon_\mathbf{k}\left(\epsilon_\mathbf{k}+2 nV_\text{11}(\mathbf{k}) \right)}$, with  $\epsilon_{\mathbf{k}}=\hbar^{2}\mathbf{k}^{2}/2m$ being the bare boson dispersion, and $W_{\mathbf{k}}=\sqrt{\epsilon_{\mathbf{k}}/\omega_{\mathbf{k}}}$; as usual, $\hat{\mathbf{r}}$ is the position operator of the impurity. The boson-boson ($V_{11}$) and impurity-boson ($V_{12}$) potentials are given by $V_{ij}(\mathbf{k})=g_{ij}+g_{ij}^{d}\left(3k_{z}^{2}/\mathbf{k}^{2}-1\right)$, where $g_{ij}=4\pi\hbar^2a_{ij}/m$ written in terms of the $s$-wave scattering length, $a_{ij}$. In addition, the long-range and anisotropic DDI is characterized by the dipolar coupling strengths $g_{ij}^{d}=\mu_{0}\mu_{i}\mu_{j}/3=4\pi\hbar^{2}\sqrt{d_{i}d_{j}}/m$ which is determined by the atomic species.

In reference~\cite{Kain2014}, Kain et al. first investigated the ground-state and dynamical properties of a single neutral impurity interacting with a dipolar condensate, followed by a perturbative study of the Fr\"ohlich model Eq.~\eqref{eqHdipFroh} by Ardila and Pohl~\cite{Ardila2018Dip} in a fully dipolar, quasi-two-dimensional set-up. Momentum-space integration of properties such as energy and quasiparticle residue introduces shifts relative to the non-dipolar scenario, and leads to anisotropic mass renormalization. Transversal and longitudinal Bogoliubov excitations exhibit different propagation behaviors in the gas as the speed of sound is direction-dependent. This also leads to different time scales associated with polaronic dressing, and thus the formation dynamics of the quasiparticle also depends on the direction~\cite{Volosniev2023}.

Purely dipolar interactions in  2D geometries were also studied by Sanchez-Baena et al.~\cite{sanchezbaena2023} using QMC methods. Thereby larger regimes of the gas parameter $x=na^{2}$ could be explored than perturbatively accessible. For a specific range of parameters, especially where the local gas parameter breaks the condition of diluteness, QMC and perturbation theory display differences as expected. Furthermore, for certain parameter ranges, polaron properties were found to be universal functions of the gas parameter, which is not a surprise as the host bath also displays such dependence in the ground-state energy~\cite{Baena22}.

Inspired by recent experiments on quantum droplets in dipolar mixtures, Bisset et al.~\cite{Bisset2020} studied the ground-state properties of a binary dipolar mixture, including in the highly imbalanced case where a mobile impurity interacts with droplets formed by the majority species. This situation draws a strong parallel with impurities immersed in helium nanodroplets~\cite{Barranco2006,navarro2012}, where the solvation energy indicates whether dopants coat the droplets or become immersed in their interior. In the context of dipolar quantum gas droplets, the polaron energy, obtained from the Fr\"ohlich model, plays the role of the solvation energy. The first-order contribution, namely the mean-field effect, tries to keep the impurity at the droplet's center as the dipole-dipole interaction increases. However, beyond a critical density the beyond-mean-field repulsive contribution exerts a force expelling the impurity away from the center. In ultracold gases, these two energy scales can be adjusted at will, allowing to tune between different configurations. Finally, the presence of the second component, even at the impurity level catalyzes droplet nucleation and the formation of dual supersolidity~\cite{Scheiermann2022} and the formation of alternating domains, creating thus immiscible double supersolids~\cite{Bland22}.

Another interesting scenario where Bose polaron formation has been studied by Ashida et al.~\cite{Ashida2018} involves an impurity immersed inside a two-component BEC. Inside the latter, the interacting impurity creates both density and spin-wave excitations which contribute to the polaronic dressing cloud. In such a setting, Ashida et al.~\cite{Ashida2018} proposed a powerful Ramsey-scheme to probe its rich polaron physics, which involves e.g. the formation of a magnetic bound state.

\subsection{Magnetic polarons}
Another setting where the Bose polaron concept has been employed is in the context of mobile dopants in antiferromagnetic Mott insulators. These systems have continued to attract significant attention due to their role in explaining unconventional superconductivity as observed for example in cuprates~\cite{Keimer2015} and heavy-fermion compounds~\cite{Stewart1984,White2015}. Polaronic physics arises in these systems when the dopant - typically a hole $\hat{h}_{\vec{k}}$ - interacts with collective spin-wave excitations (magnons) $\a_{\vec{k}}$ of the host antiferromagnet. This leads to the formation of a polaronic quasiparticle, the so-called magnetic- or spin-polaron. To model this system, Kane, Lee and Read~\cite{Kane1989} and Sachdev~\cite{Sachdev1989} have employed the linear spin-wave approximation and derived an effective Hamiltonian resembling the Fr\"ohlich model,
\begin{multline}
    \H = \sum_{\vec{k}} \omega_{\vec{k}} \ad_{\vec{k}} \a_{\vec{k}}  -\\
    - t \sum_{\vec{k},\vec{q}}  \left[ \hat{h}_{\vec{k}} \hat{h}^\dagger_{\vec{k}-\vec{q}} \left( g^{(1)}_{\vec{k},\vec{q}} \a_{\vec{q}}  + g^{(2)}_{\vec{k},\vec{q}} \ad_{-\vec{q}} \right) + \hc \right],
    \label{eqHLinSpnWvetJ}
\end{multline}
where $\omega_{\vec{k}}$ is the spin-wave dispersion, $t$ denotes the nearest-neighbor hole tunneling and $g^{(1,2)}_{\vec{k}, \vec{q}}$ are known interaction coefficients. In high-temperature superconductors the energy scale $t \sim 3 J$ significantly exceeds the typical scale $J$ associated with the spin-wave dispersion $\omega_{\vec{k}}$, rendering the problem strongly coupled.

Although Eq.~\eqref{eqHLinSpnWvetJ} formally resembles the Fr\"ohlich Hamiltonian, it notably lacks a free kinetic energy term of the impurity $\hat{h}_{\vec{k}}$. Instead, all kinetic energy associated with tunneling of the impurity is part of the interaction term and involves magnon emission or absorption. This renders the problem highly correlated, and many of the traditional solution strategies developed for polarons are not accurate. However, the self-consistent Born approximation has been extremely successful in solving the linear spin-wave Hamiltonian~Eq.~\eqref{eqHLinSpnWvetJ}, see e.g.~\cite{SchmittRink1988,Kane1989,Martinez1991,Liu1992,Reiter1994,Nielsen2021} and as recently confirmed by Diamantis and Manousakis through direct comparison to numerically exact diagrammatic Monte Carlo simulations of the same Hamiltonian \cite{Diamantis2021}. Extensions to non-zero hole concentrations (doping) have also been put forward by Chen et al.~\cite{Chen2011d}.

The ground state properties of magnetic polarons are well understood and accurately captured by the linear spin-wave model in Eq.~\eqref{eqHLinSpnWvetJ}. In particular, their dispersion minimum is located at the so-called nodal point $\vec{k}=(\pi/2,\pi/2)$ in the Brillouin zone and features an anisotropic dispersion with a heavy effective mass along the zone-boundary; The ground state energy has a characteristic scaling $E_0 = - 2 \sqrt{3} t + A t^{1/3} J^{2/3} + \mathcal{O}(J)$ in the relevant regime $t>J$ (with a non-universal constant $A$); The magnetic polaron bandwidth $W$ is rather universally given by $W \simeq 2J$ when $t>J$, independent of the bare tunneling $t$; The quasiparticle weight $Z(t/J) $ has a strong dependence on $t/J$, roughly following a fit to $(J/t)^{1/3}$ when $t$ is not too large. In combination, these properties signify the formation of a strongly renormalized quasiparticle which bares little resemblance with a free hole. 

Experimentally, magnetic polarons have been observed at low hole concentrations in copper-oxides using angle-resolved photo-emission spectroscopy (ARPES)~\cite{Wells1995,Ronning2005,Kurokawa2023}. How long the magnetic polaron picture remains adequate when the hole (i.e. impurity) concentration is increased remains presently unclear, although it is widely established that antiferromagnetic order is stronlgy suppressed beyond approximately $5\%$ hole doping in the cuprates which likely invalidates the linear spin-wave theory underlying Eq.~\eqref{eqHLinSpnWvetJ}. For a more extended discussion of magnetic polarons in the context of high-temperature superconductivity, see e.g. Chernyshev and Wood's~\cite{chernyshev2002spin} contribution in~\cite{Srivastava2003}.

More recently, magnetic polarons have also been observed directly in real-space by Koepsell et al.~\cite{Koepsell2019} using ultracold fermions in optical lattices performing quantum simulations of the Fermi-Hubbard model~\cite{Bohrdt2021PWA}. Utilizing a quantum gas microscope which takes instantaneous snapshots of quantum projective measurements on the system, three-point spin-spin-charge correlation functions were directly measured, and compared for reference to a pinned defect inside the antiferromagnet. As a result, a universal polaronic signal of reversed spin-correlations on the bonds next to the impurity were revealed, that were shown in a subsequent experiment to remain visible up to around $20\%$ hole doping~\cite{Koepsell2021}. Similar experiments were recently performed on doped holes in a triangular antiferromagnet, revealing similar magnetic polaron features in three-point spin-spin-charge correlations~\cite{Lebrat2024,prichard2023}. Extensions of the linear spin-wave theory to triangular lattices have also been developed~\cite{Chen2022,Kraats2022}.

Magnetic polarons have very rich structure beyond their equilibrium, or ground state, properties. Extensive numerical work has focused on calculating spectral functions of individual mobile dopants, often starting from the microscopic $t-J$ or Hubbard model instead of the simplified linear spin-wave polaron Hamiltonian~\eqref{eqHLinSpnWvetJ}. This led to the presently most accurate ground-state calculations beyond the linear spin-wave theory by Brunner et al.~\cite{Brunner2000} and Mishchenko et al.~\cite{Mishchenko2001} based on worm algorithm Monte Carlo simulations. These authors also extracted the spectral function using analytic continuation which revealed a characteristic vibrational excitation of the magnetic polaron at an excitation energy scaling as $t^{1/3} J^{2/3}$ when $t>J$~\cite{Mishchenko2001}. These results were further refined by time-dependent DMRG simulations by Bohrdt et al.~\cite{Bohrdt2020}. All these approaches based on the microscopic $t-J$ Hamiltonian have in common that they predict only a single vibrational excitation with a non-negligible lifetime: This is in sharp contrast to linear spin-wave calculations~\cite{Liu1992,Diamantis2021} which predict a series of long-lived resonances up to high energies, likely owing to the negligence of spin-wave non-linearities in this model~\cite{Wrzosek2021}.

A unified phenomenological theory explaining the fascinating properties of magnetic polarons is based on a picture of partons - spinons and chargons - bound together by a linear string mediating a confining force. This picture goes back to Bulaevskii et al.~\cite{Bulaevskii1968} and was further developed after the discovery of high-temperature superconductivity, most notably in Refs.~\cite{Trugman1988,Shraiman1988,Brinkman1970}. Direct connections to concepts from particle physics were later proposed by B\'eran et al.~\cite{Beran1996}. By describing the magnetic polaron as a mesonic bound state of a heavy spinon and a light chargon, several properties follow naturally: For example, the bandwidth $W \simeq 2 J$ is naturally dominated by the heavier of the two partons; The universal scaling of the ground and vibrational excitation energies with $t^{1/3} J^{2/3}$ moreover follows directly from the linear confining force between the two partons. 

Indeed, one of the hallmark predictions of the parton theory is the existence of well-defined internal excitations. In addition to vibrational modes most clearly observed in~\cite{Bohrdt2020}, rotational modes were predicted by Grusdt et al.~\cite{Grusdt2018PRX} where one parton rotates around another, endowing the magnetic polaron with internal angular momentum -- see also earlier work by Simons and Gunn~\cite{Simons1990} who found similar structures without connecting them to angular momentum yet. The existence of long-lived rotational states of magnetic polarons, with excitation energies scaling linear in $J$, was conclusively demonstrated in time-dependent numerical DMRG simulations of the $t-J$ model by Bohrdt et al.~\cite{Bohrdt2021PRL}. Extensions of the parton theory to include bi-polaronic pairs of dopants have also been developed~\cite{Shraiman1988a,Grusdt2023} and tested numerically~\cite{Bohrdt2023}, and proposed to give rise to a $d$-wave Feshbach resonance between magnetic polarons~\cite{homeier2023feshbach}.

Another central prediction of the parton theory is the existence of two separate time scales associated with the two partons forming the mesonic magnetic polaron~\cite{Beran1996,Grusdt2018PRX}. These were predicted to give rise to multiple dynamical stages of polaron formation when a localized dopant is suddenly released into a surrounding antiferromagnet~\cite{Kogoj2014,Lenarcic2014,Golez2014,Eckstein2014,Grusdt2018PRX,Bohrdt2020NJP,Hubig2020,Nielsen2022,Shen2024}. Ultracold atom experiments by Ji et al.~\cite{Ji2021} have observed dynamical formation of magnetic polarons and confirmed the existence of two distinct time-scales governing the dynamics. 

As in the case of Bose polarons, the choice of the correct model Hamiltonian is important to obtain an accurate description of the physics. This has led to various proposals how to go beyond the linear spin-wave Hamiltonian~\eqref{eqHLinSpnWvetJ}, mostly starting directly from the microscopic $t-J$ or simplified $t-J_z$ models, see for example~\cite{Chernyshev1999}. Another recent approach by Bermes et al.~\cite{bermes2024arXiv} has been to combine the parton theory, including self-interactions of the string and strong non-linearities in the vicinity of the mobile dopant, with linear spin-wave models of long wavelength magnon excitations. The resulting effective Hamiltonian describes a mesonic impurity $\hat{f}_{\vec{k},n}$ with internal ro-vibrational excitations $n=0,1,2,..$ and dispersion $\varepsilon_n(\vec{k})$ which couples to the magnons $\a_{\vec{k}}$,
\begin{multline}
    \H = \sum_{\vec{k}} \omega_{\vec{k}} \ad_{\vec{k}} \a_{\vec{k}} + \sum_{\vec{k},n} \varepsilon_n(\vec{k}) \hat{f}_{\vec{k},n}^\dagger \hat{f}_{\vec{k},n} +\\
    + \sum_{\vec{k},\vec{p},\vec{q}} \sum_{n,n'}  \hat{f}_{\vec{k},n'}^\dagger \hat{f}_{\vec{k}+\vec{p}-\vec{q},n} \left[ B^{(1)}_{n',n}(\vec{k},\vec{p},\vec{q}) \ad_{\vec{p}} \a_{\vec{q}} +  \right. \\
  \left.  +   \left( B^{(2)}_{n',n}(\vec{k},\vec{p},\vec{q}) \ad_{\vec{p}} \ad_{-\vec{q}} + \hc \right) \right].
  \label{eqHbeyLST}
\end{multline}
The coupling constants $B^{(1,2)}$ are determined from the meson wavefunction and scale as $J$~\cite{bermes2024arXiv}. Importantly, Eq.~\eqref{eqHbeyLST} includes a free kinetic energy term $\varepsilon_n(\vec{k})$, which already captures the bandwidth $\simeq J$ in the ground state, as well as beyond-Fr\"ohlich terms of comparable magnitude. This establishes a close connection to traditional Bose polarons in a BEC, including the possibility of Feshbach resonances associated with meson-magnon bound states~\cite{Cubela2023}.

Another effect that has attracted significant attention and requires beyond-linear spin-wave models of the dopant is the so-called Nagaoka effect~\cite{Nagaoka1966}. It entails that for sufficiently large $t/J \to \infty$, the kinetic term in the $t-J$, or Hubbard, Hamiltonian dominates and favors a ferromagnetic ground state. For large but finite values of $t/J \gtrsim 33$~\cite{White2001} the mobile dopant binds to a finite-size cloud of polarized spins, which grows without bounds as $t/J$ is further increased to infinity. While direct observations in the two-dimensional $t-J$ or Hubbard models remain lacking, a quantum dot experiment by Dehollain et al.~\cite{Dehollain2020} observed the effect on a four-site plaquette. Moreover, kinetic frustration arising in the doublon-doped triangular lattice favors Nagaoka-polaron formation further, which led to its direct observation in a recent ultracold atom experiment~\cite{Lebrat2024,prichard2023} - see e.g. Refs.~\cite{Davydova2023,Morera2023,schlomer2023arXiv} for recent theoretical work on this problem.

\subsection{Polarons and topology}
Finally, to demonstrate the wide applicability of the polaron concept, we discuss so-called \emph{topological polarons} arising when mobile impurities couple to topological excitations in a host many-body system. Such settings were first studied in the context of anyonic quasiparticles of fractional quantum Hall states bound to a spin-$1/2$ impurity. Grusdt et al.~\cite{Grusdt2016TP} considered the strong-coupling limit with light impurities and demonstrated that the resulting impurity-quasiparticle composite, termed topological polaron, inherits the non-trivial topology of the host many-body system. They proposed to utilize this effect as a direct means to measure the fractional many-body Chern number of the many-body system via Ramsey interferometry. 
In a follow-up work this idea was extended to topological excitations in various symmetry-protected topological systems in one dimension~\cite{Grusdt2019topoPolPRB}.

The properties of mobile impurities bound to Abelian anyons in fractional quantum Hall liquids were further analyzed and demonstrated to reflect the anyonic character of the topological excitations closely. Grass et al.~\cite{Grass2020} pointed out the fractionalization of the impurity orbital angular momentum which can be linked to the observable impurity density distribution. Furthermore they showed that the total angular momentum of several impurities reflects the anyonic statistics of the quasiparticles bound to the impurities. Subsequently, generalizations to non-Abelian anyons bound to impurities in the context of the Moore-Read Pfaffian state were provided~\cite{Baldelli2021}. In addition to these studies which are largely based on trial states of the host fractional quantum Hall system, a strong-coupling Born Oppenheimer type ansatz was used by Munoz de las Heras et al.~\cite{MunozdelasHeras2020} who also demonstrated that impurities can serve as quantitative probes of fractional charge and statistics of the quasiparticles they bind to. This confirmed earlier arguments by Zhang et al.~\cite{Zhang2014} coming to similar conclusions.

A more direct analog of polaron physics with a topologically non-trivial bath was studied by Camacho-Guardian et al.~\cite{CamachoGuardian2019}. They considered a mobile impurity interacting with the particle-hole excitations of a Chern insulator constituted by fermions. Even if the impurity is merely dressed by excitations of the topologically non-trivial bath, Camacho-Guardian et al.~\cite{CamachoGuardian2019} demonstrated that it inherits topological properties from the latter. Specifically, the dressing leads to a polaronic Hall conductivity, i.e. a transverse drag force exerted on the polaronic cloud. This study, though focusing on integer Chern insulating baths, stands as a prototypical example of how mobile impurities couple to charge-neutral collective excitations (magneto rotons) in a larger class of systems including fractional Chern insulators and fractional quantum Hall states. Another recent proposal involved to binding of a mobile impurity to topological solitons in ultracold atomic gases, or optical settings~\cite{Mostaan2022}.

\section{Outlook}
\label{secOutlook}
We close this review by providing a discussion of open questions in the field that the authors feel are of particular relevance and should be addressed in future research. Although this review likely manifests to the extensive and impressive work that has already been done on Bose polarons, in some ways the field is only just in its infancy. 

For example, no measurements of the renormalized polaron mass exist beyond one-dimensional systems, despite being a hallmark signature of strong polaronic dressing. We view this as an important future task for experiments, that can shed additional light onto various theoretical approaches that have all been successfully been employed to model polaron spectra. As a direct extension, momentum dependent polaron spectroscopy would be highly desirable to achieve, in order to reveal possible deviations from a parabolic quasiparticle dispersion and study in more detail the subsonic-to-supersonic transition and the onset of Cherenkov-type radiation. Experimentally, these quests require low temperatures, clean experiments without disorder and box-like trapping potentials to avoid inhomogeneous broadening. As discussed in a previous section, measurements of the effective mass may be simpler, requiring well-controlled polaron dynamics. 

Another key question concerns the fate of Bose polarons upon increasing the impurity concentration. This leads us squarely into the field of Bose-Bose or Fermi-Bose mixtures that has exciting prospects such as the ability to control electron-electron interactions in solids or the mediation of attractive pairing interactions in ultracold atomic mixtures. The recent advances in realizing Bose polarons of exciton-polariton condensates in semiconductors~\cite{Tan2023} are particularly exciting developments in this regard. Another example along these lines is the decades-old problem of high-temperature superconductivity, in which the fate of the magnetic polaron upon doping remains to be settled, and theories involving bipolarons - though certainly not the main contenders for explaining high-temperature superconductivity - remain being debated.

As far as ultracold atom experiments are concerned, the exploration of few-body effects in the vicinity of Feshbach-resonances has only just started. In addition to possibly Efimov-type states with intrinsic three-body correlations, the interplay of strong impurity-boson and boson-boson interactions has been predicted by multiple groups to give rise to additional meta-stable branches in polaron spectra starting from impurity-boson bound states~\cite{christianen2024phase,Mostaan2023} that are awaiting experimental detection. Key challenges in this regard are the limited life-times of impurity atoms in the vicinity of a Feshbach resonance that need to be overcome. 

Another very promising development is the increasing amount of coherent control of individual atoms by optical tweezers, and the ability to perform textbook-style quantum projective measurements in quantum gas microscopy experiments. This opens the door to study polaronic dressing clouds on an unprecedented microscopic level, initialize interesting quantum quench experiments and study far-from equilibrium polaron dynamics, or even to perform fully coherent braiding experiments of topological excitations. Such studies will without doubt challenge theorists in providing ever-more accurate models of Bose polarons in various settings.

The list of possible future directions does not stop here, of course. By replacing the bosonic quantum fluid with ever more interesting systems -- for example a fermionic mixture near a BEC-to-BCS crossover~\cite{nishida2015polaronic} -- impurities in increasingly more difficult environments can be investigated. Here, the impurities can serve as probes of the surrounding many-body physics, but when the impurity-boson interactions become sizable even richer emergent few-body physics can be expected to arise. This line of thinking also plays an increasingly important role in the study of strongly correlated quantum matter, where exotic quantum spin liquids can be doped by mobile charge carriers. Instead of the celebrated resonating valence bond (RVB) picture proposed by Anderson~\cite{Anderson1987} where spin and charge degrees of freedom form featureless constituents, a trend to models with much richer bound states of these constituents can be recognized. As the story of Bose polarons - starting from featureless impurities modeled by the Fr\"ohlich Hamiltonian - teaches us, once interesting bound states form we can expect near-resonant interactions to arise when the bound states disappear, associated with interesting polaronic and universal few-body physics.

\section*{Acknowledgements}
We are indebted to many colleagues and friends with whom we had the pleasure to collaborate on research involving polaron physics, among which we would like to highlight in particular:
D. Abanin, J. Artl, G. Astrakharchik, P. Bermes, I. Bloch, A. Bohrdt, G. Bruun, A. Camacho, J.P. Christ, I. Cirac, S. Giorgini, N. Goldman, K. Hazard, T. Hilker, L. Homeier, C. Hubig, A. \.{I}mamo\u{g}lu, K. Jachymski, M. Knap, J. Koepsell, H. Lange, J. Levinsen, S. Mistakidis, M. Parish, T. Pohl, A. Rubtsov, L. Santos, H. Schl\"omer, P. Schmelcher, R. Schmidt, K. Seetharam, A. Shashi, Y. Shchadilova, T. Shi, Y. Wang, N. Yao, Z. Zhu.
We would like to thank Christoph Eigen, Tilman Enss, Jesper Levinsen, Pietro Massignan, Pascal Naidon and Martin Zwierlein for providing valuable feedback on our manuscript. 
LAPA acknowledges financial support from PNRR MUR project PE0000023-NQSTI and the Deutsche Forschungsgemeinschaft (DFG, German Research Foundation) under Germany’s Excellence Strategy– EXC-2123 QuantumFrontiers– 390837967, and FOR 2247.
F.G. and N.M. acknowledge funding by the Deutsche Forschungsgemeinschaft (DFG, German Research Foundation) under Germany's Excellence Strategy -- EXC-2111 -- 390814868. F.G. acknowledges funding from the European Research Council (ERC) under the European Union’s Horizon 2020 research and innovation program (Grant Agreement no 948141) — ERC Starting Grant SimUcQuam.

\bibliography{ReviewReferences}

\end{document}